\documentclass{aa}
\usepackage{graphicx}
\usepackage[varg]{txfonts}
\usepackage{epsfig}
\usepackage{natbib}
\usepackage{color}
\usepackage{soul}
\usepackage{longtable,lscape}
\definecolor{darkmagenta}{rgb}{0.55, 0.0, 0.55}
\definecolor{fashionfuchsia}{rgb}{0.96, 0.0, 0.63}
\definecolor{desertsand}{rgb}{0.93, 0.79, 0.69}
\definecolor{darkcoral}{rgb}{0.8, 0.36, 0.27}
\definecolor{kellygreen}{rgb}{0.3, 0.73, 0.09}
\definecolor{orange}{rgb}{1.0, 0.68, 0.10}
\definecolor{blue-violet}{rgb}{0.54, 0.17, 0.89}
\bibpunct{(}{)}{;}{a}{}{,} 
\begin{document}
\title{Discovery of new members of the nearby young stellar association in Cepheus
\thanks{Based on observations collected at the Isaac Newton Telescope (INT) operated on the island of La Palma by the Isaac Newton Group in the Spanish Observatorio del Roque de Los Muchachos of the Instituto de Astrof\'{\i}sica de Canarias, the 2.2-m telescope of the German-Spanish Astronomical Centre, Calar Alto (Almer\'{\i}a, Spain), operated by the Max-Planck-Institute for Astronomy, Heidelberg, jointly with the Spanish National Commission for Astronomy, and the SOPHIE spectrograph on the 1.93-m telescope at Observatoire de Haute-Provence (CNRS), France (program 09A.PNPS.GUIL).}
\fnmsep\thanks{Tables~\ref{Tab:List_Candidates} and \ref{Tab:AP_CepSurv}--\ref{Tab:VB_Gaia} are available at the CDS via anonymous FTP to 
		   {\tt cdsarc.u-strasbg.fr (130.79.128.5)} or via 
		   {\tt http://cdsweb.u-strasbg.fr/cgi-bin/qcat?J/A+A/}.}
}
\author{A. Klutsch \inst{1,2,4}, A. Frasca\inst{2}, P. Guillout\inst{3}, D. Montes\inst{4},  F.-X. Pineau\inst{3}, N. Grosso\inst{5} \and B. Stelzer \inst{1,6}
	    }

\offprints{A. Klutsch \\
\email{klutsch@astro.uni-tuebingen.de}\\}

\institute{Institut f\"ur Astronomie und Astrophysik, Eberhard Karls Universit\"at, Sand 1, D-72076 T\"ubingen, Germany
\and
INAF-Osservatorio Astrofisico di Catania, via S. Sofia 78, I-95123 Catania, Italy
\and
Observatoire Astronomique, Universit\'e de Strasbourg \& CNRS, UMR 7550, 11 rue de l'Universit\'e, F-67000 Strasbourg, France
\and
   Departamento de F\'{\i}sica de la Tierra y Astrof\'{\i}sica and IPARCOS-UCM (Instituto de F\'{\i}sica de Part\'{\i}culas y del Cosmos de la UCM),
Facultad de Ciencias F\'{\i}sicas, Universidad Complutense de Madrid, E-28040 Madrid, Spain
\and
Aix Marseille Univ, CNRS, CNES, LAM, Marseille, France
\and
INAF--Osservatorio Astronomico di Palermo, Piazza del Parlamento 1, I-90134 Palermo, Italy
}
\date{Received 29 November 2019 / Accepted 07 March 2020}
\abstract 
{Young field stars are hardly distinguishable from older ones because their space motion rapidly mixes them with the stellar population of the Galactic plane. Nevertheless, a careful target selection allows for young stars to be spotted throughout the sky.}
{We aim to identify additional sources associated with the four young comoving stars that we discovered towards the CO~Cepheus void and to provide a comprehensive view of the Cepheus association.} 
{Based on multivariate analysis methods, we have built an extended sample of $193$ young star candidates, which are the optical and infrared counterparts of ROSAT All-Sky Survey and XMM-Newton X-ray sources. From optical spectroscopic observations, we measured their radial velocity with the cross-correlation technique. We derived their atmospheric parameters and projected rotational velocity with the code {\tt ROTFIT}. We applied the subtraction of inactive templates to measure the lithium equivalent width, from which we infer their lithium abundance and age. Finally, we studied their kinematics using the second \emph{Gaia} data release.}
{Our sample is mainly composed of young or active stars and multiple systems. We identify two distinct populations of young stars that are spatially and kinematically separated. Those with an age between $100$ and $300$~Myr are mostly projected towards the Galactic plane. In contrast, $23$ of the $37$ sources younger than $30$~Myr are located in the CO~Cepheus void, and $21$ of them belong to the stellar kinematic group that we previously reported in this sky area. We report a total of $32$ bona fide members and nine candidates for this nearby (distance $=157\pm 10$~pc) young (age $= 10$--$20$~Myr) stellar association. According to the spatial distribution of its members, the original cluster is already dispersed and partially mixed with the local population of the Galactic plane.}
{}
\keywords{Stars: pre-main sequence -- Stars: fundamental parameters -- Stars: kinematics and dynamics -- X-rays: stars}
\titlerunning{Properties of the young stellar association in Cepheus}
\authorrunning{A. Klutsch et al.}
\maketitle

\section{Introduction}
\label{Sec:Intro}

The ``natural'' birth sites of stars are young open clusters and star-forming regions (SFRs) which are often tightly associated with emission nebulae and molecular clouds \citep[e.g.,][]{Piskunov2008, Zinnecker2008}. However young stars are also found in wide regions around SFRs and also in the field, apparently unrelated to any of the known star-forming sites. Such populations of young stars are composed of both classical T\,Tauri stars (CTTSs) and weak-line T\,Tauri stars (WTTSs). There is evidence of different space distributions for CTTSs and WTTSs, the former  located near the cloud cores while the latter are spread all around the SFR \citep[e.g.,][]{Alcala1997}. Moreover, WTTSs are on average older than CTTSs and have already dissipated their accretion disks \citep{2007A&A...473L..21B}. These young stars may form substructures that can be located up to tens of parsecs away from the SFR's core \citep[e.g.,][]{2015A&A...578A..35M}. 

\medskip
In the 1990s, various scenarios for explaining the presence of dispersed young stars were proposed. For those located in the outskirts of SFRs or in the space surrounding them, the simplest explanation was that they had drifted by thermal velocity dispersion of gas within star-forming sites. On the other hand, those far away from any conventional SFRs should have formed locally from cloudlets in turbulent giant molecular clouds \citep{Feigelson96}. However neither of these theories could explain the presence of T\,Tauri stars (TTSs) with high space velocity, the so-called ``runaway'' TTSs \citep{Sterzik95}. Their ejection could have occurred during the dynamical evolution of young multiple systems and come from close encounters with other members of their parent cloud \citep{SD95, SD98, GB96}.

\medskip
The picture of star formation in our Galaxy is still not well-defined and new important details are being added by the recent large spectroscopic surveys and astrometric space missions, such as \emph{Gaia}. One of the most relevant results of the \emph{Gaia}-ESO survey \citep{2012Msngr.147...25G, 2013Msngr.154...47R} is the discovery of two kinematically distinct populations (A and B) in the field of both the \object{Gamma Velorum} \citep{Jeffries2014} and \object{NGC\,2547} \citep{Sacco2015} clusters. The properties of sources belonging to Gamma Velorum~B and NGC\,2547~B are fairly similar in terms of age and kinematics. Since these two clusters are close to each other ($\approx 6\degr$ on the sky corresponding to about 40\,pc at their distance) and both are in the field of the Vela OB2 association, the B populations of these two clusters might form an extended low-mass population in the Vela OB2 association. 

\medskip
Over the past two decades, several nearby ($30$--$150$\,pc) young ($5$--$70$\,Myr) associations were identified, mostly in the southern hemisphere \citep{ZS04}. \citet{Torres06,Torres08} found many of them and their members during the SACY survey. The use of a Bayesian analysis and, subsequently, the BANYAN $\Sigma$ multivariate Bayesian algorithm contributed significantly to the continued identification of new members, mainly low-mass stars and brown dwarfs \citep{2013ApJ...762...88M, 2014ApJ...783..121G, 2015ApJS..219...33G, 2015ApJ...798...73G, 2018ApJ...856...23G}. The first release of the \emph{Gaia} mission \citep{2016A&A...595A...1G} improved the astrometric accuracy for the \emph{Tycho} sources. This has made it possible to search for new comoving stars \citep{2017AJ....153..257O}, as well as stellar kinematic groups and clusters \citep{2018ApJ...863...91F,2018ApJ...860...43G}. Based on the very accurate astrometry from the second data release of the \emph{Gaia} mission \citep[\emph{Gaia} DR2,][]{GaiaDR2}, \citet{2018ApJ...862..138G} discovered a considerable number of likely new members in these young associations.

\medskip
However, it remains difficult to recognize the young stars without a circumstellar disk in the field among the Galactic plane stellar population. Neither their global photometric properties nor the presence of nearby gas differentiates them from older stars. An efficient methodology for identifying the young stars is through the use of large X-ray surveys because the stellar X-ray sources in the ROSAT catalog are mainly stars younger than $1$\,Gyr \citep[e.g.,][]{Motch1997a}. \citet{Guillout98a} showed that this stellar population can be used as a tracer of young local structures, like the late-type stellar population in the Gould Belt \citep[][]{Guillout98b}. To this end, \citet{Guillout99} cross-correlated the ROSAT All-Sky Survey (RASS) with the \emph{Tycho} catalog building the first large dataset ($\approx 14,000$ objects) of late-type stellar X-ray sources, the so-called \emph{RasTyc} sample. 

\medskip
Spectroscopic surveys of northern \emph{RasTyc} sources were performed by \citet[][Paper~I]{Guillout09} and \citet[][Paper~III]{Frasca2018}. In Paper~I, we identified five young field stars in the optically bright ($V<9.5$\,mag) sample. Afterwards, BD+44\,3670 and BD+45\,598 were recognized as members of the Columba association \citep[$\sim$$30$\,Myr;][]{Zuckerman2011} and of the $\beta$~Pic~moving group \citep[$\sim$$12$\,Myr;][]{Moor2011}, respectively. \citet{Klutsch08} noticed that an almost uniform spatial density of young stars in the optically bright \emph{RasTyc} sample. At the same time, the spatial density in the northern hemisphere (Paper~I) is about one order of magnitude lower, on average, than that in the southern hemisphere (the SACY survey). This is consistent with the significant asymmetry in the all-sky \emph{RasTyc} map with respect to the Galactic plane, as reported by \citet{Guillout98a}, and is likely related to the structure of the Gould Belt. A higher fraction of young stars was found in the optically faint sample (Paper~III). This may be due to the larger distances involved, on average, in the faint survey, which facilitated the detection of many more intrinsically brighter X-ray (more active and younger) sources when compared to the bright sample. A larger contribution from the Gould Belt, which is more distant in the northern hemisphere, can also explain these differences.

\medskip
\citet{Klutsch08} analyzed the early spectroscopic observations of optically faint \emph{RasTyc} sources. This led to the discovery of an over-density of stellar X-ray sources near the Cepheus-Cassiopeia complex. We refer the reader to the reviews on SFRs in the Cassiopeia and Cepheus constellations of \citet{Kun2008b} and \citet{Kun2008}, respectively. Although this sky area is rich in CO molecular regions \citep{Dame01} and dark clouds \citep{Dobashi05, Kiss2006}, \citet[][Paper~II]{Guillout10a} identified four comoving TTSs towards a region devoid of interstellar matter that is denoted as the fourth void in the Cepheus Flare region by \citet{Kiss2006}. Moreover, \citet{Tachihara05} had already reported the discovery of 16 WTTSs in this region. In Paper~III, we found four other likely members of this group based on their spectral characteristics, position in the HR diagram, and kinematic properties. The group~38 in \citet{2017AJ....153..257O} and \citet{2018ApJ...863...91F} is composed of seven stars all belonging to the young association analyzed in the present paper and formerly discovered by our group (\citealt{Klutsch08}; Paper~II), which already included four of their seven stars; here, we extend the census of this young association.

\medskip
The main aim of the present work is to identify other young stars surrounding the four comoving TTSs in Cepheus reported in Paper~II, and to characterize their physical and kinematic properties, including the WTTSs provided by \citet{Tachihara05}. In Sect.~\ref{Sec:SelecDataObs} we detail our procedure to build a sample of ``young star candidates'', based on the data available in 2009, when it was created. We also describe our campaign of optical spectroscopic observations. Section~\ref{Sec:Analysis} presents our analytical methods and results thereof. In Sect.~\ref{Sec:Results} we mainly focus the discussion on the properties of the selected sources, on the spatial distribution and kinematics of the young stars belonging to the young association towards the CO Cepheus void, and on its reliability. Finally, we outline our conclusions and perspectives in Sect.~\ref{Sec:Conc}. We present the extraction of spectra with a double Gaussian profile in Appendix~\ref{appendix:Extraction_2_blended_spectra}. We also provide further information on multiple systems (Appendix~\ref{appendix:multiple_targets}), low-mass stars (Appendix~\ref{appendix:Mtype_stars}), and a few specific targets (Appendix~\ref{appendix:Info_targets}).

\section{Sample and observations}
\label{Sec:SelecDataObs}

\subsection{Sample selection}
\label{Sec:Sample}

The greatest difficulty in finding young stars in the field is the selection of a suitable sample of candidates. The best manner for identifying such a population is to excavate into an extensive and comprehensive sample of stellar X-ray sources, such as the \emph{RasTyc} sample \citep[Papers~I and III]{Guillout99}. 

\medskip
To this end, we picked out all counterparts of the X-ray sources by cross-matching the RASS catalog \citep{Voges99, Voges00} with the Two Micron All-Sky Survey catalog \citep[2MASS,][]{2MASS06} using a likelihood ratio approach and an original way to estimate the rate of spurious associations \citep{Pineau08a, Pineau08b, Pineau11, Pineau09}. We discarded all the matches having an angular separation larger than $15\arcsec$ and a probability of identification less than~$0.7$. We then correlated the remaining sources with the Guide Star Catalog~II \citep[GSC\,II,][]{Lasker08}. We kept all the optical sources at an angular distance not exceeding $5\arcsec$ from 2MASS sources and $15\arcsec$ from RASS ones. 

\begin{figure}[t]
\centering
\includegraphics[width=8.1cm, bb=50 40 370 340, clip=true]{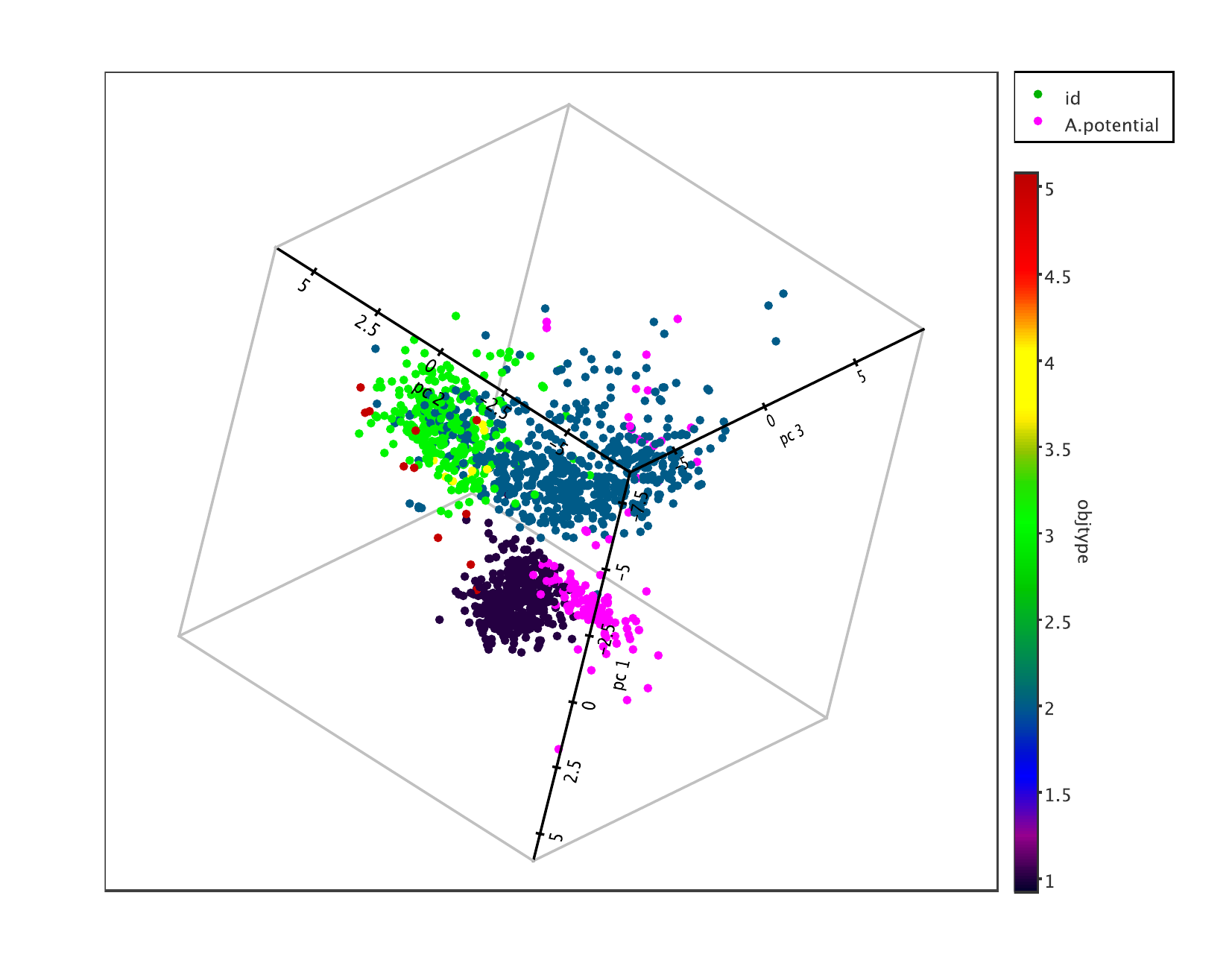}
\caption{Result of the principal component analysis  applied to RASS X-ray sources. The 3D scatter plot displays the three first components, adopting different colors for quasars (green dots), galaxies (blue dots), and stars (other dots). Such a method optimizes the disentangling of various classes of X-ray emitters. The pink dots denote the locus of a priori young stars selected using our learning sample (see text).}
\label{Fig:YoungStarCandidates_ACP}
\end{figure}

We proceeded in a similar way for the XMM-Newton X-ray sources of the 2XMMi catalog \citep{Watson09}. This turned out to be quite inconclusive because we only selected three stars. However, we note that the source \object{1RXS\,184257} \citep[=~\object{[KP93]~2$-$43} in][]{Kun1993}, which is a well-known young visual binary, is part of the selection. That makes us confident in the effectiveness of our approach, even for optically faint sources.

\medskip
We then applied a multivariate analysis on these two X-ray selected datasets. A principal component analysis, which takes measurements errors \citep{Pineau08b} on $19$ parameters into account, was used to build up linear combinations of flux ratios from X-ray to near-infrared (NIR) wavelength domain, X-ray hardness ratios, and color indexes. Applying a mean shift procedure on the three first principal components, we were able to disentangle the stellar population from the extragalactic component (galaxies and quasars) that also emits X-rays (Fig.~\ref{Fig:YoungStarCandidates_ACP}). We followed the guidelines of \citet{Pineau11} to identify the local maximum of the population of well-known young stars in this three-dimensional space. We therefore built a reliable learning sample composed of sources whose object type is either ``young stellar object'', ``T~Tau-type star'', or ``pre-main-sequence star'' in the Simbad database. We finally labeled all the likely stellar X-ray sources located in the latter region as our young star candidates. The objects are distributed all over the sky. 

\medskip
This work is focused on the identification of young stars towards the CO Cepheus void and its surroundings. Therefore, we selected all the young stars located in a $30\degr$-wide region encompassing the four comoving TTSs discovered in Paper~II in that region and we then restricted our sample to the population of late-type ($B$--$V > 0.5$~mag) and faint ($V > 9$~mag) stars. Relying on the photometric distance estimate of the group (Paper~II), we retain all the candidates within $170$~pc of the Sun and with X-ray luminosity L$_{X}\geqslant10^{29}$~erg~s$^{-1}$ \citep{Guillout98b}. This X-ray emission threshold picks the majority of 1~Myr-old stars with spectral type G and K \citep{2005ApJS..160..401P}. The choice of these selection criteria and of this vast sky area allows us \textit{i)} to look for additional bona fide members both in a restricted region around the Cepheus void and far away (i.e., to search for any runaway WTTSs), and \textit{ii)} to characterize the population of stellar X-ray sources in an under-researched sky area. The $162$ strongest stellar X-ray sources, which are also a priori the youngest stars selected, form our list of prime targets. 

The low-resolution spectroscopic survey by \citet{Tachihara05} includes $14$ sources common to our sample, as well as one analyzed in Paper~II (i.e., \object{TYC\,4500-1478-1}). We decided to add the remaining $31$ stars from \citet{Tachihara05} to our sample, which leads to a total of $193$ candidates. Table~\ref{Tab:List_Candidates} lists their basic data, together with that of the eight young stars reported in Papers~II and III (labeled as G1--G4 and F1--F4, respectively). 

\subsection{Photometric and astrometric data}
\label{subsection:phot_astro}

The selection of young star candidates to be observed spectroscopically was performed before the launch of \emph{Gaia}. At that time, only two of the $193$ pre-selected candidates had reliable \emph{Hipparcos} parallaxes, while 102 sources, which are included in the \emph{Tycho} catalog \citep{Perryman1997}, had unreliable parallaxes and proper motions with a fair accuracy. We therefore needed multi-band photometric data (from optical to infrared wavelengths) to select the young star candidates with the multivariate analysis described in Sect.~\ref{Sec:Sample}. However, meanwhile, the \emph{Gaia} DR2 catalog reports very accurate parallaxes and proper motions for nearly all the sources investigated in the present paper, which allows us to study their kinematics and evolutionary status. 

\medskip
Since we only selected stellar X-ray sources with a 2MASS counterpart, their photometry in the NIR wavelength domain is homogeneous. The $J$, $H$, and $K_{\rm s}$ magnitudes of the targets are in the range $7$--$12$~mag and have generally a high level of accuracy (in more than $96$\,\% of cases). For seven sources, at least one of NIR magnitudes is an upper limit or a poor-quality value in the final release of 2MASS data.

\begin{figure}[!t]
\centering 
\hspace{-0.5cm}
\includegraphics[width=9.2cm]{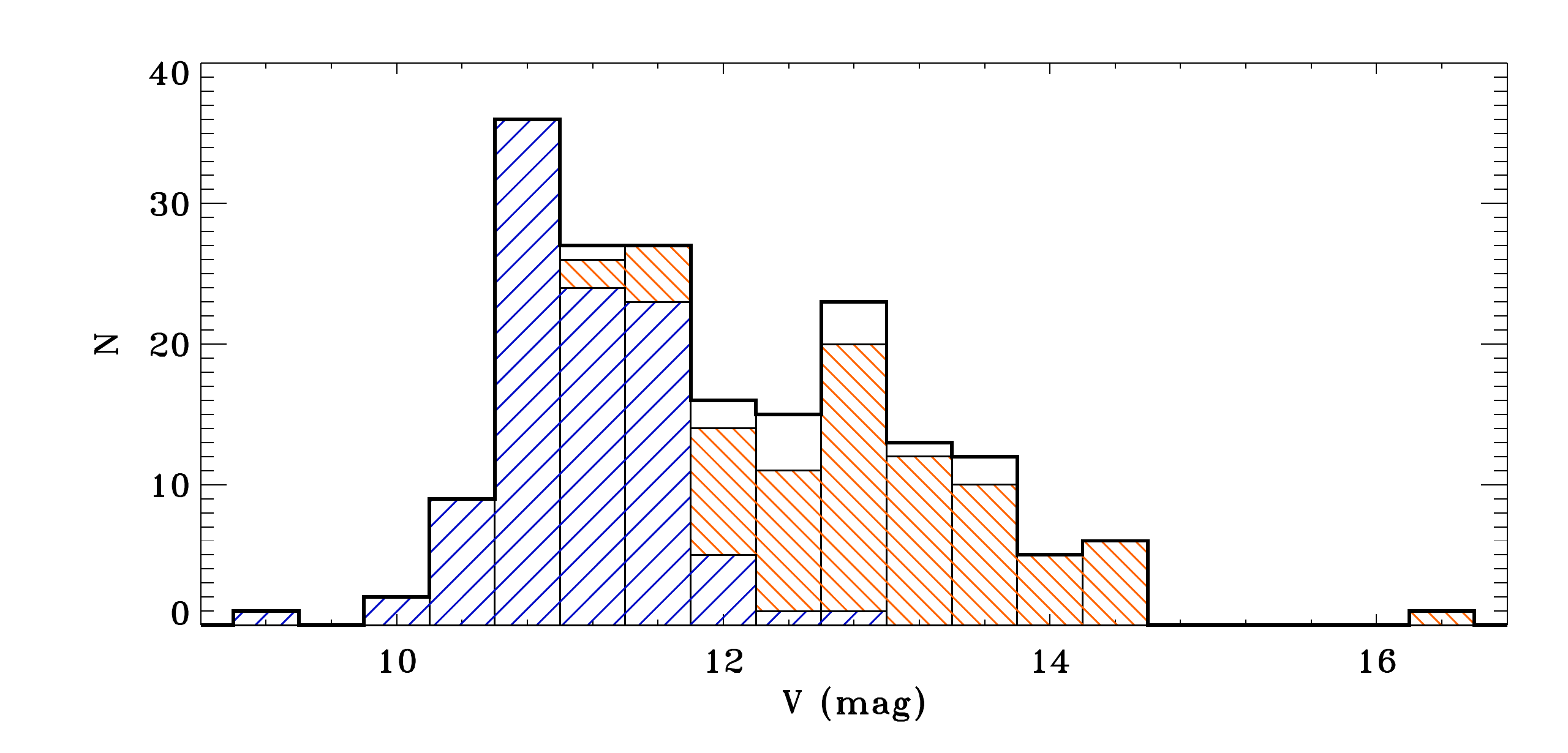}
\vspace{-0.4cm}
\\
\hspace{-0.5cm}
\includegraphics[width=9.2cm]{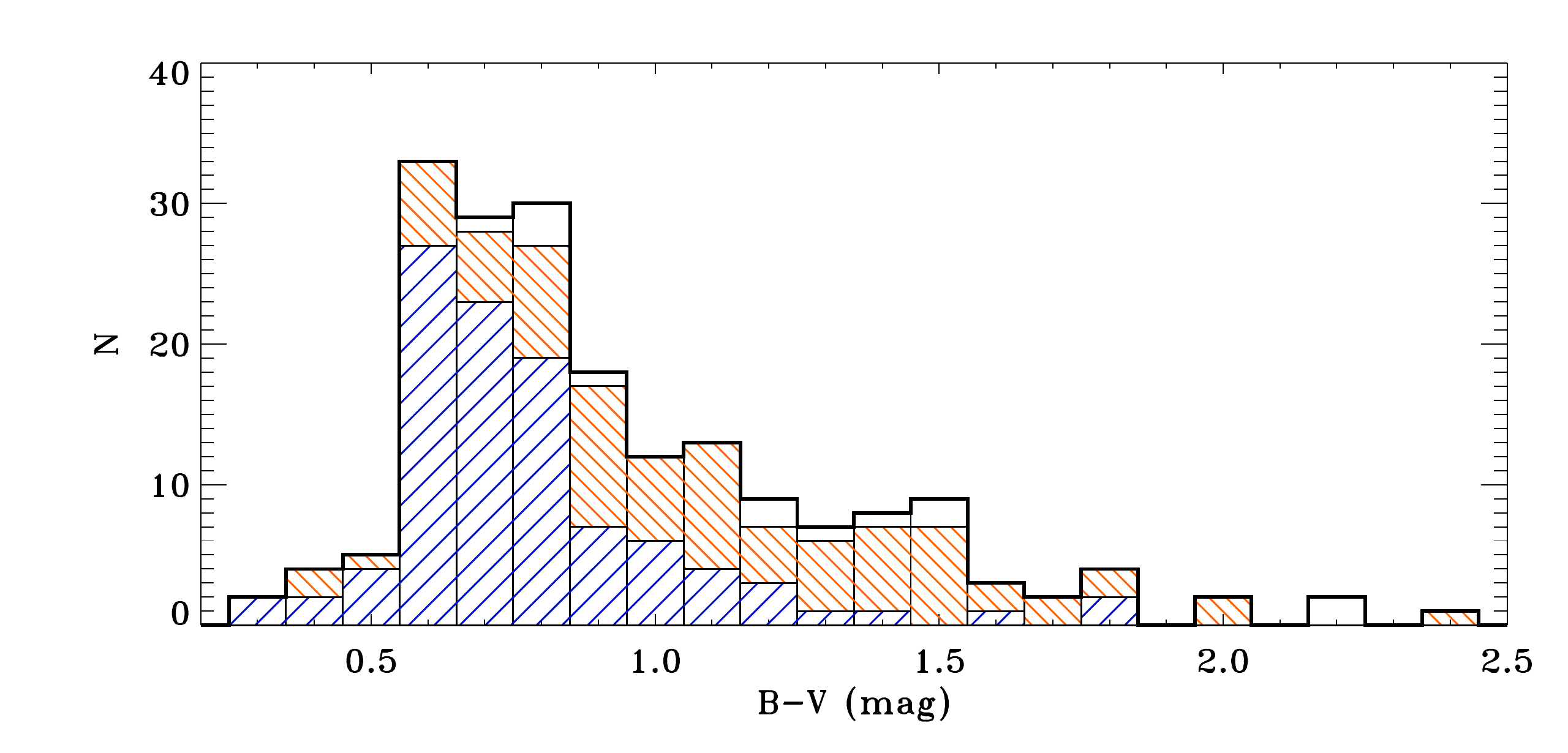}
\caption{Distributions of $V$ magnitude \textit{(upper panel)} and $B$--$V$ color index \textit{(lower panel)} for our targets. The blue and red stripes correspond to the \emph{Tycho} and USNO-B1.0 magnitudes converted into the Johnson $V$ magnitudes based on the relations of \citet{Hog2000} and \citet{LS2005}, respectively (see text for details). The white areas mark the $V$ photographic (GSC\,II or YB6) ones.} 
\label{Fig:BV_photometry_color}
\end{figure}

\medskip
We encountered difficulties with the homogenization of photometry in optical bands. Indeed, when we started our target selection, no catalog provided Johnson $B$ and $V$ magnitudes up to $V=14.5$~mag. The completeness of the \emph{Tycho}-2 catalog \citep{Hog2000} is to about 90~\% for sources brighter than $V$\,$\sim$$11.5$~mag, although it also contains fainter stars. For the non-\emph{Tycho} sources fainter than $V=12$~mag, only the photographic magnitudes were available and reliable. For the $102$ sources with an entry in the \emph{Tycho}-2 catalog, the $V_{T}$ magnitude ranges from $9$ to $13$~mag (Fig.~\ref{Fig:BV_photometry_color}). We converted the $B_{T}$ and $V_{T}$ magnitudes into Johnson $B$ and $V$ magnitudes following the transformation given in the introduction of the \emph{Tycho}-2 catalog: $B = B_{T} - 0.24 (B_{T} - V_{T})$ and $V = V_{T} - 0.09 (B_{T} - V_{T})$. 

For $79$ of the remaining sources with both B$_{J}$ and R$_{F}$ photographic magnitudes in the USNO-B1.0 catalog \citep{USNO2003}, we estimated the $V$ magnitude using the formula proposed by \citet{LS2005}: $V=B_{J} - 0.46 (B_{J} - R_{F})$. We used the $B_{J}$ and $R_{F}$ magnitudes extracted from the second Palomar Observatory Sky Survey (POSS) scans in the IIIa-J and IIIa-F passbands, respectively. All these sources are fainter than $V=11$~mag. Additional magnitudes in the Johnson $B$ band were retrieved either from the GSC\,II catalog or from the unpublished Yellow-Blue 6 catalog (YB6 -- USNO). The latter is available through the NOMAD catalog \citep{NOMAD2004}. 

In Table~\ref{Tab:List_Candidates}, we mainly list the $B$ and $V$ magnitudes in the Johnson-Cousins system ($\sim$$88$\,\% and $\sim$$94$\,\% of sources). If one of these values are not available, we substitute them by the $B_{J}$ or $V$ photographic passbands taken from GSC\,II (Fig.~\ref{Fig:BV_photometry_color}). 

\onllongtab{
\begin{landscape}
{\scriptsize
\begin{longtable}{lllllccrcrcrrrcc}
\caption{Main parameters of the young stars candidates available in 2009, when the sample was created.}
\label{Tab:List_Candidates}\\
\hline
\hline
\noalign{\smallskip}
\multicolumn{1}{c}{\#} & \multicolumn{1}{c}{X-ray name} & \multicolumn{1}{c}{IR name} & GSC\,II  name & Other name & $\alpha$ (2000) &  $\delta$ (2000) & \multicolumn{1}{c}{$B$} & Ref. & \multicolumn{1}{c}{$V$} & Ref.  & \multicolumn{1}{c}{$J$} & \multicolumn{1}{c}{$H$} & \multicolumn{1}{c}{$K_{\rm s}$} & $B$--$V$ $^{a}$ & $V$--$J$ $^{b}$ \\
& \multicolumn{1}{c}{1RXS} & \multicolumn{1}{c}{2MASS} &  &  & (h m s) & ($\degr ~\arcmin ~\arcsec$) & (mag) &  & (mag) & & (mag) & (mag) & (mag) & (mag) & (mag) \\
\noalign{\smallskip}
\hline
\noalign{\smallskip}
\endfirsthead
\caption{Continued.} \\
\hline
\hline
\noalign{\smallskip}
\multicolumn{1}{c}{\#} & \multicolumn{1}{c}{X-ray name} & \multicolumn{1}{c}{IR name} & GSC\,II  name & Other name & $\alpha$ (2000) &  $\delta$ (2000) & \multicolumn{1}{c}{$B$} & Ref. & \multicolumn{1}{c}{$V$} & Ref.  & \multicolumn{1}{c}{$J$} & \multicolumn{1}{c}{$H$} & \multicolumn{1}{c}{$K_{\rm s}$} & $B$--$V$ $^{a}$ & $V$--$J$ $^{b}$  \\
& \multicolumn{1}{c}{1RXS} & \multicolumn{1}{c}{2MASS} &  &  & (h m s) & ($\degr ~\arcmin ~\arcsec$) & (mag) &  & (mag) & & (mag) & (mag) & (mag) & (mag) & (mag) \\
\noalign{\smallskip}
\hline
\noalign{\smallskip}
\endhead
\noalign{\smallskip}
\hline
\endfoot
\noalign{\smallskip}
\hline
\endlastfoot
& \multicolumn{15}{l}{Stars cross-identified with a source of the RASS catalogs:}\\
\noalign{\smallskip}
\hline             
\noalign{\smallskip}
1		&	J000221.0+552519	&	J00022092+5525145	&	NAHS000445	&		&	00 02 20.929	&	+55 25 14.55	&	13.58	&	G$_{J}$	&	12.64	&	G$_{V}$	&	9.828	&	9.056	&	8.859	&	(0.94)	&	(2.81)	\\
2	$^{\ddag}$	&	J000806.3+475659	&	J00080642+4757025	&	NAH9000368	&		&	00 08 06.428	&	+47 57 02.50	&	14.53	&	G$_{B}$	&	12.69	&	U	&	8.523	&	8.000	&	7.677	&	1.84	&	4.17	\\
3	$^{\ddag}$	&	J002158.0+491245	&	J00215781+4912379	&	NAH1000233	&	GJ 3030	&	00 21 57.815	&	+49 12 37.96	&	13.43	&	G$_{J}$	&	12.75	&	U	&	9.139	&	8.454	&	8.205	&	(1.14)	&	3.61	\\
4	 $^{\ast,}$ $^{\star}$	&	J002827.8+785750	&	J00282798+7857426	&	NAG0000520	&	TYC 4500-730-1	&	00 28 27.987	&	+78 57 42.70	&	11.20	&	T	&	10.61	&	T	&	9.251	&	8.916	&	8.868	&	0.59	&	1.36	\\
5		&	J002854.1+502226	&	J00285391+5022330	&	NAIJ018308	&	GJ 3036	&	00 28 53.920	&	+50 22 33.02	&	14.09	&	G$_{J}$	&	13.16	&	U	&	8.847	&	8.289	&	7.991	&	[0.93]	&	4.32	\\
6	 $^{\ast,}$ $^{\star,}$ $^{\ddag}$	&	J003904.2+791912 \footnote{This source is the companion to \object{TYC 4500-1549-1}, which turns out to be the brighter component of a visual binary reported by \citet{Tachihara05}. See Appendix~\ref{appendix:galaxy} for details.}	&	J00390355+7919191	&	NAG3000177	&		&	00 39 03.554	&	+79 19 19.11	&	14.66	&	G$_{J}$	&	13.46	&	G$_{V}$	&	10.479	&	9.837	&	9.627	&	(1.20)	&	(2.98)	\\
7	 $^{\ast,}$ $^{\ddag}$	&	J005300.8+682125	&	J00530041+6821225	&	NAZ0000392	&	GSC 04296-00542	&	00 53 00.419	&	+68 21 22.52	&	12.80	&	G$_{J}$	&	12.01	&	G$_{V}$	&	10.088	&	9.468	&	9.367	&	(0.79)	&	(1.92)	\\
\end{longtable}
}
{\small
\begin{tabular}{l}
\textbf{Notes associated to the recno.} $^{\ast}$ Selected by both this study and the work of \citet{Tachihara05}; $^{\star}$ Towards the CO Cepheus void; $^{\ddag}$ Multiple system (VB and SB). \\
\begin{minipage}{23.2cm}
\textbf{Notes.} \\ $^{a}$ The $B$--$V$ color indexes are mainly in the Johnson-Cousins system. The bracketed values are the GCS\,II photographic color indexes, while the square-bracketed ones are the result of a mixture of both systems. \\$^{b}$ The $V$--$J$ color index values are put in parentheses if the $V$ magnitude is not in the Johnson-Cousins system or if the $J$ magnitude is an upper limit or a poor quality value. 
\end{minipage} \\
\textbf{References.} T =  \citet{Hog2000}; U =  \citet{USNO2003}; Y =  \citet{NOMAD2004}; G = \citet{Lasker08} where the suffix corresponds to their $B$, $j$, or $V$-band magnitude. \\
\end{tabular}
}
\end{landscape}
} 

\medskip
However, for the analysis of the spectral energy distribution (Sect.~\ref{subsection:SED}) we have updated the magnitudes with more accurate values of $B$ and $V$ retrieved from the AAVSO Photometric All Sky Survey (APASS, \citealt{APASS}) catalog. The $V$ magnitudes are also available in the TASS Mark IV patches photometric catalog \citep{TASS2006}. For most sources we retrieved the $I_{\rm C}$ magnitudes from the TASS catalog or, if not available, from the catalog of \citet{2014MNRAS.443..725G}. In this way, the dataset of the magnitudes is homogeneous and avoids a mixture of photometric and photographic estimates. We found $177$ (plus three whose $V$ magnitude is not provided) and $166$ sources in common with the APASS and TASS catalogs, respectively. The different datasets are consistent with the $V$ magnitudes in Table~\ref{Tab:List_Candidates}, with a Pearson's correlation coefficient (PCC) of $\sim$$0.967$. We also note that the $V$ magnitudes in APASS and TASS catalogs are fully consistent with each other (PCC $\sim 0.992$). In the case of the bright sources ($V \leqslant13$~mag), we obtained a good agreement between the various magnitudes and the spread increases towards the faintest objects, as expected in such a comparison. Regarding the faint objects ($V > 13$~mag), the $V$ magnitudes in Table~\ref{Tab:List_Candidates} tend to be overestimated by $0.5$~mag. This corresponds to the typical accuracy of the \citet{LS2005} relation at this brightness, which might be subject to systematic errors and other effects. 

\medskip
The photometric standard errors of the \emph{Tycho} sources with $9 <V_{T}<12$ are typically $0.15$ and $0.1$ mag in $B$ and $V$ passbands, respectively, while they are $\sim$$0.25$ mag in both bands for the faintest sources. The tabulated errors of the GCS\,II sources are $\sim$$0.3$ mag for the $V$ magnitudes and $\sim$$0.4$ mag for both the $B$ and $B_{J}$ magnitudes. The error on the optical magnitudes for the sources fainter than $V=12$~mag is considered to be at least of $0.5$ mag. For our targets, the typical error for the APASS photometry is instead $0.05$\,mag.

\medskip
Five of the six sources with a $B$--$V$ color index below our threshold of $0.5$~mag come from the inclusion of the \citet{Tachihara05} sample. The use of APASS magnitudes leads to the same statement for sources \#176 and \#178 only. Regarding the remaining source (\#102), its color index of $0.546\pm0.276$~mag in the APASS catalog fulfills our initial selection criterion.

\subsection{Observations and data reduction}
\label{Sec:Obs}

To study the physical and kinematic properties of all these young star candidates, we conducted observing campaigns using three spectrographs (Table~\ref{Tab:Inst_Sum}). We acquired both intermediate- and high-resolution optical spectra, sharing our candidates in an optimized way, depending on their brightness. 

\medskip
We observed the brightest ($V\leqslant12$\,mag) targets with two instrumental setups. 
From $26$ to $27$ September 2009, we used the cross-dispersed \'echelle SOPHIE\footnote{http://www.obs-hp.fr/guide/sophie/sophie-eng.shtml} spectrograph \citep{Perruchot08, Bouchy09} mounted on the 1.93-m telescope of the Observatoire de Haute Provence (OHP, France). The EEV-4482 detector is a thinned, back-illuminated, anti-reflection coated $4$k$\times 2$k $15$-$\mu$m-pixel CCD cooled at $-100\degr$C, with fast-readout mode. We chose the high-efficiency mode equipped with a double-fiber scrambler to homogenize and stabilize the illumination of the spectrograph entrance. This allowed us to record the $39$ spectral orders covering effectively the wavelength range $3872$--$6943$~\AA, with a resolution of about $39$,$000$ at $5500$~\AA. The data were treated using the automatic data-reduction pipeline, adapted from the HARPS software.\footnote{http://www.eso.org/sci/facilities/lasilla/instruments/harps/}

From $2$ to $5$ October 2009 and on $12$ December 2009, we completed our survey of the bright targets. The \'echelle spectra were taken with the fibre optics Cassegrain \'echelle spectrograph \citep[FOCES,][]{1998A&AS..130..381P} at the 2.2-m telescope of the German-Spanish Calar Alto Observatory (CAHA, Sierra de Los Filabres, Spain). We adopted the unique-fiber mode and chose the Site\#1d detector, a 2048$\times$2048 24\,$\mu$m-pixel CCD. We used the configuration with the $400\,\mu$m-slit width, leading to a resolution of about $28$,$000$ and covering the wavelength range from $3800$ to $10000$~\AA, in a total of $100$ orders. Using tasks of the {\sc Echelle} package within the {\tt IRAF} environment, we reduced the data following the standard steps of background subtraction, division by a flat-field spectrum given by a halogen lamp, wavelength calibration using the emission lines of the Th-Ar arc lamp, and normalization to the continuum through a polynomial fit. 

\begin{table}[!t]
\caption{Instrumental setup summary.}
\smallskip
\begin{center}
{
\footnotesize
\vspace{-0.4cm}
\begin{tabular}{llccc}
\hline
\hline
\noalign{\smallskip}
Telescope  & Inst. & Spectra & Spectral region & $\lambda/\Delta\lambda$ \\ 
\noalign{\smallskip}
\hline
\noalign{\smallskip}
INT 2.5-m & IDS & 1D & $5800$--$7200~\AA$ & ~~9,200 \\ 
CAHA 2.2-m & FOCES & \'echelle & $3800$--$10000~\AA$ & 28,000 \\ 
OHP 1.93-m & SOPHIE & \'echelle & $3872$--$6943~\AA$ & 39,000 \\
\noalign{\smallskip}
\hline 
\end{tabular}
\vspace{-0.4cm}
\label{Tab:Inst_Sum}
}
\end{center}
\end{table}

\onllongtab{
\begin{landscape}
{\scriptsize
\begin{longtable}{r@{\hspace{0.05cm}}l@{\hspace{0.2cm}}lcr@{\hspace{0.05cm}}c@{\hspace{0.05cm}}lr@{\hspace{0.05cm}}c@{\hspace{0.05cm}}lcr@{\hspace{0.05cm}}c@{\hspace{0.05cm}}lr@{\hspace{0.05cm}}c@{\hspace{0.05cm}}lr@{\hspace{0.05cm}}c@{\hspace{0.05cm}}lr@{\hspace{0.05cm}}c@{\hspace{0.05cm}}ll}
\caption{Radial and rotational velocities, atmospheric parameters, and lithium equivalent widths derived for our targets during the survey.}
\label{Tab:AP_CepSurv}\\
\hline
\hline
\noalign{\smallskip}
\multicolumn{2}{c}{\#$^{\star}$}	& Instrument	& H.J.D.					& \multicolumn{3}{c}{$RV$}			& \multicolumn{3}{c}{$v\sin i$}		& SpT & \multicolumn{3}{c}{$T_{\rm eff}$}	& \multicolumn{3}{c}{$\log g$}			& \multicolumn{3}{c}{$[Fe/H]$}	& \multicolumn{3}{c}{$W_{\rm Li}$}	& Note	\\
			&			&			& {\scriptsize [2400000+]}		& \multicolumn{3}{c}{[km\,s$^{-1}$]}	& \multicolumn{3}{c}{[km\,s$^{-1}$]}	&         & \multicolumn{3}{c}{[K]}			& \multicolumn{3}{c}{[cm$^{2}$ s$^{-1}$]}	& \multicolumn{3}{c}{[dex]}	& \multicolumn{3}{c}{[m$\AA$]}	& 		\\
\noalign{\smallskip}
\hline
\noalign{\smallskip}
\endfirsthead
\caption{Continued.} \\
\hline
\hline
\noalign{\smallskip}
\multicolumn{2}{c}{\#$^{\star}$}	& Instrument	& H.J.D.					& \multicolumn{3}{c}{$RV$}			& \multicolumn{3}{c}{$v\sin i$}		& SpT & \multicolumn{3}{c}{$T_{\rm eff}$}	& \multicolumn{3}{c}{$\log g$}			& \multicolumn{3}{c}{$[Fe/H]$}	& \multicolumn{3}{c}{$W_{\rm Li}$}	& Note	\\
			&			&			& {\scriptsize [2400000+]}		& \multicolumn{3}{c}{[km\,s$^{-1}$]}	& \multicolumn{3}{c}{[km\,s$^{-1}$]}	&         & \multicolumn{3}{c}{[K]}			& \multicolumn{3}{c}{[cm$^{2}$ s$^{-1}$]}	& \multicolumn{3}{c}{[dex]}	& \multicolumn{3}{c}{[m$\AA$]}	& 		\\
\noalign{\smallskip}
\hline
\noalign{\smallskip}
\endhead
\noalign{\smallskip}
\hline
\endfoot
\noalign{\smallskip}
\hline
\endlastfoot
\noalign{\smallskip}
1			&		&	IDS		&	55079.6090728		&	$2.2$	&	$\pm$	&	$0.3$	&	$17$		&	$\pm$	& 	$9$		&	G1.5V	&	$5827$	&	$\pm$ 	&	$61$		&	$4.07$	&	$\pm$ 	&	$0.12$	&	$-0.12$	&	$\pm$ 	&	$0.11$	&	$72$	&	$\pm$	&	$15$	&		\\
2			&		&	IDS		&	55079.6255449		&	$-75.7$		& $\pm$		&	$5.6$		&	\multicolumn{3}{c}{\dots}		&	\dots		&	\multicolumn{3}{c}{\dots}	&	\multicolumn{3}{c}{\dots}	&	\multicolumn{3}{c}{\dots}	&	\multicolumn{3}{c}{\dots}		&	$RV1$ of an SB2 system with two M-type stars (M3.5V)	\\
2			&		&	IDS		&	55079.6255449		&	$10.6$		& $\pm$		&	$8.4$		&	\multicolumn{3}{c}{\dots}		&	\dots		&	\multicolumn{3}{c}{\dots}	&	\multicolumn{3}{c}{\dots}	&	\multicolumn{3}{c}{\dots}	&	\multicolumn{3}{c}{\dots}		&	$RV2$ of an SB2 system with two M-type stars (M3.5V)	\\
3			&		&	IDS		&	55079.5934565		&	$-14.3$	&	$\pm$	&	$2.0$	&	$15$		&	$\pm$	&	$6$		&	M4V		&	$3397$	&	$\pm$	&	$50$		&	$4.56$	&	$\pm$	&	$0.15$	&	$-0.15$	&	$\pm$	&	$0.11$	&	$18$	&	$\pm$	&	$41$	&		\\
4			&		&	SOPHIE	&	55101.4586116		&	\multicolumn{3}{c}{\dots}	&		\multicolumn{3}{c}{\dots}	&	\dots		&	\multicolumn{3}{c}{\dots}	&	\multicolumn{3}{c}{\dots}	&	\multicolumn{3}{c}{\dots}	&	\multicolumn{3}{c}{\dots}	&	W\,UMa-type star	\\
5			&		&	IDS		&	55080.6417659		&	$3.1$	&	$\pm$	&	$2.4$	&	$38$		&	$\pm$	&	$3$		&	M4V		&	$3397$	&	$\pm$	&	$50$		&	$4.58$	&	$\pm$	&	$0.16$	&	$-0.15$	&	$\pm$	&	$0.11$	&	$9$	&	$\pm$	&	$31$	&		\\
5			&		&	IDS		&	55083.6659514		&	$3.0$	&	$\pm$	&	$1.3$	&	$15$		&	$\pm$	&	$4$		&	M5V		&	$3339$	&	$\pm$	&	$70$		&	$4.57$	&	$\pm$	&	$0.18$	&	$-0.14$	&	$\pm$	&	$0.11$	&	$45$	&	$\pm$	&	$59$	&		\\
6			&	c1	&	IDS		&	55084.6707247		&	$-108.0$	&	$\pm$	&	$4.5$	&	\multicolumn{3}{c}{\dots}	&	\dots		&	\multicolumn{3}{c}{\dots}	&		\multicolumn{3}{c}{\dots}	&		\multicolumn{3}{c}{\dots}	&	\multicolumn{3}{c}{\dots}	&	$RV1$ of the SB2 system, strong Li line, Target = TYC 4500-1549-1	\\
\end{longtable}
}
{\small
\begin{tabular}{c}
\textbf{$^{\star}$} During one IDS observation, several apertures might be visible on the cuts perpendicular to the spectral direction. In such a case, \\ we ranked the sources in reverse order to their maximum intensity (i.e., ``c1'' = the brightest one).\\
\end{tabular}
}
\end{landscape}
}

\onllongtab{
\begin{landscape}
{\scriptsize
\begin{longtable}{lr@{\hspace{0.05cm}}c@{\hspace{0.05cm}}l@{\hspace{0.15cm}}r@{\hspace{0.05cm}}c@{\hspace{0.05cm}}l@{\hspace{0.15cm}}r@{\hspace{0.05cm}}c@{\hspace{0.05cm}}l@{\hspace{0.15cm}}r@{\hspace{0.05cm}}c@{\hspace{0.05cm}}lr@{\hspace{0.05cm}}c@{\hspace{0.05cm}}l@{\hspace{0.15cm}}r@{\hspace{0.05cm}}c@{\hspace{0.05cm}}l@{\hspace{0.15cm}}r@{\hspace{0.05cm}}c@{\hspace{0.05cm}}lr@{\hspace{0.05cm}}c@{\hspace{0.05cm}}l@{\hspace{0.15cm}}r@{\hspace{0.05cm}}c@{\hspace{0.05cm}}l@{\hspace{0.15cm}}r@{\hspace{0.05cm}}c@{\hspace{0.05cm}}lr@{\hspace{0.05cm}}c@{\hspace{0.05cm}}lcl}
\caption{Galactic positions and space velocities of the targets, along with the astrometry in \emph{Gaia} DR2, and the average $RV$ values and lithium equivalent widths obtained during our survey.}
\label{Tab:EWLi_Kinematics_CepSurv}\\
\hline
\hline
\noalign{\smallskip}
   \multicolumn{1}{c}{\#$^{\star}$}	& \multicolumn{3}{c}{$\pi$}	& \multicolumn{3}{c}{$\mu_{\alpha} \cos\delta$} 	& \multicolumn{3}{c}{$\mu_{\delta}$} 		& \multicolumn{3}{c}{$\langle RV \rangle$}	& \multicolumn{3}{c}{X} 		& \multicolumn{3}{c}{Y} 		& \multicolumn{3}{c}{Z} 		& \multicolumn{3}{c}{U$_{\sun}$} 	& \multicolumn{3}{c}{V$_{\sun}$} 	& \multicolumn{3}{c}{W$_{\sun}$} 	& \multicolumn{3}{c}{$\langle W_{\rm Li} \rangle$}	&	A$_{v}$	& Note	\\
\noalign{\smallskip}
							& \multicolumn{3}{c}{[mas]}	& \multicolumn{3}{c}{[mas\,yr$^{-1}$]}			& \multicolumn{3}{c}{[mas\,yr$^{-1}$]}	& \multicolumn{3}{c}{[km s$^{-1}$]}			& \multicolumn{3}{c}{[pc]}	 	& \multicolumn{3}{c}{[pc]}		& \multicolumn{3}{c}{[pc]}		 & \multicolumn{3}{c}{[km s$^{-1}$]}	& \multicolumn{3}{c}{[km s$^{-1}$]}	& \multicolumn{3}{c}{[km s$^{-1}$]}	& \multicolumn{3}{c}{[m$\AA$]}	& [mag]	& 	\\
\noalign{\smallskip}
\hline
\noalign{\smallskip}
\endfirsthead
\caption{Continued.} \\
\hline
\hline
\noalign{\smallskip}
   \multicolumn{1}{c}{\#$^{\star}$}	& \multicolumn{3}{c}{$\pi$}	& \multicolumn{3}{c}{$\mu_{\alpha} \cos\delta$} 	& \multicolumn{3}{c}{$\mu_{\delta}$} 		& \multicolumn{3}{c}{$\langle RV \rangle$}	& \multicolumn{3}{c}{X} 		& \multicolumn{3}{c}{Y} 		& \multicolumn{3}{c}{Z} 		& \multicolumn{3}{c}{U$_{\sun}$} 	& \multicolumn{3}{c}{V$_{\sun}$} 	& \multicolumn{3}{c}{W$_{\sun}$} 	& \multicolumn{3}{c}{$\langle W_{\rm Li} \rangle$}	&	A$_{v}$	& Note	\\
\noalign{\smallskip}
							& \multicolumn{3}{c}{[mas]}	& \multicolumn{3}{c}{[mas\,yr$^{-1}$]}			& \multicolumn{3}{c}{[mas\,yr$^{-1}$]}	& \multicolumn{3}{c}{[km s$^{-1}$]}			& \multicolumn{3}{c}{[pc]}	 	& \multicolumn{3}{c}{[pc]}		& \multicolumn{3}{c}{[pc]}		 & \multicolumn{3}{c}{[km s$^{-1}$]}	& \multicolumn{3}{c}{[km s$^{-1}$]}	& \multicolumn{3}{c}{[km s$^{-1}$]}	& \multicolumn{3}{c}{[m$\AA$]}	& [mag]	&	\\
\noalign{\smallskip}
\hline
\noalign{\smallskip}
\endhead
\noalign{\smallskip}
\hline
\endfoot
\noalign{\smallskip}
\hline
\endlastfoot
\noalign{\smallskip}
1		&	$0.3580$		&	$\pm$	&	$0.0294$	&	$-4.088$		&	$\pm$	&	$0.043$		&	$-3.706$		&	$\pm$	&	$0.041$		&	$2.2$	&	$\pm$	&	$0.3$	&	$1210$	&	$\pm$	&	$100$	&	$2495$	&	$\pm$	&	$205$	&	$-330$	&	$\pm$	&	$27$		&	$-41.6$	&	$\pm$	&	$4.0$	&	$35.6$	&	$\pm$	&	$2.0$	&	$-48.5$	&	$\pm$	&	$4.1$	&	$72$		&	$\pm$	&	$15$		&	$1.76$		&		\\
2		&	$53.6844$	&	$\pm$	&	$0.0763$	&	$-123.729$	&	$\pm$	&	$0.114$		&	$4.509$		&	$\pm$	&	$0.071$		&	\multicolumn{3}{c}{~~~\dots}	&	$7.762$	&	$\pm$	&	$0.017$	&	$16.297$	&	$\pm$	&	$0.024$	&	$-4.598$	&	$\pm$	&	$0.016$	&	\multicolumn{3}{c}{\dots}	&	\multicolumn{3}{c}{\dots}	&	\multicolumn{3}{c}{\dots}	&	\multicolumn{3}{c}{\dots}	&	\dots	&	SB2	\\
3		&	$33.8287$	&	$\pm$	&	$0.0602$	&	$208.768$	&	$\pm$	&	$0.097$		&	$-35.025$		&	$\pm$	&	$0.068$		&	$-14.3$	&	$\pm$	&	$2.0$	&	$13.498$	&	$\pm$	&	$0.028$	&	$25.394$	&	$\pm$	&	$0.046$	&	$-6.840$	&	$\pm$	&	$0.025$	&	$19.7$	&	$\pm$	&	$0.9$	&	$-26.1$	&	$\pm$	&	$1.7$	&	$-3.4$	&	$\pm$	&	$0.5$	&	$18$		&	$\pm$	&	$41$		&	$0$		&	VB	\\
4		&	$5.7787$		&	$\pm$	&	$0.0258$	&	$-12.682$		&	$\pm$	&	$0.045$		&	$0.420$		&	$\pm$	&	$0.033$		&	\multicolumn{3}{c}{~~~\dots}	&	$87.57$	&	$\pm$	&	$0.39$	&	$141.29$	&	$\pm$	&	$0.63$	&	$48.11$	&	$\pm$	&	$0.22$	&	\multicolumn{3}{c}{\dots}	&	\multicolumn{3}{c}{\dots}	&	\multicolumn{3}{c}{\dots}	&	\multicolumn{3}{c}{\dots}	&	\dots	&	W\,UMa-type star	\\
5		&	$74.2071$	&	$\pm$	&	$0.9587$	&	$447.857$	&	$\pm$	&	$1.179$		&	$132.925$	&	$\pm$	&	$0.941$		&	$3.05$	&	$\pm$	&	$0.93$	&	$6.434$	&	$\pm$	&	$0.083$	&	$11.485$	&	$\pm$	&	$0.148$	&	$-2.879$	&	$\pm$	&	$0.037$	&	$24.0$	&	$\pm$	&	$0.6$	&	$-12.5$	&	$\pm$	&	$0.8$	&	$12.9$	&	$\pm$	&	$0.2$	&	$27$		&	$\pm$	&	$26$		&	$0.49$		&		\\
\end{longtable}
}
{\small
\begin{minipage}{22cm} 
\begin{tabular}{l}
\textbf{Note.} \\ 
$^{\star}$ If we put two sources on the slit during one IDS observation, the brightest and faintest sources are ``c1'' and ``c2'', respectively.  \\ 
$^{\dagger}$ Astrometry from the URAT Parallax Catalog \citep[UPC -- ][]{Zacharias2015,Finch2016}.\\
$^{\ddagger}$ Radial velocity and lithium equivalent width from Paper~III.\\
$^{\ast}$ Lithium equivalent width from Paper~III. Radial velocity near the conjunction reported in Paper~III (i.e., at HJD = $2455077.6191$).\\
$^{\bigtriangleup}$ We used the $RV$ measurement quoted in \emph{Gaia}~DR2 to compute the space-velocity components. \\
\end{tabular}
\end{minipage}
}
\end{landscape}
}

\onllongtab{
\begin{landscape}
{\scriptsize
\begin{longtable}{rlcrrrrrrr}
\caption{Comoving companions of our targets whose names are in bold. We reported the astrometry, $G$ magnitude and radial velocity $RV_{Gaia}$ released in the \emph{Gaia}~DR2 catalog. We derived their angular separation~($\rho$), position angle ($\theta$), and projected physical separation ($s$) with regard to the brightest star, along with the associated errors.}
\label{Tab:VB_Gaia}\\
\hline
\hline
\noalign{\smallskip}
  \# &
  \multicolumn{1}{c}{\emph{Gaia} DR2 name} &
  \multicolumn{1}{c}{$\pi$} &
  \multicolumn{1}{c}{$\mu_{\alpha} \cos \delta$} &
  \multicolumn{1}{c}{$\mu_{\delta}$} &
  \multicolumn{1}{c}{Gmag} &
  \multicolumn{1}{c}{RV$_{Gaia}$} &
  \multicolumn{1}{c}{$\rho$} &
  \multicolumn{1}{c}{$\theta$} &
  \multicolumn{1}{c}{$s$} \\
&  & \multicolumn{1}{c}{[mas]} & \multicolumn{1}{c}{[mas\,yr$^{-1}$]} & \multicolumn{1}{c}{[mas\,yr$^{-1}$]} & \multicolumn{1}{c}{[mag]} & \multicolumn{1}{c}{[km\,s$^{-1}$]} & \multicolumn{1}{c}{[$\arcsec$~$\pm$~mas]} & \multicolumn{1}{c}{[$\degr$~$\pm$~$\arcsec$]} & \multicolumn{1}{c}{[AU]}\\
\noalign{\smallskip}
\hline
\noalign{\smallskip}
\endfirsthead
\caption{Continued.} \\
\hline
\hline
\noalign{\smallskip}
  \# &
  \multicolumn{1}{c}{\emph{Gaia} DR2 name} &
  \multicolumn{1}{c}{$\pi$} &
  \multicolumn{1}{c}{$\mu_{\alpha} \cos \delta$} &
  \multicolumn{1}{c}{$\mu_{\delta}$} &
  \multicolumn{1}{c}{Gmag} &
  \multicolumn{1}{c}{RV$_{Gaia}$} &
  \multicolumn{1}{c}{$\rho$} &
  \multicolumn{1}{c}{$\theta$} &
  \multicolumn{1}{c}{$s$} \\
&  & \multicolumn{1}{c}{[mas]} & \multicolumn{1}{c}{[mas\,yr$^{-1}$]} & \multicolumn{1}{c}{[mas\,yr$^{-1}$]} & \multicolumn{1}{c}{[mag]} & \multicolumn{1}{c}{[km\,s$^{-1}$]} & \multicolumn{1}{c}{[$\arcsec$~$\pm$~mas]} & \multicolumn{1}{c}{[$\degr$~$\pm$~$\arcsec$]} & \multicolumn{1}{c}{[AU]}\\
\noalign{\smallskip}
\hline
\noalign{\smallskip}
\endhead
\noalign{\smallskip}
\hline
\endfoot
\noalign{\smallskip}
\hline
\endlastfoot
3 &   \emph{Gaia} DR2 394177942214683520 & $33.8287 \pm 0.0602$~~ & $208.768 \pm 0.097$ & $-35.025 \pm 0.068$ & $11.8573 \pm 0.0010$ & $-12.39 \pm 0.56$ & \dots & \dots & \dots \\
   &   \textbf{\emph{Gaia} DR2 394177946510277248} & $33.7578 \pm 0.1084$~~ & $209.999 \pm 0.175$ & $-29.251 \pm 0.105$ & $14.3724 \pm 0.0009$ & \dots & $2.27\pm 0.125$ & $300.89\pm 47.17$ & $67.2\pm0.1$ \\
\noalign{\smallskip}
\hline
\noalign{\smallskip}
6 &   \emph{Gaia} DR2 564707973733132800 & $6.4739 \pm 0.1404$ & $24.376 \pm 0.307$ & $-3.003 \pm 0.209$ & $11.6041 \pm 0.0081$ & \dots & \dots & \dots & \dots \\
   &   \textbf{\emph{Gaia} DR2 564707973733132032} & $6.7211 \pm 0.0321$ & $23.318 \pm 0.061$ & $-2.834 \pm 0.046$ & $12.9056 \pm 0.0009$ & $-8.34 \pm 2.37$ & $12.01 \pm 0.127$ & $322.07 \pm 3.02$ & $1856 \pm 9$ \\
   &   \emph{Gaia} DR2 564802050695978752 & $6.9090 \pm 0.0928$ & $23.770 \pm 0.156$ & $-3.458 \pm 0.135$ & $17.1583 \pm 0.0013$ & \dots & $205.48 \pm  0.185$ & $358.53 \pm 0.05$ & $31740 \pm 160$ \\
\noalign{\smallskip}
\hline
\noalign{\smallskip}
 18  &   \textbf{\emph{Gaia} DR2 575921721024508928} & $6.9947 \pm 0.0256$ & $47.790 \pm 0.048$ & $-54.249 \pm 0.042$ & $12.6970 \pm 0.0012$ & \dots & \dots & \dots & \dots \\
       &   \emph{Gaia} DR2 575921721024509312 & $7.0859 \pm 0.0239$ & $47.536 \pm 0.048$ & $-54.317 \pm 0.039$ & $14.2886 \pm 0.0005$ & \dots & $5.84 \pm 0.031$ & $224.01 \pm 2.03$ & $835.1 \pm 3.1$ \\
\noalign{\smallskip}
\hline
\noalign{\smallskip}
25 &   \textbf{\emph{Gaia} DR2 488038959210617088} & $9.2418 \pm 0.0296$ & $-55.973 \pm 0.044$ & $-47.168 \pm 0.044$ & $11.9592 \pm 0.0005$ & $-24.49 \pm 0.67$ & \dots & \dots & \dots \\
\end{longtable}
}
\end{landscape}
}

\medskip
The optically faint candidates ($12$$<$$V$$<$$14.5$ mag) were observed with the Intermediate Dispersion Spectrograph (IDS)\footnote{http://www.ing.iac.es/Astronomy/instruments/ids/} at the 2.5-m Isaac Newton Telescope (INT) of the Observatorio del Roque de los Muchachos (La Palma, Canary Islands, Spain). We acquired long-slit spectra during several observing runs that took place in $4$--$10$ September $2009$, $23$--$28$ November $2010$, $24$--$27$ October $2012$, $27$ March to $2$ April $2013$, $16$--$20$ November $2013$, and $18$--$20$ February $2014$. During the observing night of $12$ to $13$ December $2009$, some of our targets were also observed in queue mode. We used a slit width of $0\farcs95$ and the H1800V dispersion grating with the $235$~mm Camera ($2148$$\times$$4200$ EEV10a CCD detector until 2010 and $2k$$\times$$4k$ RED+2 CCD detector since $2012$) and with slow-readout mode. We chose the central wavelength of $6500$~\AA. As the outer regions of the dispersed light beam are severely vignetted by the camera optics, only $2070$ of the pixels (roughly from pixel number $1000$ to $3070$ in the direction of the dispersion) are clear and unvignetted regarding the H1800V grating. This allows for efficient coverage of the wavelength range $6200$--$6800$~\AA, with a resolving power of about $9200$. For most of our targets and standard stars, we set the slit position angle to the current parallactic angle. Using the typical {\tt IRAF} tasks, all spectra were de-biased, flat-fielded, distortion-corrected, wavelength calibrated (using the emission lines of the Cu-Ar and Cu-Ne lamps), and finally normalized to the continuum. When a source was already reported as a visual binary in the literature or was identified as such during our observing runs (i.e., with a small angular separation), we aligned the slit on its position angle to simultaneously acquire the spectra of both sources. In this way we made sure to observe the optical counterpart(s) of the unresolved X-ray source. We subsequently examined whether or not the two stars form a physical binary system. When their seeing profiles are partially overlapped on the CCD, we extracted the two spectra by means of a {\tt IDL} code similar to that of \citet{Frasca1997}, rather than the {\tt IRAF} task {\tt APALL} (see Appendix~\ref{appendix:Extraction_2_blended_spectra} for details). The use of a double Gaussian profile optimizes the spectrum extraction of the fainter source, which may be heavily contaminated by the brighter one.

\medskip
We removed the telluric water vapor lines only in the spectral region around the H$\alpha$ line of every spectrum, in all instrumental setups. We applied a method similar to that described by \citet{Frasca2000} using telluric templates taken with the same instrumentation as the targets. We show a few results for the case of SOPHIE spectra (Fig.~\ref{Fig:Telluric_Removal_Sophie}) to illustrate its effectiveness. 

\medskip
Owing to bad weather conditions during our IDS observing runs, we were unable to obtain a spectrum for $35$ of our targets. All the unobserved targets have a lower priority because their Galactic latitude is higher than $25 \degr$ or they were not classified as WTTSs by \citet{Tachihara05}. Figure~\ref{Fig:HaLiSpec_CepSurv} shows the spectra acquired in the region around the \ion{Li}{i} $\lambda$6707.8 lines. 

\section{Analysis and results}
\label{Sec:Analysis}

\subsection{Radial velocity and multiplicity}
\label{Sec:Analysis_RV}

We measured the heliocentric radial velocities ($RV$s) using the cross-correlation technique. We distinguished spectroscopic systems from single stars (or single-lined binaries) based on the number of peaks visible in the cross-correlation function (CCF).  

\medskip
Regarding the IDS data, we cross-correlated each long-slit spectrum of our targets with that of RV-standard stars observed during the same night. For this purpose, we have upgraded the {\tt IDL} procedure described in \citet{Klutsch08} and Paper~I. We masked the spectral range around the H$\alpha$ line because of its strong wings and the possible chromospheric emission in the core that can considerably broaden and distort the CCF peak. For the same reason we discarded the spectral regions strongly affected by telluric absorption lines. When only one CCF peak was detected, the code performs an additional analysis to report any asymmetry of the CCF profile, possibly interpreted as an SB2 system observed near the conjunction. Such a procedure is only relevant when the masses of the two stars differ significantly. We computed the errors of the pixel shift from the center of the Gaussian function fitting the CCF. We computed the $RV$ errors as a function of the fitted peak height and the antisymmetric noise through the $R$ factor, as described in \citet{TonryDavis1979}.

For the FOCES data we derived the $RV$ values with the {\tt IRAF} task {\tt FXCOR}, by cross-correlating each \'echelle order of the target spectra with the corresponding one of the most similar RV-standard star, which was acquired during the same run. We discarded the orders heavily contaminated by telluric lines or those including very broad lines and activity indicators, such as the \ion{Ca}{ii} H~\&~K lines, the Balmer series (H$\alpha$ and H$\beta$), and the \ion{Ca}{ii} IR triplet lines. We fitted the entire CCF peak with a Gaussian to measure more precisely the $RV$ values, which were obtained as the weighted average of all the individual $RV$ measurements computed by {\tt FXCOR} per each \'echelle order. The resulting $RV$ uncertainty is the standard error of the weighted mean, $\sigma_{RV}$.  

The $RV$ measurements derived from the target spectra acquired with the SOPHIE spectrograph were performed with the cross-correlation analysis that is included in the automatic data-reduction pipeline developed by the OHP staff. We used the most appropriate mask depending on the $B$--$V$ color index (or spectral type) of the target, among the five available ones. This procedure failed for some of the multiple systems and fast rotators. From a first guess of their spectral type (Sect.~\ref{Sec:Analysis_AP}), we selected the most appropriate synthetic spectrum from the POLLUX database \citep{Palacios2010} to be used as template for computing the CCF, which was automatically fitted by a procedure similar to that used for IDS data. Taking advantage of the high resolution of these spectra, we improved the code to detect and fit the CCF peak of a companion even having a much lower brightness in a stellar binary system, that is, with a low mass-ratio. This requires careful analysis of the CCF residues after subtracting the best fit, especially when the two peaks are severely blended (i.e., one CCF peak with an asymmetric shape). 

Table~\ref{Tab:AP_CepSurv} contains all our $RV$ measurements and associated errors, with one entry for each acquired spectrum and, in case of multiple systems, for each component. 

\medskip
Since our {\tt IDL} procedure and the task {\tt FXCOR} use the RV-standard stars as templates, we need an accurate $RV$ value for them. So we used the $RV$ measurements available from the \'Elodie archive,\footnote{http://atlas.obs-hp.fr/elodie/} which were derived from high-resolution spectra. Table~\ref{tab:StandardStars} lists all these values that agree with those in \citet{Nidever2002} or other compilations of $RV$ measurements \citep[e.g.,][]{Gontcharov2006, Kharchenko2007}.

\medskip
For multiple systems, we adopted an independent Gaussian fit for each CCF peak whenever they are not blended (i.e., far from the conjunctions). We used instead a multiple-Gaussian fit algorithm when the spectral lines are partially blended. Moreover, we automatically looked for any significant peaks in the CCF and fitted them with single, double, or multiple functions (i.e., up to four Gaussian profiles), using our {\tt IDL} code that computes the CCF and treats the errors as {\tt FXCOR} does. We refer the reader to Paper~I for details on  this code and its performances.

\medskip
Among the $47$ sources observed at multiple epochs, eleven sources display a large variation in radial velocity (Table~\ref{Tab:AP_CepSurv}). We classified those with $RV$ change significantly larger than the $RV$ errors as single-lined binaries ({\tt SB1}) and those with smaller $RV$ variation as possible SB1 systems ({\tt SB1?}). We discovered that $44$ of the surveyed sources are multiple spectroscopic systems, which are distributed as follow: nine SB1 ($4.8$\,\%), $28$ double-lined (SB2) plus four likely SB2 ($17.2$\,\%), and three triple (SB3;  $1.6$\,\%) systems (see Table~\ref{tab:MultipleSyst}). This corresponds to $20$\,\% of the overall sample (i.e., our targets, plus the $20$ sources listed in Table~\ref{Tab:IDS_companion_slit} and the eight sources from Papers~II and III) and to $24$\,\% of the sources with new observations presented here, as listed in Table~\ref{Tab:AP_CepSurv}. For the $17$ sources classified as possible spectroscopic systems (Table~\ref{tab:MultipleSyst}), additional observations are needed to draw firm conclusions on their multiplicity.   

\begin{figure}[!t]
\begin{center}
\includegraphics[width=8.8cm]{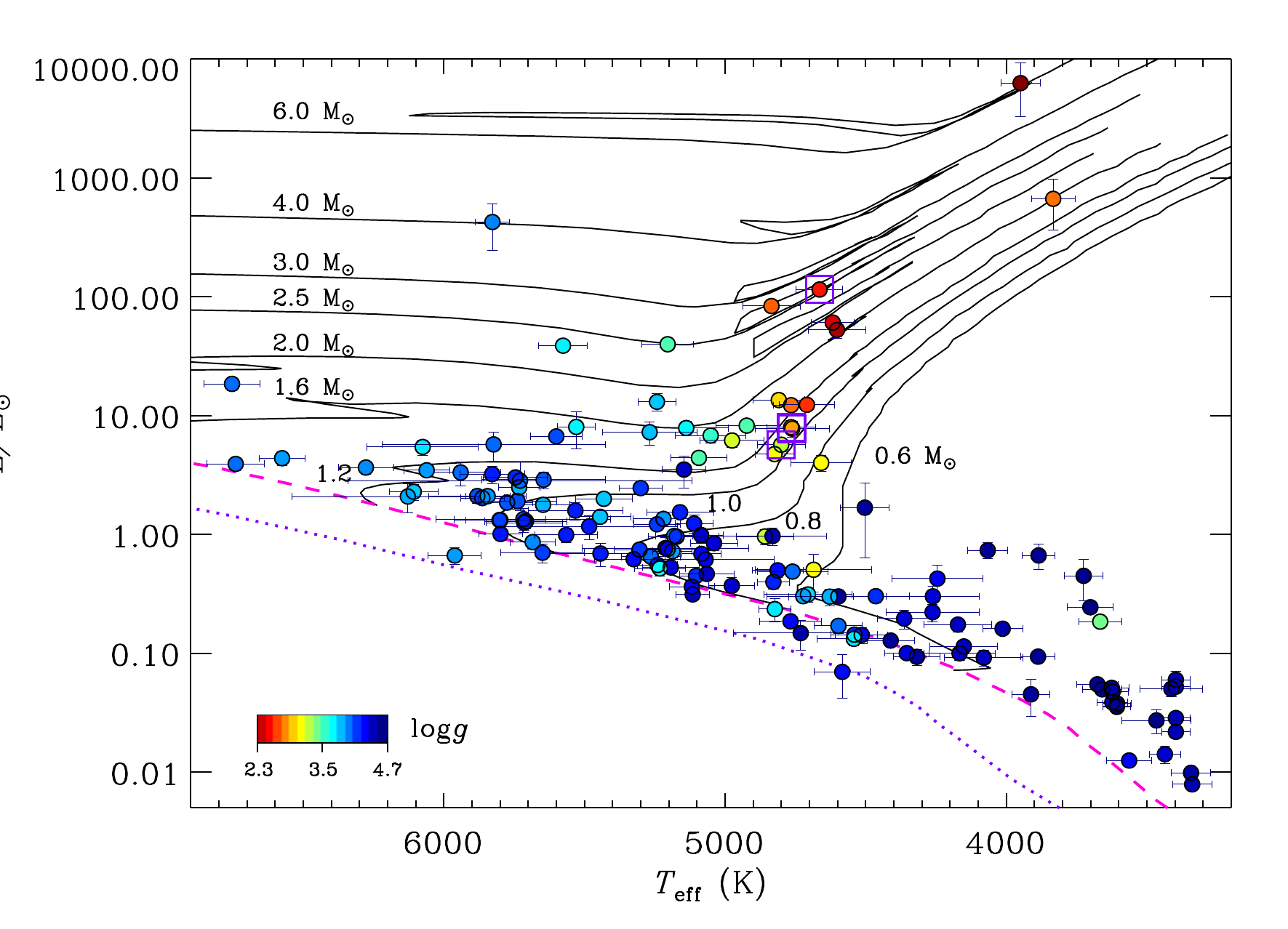}
\caption{HR diagram of targets with known APs and parallaxes. The symbols are color coded by $\log g$ derived from spectra (Sect.~\ref{Sec:Analysis_AP}). The four lithium-rich sub-giant or giant stars are enclosed in open squares (Sect.~\ref{Sec:Analysis_Li}). The evolutionary tracks from the ZAMS of \citet{Girardi2000} are shown as solid lines with the labels representing their masses. The dashed and dotted lines display the ZAMS with solar metallicity ($Z=0.019$) and $Z=0.001$ by the same authors, respectively.}
\label{Fig:HR}
 \end{center}
\end{figure}

\subsection{Spectral classification and projected rotational velocity}
\label{Sec:Analysis_AP}

We applied the {\tt IDL} code {\tt ROTFIT} \citep{Frasca2006} to perform the physical characterization of SB1 systems and single stars, that is, those with only one peak in the CCF profile. This code searches for the best combination of parameters by comparing the target spectrum with reference star spectra that are rotationally broadened until a minimum $\chi^2$ was reached. As in our previous works (Papers~I and III), we determined the spectral type (SpT), effective temperature ($T_{\rm eff}$), gravity ($\log g$), and metallicity ([Fe/H]) using as templates a library of $270$ high-resolution spectra of low-activity and slowly rotating stars retrieved from the \'Elodie archive \citep{Moultaka2004}. For these stars we took the atmospheric parameters (APs) from the PASTEL catalog \citep{Pastel_Cat2010}. We report these results in Table~\ref{Tab:AP_CepSurv} in which any source observed at several epochs\footnote{whether with different instruments (usually for the bright sources) or with the same instrumental setting (in case of the faint ones).} has multiple entries. The typical accuracy on the spectral type is about one subclass. We refer the reader to Appendix~\ref{appendix:Mtype_stars} for a description of our procedures dedicated to the analysis of the M-type stars.

\medskip
To minimize possible effects on projected rotational velocities ($v \sin i$) due to different resolutions and instrumental setups between the target spectra (acquired with both IDS and FOCES) and the \'Elodie ones we also used a smaller library of template spectra taken with these spectrographs during our observing runs. Table~\ref{tab:StandardStars} lists them as well as all their relevant information. Analogous to the treatment of the full library of \'Elodie templates, the rotation velocity of each template is progressively increased by convolving its spectrum with a rotational profile of a given $v \sin i$. The best match of the template spectrum with the target spectrum, found by $\chi^2$ minimization, gives us the value of $v \sin i$. We note that most $v \sin i$ values derived with the \'Elodie templates are slightly underestimated compared to those obtained when the spectra are taken with the same instrumental setup. This is probably caused by the resampling of the \'Elodie spectra on the points of the IDS or FOCES ones. We thus adopted the $v \sin i$ values obtained with the template acquired with the same spectrograph as the target (Table~\ref{Tab:AP_CepSurv}).  At the IDS, FOCES, and SOPHIE resolutions, any $v\sin i$ determination is reliable only if the $v\sin i$ value is larger than $15$~km\,s$^{-1}$,~$5$~km\,s$^{-1}$, and $2$~km\,s$^{-1}$, respectively. We determined these lower limits with simulations similar to those made by \citet{Frasca2015}.

\begin{figure}[!t]
\centering 
\hspace{-0.5cm}
\includegraphics[width=9.4cm]{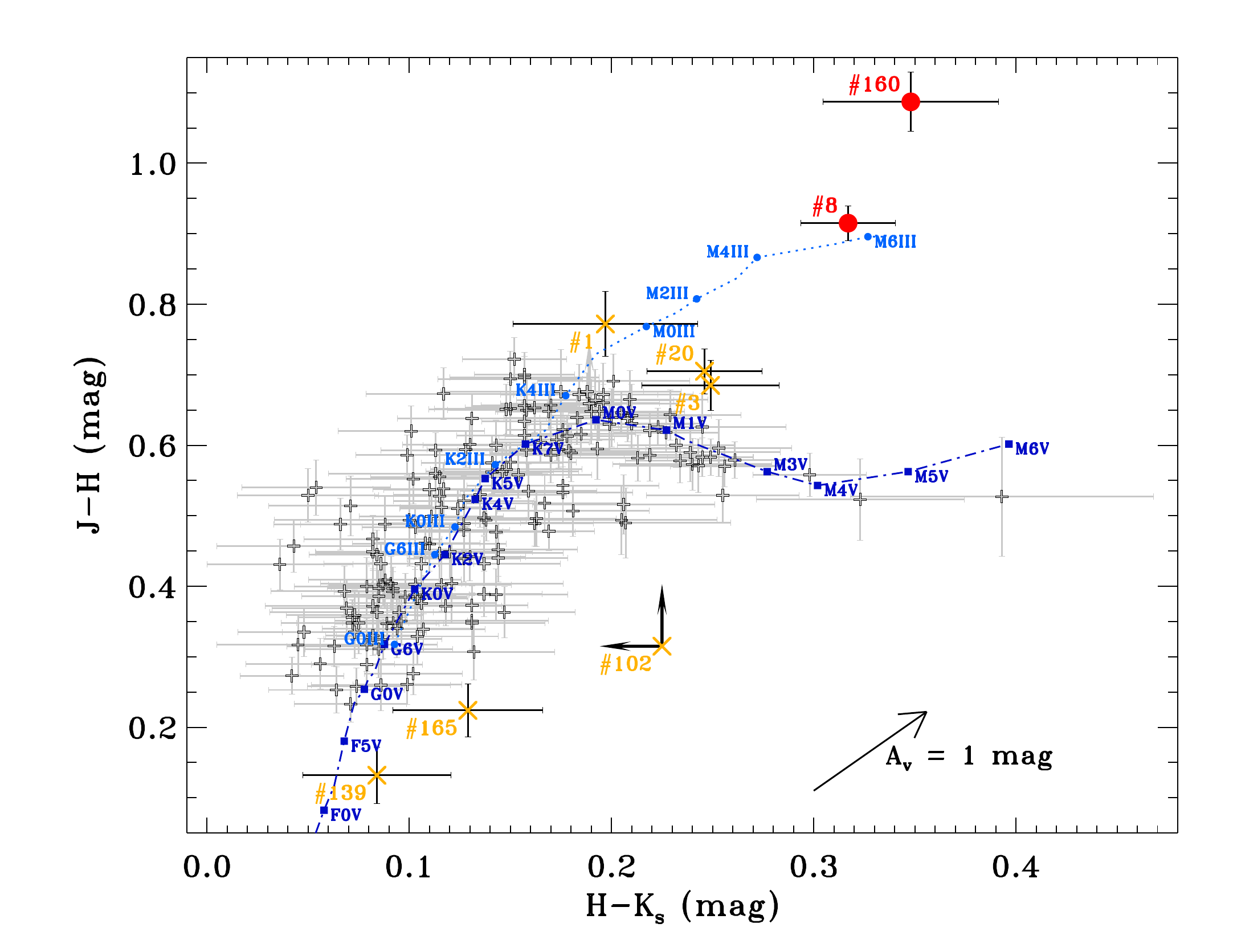}
\caption{
($H$--$K_{\rm s}$, $J$--$H$) color-color diagram of our targets. We also display the intrinsic color tracks of main-sequence (MS; dash-dotted line) and giant (dotted line) stars from \citet{Bessell1988}, which we converted into the 2MASS photometric system using the \citet{Carpenter2001} relation. 
Eight sources lie outside the locus of MS stars: two sources likely suffering from extinction (red circles) and six sources with possibly small extinction or poor-quality infrared magnitudes (orange crosses). We also overplot their error bars, except for source \#102 because its $J$ and $K_{\rm s}$ magnitudes are upper limits in 2MASS. The arrow indicates the reddening vector for $A_{V} =1$~mag. For a better readability, the source \#120 with $H$--$K_{\rm s} = -0.152\pm0.045$~mag does not appear.}
\label{Fig:2mass_photometry_color}
\end{figure}

\begin{figure*}[!t]
\centering 
\includegraphics[width=9.5cm]{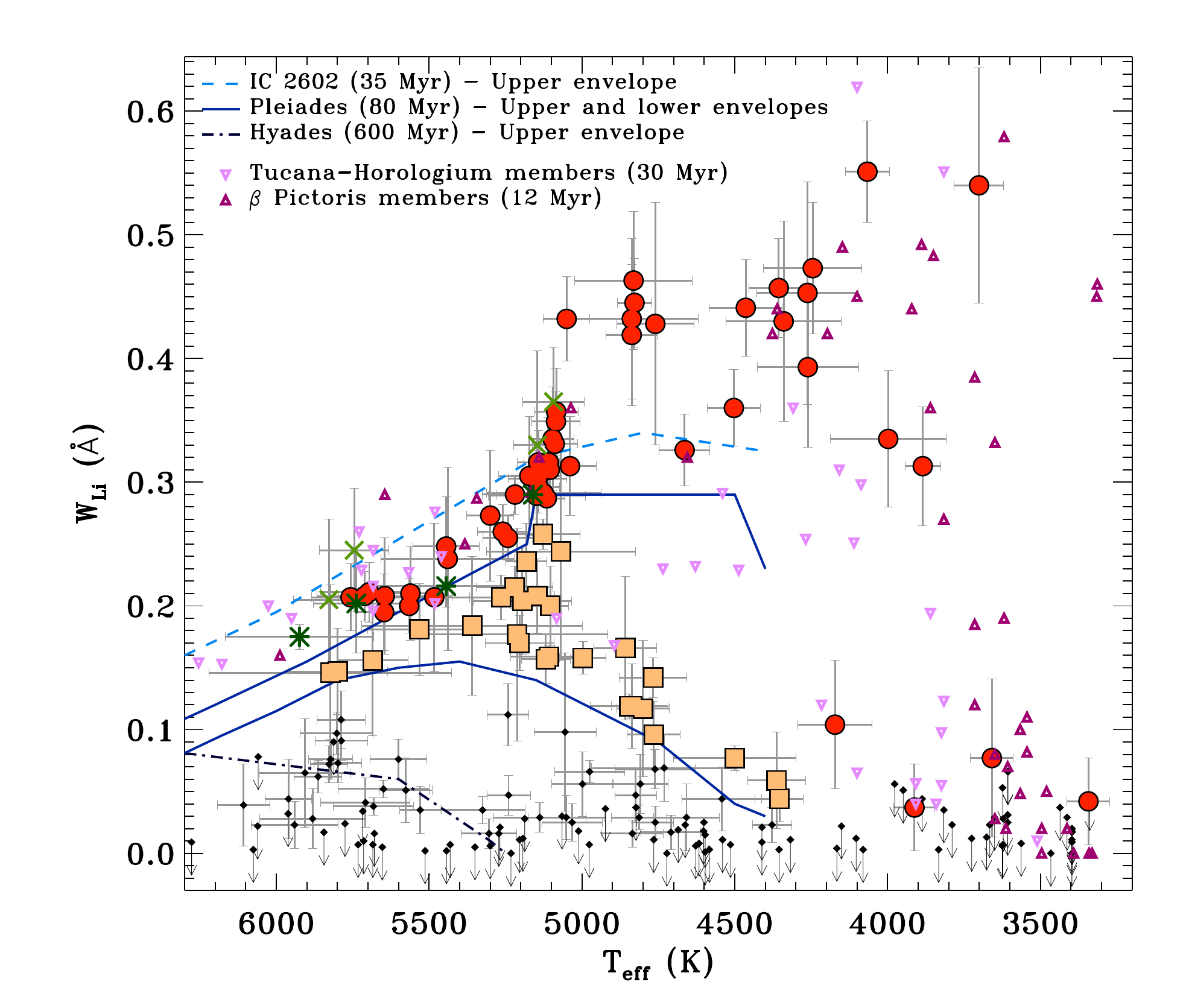}
\hspace{-1.cm}
\includegraphics[width=9.5cm]{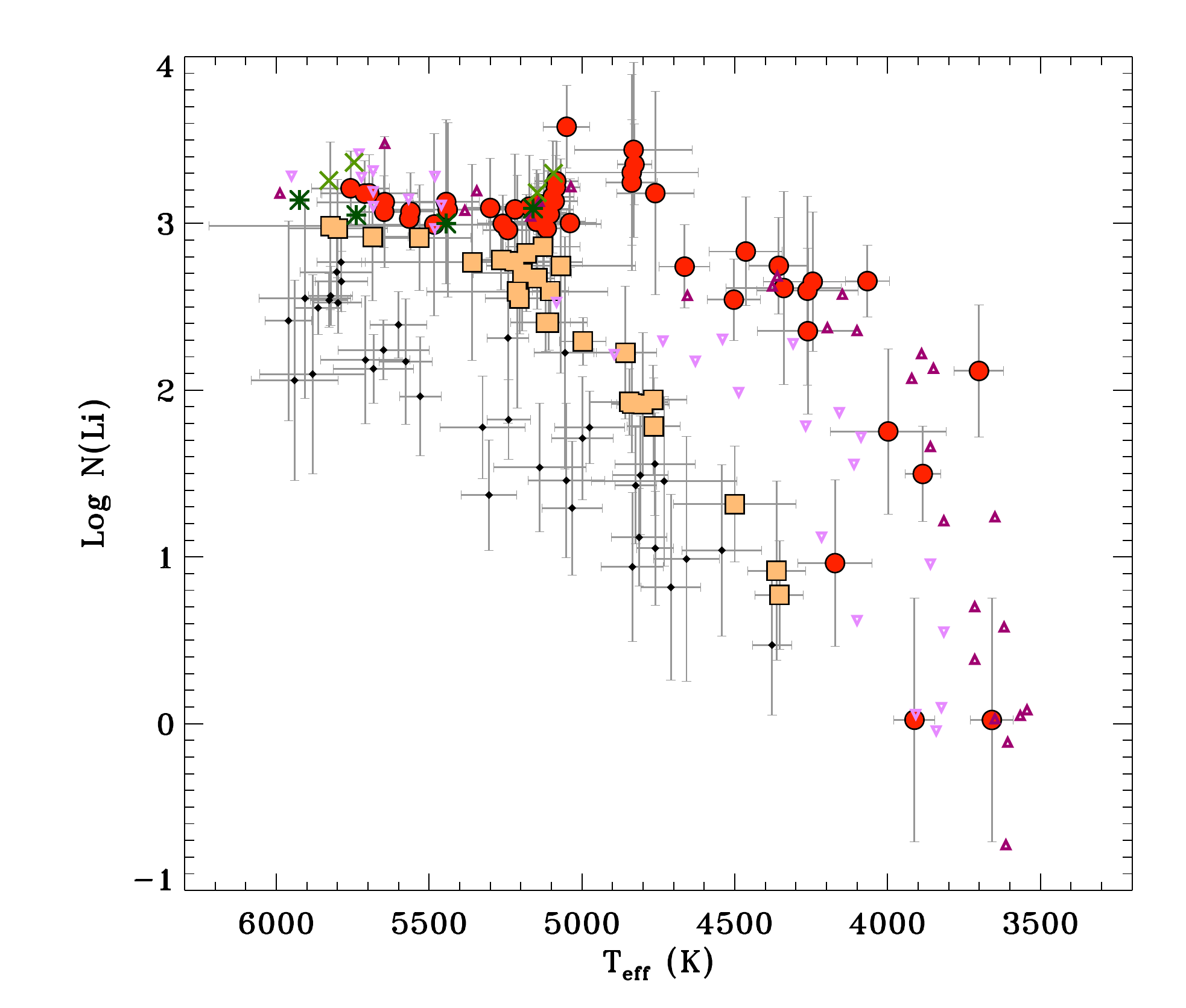}
\caption{\textit{Left panel}: \ion{Li}{i} $\lambda$6707.8 equivalent widths as a function of effective temperatures for our targets. Since each symbol results from the analysis of a unique spectrum (Table~\ref{Tab:AP_CepSurv}), any source observed at multiple epochs is marked several times. The arrow symbols indicate the upper limits on lithium equivalent width. The lines of different color and style mark the boundaries for three young clusters. \textit{Right panel}: lithium abundances versus effective temperatures. We disregard the upper limits. In both panels, we display the {PMS-like (big filled circles), Pleiades-like (filled squares), and older (filled diamonds) stars (see Sect.\,\ref{Sec:Analysis_Li} for details)}. The four comoving TTSs towards the CO Cepheus void (Paper~II) and the additional members (Paper~III) are highlighted with asterisks and crosses, respectively. For comparison, we also overplot the locus of the sources belonging to two young (age $= 12$--$30$\,Myr) associations listed by \citet{daSilva2009} with different small triangles.}
\label{Fig:EWLI_ALI_Teff}
\end{figure*}

\subsection{Stellar luminosity and extinction}
\label{subsection:SED}

We analyzed the spectral energy distribution (SED) of our targets with the aim of checking the spectroscopic atmospheric parameters, of estimating the extinction, $A_V$, and of deriving the stellar luminosities. The $BVI_{\rm C}JHK_{\rm s}$ SEDs were constructed using the optical and NIR photometric data described in Sect.~\ref{subsection:phot_astro}. We completed them with mid-infrared photometry provided by the WISE All-Sky survey \citep{Wright2010, Cutri2012}. 

\medskip
Figure~\ref{Fig:HR} shows the Hertzsprung-Russell (HR) diagram for the targets with accurate parallaxes (Table~\ref{Tab:EWLi_Kinematics_CepSurv}) and APs derived in the present study (Table~\ref{Tab:AP_CepSurv}). For $187$ of our targets we made use of the \emph{Gaia}~DR2 parallax, while we adopted the values of 18.4\,$\pm$\,2.7\,mas and 36.0\,$\pm$\,6.9\,mas for the sources \#$15$ and \#$101$, respectively, as quoted in the URAT Parallax Catalog \citep[][]{Zacharias2015,Finch2016}. No parallax is available for the remaining four targets. Evidently, the values of $\log g$ found by our spectroscopic analysis are fully consistent with the evolutionary status in the HR~diagram, demonstrating the reliability of APs derived by {\tt ROTFIT} (Sect.~\ref{Sec:Analysis_AP}).  Most targets are located close to (or slightly above) the zero-age main sequence (ZAMS) and show $\log g$$>$$3.5$. These are good young star candidates. As for the $16$ sub-giant or giant stars, the X-ray emission can be related to their evolutionary status or to the presence of an unseen companion, which forces the evolved star into fast rotation.

\medskip
We adopted the grid of NextGen low-resolution synthetic spectra, with $\log g$ in the range $3.5$--$5.0$ and solar metallicity by \citet{Hau99a}, to fit the optical-NIR portion (from $B$ to $H$ band) of the SEDs, in the same way as in Paper~III. We fixed $T_{\rm eff}$ and $\log g$ of each target to the values found with {\tt ROTFIT} (Table~\ref{Tab:AP_CepSurv}) and let the angular stellar diameter and the extinction $A_V$ vary until a minimum $\chi^2$ was reached. For the stars with a known distance, this also provides us with a measure of the stellar radius and luminosity that was obtained by integrating the best-fit model spectrum. As listed in Table~\ref{Tab:EWLi_Kinematics_CepSurv}, we found low extinction values ($A_V<0.5$\,mag), with the exception of the more distant targets, which are mostly sub-giant or giant stars.

\medskip
In line with our selection criteria, most of our targets are consistent with late-type stars near the main sequence (Fig.~\ref{Fig:HR}), with no or low reddening (Table~\ref{Tab:EWLi_Kinematics_CepSurv}). These results fully agree with the distribution of our targets in the infrared color-color diagram (Fig.~\ref{Fig:2mass_photometry_color}). We found nine sources that lie outside the locus of main-sequence (MS) stars. For a better readability the source \#120 with $H$--$K_{\rm s} = -0.152\pm0.045$~mag is not shown. Sources \#$8$ and \#$160$ likely suffer from extinction, as they lie at the upper right corner in the color-color diagram ($J$--$H > 0.9$~mag and $H$--$K_{\rm s} > 0.3$~mag). These color excesses E($J$--$H$) and E($H$--$K_{\rm s}$) allow us to estimate their extinction $A_V$ by means of the mid-infrared extinction law of \citet{2019ApJ...877..116W} and the standard optical one $R_\mathrm{V}=3.1$. We classified the source \#$8$ as a K5 giant star with a temperature $T_{\rm eff}=3833\pm\,78$\,K and a surface gravity $\log g = 2.84 \pm 0.44$, while its infrared colors place it near the position of an M6 giant star, implying an extinction of $\sim$$2$~mag. Moreover, this source is at least $2$~mag fainter than giants having a similar $J$--$K_{\rm s}$ color index in the LSPM-North catalog \citep{LS2005}. This is consistent with a high reddening, as also indicated by the SED analysis. Source \#$160$ is a known member of the L1251 cloud \citep{Kun1993} and classified as a WTTS with spectral type K5 \citep{Kun2009}, implying an extinction of $\sim$$3$~mag.
The other sources outside the locus of MS stars include two visual binary candidates (\#$102$ and \#$120$, see Table~\ref{Tab:VB_Gaia}), one spectroscopic system (\#$20$), and four possible single stars (\#$1$, \#$3$, \#$139$, and \#$165$). We note that the entire photometry of the source \#$139$ could be affected by the extended halo of light of the nearby B5~binary \object{V447~Cep} (see Appendix~\ref{appendix:Id139}).

\begin{figure}  
\begin{center}
\includegraphics[width=8.8cm]{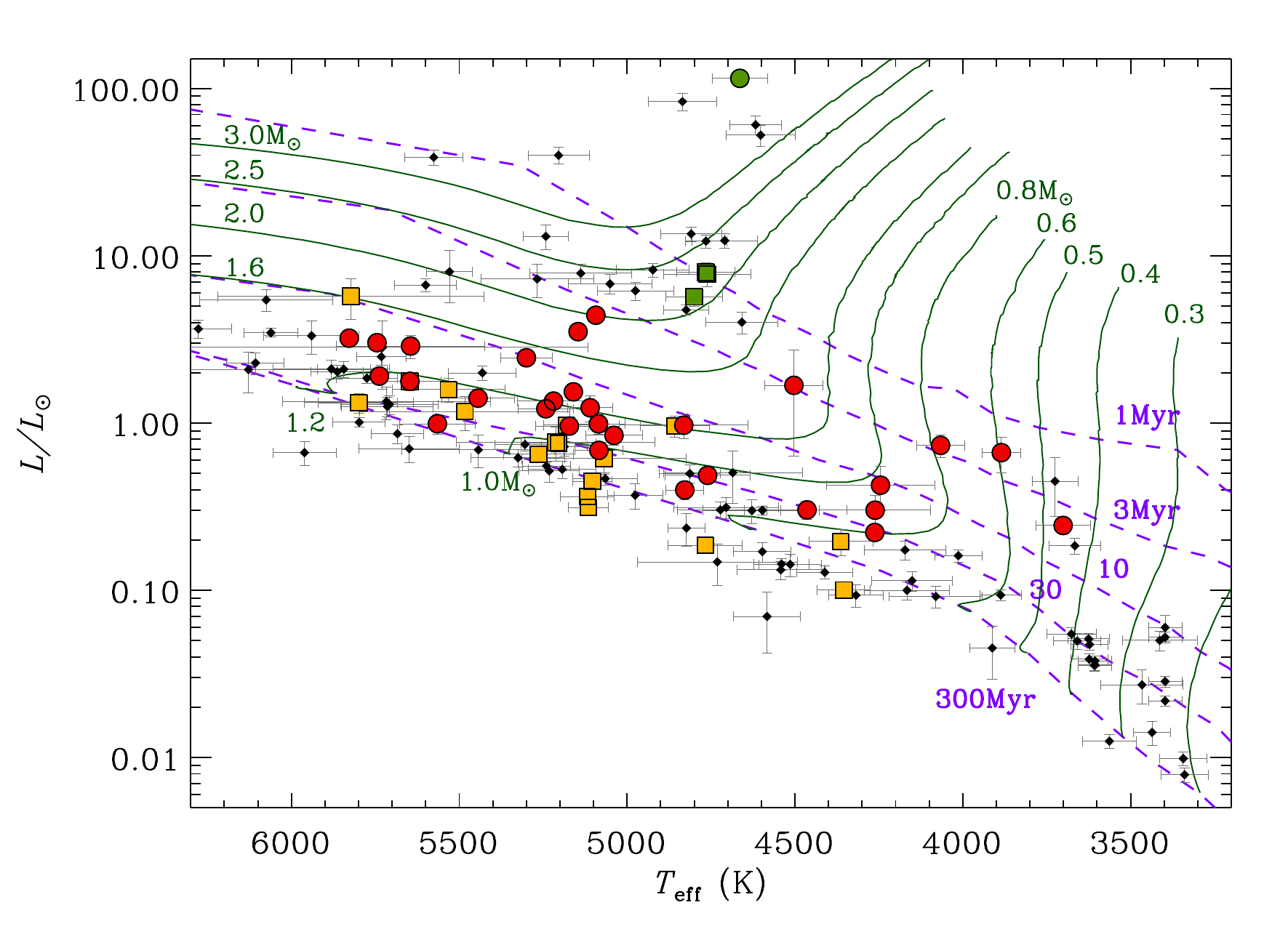}
\caption{HR diagram of our targets with known APs and parallaxes. We display the young stars selected as PMS-like (red circles) and Pleiades-like (orange squares) sources, along with the lithium-rich sub-giant or giant stars (green symbols) and older (black diamonds) sources. The pre-main-sequence evolutionary tracks of \citet{Siess2000} are shown as solid lines with the labels representing their masses. The dashed lines denote the isochrones at ages of $1$, $3$, $10$, $30$, and $300$\,Myr.}
\label{Fig:HR_Siess}
 \end{center}
\end{figure}

\subsection{Lithium content and age}
\label{Sec:Analysis_Li}

Lithium is a fragile element that is progressively depleted in the deep convective envelopes of late-type stars where it is brought at temperatures of about $3\times10^6$\,K. Therefore, the strength of the \ion{Li}{i} $\lambda$6707.8\,\AA~line can be used as an age proxy for dwarfs with spectral type later than mid-G; for these stars, a high lithium abundance is a clear sign of youth \citep[e.g.,][]{Soderblom1998}. 

\medskip
In all instrumental setups used in this study, the lithium line and the nearby \ion{Fe}{i} $\lambda$6707.4\,\AA~line are blended, at least partly. However, when applying the spectral subtraction technique \citep[see, e.g.,][]{Herbig1985, Barden1985, Frasca1994, Montes1995}, this contamination is automatically corrected because the best-fitted lithium-poor template, which is rotationally broadened by {\tt ROTFIT} to the $v\sin i$ of the target, faithfully reproduces the spectral behavior around the lithium line (Fig.~\ref{Fig:HaLiSpec_CepSurv}). We measured the \ion{Li}{i} equivalent width, $W_{\rm Li}$, for the whole sample of single stars and SB1 systems (Table~\ref{Tab:AP_CepSurv}). We then derived the lithium abundance, $\log N$(Li), based on the \citet{PavMag96} calculations. 

\medskip
Following the classification of Papers~I and III, we defined as Pleiades-like the stars lying between the lower and upper envelopes of the Pleiades cluster in the  $T_{\rm eff}$--$W_{\rm Li}$ diagram (filled squares in Fig.~\ref{Fig:EWLI_ALI_Teff}). Since the lower envelope of the Pleiades nearly coincides with the upper envelope of the 300-Myr-old UMa cluster \citep{Soderblom1993b}, the age of Pleiades-like objects should range between $100$ and $300$\,Myr. Analogously, we have considered as PMS-like the stars lying above the Pleiades upper envelope (filled circles in Fig.~\ref{Fig:EWLI_ALI_Teff}). 

\medskip
In the literature, a significant scattering of $\log N$(Li) was found for stars belonging to the same cluster. This can be due to a real age spread within the cluster or to further parameters affecting the lithium depletion, such as stellar rotation \citep[e.g.,][and references therein]{Jeffries2017,Bouvier2018}. For a field star, this prevents any direct conversion of $\log N$(Li) into~age. 

However, a relative age estimate can be obtained by comparing the position in the $T_{\rm eff}$--$\log N$(Li) diagram with that of stars with a well-defined age and similar APs. We therefore overplot the envelopes of clusters IC\,2602, Pleiades, and Hyades \citep[see, e.g.,][]{Soderblom1993, Soderblom1993b, Jeffries2000}, along with the locus of known members of the $\beta$~Pictoris and Tucana-Horologium associations whose quoted ages are about $12$ and $30$\,Myr, respectively \citep{daSilva2009}. As a result, most PMS-like stars are distributed similarly to the $\beta$~Pictoris members and slightly above those of the Tucana-Horologium association (Fig.~\ref{Fig:EWLI_ALI_Teff}).

We got an alternative estimate of the ages of our targets by comparing their position in the HR diagram with pre-main-sequence isochrones from \citet{Siess2000}, as shown in Fig.~\ref{Fig:HR_Siess}. Most PMS-like stars are located between the isochrones at $10$ and $30$\,Myr, while six sources (\#6\,c2, \#189, \#191\,c1\,\&\,c2, F3, F4) are found between the isochrones at $1$ and $10$\,Myr (Fig.~\ref{Fig:HR_Siess}). This agrees with the young age assumed in Paper~II. In contrast, the Pleiades-like stars are mostly below the isochrone at $30$\,Myr. We are finally able to confirm that all the PMS-like and Pleiades-like sources are likely young stars, with the exception of four sources (\#109, \#133, \#159, \#185\,c1), which are likely lithium-rich sub-giant or giant stars (Figs.~\ref{Fig:HR} and \ref{Fig:HR_Siess}).

\begin{figure}[!t]
\centering 
\hspace{-0.4cm}
\includegraphics[width=8.9cm]{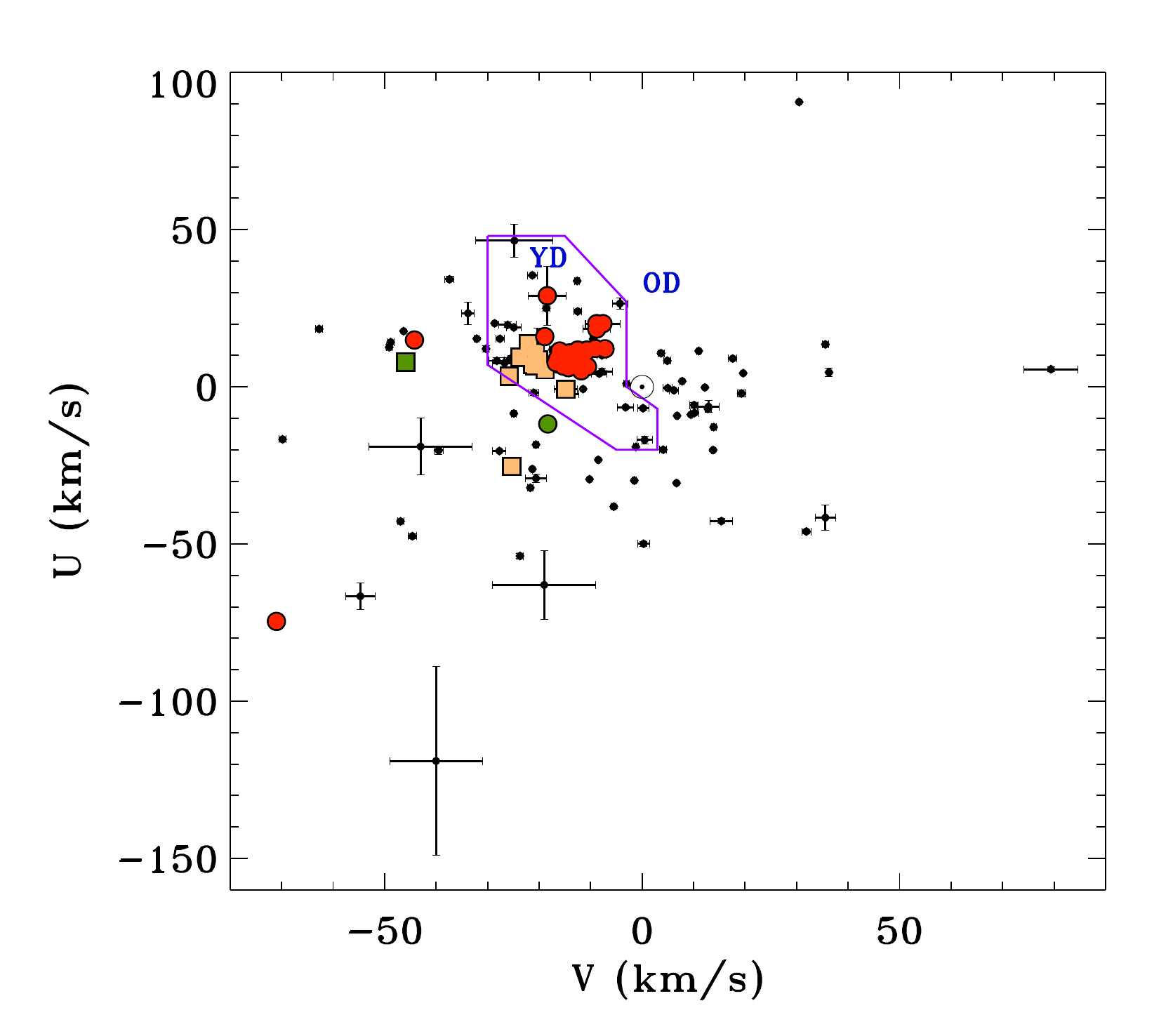}
\caption{($U$, $V$) plane of targets with astrometry and $RV$ value (Table~\ref{Tab:EWLi_Kinematics_CepSurv}). The solid line marks the separation between the young- (YD; age $<2$\,Gyr) and old-disk (OD) populations in the solar neighborhood, according to \citet{Eggen1996}. We use the same symbols as in Fig.~\ref{Fig:HR_Siess}.}
\label{Fig:Kinematic_UV_all}
\end{figure}

\begin{figure*}[!t]
\centering 
\hspace{-0.7cm}
\vspace{-0.4cm}
\includegraphics[width=6.5cm]{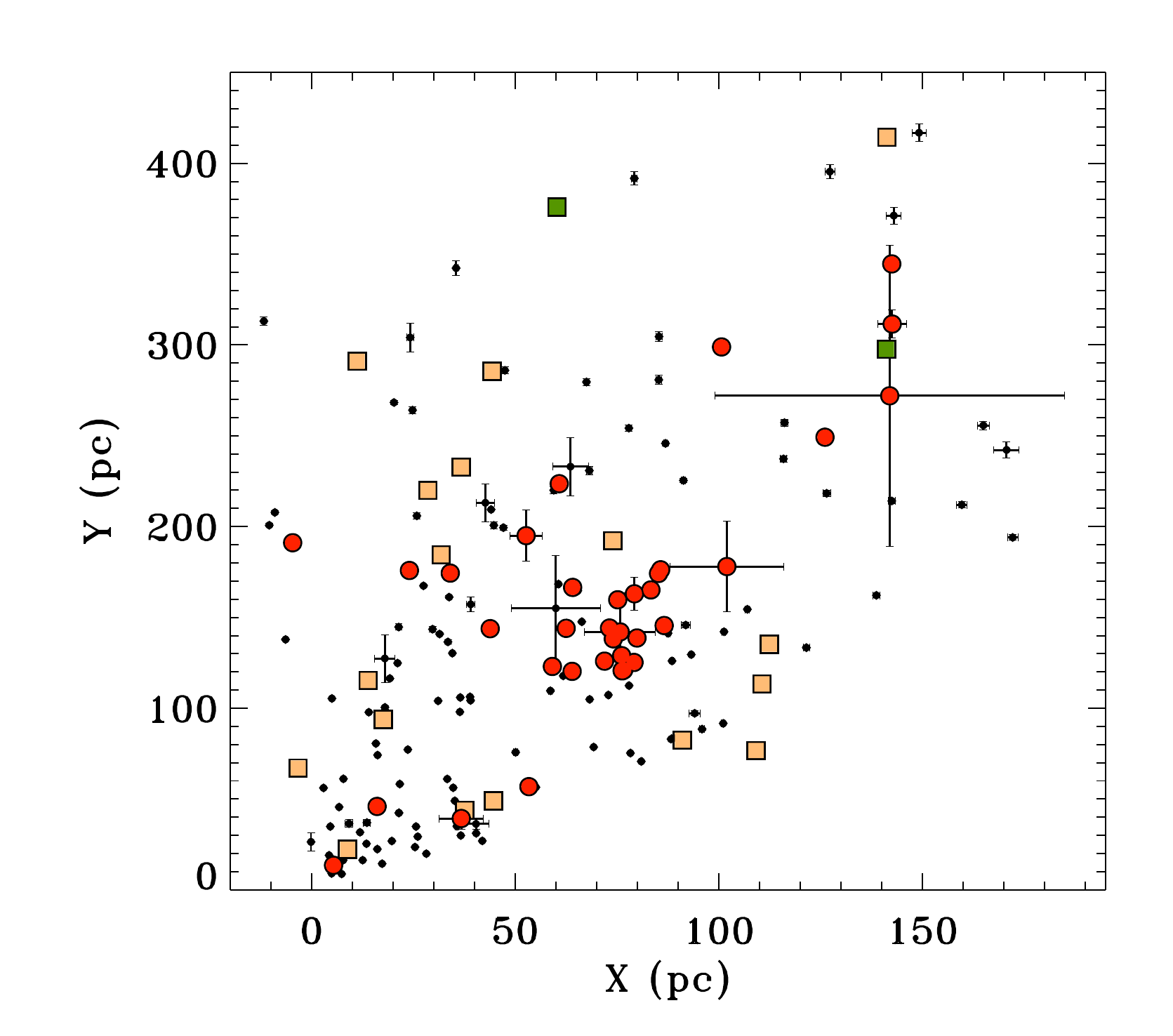}
\hspace{-0.7cm}
\includegraphics[width=6.5cm]{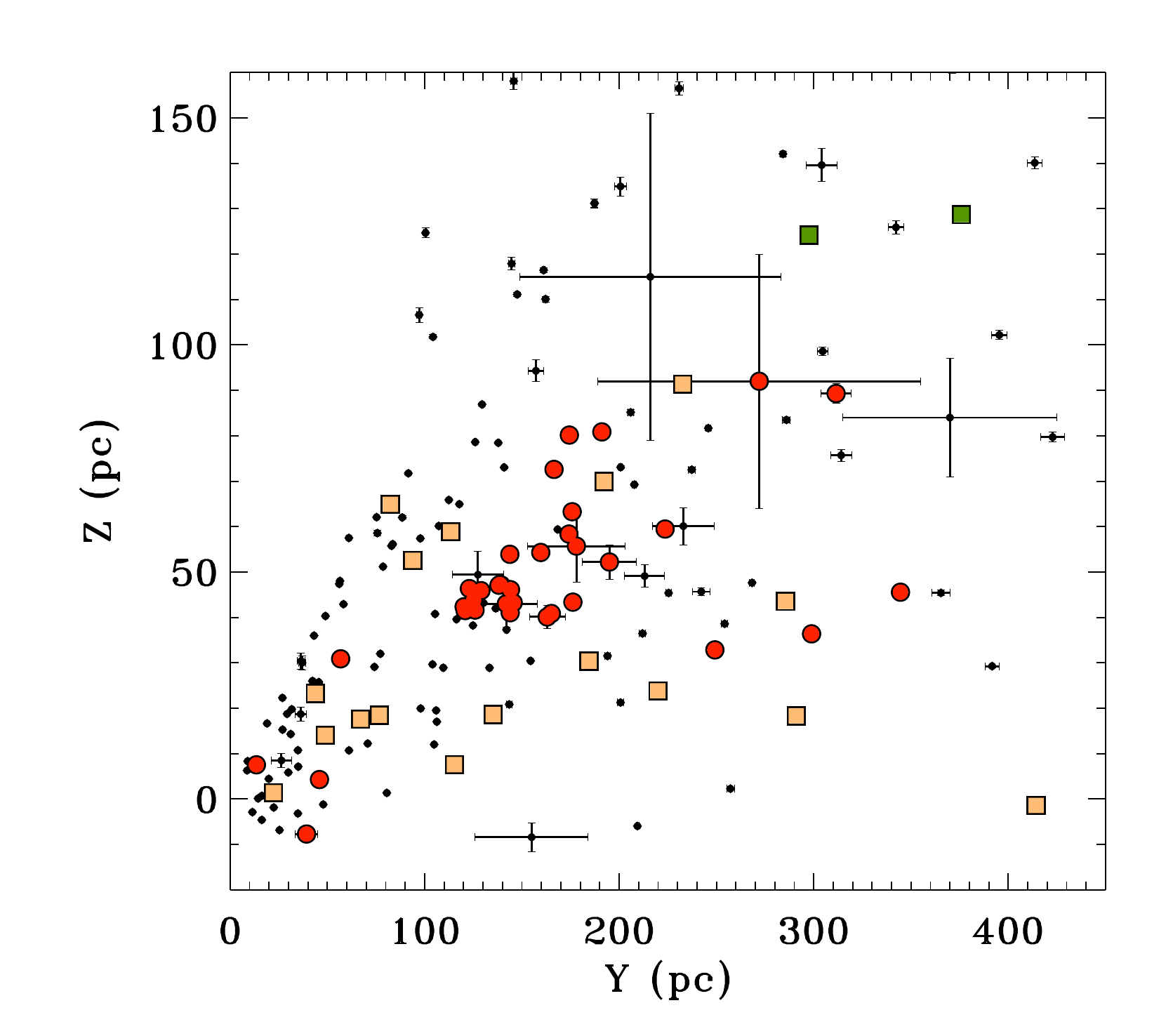}
\hspace{-0.7cm}
\includegraphics[width=6.5cm]{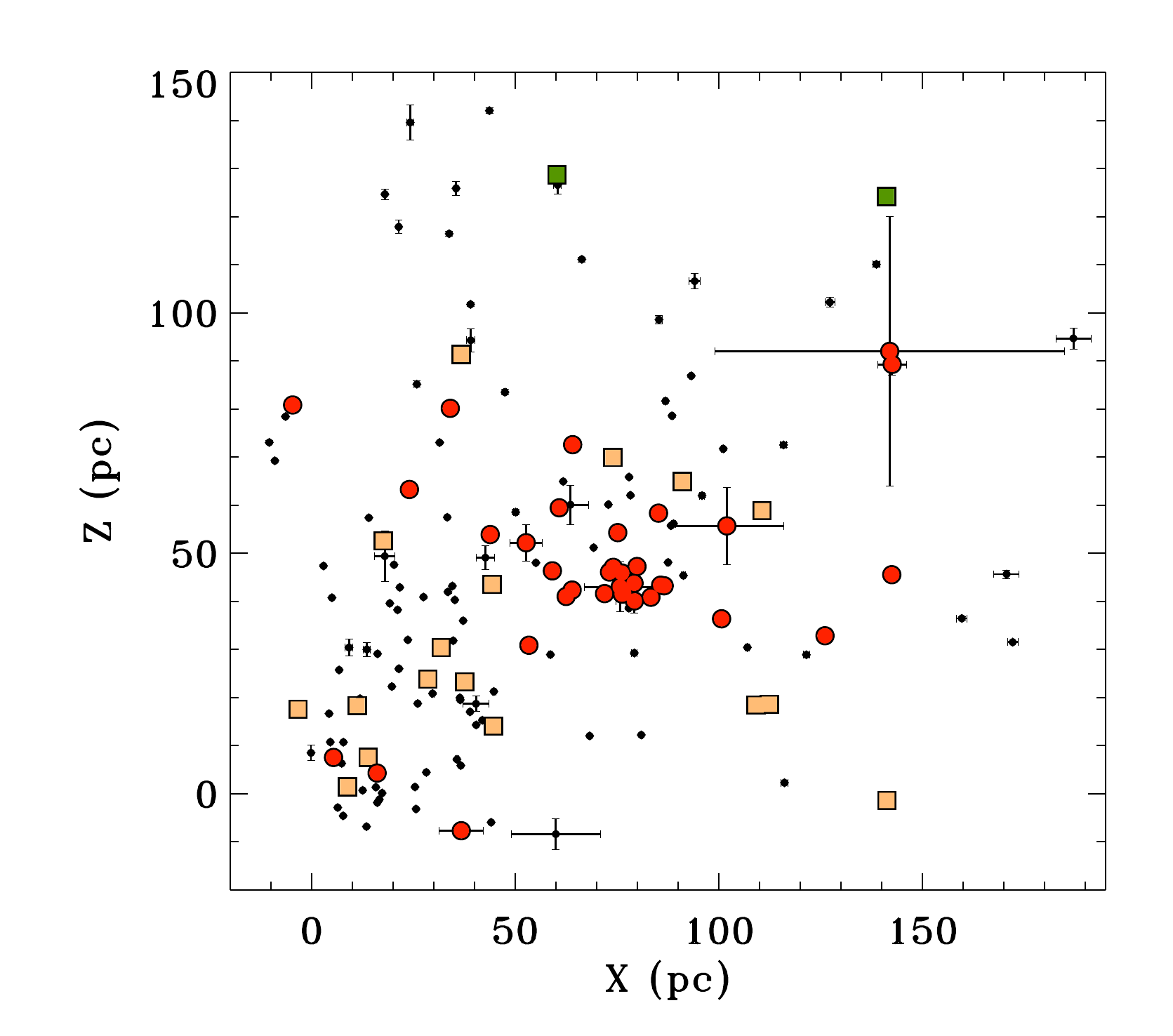}\\
\hspace{-0.7cm}
\vspace{-0.4cm}
\includegraphics[width=6.5cm]{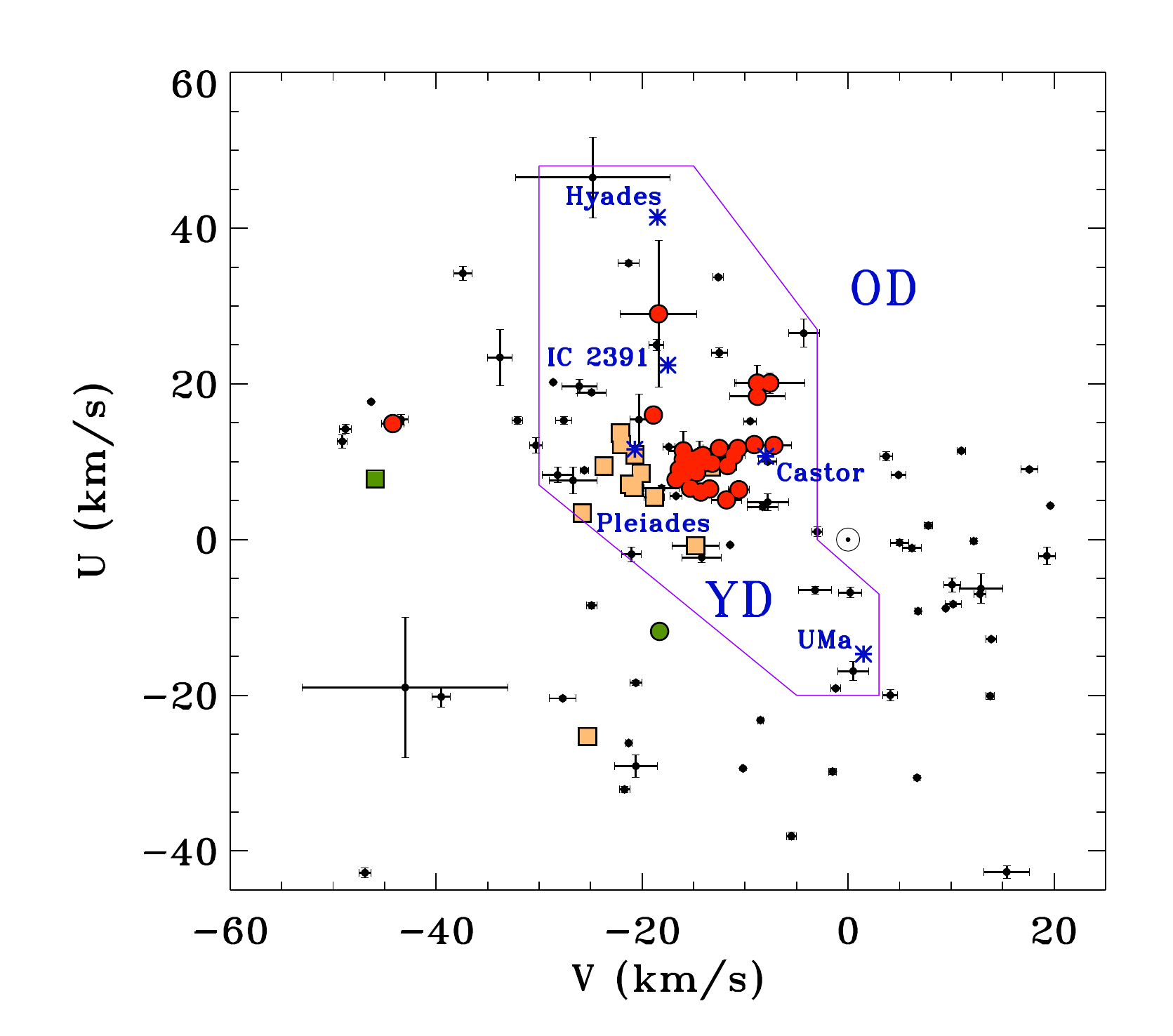}
\hspace{-0.7cm}
\includegraphics[width=6.5cm]{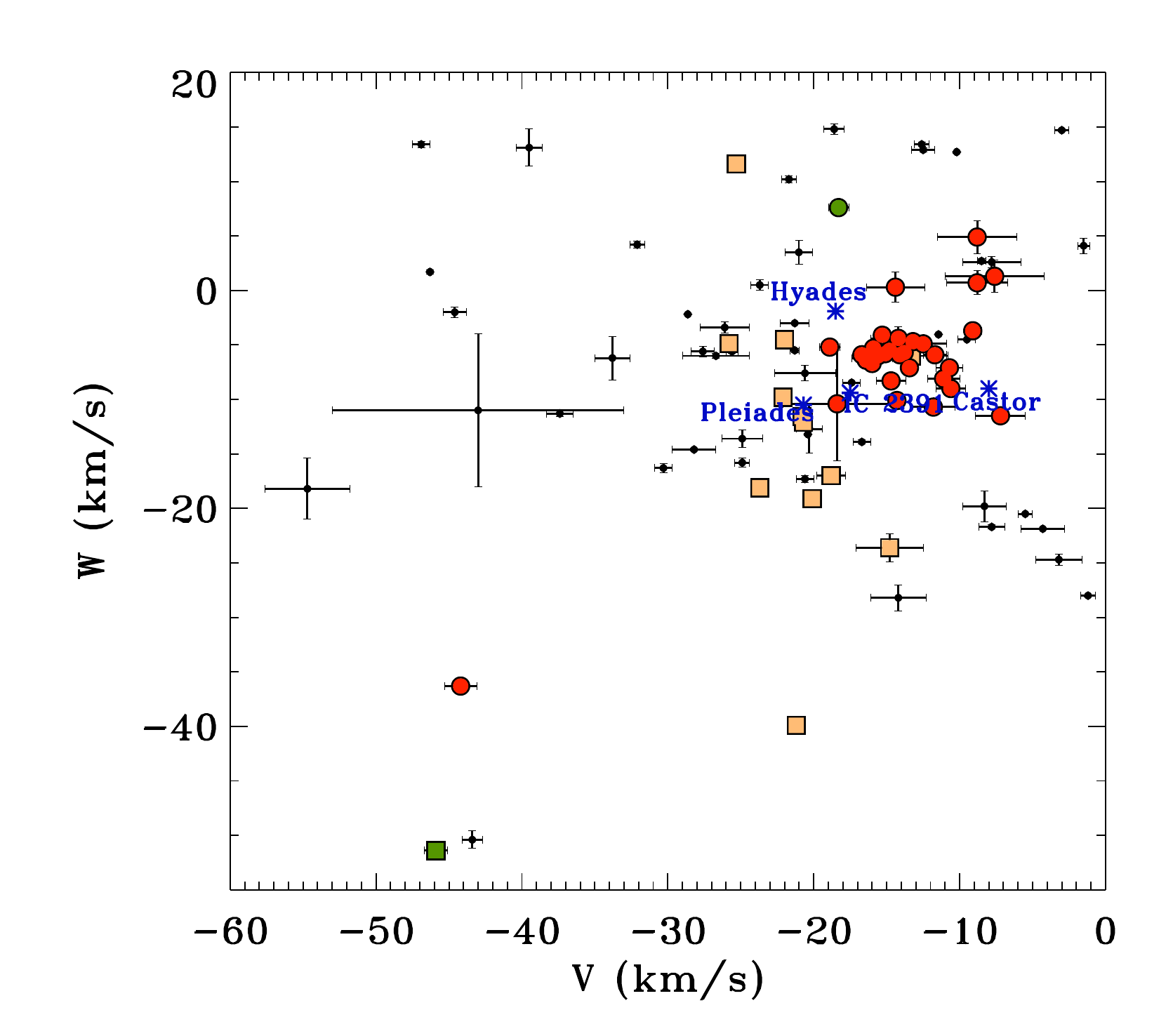}
\hspace{-0.7cm}
\includegraphics[width=6.5cm]{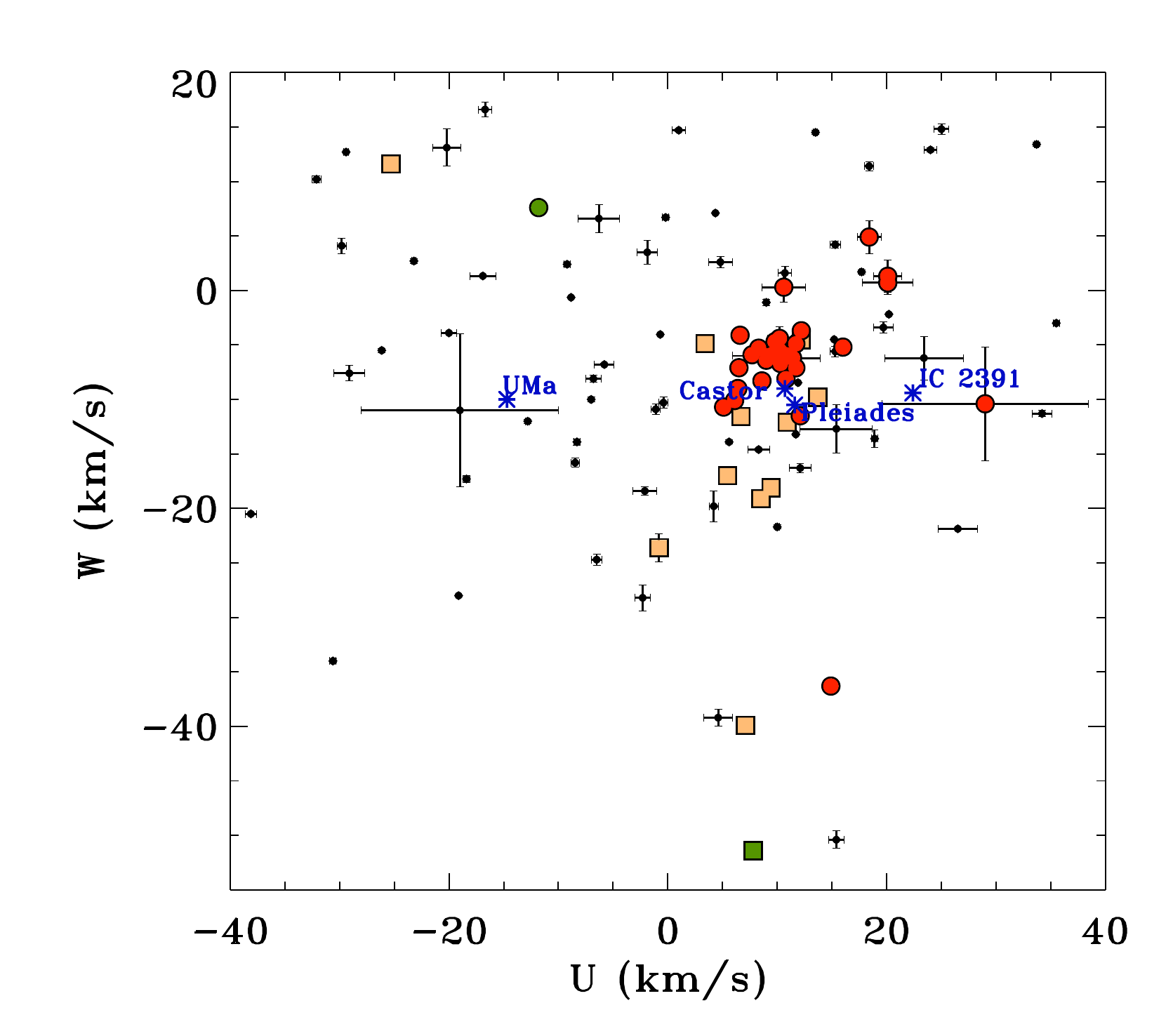}\\
\hspace{-0.7cm}
\includegraphics[width=6.5cm]{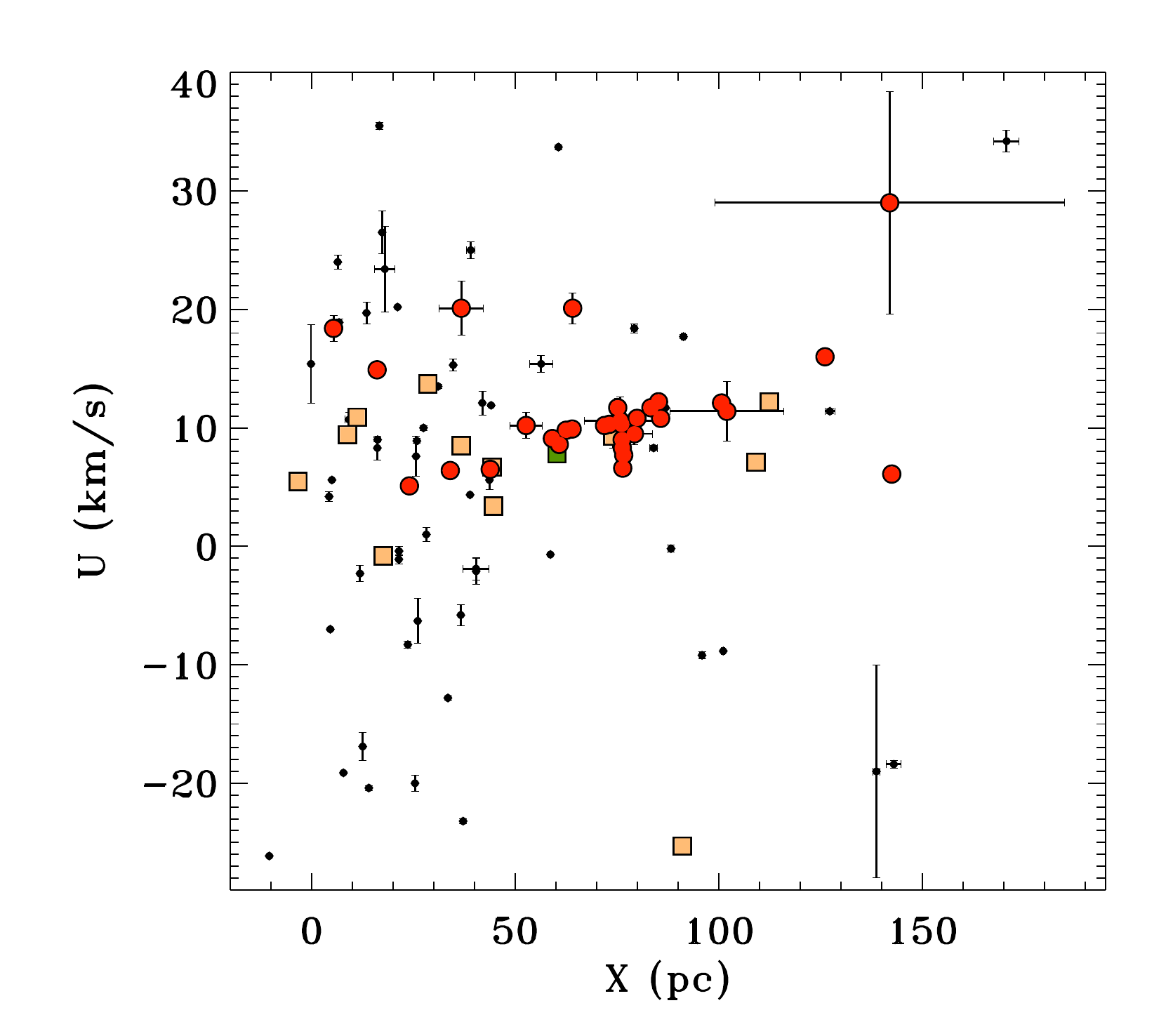}
\hspace{-0.7cm}
\includegraphics[width=6.5cm]{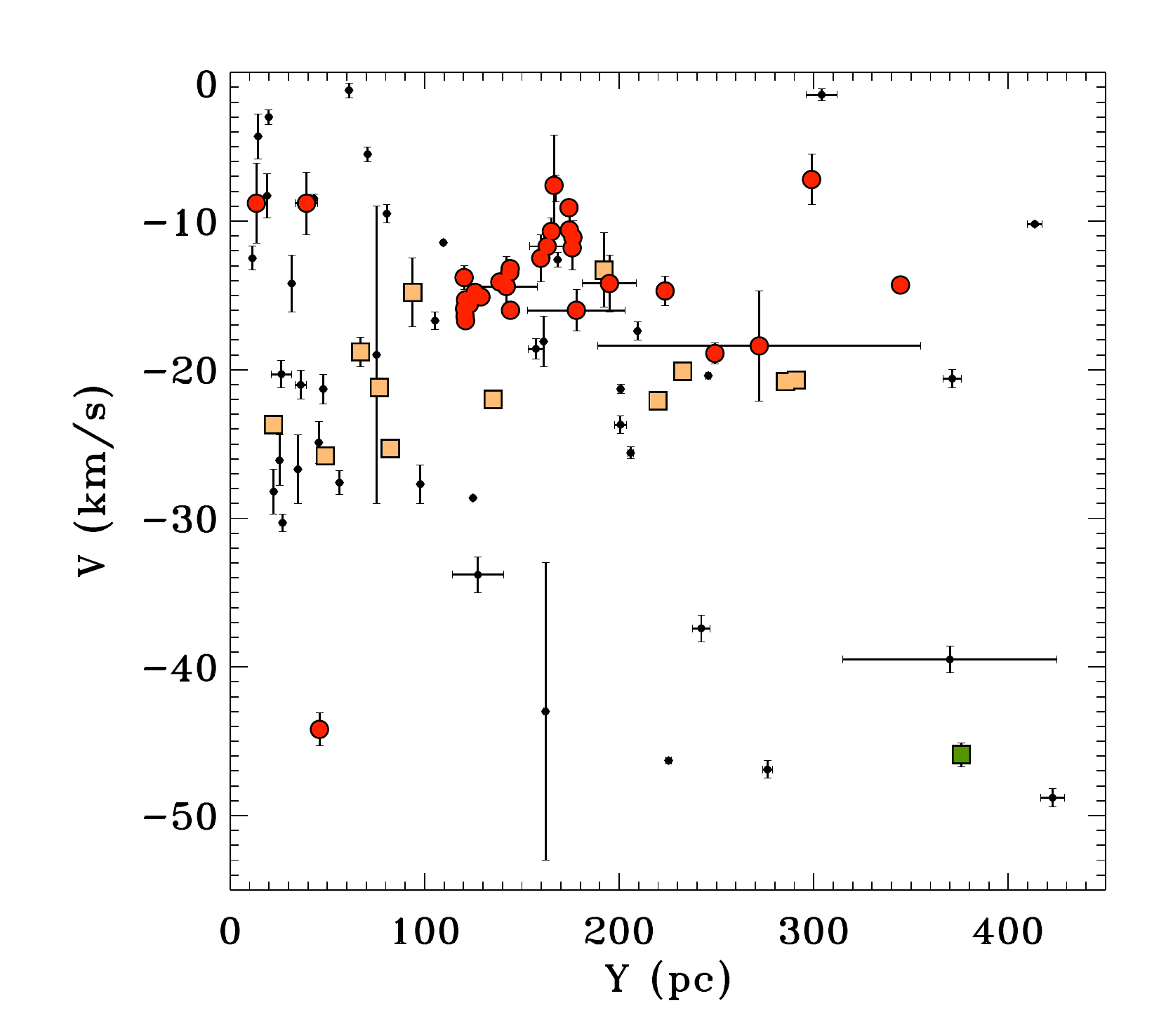}
\hspace{-0.7cm}
\includegraphics[width=6.5cm]{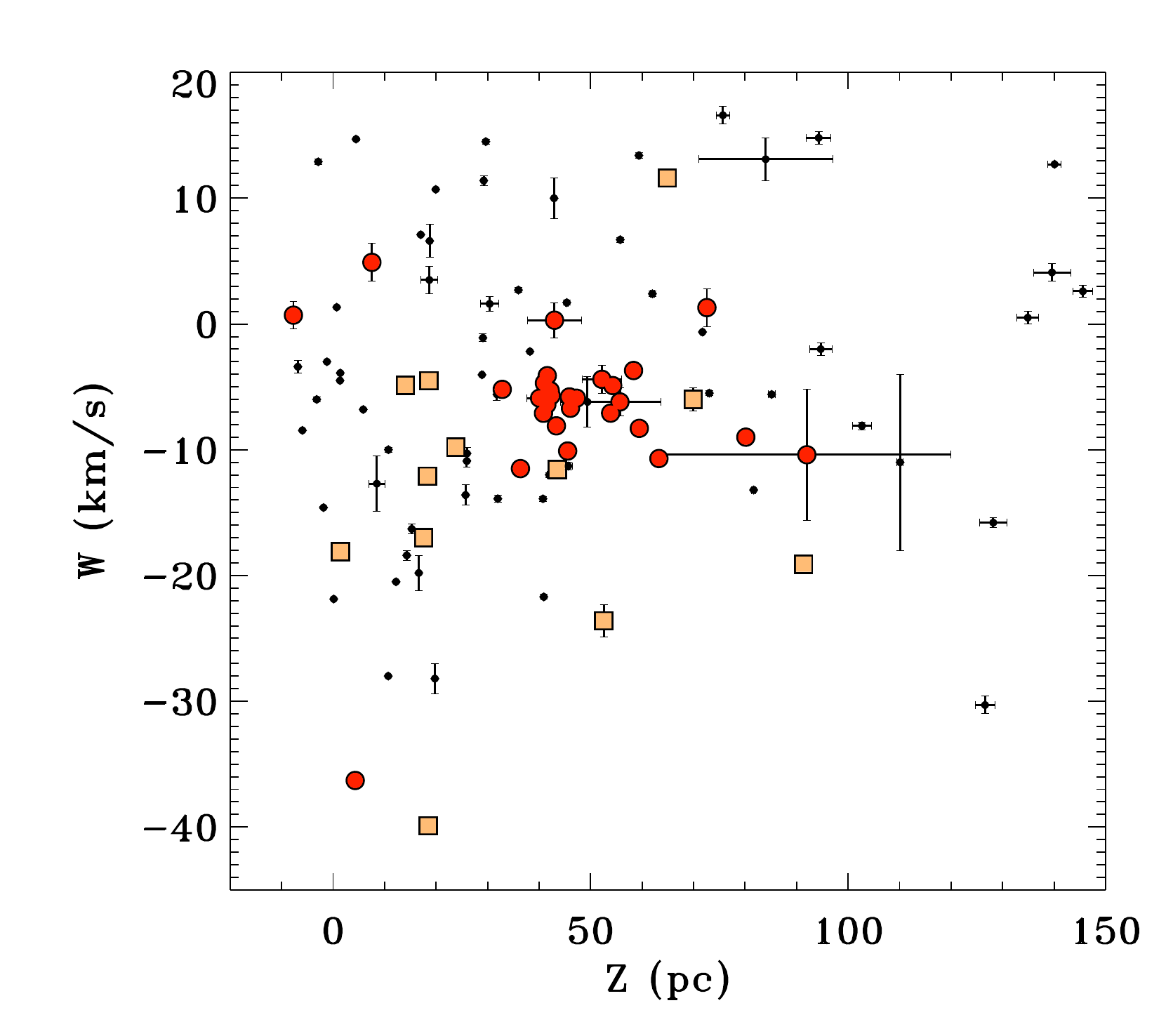}\\
\caption{Distribution of our targets in Galactic coordinates (\textit{upper panels}) and space-velocity components (\textit{middle panels}){, along with the diagram of position versus velocity pointing in a given direction (\textit{lower panels})}. We restricted the parameter ranges on all diagrams to better see the distribution of the youngest sources. We use the same symbols as in Fig.~\ref{Fig:HR_Siess}. In the middle panels, we also display the mean position of the IC\,2391 ($\sim$$50$\,Myr) and Hyades ($\sim$$600$\,Myr) superclusters, Pleiades ($\sim$$100$\,Myr) and Castor ($\sim$$200$\,Myr) moving groups, and Ursa Major (UMa) group ($\sim$$300$\,Myr). In the ($U$, $V$) plane, the solid line delimits the area of young- and old-disk populations in the solar neighborhood \citep{Eggen1996}.}
\label{Fig:Kinematic_XYZUVW_young}
\end{figure*}

\subsection{Calculation of spatial-kinematic coordinates}
\label{sec:Kin}

To study the kinematics of our targets in the $XYZUVW$ space, we combined the average radial velocities derived in Sect.~\ref{Sec:Analysis_RV} with the sky positions, proper motions, and parallaxes reported in \emph{Gaia}~DR2. For sources \#$15$ and \#$101$, we actually derived their kinematics by making use of the sky positions from the 2MASS catalog and the astrometry from the URAT Parallax Catalog. We then computed the Galactic positions ($X$,$Y$,$Z$) and heliocentric space-velocity components ($U$,$V$,$W$) in the left-handed coordinate system.\footnote{$X$ and $U$ are pointing towards the Galactic anti-center; $Y$ and $V$ in the Galactic rotation direction; $Z$ and $W$ towards the North Galactic pole.} All these values are listed in Table~\ref{Tab:EWLi_Kinematics_CepSurv}.

\medskip
As shown in Fig.~\ref{Fig:Kinematic_UV_all}, the ($U$,$V$) plane of targets with astrometry and $RV$ measurement (Table~\ref{Tab:EWLi_Kinematics_CepSurv}) shows that most PMS-like and Pleiades-like stars are located in the young-disk (YD) population in the solar neighborhood, according to \citet{Eggen1996}. However we found three PMS-like (\#35, \#146, \#185c1) and two Pleiades-like (\#46, \#109) in the region populated by old-disk stars. While the sources \#109 and \#185c1 are Li-rich giant stars, additional observations are needed to confirm any possible variation in radial velocities for the others. Our observations reveal that the sources \#35, \#46, \#109 are possible SB2 systems due to a double-peaked H$\alpha$ profile or their CCF shape (i.e., slightly asymmetric or with a possible small peak below our detection limit; Table~\ref{Tab:AP_CepSurv}). Regarding source \#179, we do not exclude a problem in the wavelength calibration. From our $RV$ value of $-147.8\pm0.8$~km\,s$^{-1}$, this source might be located in the old-disk (OD) area. Thus this $RV$ value must be used with care, especially because this source is not a high-proper-motion star and its $RV$ value in \emph{Gaia}~DR2 is of $-9.74\pm2.88$~km\,s$^{-1}$, which we use afterwards (Table~\ref{Tab:EWLi_Kinematics_CepSurv}). Finally, we note that the signal-to-noise of the spectrum acquired for source \#146 is not high enough to obtain an accurate $W_{\rm Li}$ value.

\medskip
Figure~\ref{Fig:Kinematic_XYZUVW_young} shows the distribution of our targets in various 2D planes of the $XYZUVW$ space restricting the parameter ranges to the locus of the young sources. We used the probabilistic approach described in \citet{2014A&A...567A..52K} to assess their membership to the five young stellar kinematic groups denoted in Fig.\,\ref{Fig:Kinematic_XYZUVW_young}. Their ages are adopted from Table~1 of \citet{Montes2001}. In the kinematic velocity space, the populations of PMS-like and Pleiades-like stars are located in well-distinguishable regions. On the one hand, the heliocentric space-velocity components of a compact group of PMS-like stars are marginally consistent with those of the Castor moving group. Nevertheless, such a link is hardly compatible because of the age difference between these two groups of young stars, but above all because the PMS-like stars are only weakly dispersed in the $XYZUVW$ space. 
On the other hand, $75$\,\% of the $12$ Pleiades-like single stars with an $RV$ value in Table~\ref{Tab:EWLi_Kinematics_CepSurv} have membership probability to the Pleiades moving group larger than $40$\,\%. These nine sources are \#16, \#22, \#102, \#112, \#123, \#125, \#129, \#151, and \#179. Our procedure disregarded two sources  (\#31c2, \#86) in the YD~area due to a space-velocity component $W$ lower than $-23$\,km\,s$^{-1}$, along with source \#46 located in the OD~area (Fig.~\ref{Fig:Kinematic_UV_all}). These three sources probably exhibit some variation in radial velocity.

\section{Discussion}
\label{Sec:Results}

\subsection{Properties of the sample}
\label{ssec:Sample_candidates}

Our study confirms that the multivariate analysis allows for the optimization of the disentangling of the stellar population from X-ray emitting extragalactic components. In fact, only one (\#99) of the 162 sources is wrongly classified as a star due to a rather singular spatial configuration of the different sources (Fig.~\ref{Fig:2MASX_1922_6739_field}) and the photometric properties of this galaxy that are similar to late M-type stars (Appendix~\ref{appendix:galaxy}). The main stellar populations in our sample are young objects, multiple systems (Appendix~\ref{appendix:multiple_targets}), and low-mass stars (Appendix~\ref{appendix:Mtype_stars}). This latter group includes $36$ M-type dwarfs (i.e., from M0 to M4.5; see Table~\ref{Tab:M_SpT}).

\medskip
We selected young stars more efficiently than in our previous works. This may be linked to a sampling of more distant sources in comparison to previous studies, as explained by the Gould Disk scenario \citep{Guillout98b}. It could also be related to the sky region considered here in which stars younger than $30$~Myr are found at three different distance scales: the visual binary V368\,Cep and its comoving companion NLTT\,56725 \citep[$\sim$$20$~pc;][]{Makarov07}, the Cepheus association (Table~\ref{Tab:Prop_Cepheus}), and the Cepheus Flare region \citep{Kun2008b,Kun2008}.

\medskip
We identified $59$ lithium-rich sources. Most of them prove to be young but four sub-giant or giant stars also display a strong lithium line in absorption (Sect.\,\ref{Sec:Analysis_Li}). This is consistent with {the discovery of several lithium-rich giants from our previous studies of the stellar X-ray population (Papers~I and III). We classified $18$ targets as Pleiades-like sources (age $=100$--$300$\,Myr) and $37$ as PMS stars (age $=10$--$30$\,Myr). These two distinct stellar populations are found in different sky areas (Fig.~\ref{Fig:Distribution_all}). While the former are mainly located towards the Galactic plane, the latter are mostly projected in front of the CO Cepheus void, in the Cepheus Flare region. 

\medskip
During our survey of 186 sources, we identified $44$ ($24$\,\%) spectroscopic multiple systems. This fraction is $30$\,\% lower with respect to Papers~I and~III. Nine and six of them also display a strong and very strong lithium line (Table~\ref{tab:MultipleSyst}), respectively. Using \emph{Gaia} DR2 data, we found that $40$ of our targets and six of eight young stars in Papers~II and III have comoving companions with an angular separation ranging from a few arcsecond to a few arcmin (Table~\ref{Tab:VB_Gaia}). Five of these visual binaries ($\#6$, $\#18$, $\#119$, $\#131$, G4) also have one component classified as an SB2 system. They are, therefore, likely hierarchical triple systems. 

\begin{figure}[!t]
 \centering
 \includegraphics[width=8.9cm]{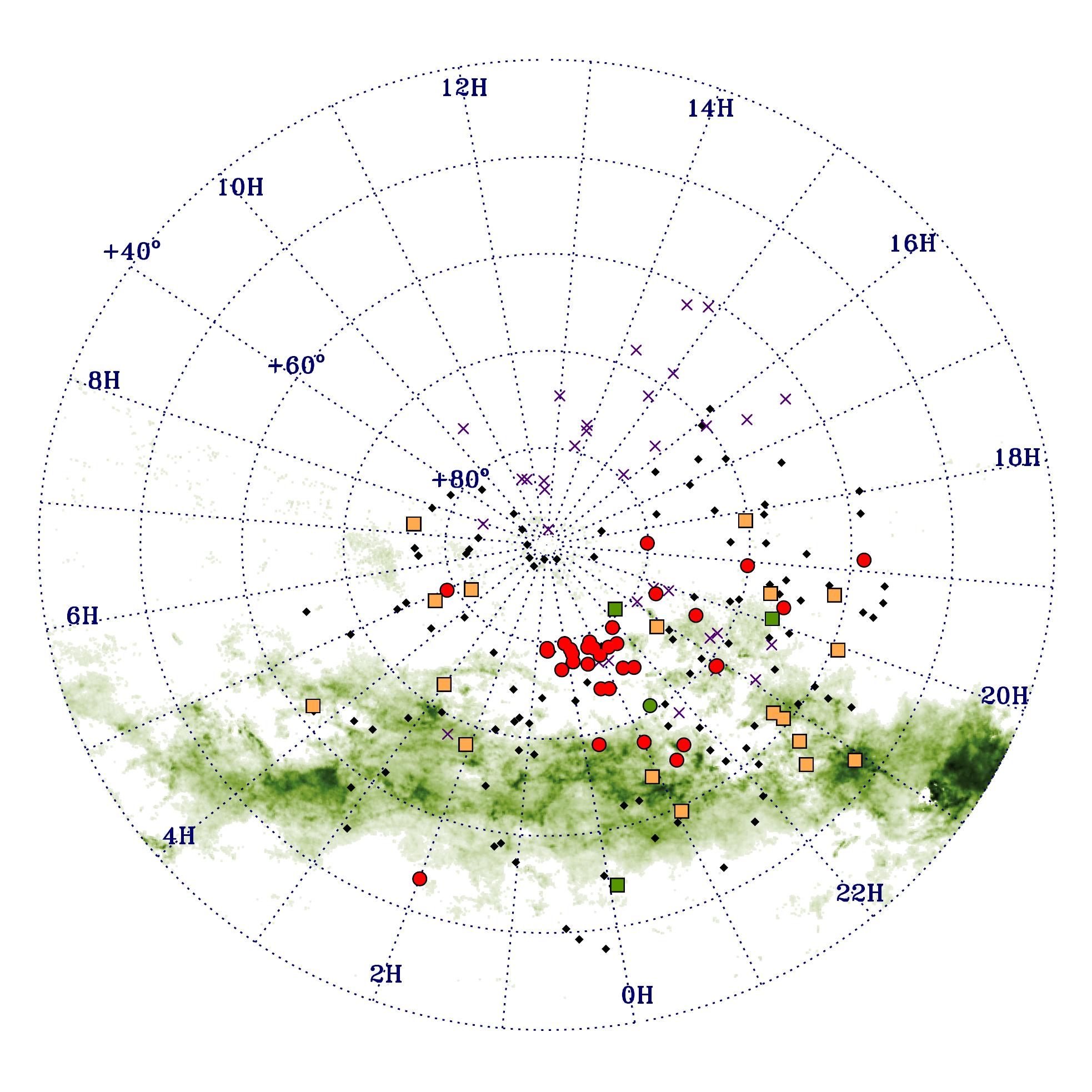}
 \caption{Spatial distribution of our targets, overplotted on the extinction (A$_{v}$) map of \citet{Dobashi05}. In addition to the symbols defined in Fig.~\ref{Fig:HR_Siess}, we denote the unobserved sources with crosses.}
  \label{Fig:Distribution_all}
\end{figure}

\begin{figure*}[t]
 \centering
\includegraphics[width=16.3cm]{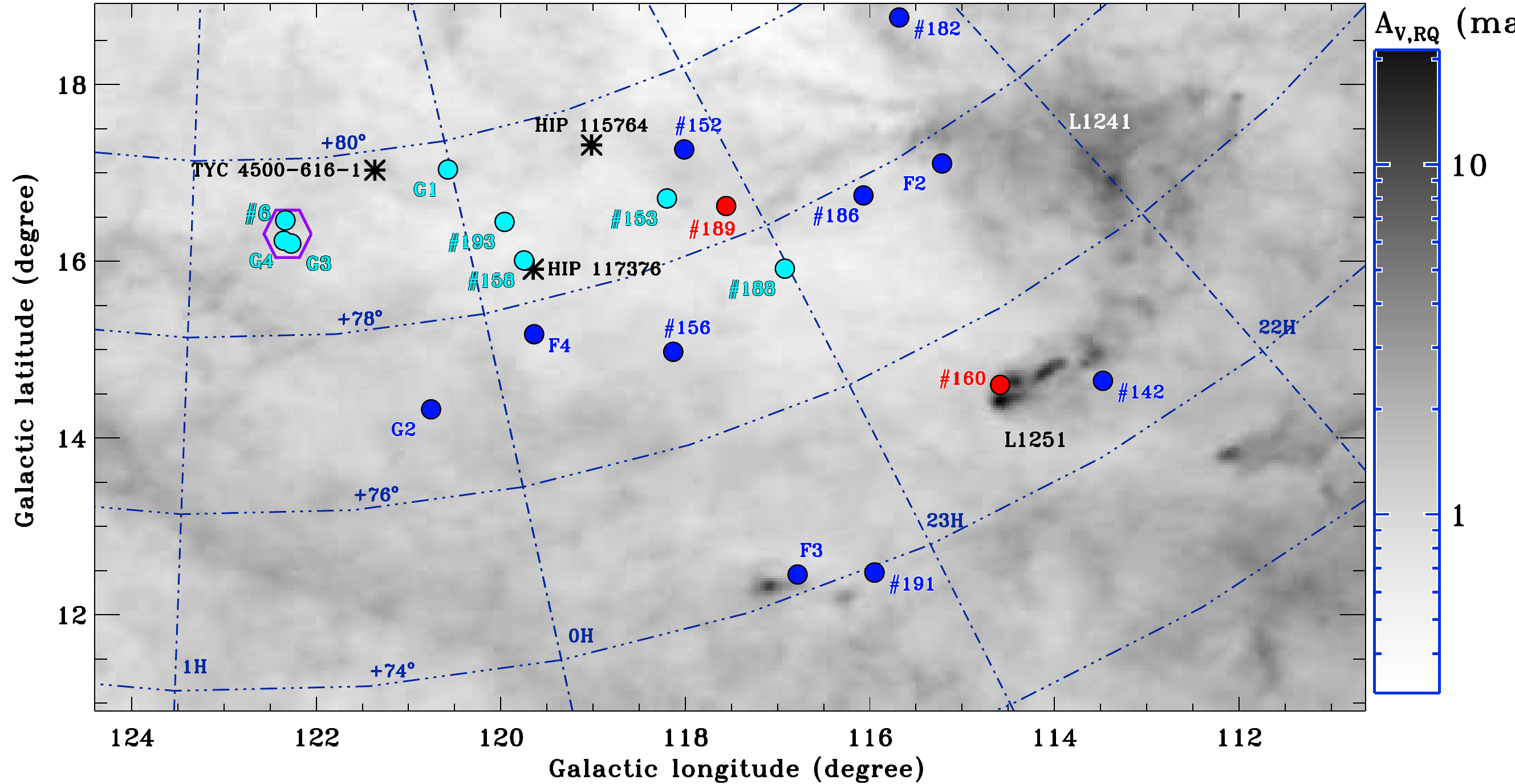}
\caption{Spatial distribution of the $20$ PMS stars (circles) in the sky area towards the CO~Cepheus void, overplotted on the visible extinction A$_{V,RQ}$ map of \citet{2016A&A...586A.132P}.$^{\ref{planck-map}}$ We display the Cep\,III members as blue circles and those still gravitationally bound (i.e., Cep\,II that is a subgroup of Cep~III, see Sect.~\ref{ssec:Stat_candidates}) as cyan circles, while the non-members are shown in red. The big hexagon encloses the Cep\,I members. The asterisks mark the locus of the three additional members reported in \citet{2017AJ....153..257O} and \citet{2018ApJ...863...91F}, as described in Sect.~\ref{ssec:Origin_Cepheus}.}
\label{Fig:Distribution_Cep}
\end{figure*}

\begin{figure*}[!t]
\centering 
\hspace{-0.7cm}
\vspace{-0.4cm}
\includegraphics[width=6.5cm]{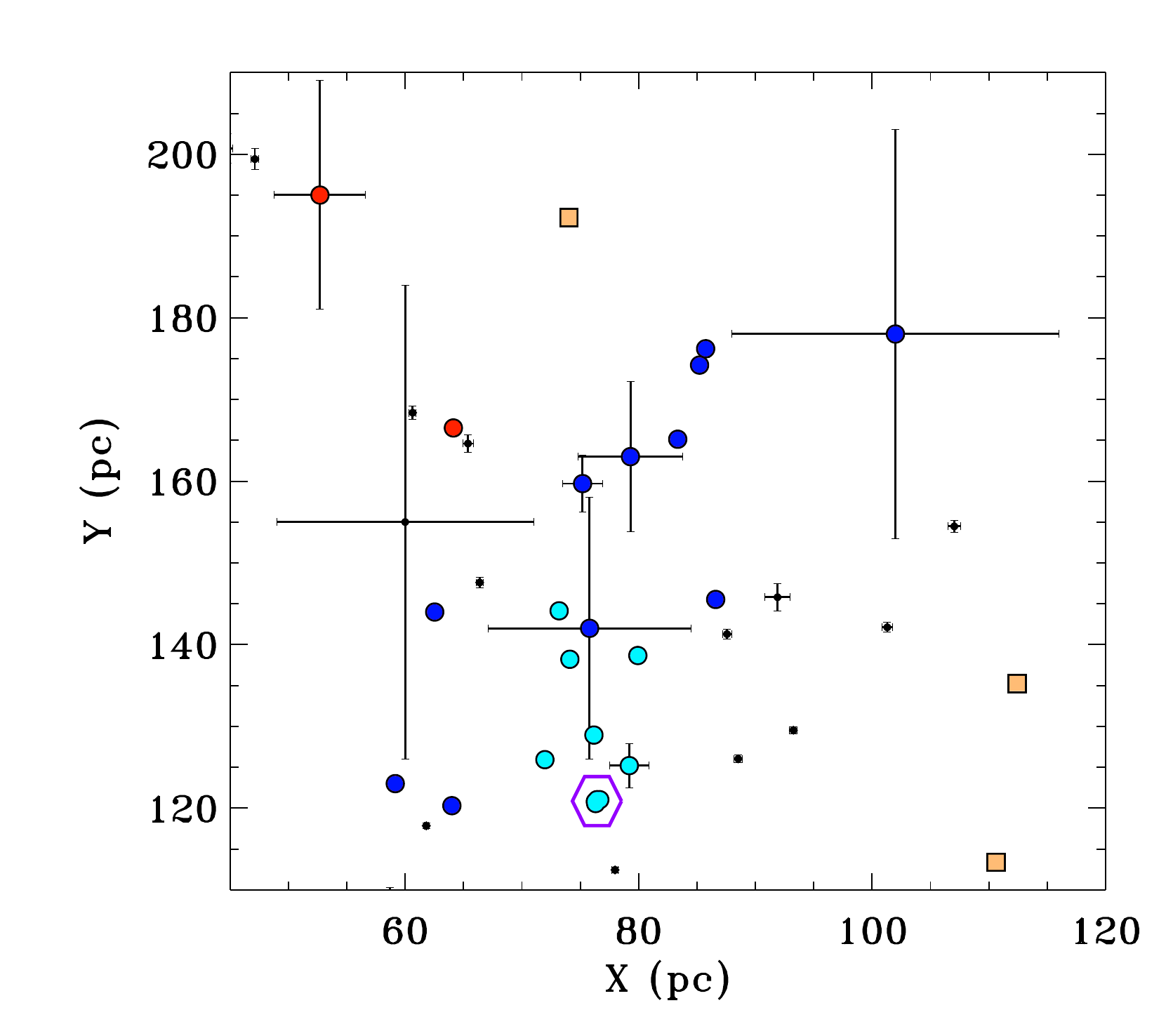}
\hspace{-0.7cm}
\includegraphics[width=6.5cm]{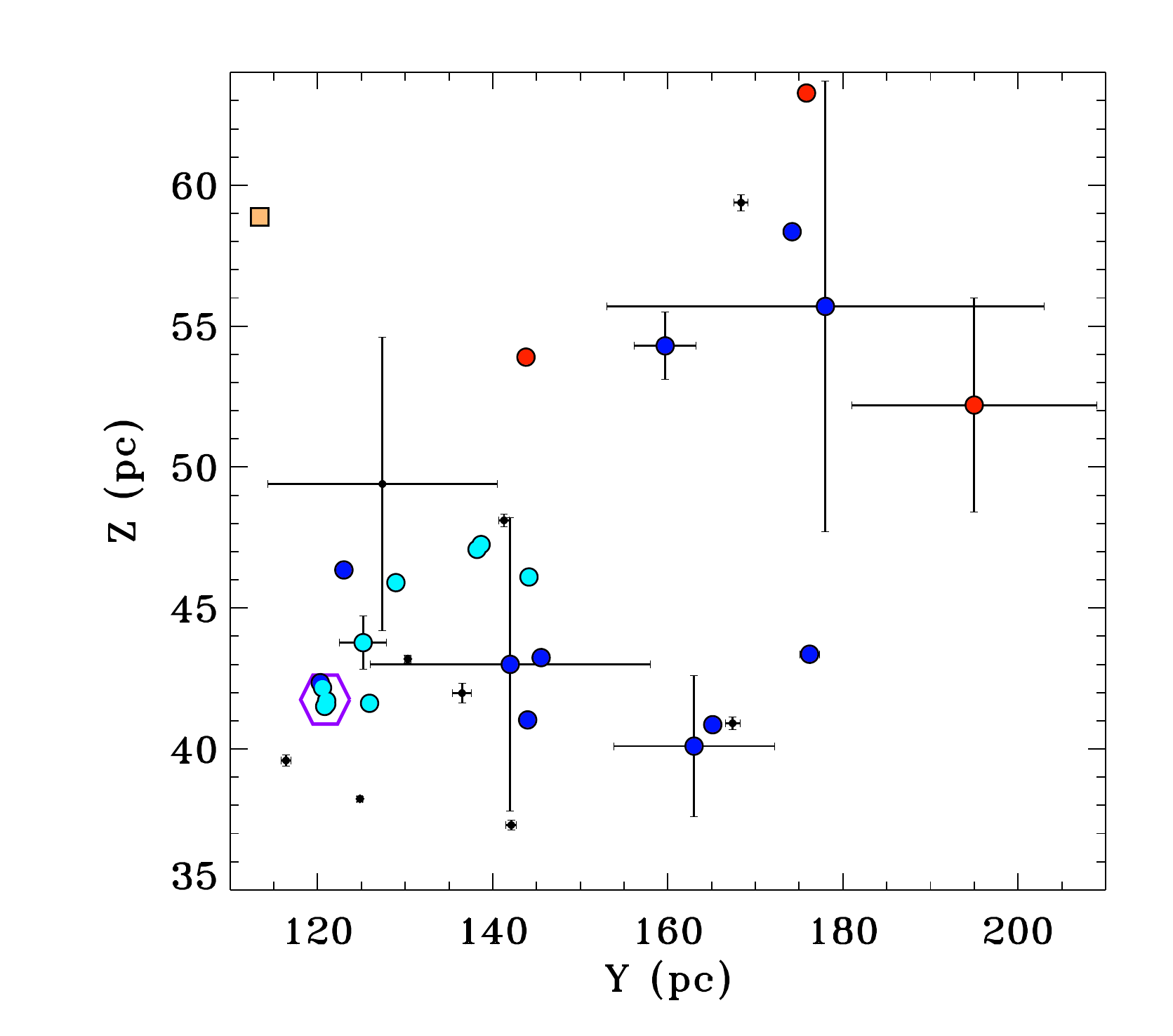}
\hspace{-0.7cm}
\includegraphics[width=6.5cm]{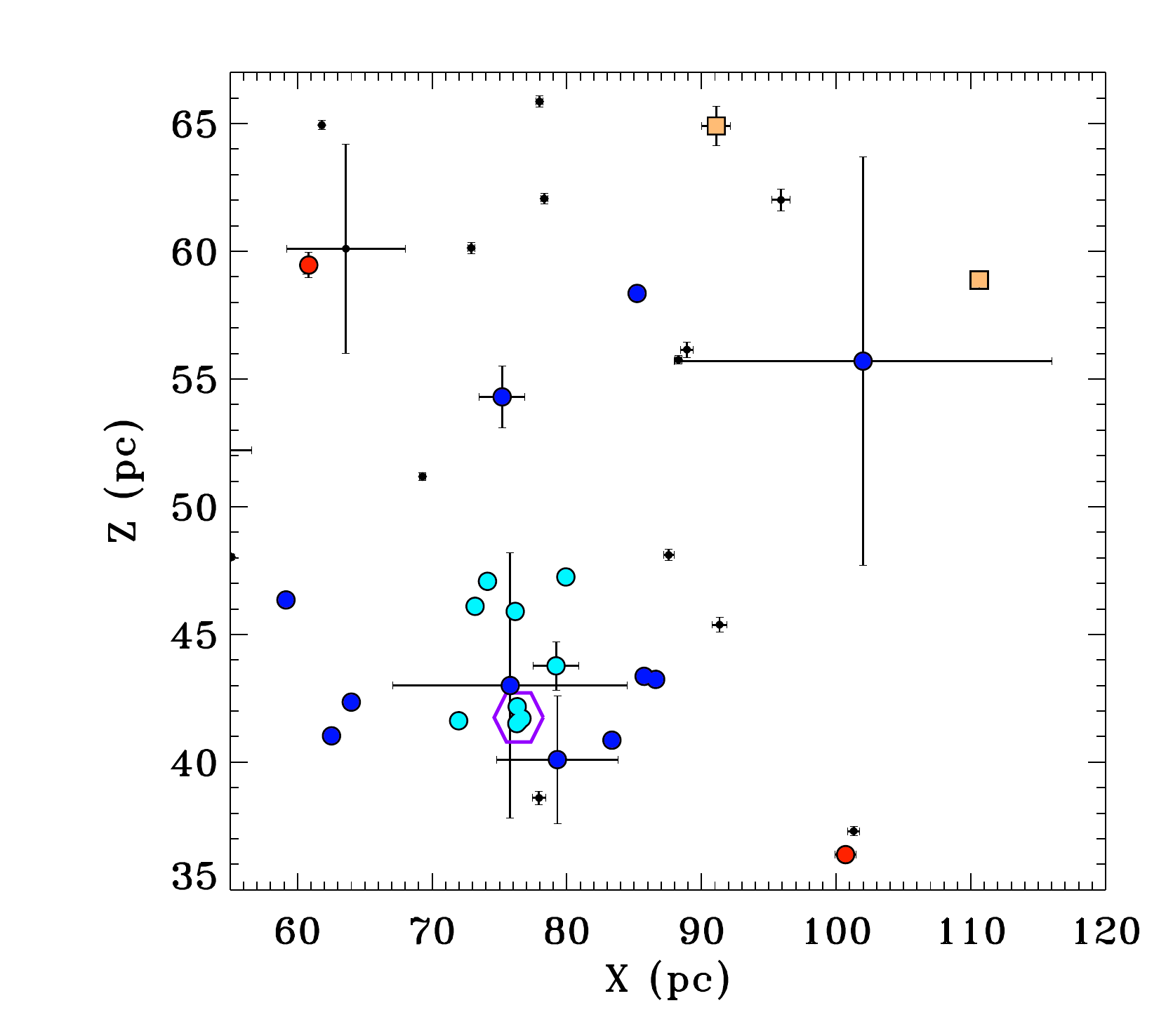}\\
\hspace{-0.7cm}
\vspace{-0.4cm}
\includegraphics[width=6.5cm]{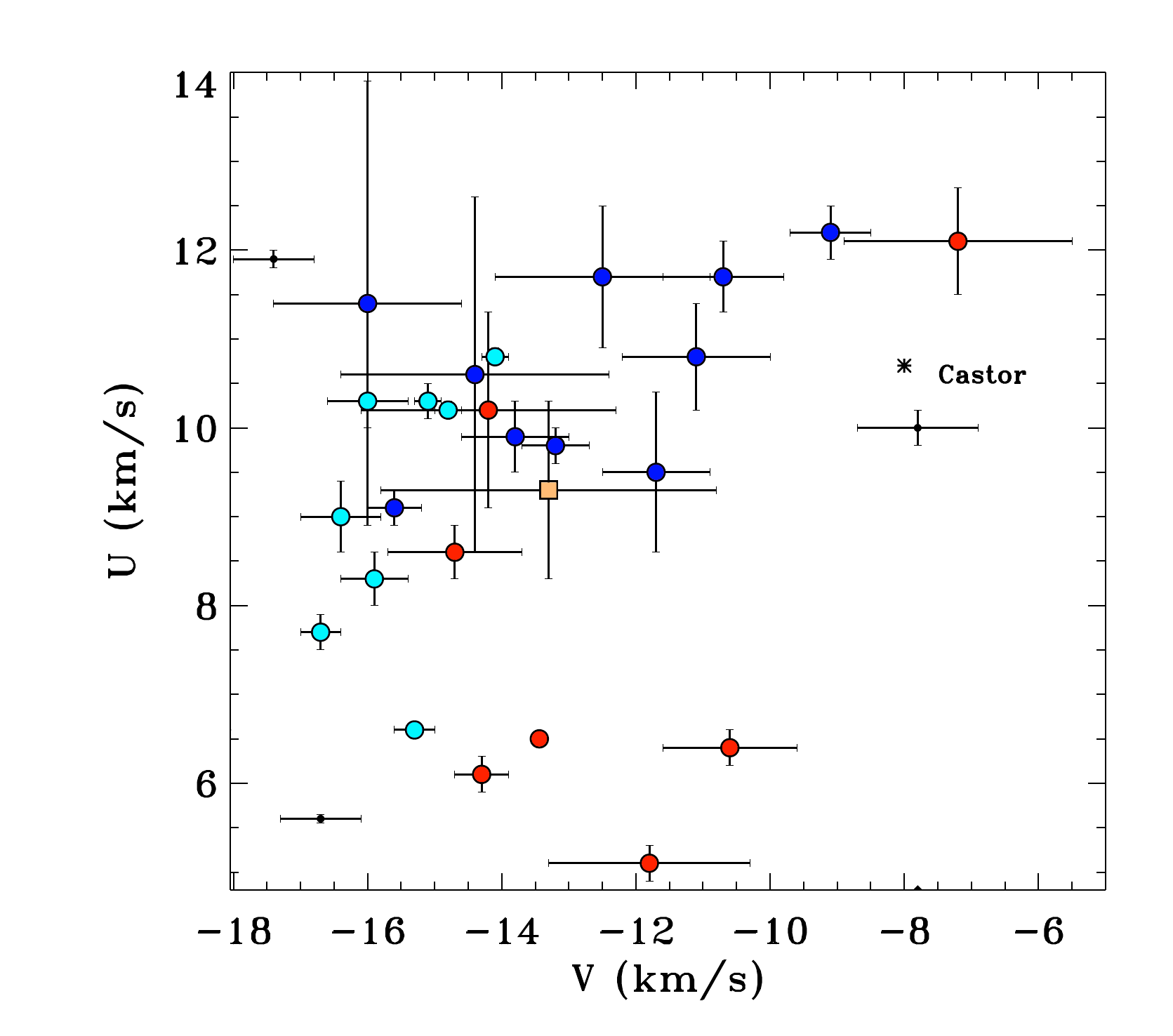}
\hspace{-0.7cm}
\includegraphics[width=6.5cm]{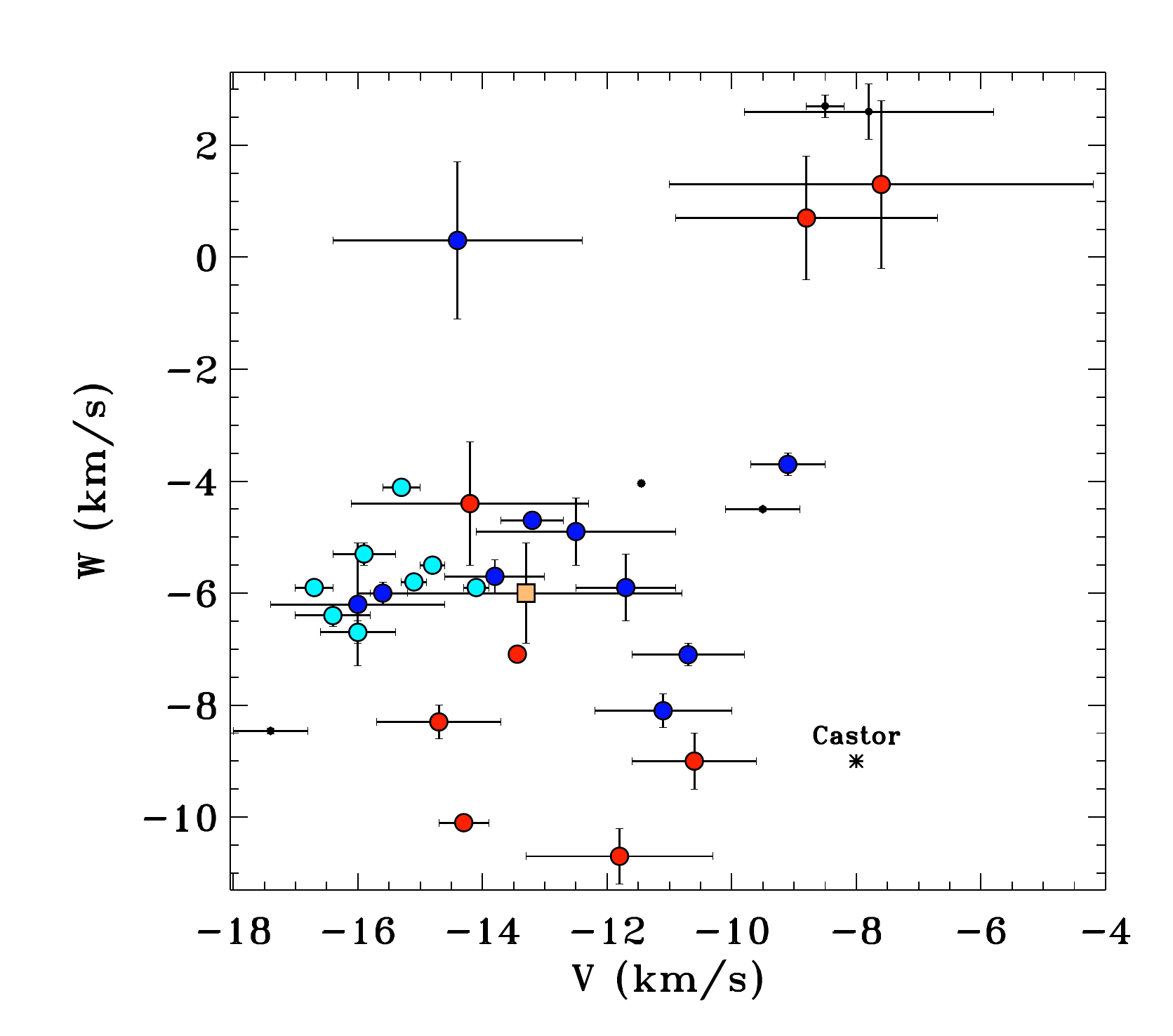}
\hspace{-0.7cm}
\includegraphics[width=6.5cm]{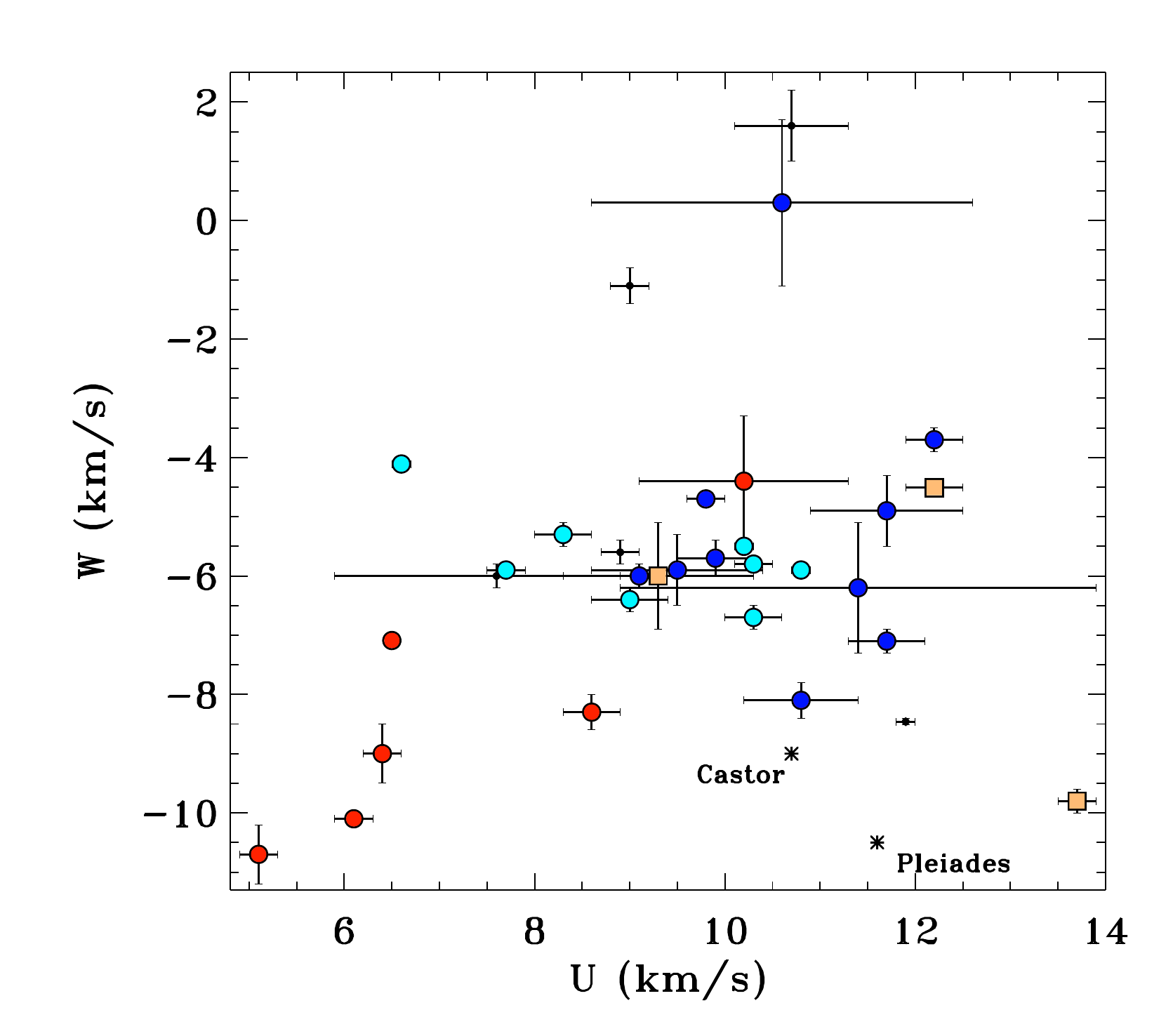}\\
\hspace{-0.7cm}
\includegraphics[width=6.5cm]{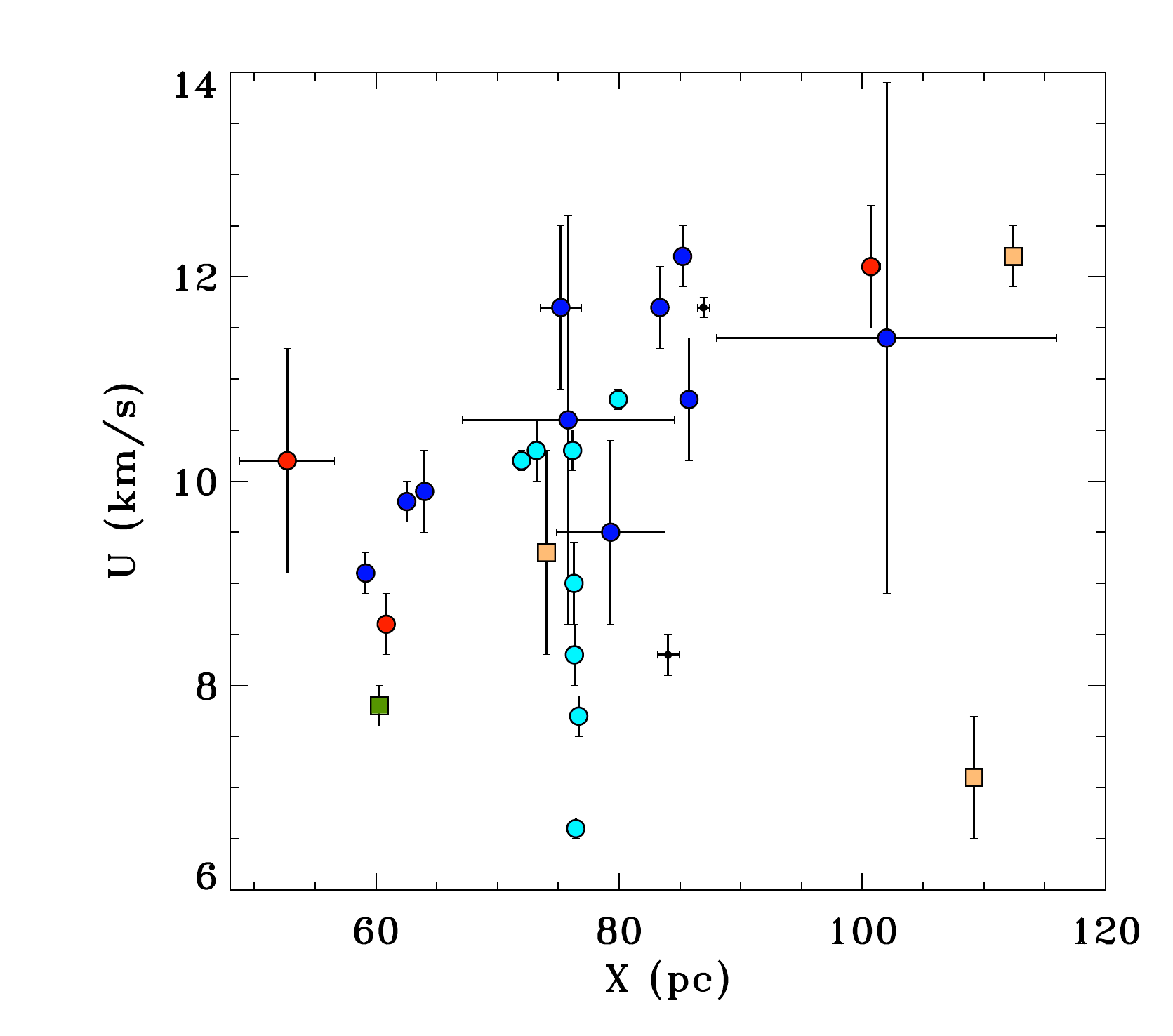}
\hspace{-0.7cm}
\includegraphics[width=6.5cm]{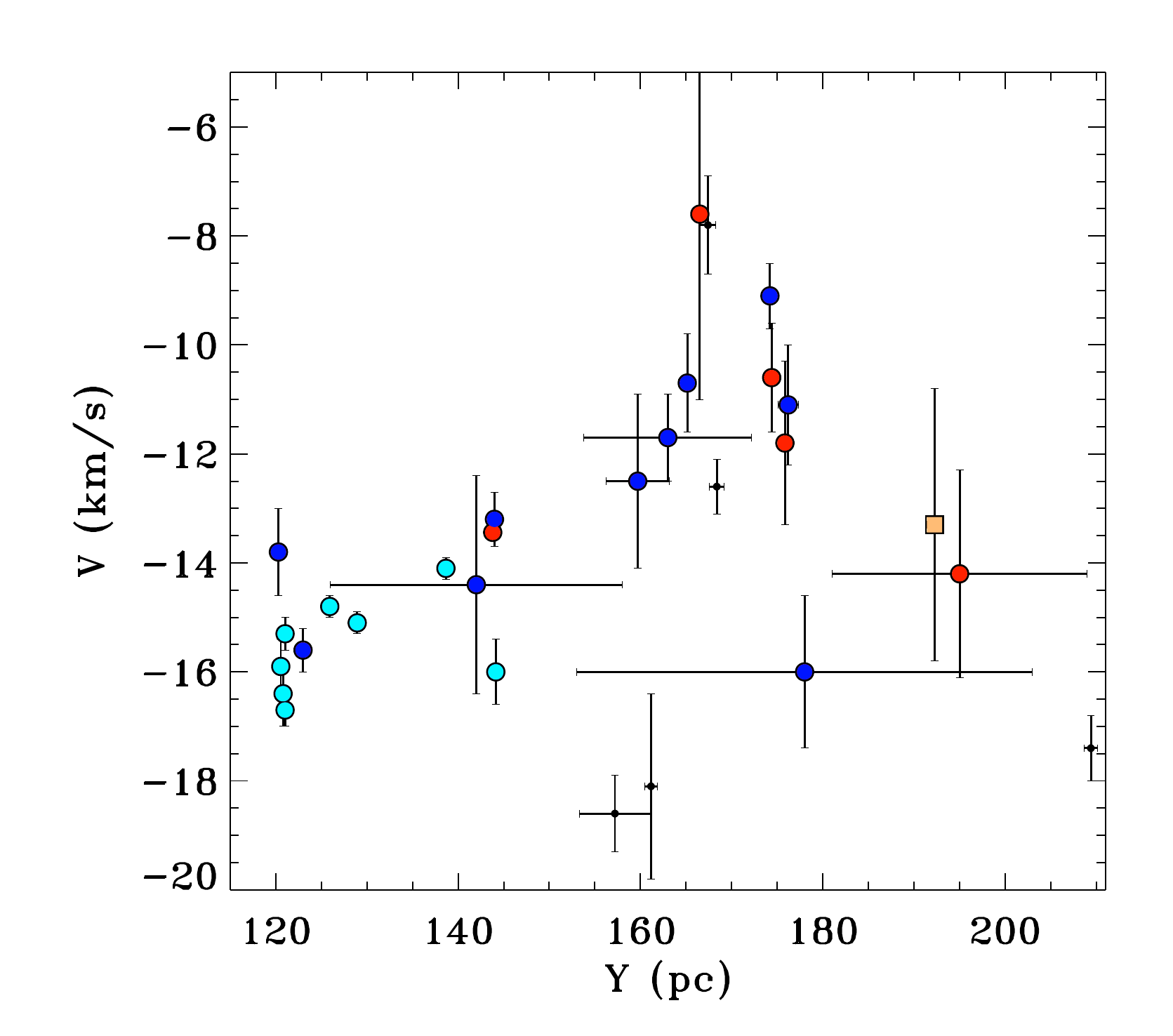}
\hspace{-0.7cm}
\includegraphics[width=6.5cm]{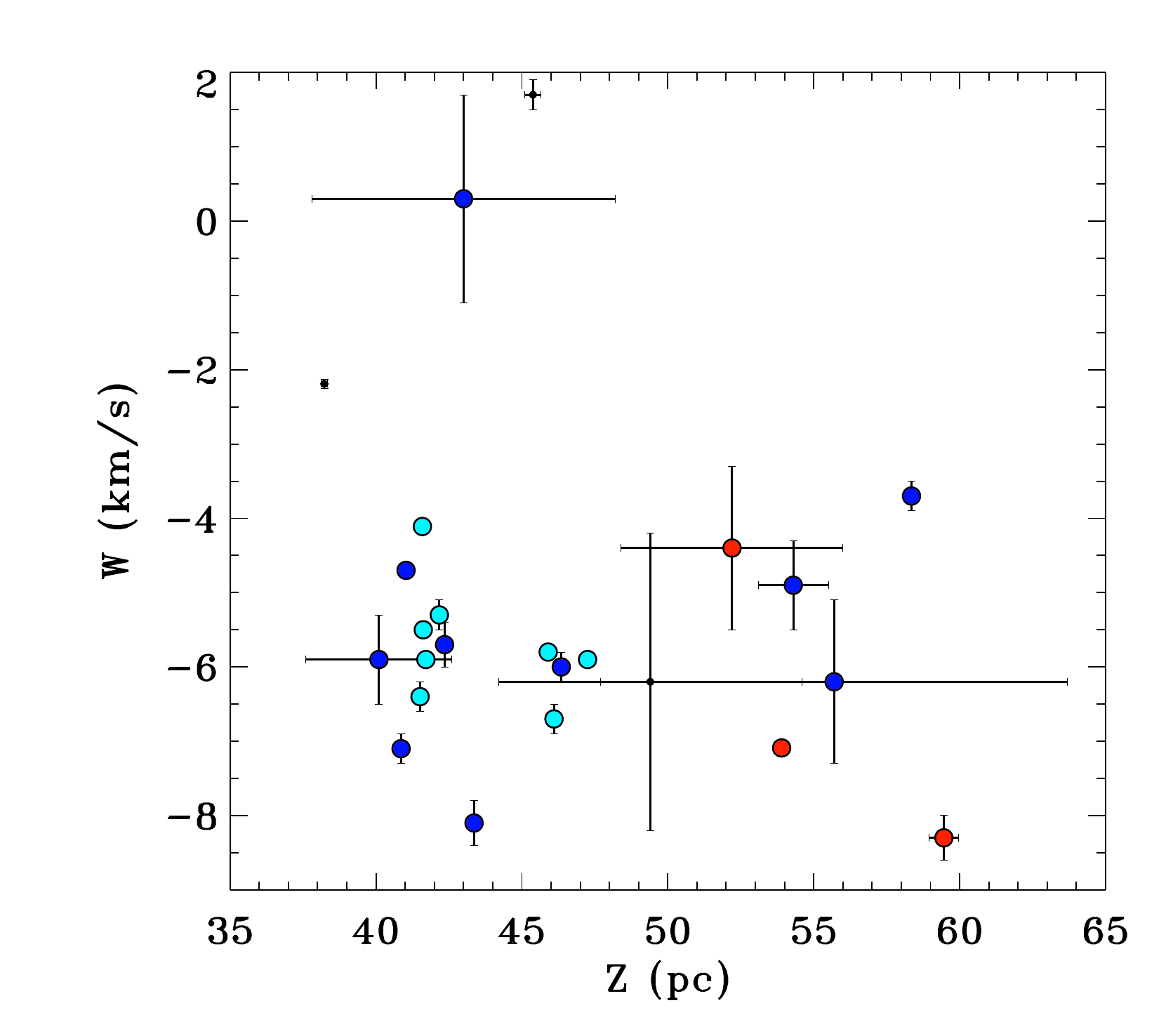}
\caption{Same diagrams as in Fig.\,\ref{Fig:Kinematic_XYZUVW_young} but covering only the region populated by the PMS-like stars towards the CO Cepheus void. In all the panels, we display the stars composing Cep\,III and its subgroup (i.e., Cep\,II) as blue and cyan circles, respectively. The red circles, the orange squares, and black diamonds correspond to the PMS-like that are non-members of Cep\,III, Pleiades-like stars, and older sources, respectively. In the upper panels, the big open hexagon marks the locus of the smaller concentration (i.e., Cep\,I).}
\label{Fig:Kinematic_XYZUVW_cep}
\end{figure*}

\begin{figure*}[!t]
\centering 
\hspace{-0.8cm}
\vspace{-0.3cm}
\includegraphics[width=6.5cm]{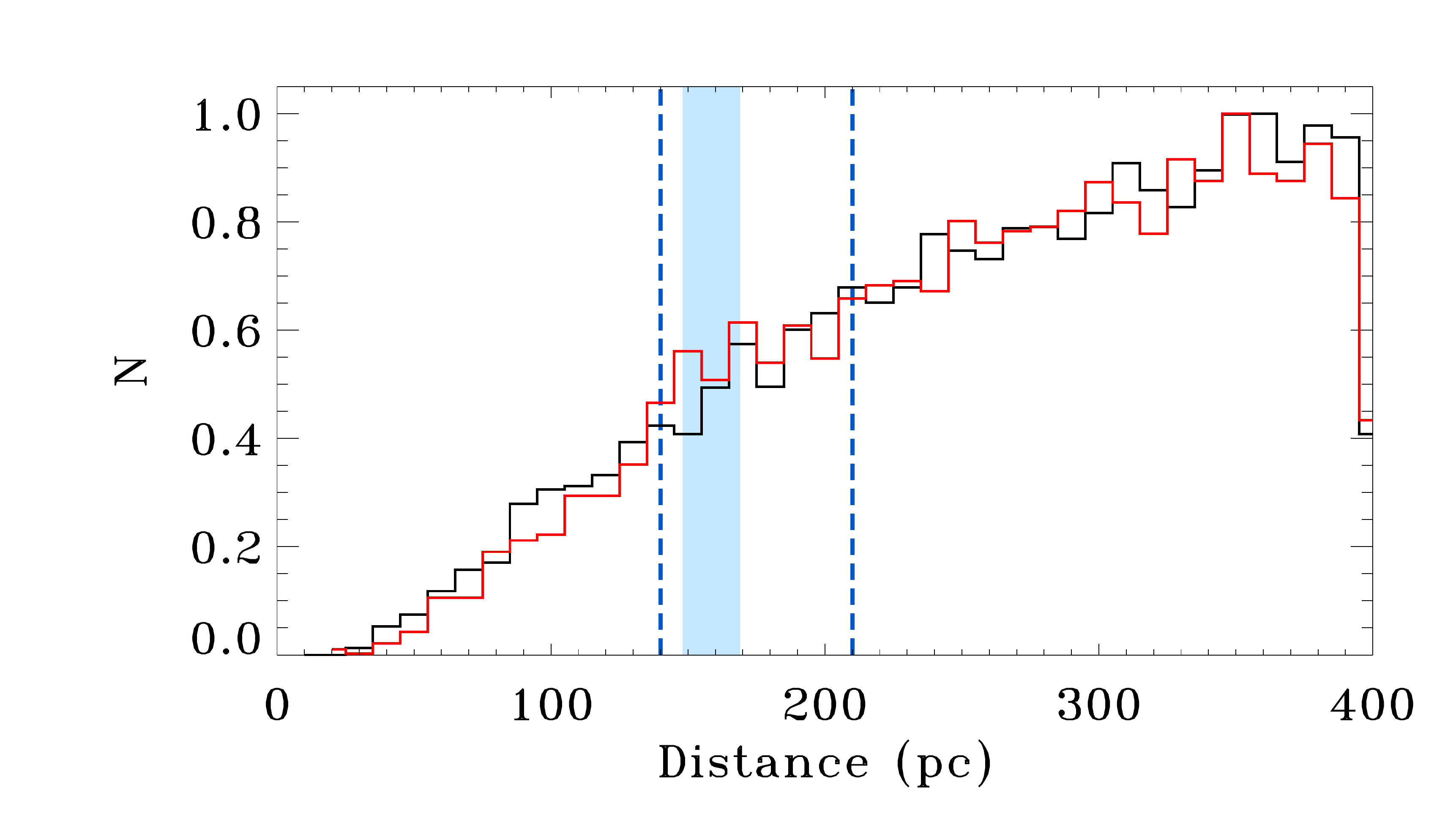}
\hspace{-0.7cm}
\includegraphics[width=6.5cm]{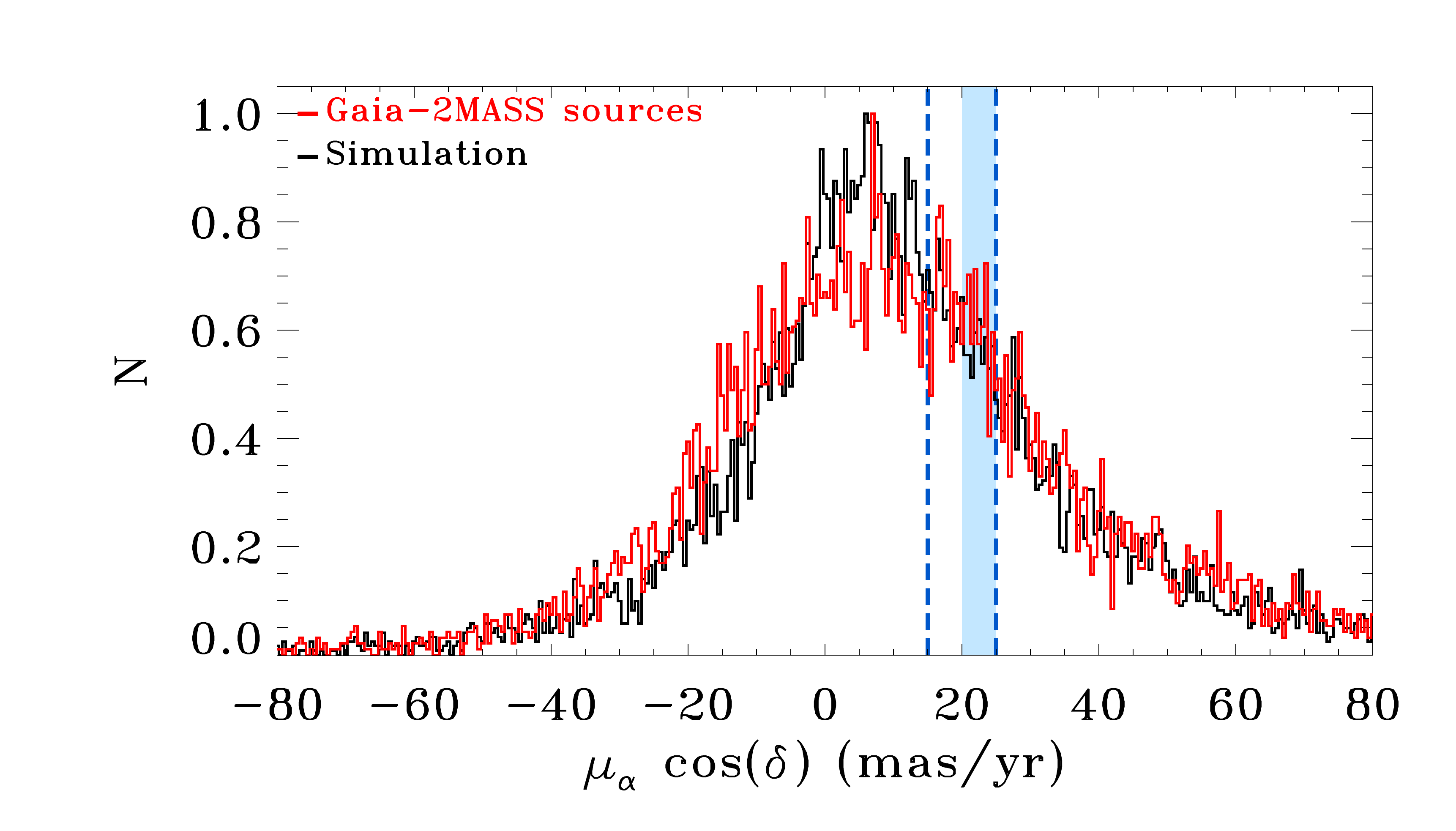}
\hspace{-0.7cm}
\includegraphics[width=6.5cm]{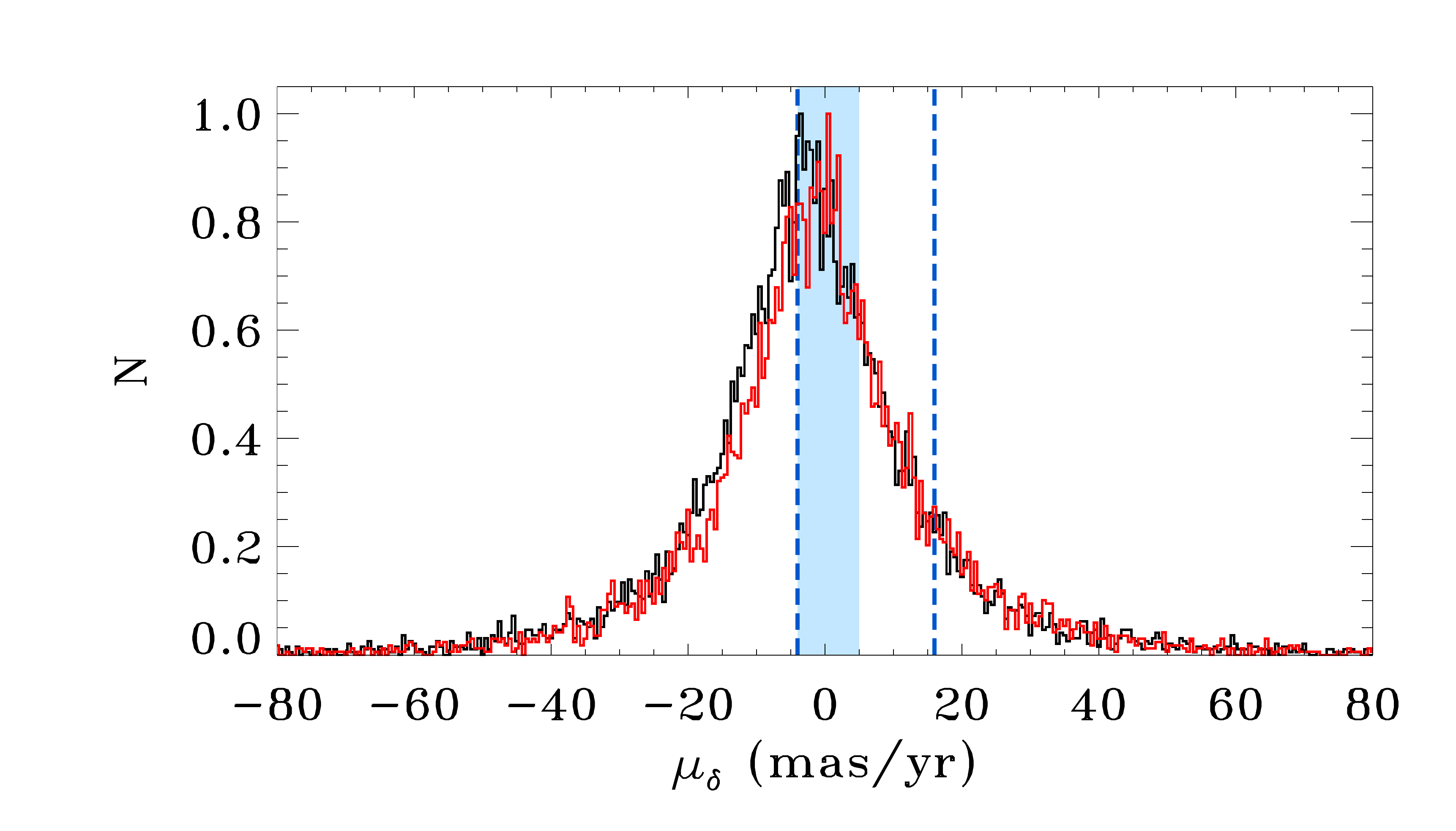}\\
\hspace{-0.8cm}
\includegraphics[width=6.5cm]{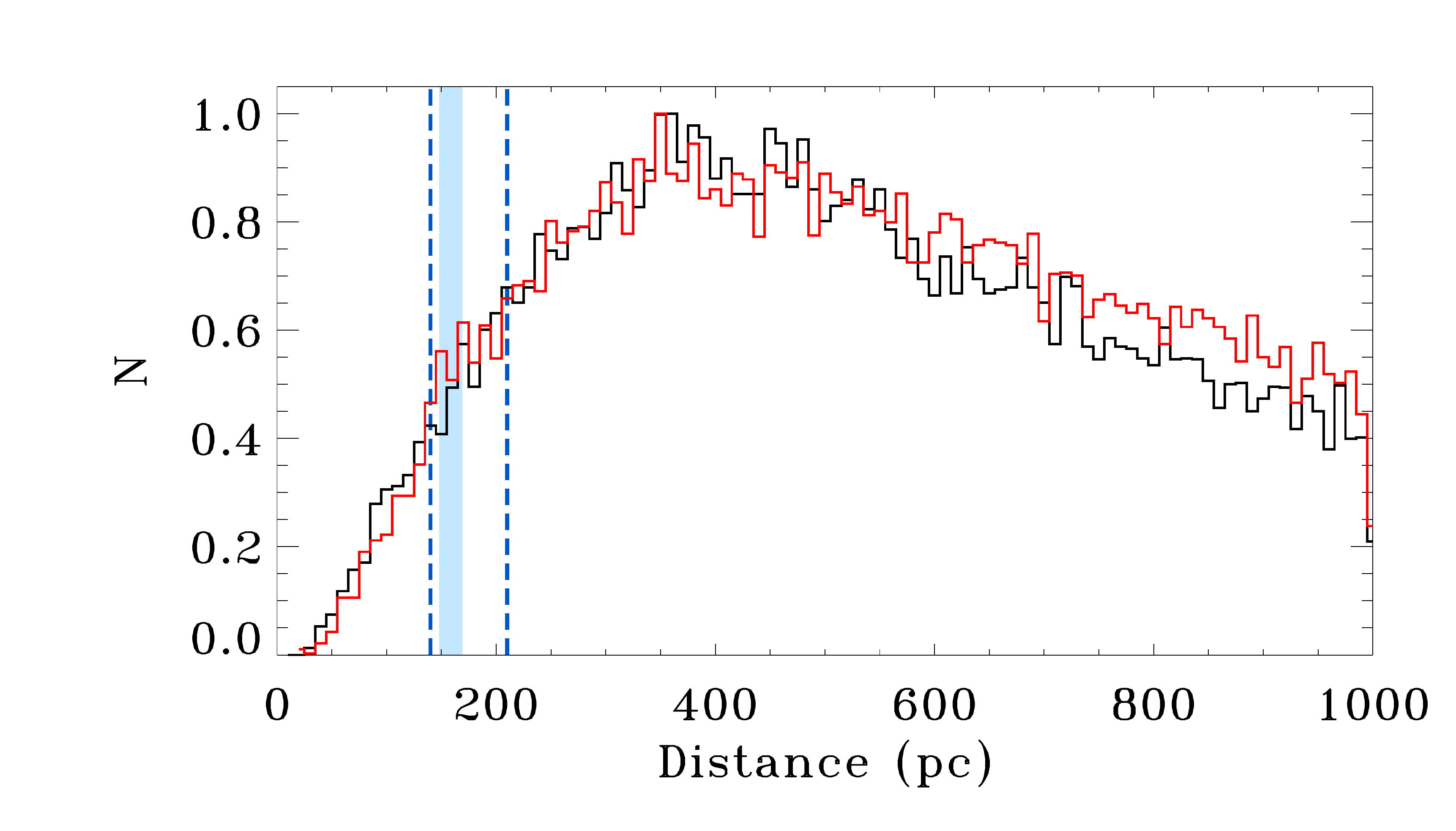}
\hspace{-0.7cm}
\includegraphics[width=6.5cm]{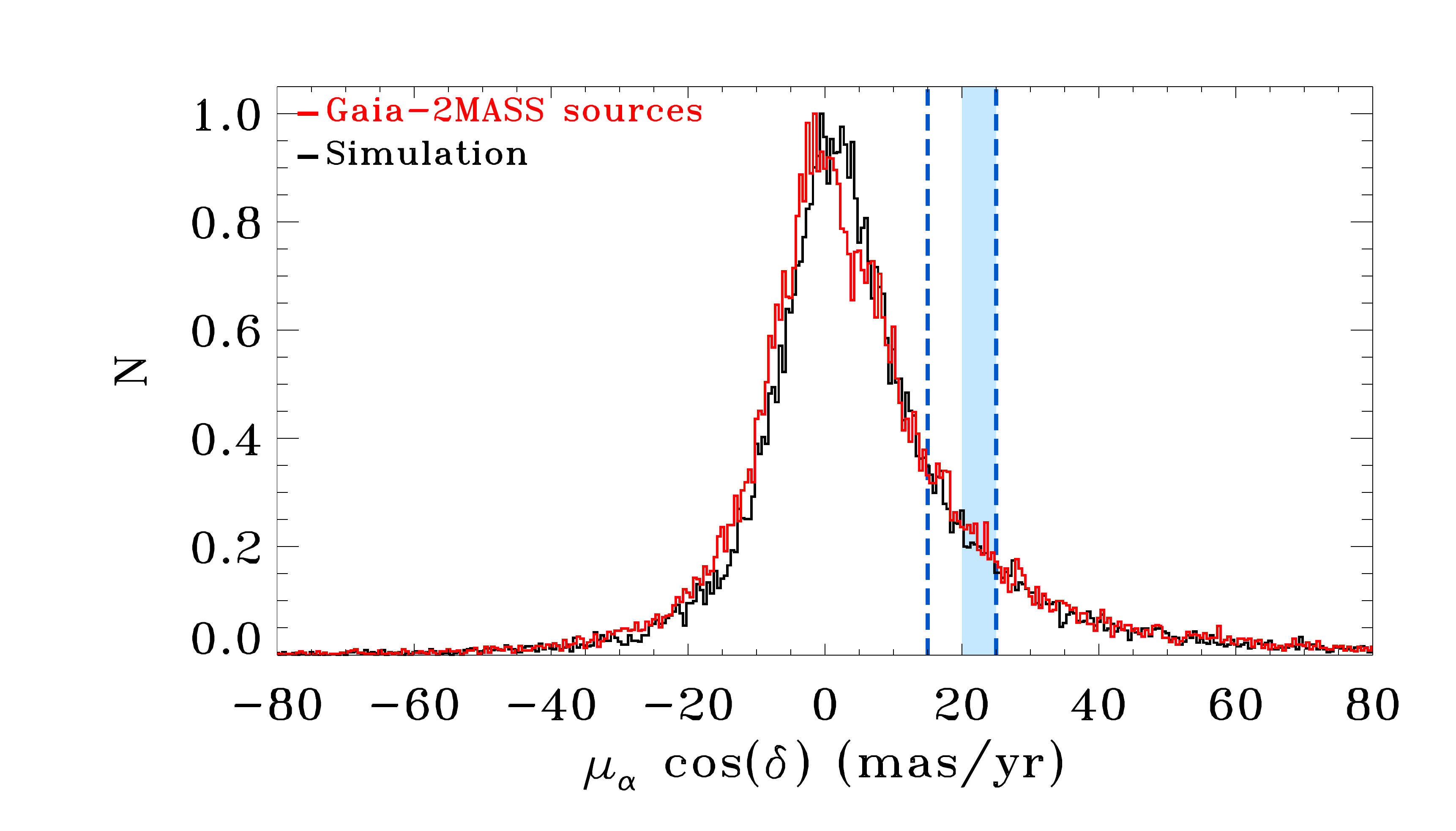}
\hspace{-0.7cm}
\includegraphics[width=6.5cm]{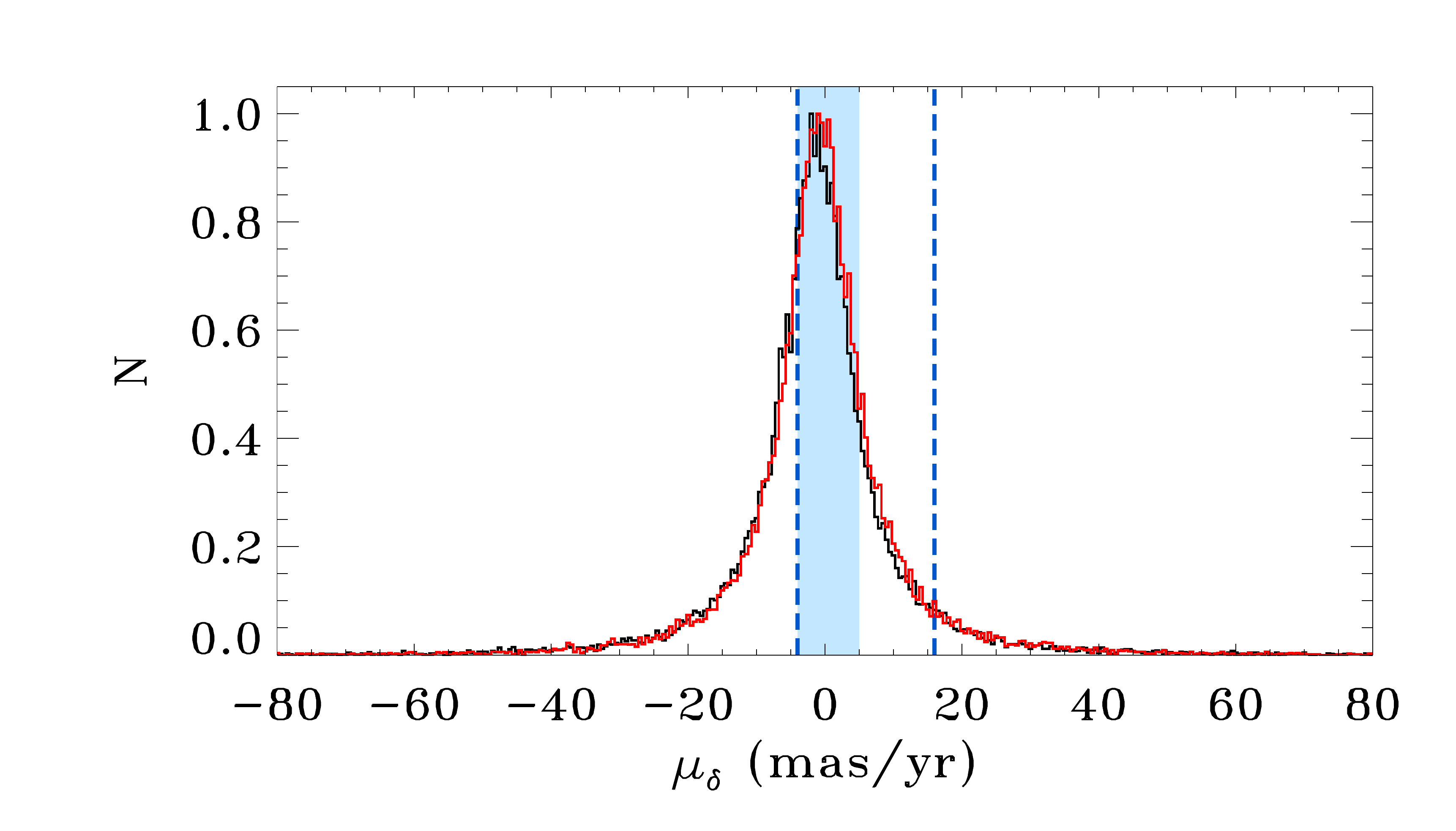}\\
\caption{Comparison of the distribution of distance (\textit{left panels}), and proper motion in right ascension (\textit{middle panels}) and in declination (\textit{right panels}) between the \emph{Gaia}-2MASS sources and the simulation with the Besan\c con model (Sect.\,\ref{ssec:Comp_Simulation}). We normalized the number of stars in each bin, dividing by the maximum number of sources obtained for each histogram. For these plots, we only considered all the sources having a distance lower than $400$\,pc (\textit{upper panels}) and $1000$\,pc (\textit{lower panels}). In each panel, the blue-shaded area denotes the region populated by the likely members of Cep\,II, while the dashed lines correspond to those belonging to Cep\,III.}
\label{Fig:Comp_dist_sources_vs_simul}
\end{figure*}

\subsection{Young stars towards the CO Cepheus void}
\label{ssec:Stat_candidates}

The possible detection of a new young association in Cepheus (age $=10$--$20$\,Myr; Paper~II) is supported by our identification of $20$ PMS stars (plus three small-separation companions observed during our spectroscopic campaigns) in the same region (Fig.~\ref{Fig:Distribution_Cep}) and our discovery of their eight comoving companions (Table~\ref{Tab:VB_Gaia}). These are overplotted on the visible extinction A$_{V,RQ}$ map of \citet{2016A&A...586A.132P},\footnote{\label{planck-map}The map results from the renormalization of the visible extinction one based on the analysis of quasi-stellar objects observed in the Sloan Digital Sky Survey. This A$_{V,RQ}$ map is available from the \emph{Planck} Legacy Archive (http://pla.esac.esa.int/pla/\#home) and corresponds to the file: \\COM\_CompMap\_Dust-DL07-AvMaps\_2048\_R2.00.fits[AV\_RQ].} derived from the thermal dust emission modeling presented by \citet{2007ApJ...657..810D}. The PMS stars are highly clumped near the Galactic coordinates ($X,Y,Z$) $=$ ($76$, $130$,~$43$)~pc (Fig.~\ref{Fig:Kinematic_XYZUVW_young}). To better see their distribution, we zoomed in this smaller region (Fig.~\ref{Fig:Kinematic_XYZUVW_cep}). 

\begin{table}
\caption{Properties of the three groups of PMS stars in Cepheus. We give for each group the number of targets and their multiplicity.}
\smallskip
\begin{center}
{\scriptsize
\vspace{-0.4cm}
\begin{tabular}{ll@{\hspace{0.2cm}}r@{\hspace{0.05cm}}c@{\hspace{0.05cm}}l@{\hspace{0.2cm}}r@{\hspace{0.05cm}}c@{\hspace{0.05cm}}l@{\hspace{0.2cm}}r@{\hspace{0.05cm}}c@{\hspace{0.05cm}}l}
\hline
\hline
\noalign{\smallskip}
Parameter & Unit & \multicolumn{3}{c}{Cep\,I}& \multicolumn{3}{c}{Cep\,II}& \multicolumn{3}{c}{Cep\,III} \\
\hline
\noalign{\smallskip}
$\langle l \rangle$ & [$\degr$] 						& $122.31$ & $\pm$ & $0.04$ 			& $120.7$ & $\pm$ & $2.0$ 	& $118.7$ & $\pm$ & $2.8$\\
$\langle b \rangle$ & [$\degr$] 						& $16.31$ & $\pm$ & $0.14$ 			& $16.37$ & $\pm$ & $0.33$ 	& $15.7$ & $\pm$ & $1.7$\\
$\langle \mu_\alpha\cos\delta\rangle$ & [mas\,yr$^{-1}$] 	& $23.88$ & $\pm$ & $0.29$ 			& $22.63$ & $\pm$ & $0.25$ 	& $20.35$ & $\pm$ & $0.34$\\
$\langle \mu_\delta\rangle$ & [mas\,yr$^{-1}$] 			& $-3.11$ & $\pm$ & $0.25$ 			& $-0.31$ & $\pm$ & $0.51$ 	& $1.98$ & $\pm$ & $0.49$\\
$\langle RV \rangle$ & [km\,s$^{-1}$] 				& $-9.05$ & $\pm$ & $0.24$ 			& $-8.98$ & $\pm$ & $0.3$ & 	$-8.30$ & $\pm$ & $0.43$\\
$\langle\pi\rangle$ & [mas] 						& $6.713$ & $\pm$ & $0.016$ 			& $6.439$ & $\pm$ & $0.054$ 	& $6.044$ & $\pm$ & $0.094$\\
$\langle${\rm Trigo. distance}$\rangle$ & [pc] 			& $148.97$ & $\pm$ & $0.36$ 			& $155.7$ & $\pm$ & $1.3$ 	& $167.7$ & $\pm$ & $2.9$\\
$\langle X\rangle$ & [pc] 							& $76.43$ & $\pm$ & $0.19$ 			& $76.0$ & $\pm$ & $0.5$ 	& $75.9$ & $\pm$ & $1.0$\\
$\langle Y\rangle$ & [pc] 							& $120.86$ & $\pm$ & $0.29$ 			& $128.5$ & $\pm$ & $1.4$	& $141.7$ & $\pm$ & $2.7$\\ 
$\langle Z\rangle$ & [pc] 							& $41.75$ & $\pm$ & $0.13$ 			& $43.9$ & $\pm$ & $0.4$ 	& $45.1$ & $\pm$ & $0.8$\\
$\langle U_{\sun}\rangle$ & [km\,s$^{-1}$] 			& $7.90$ & $\pm$ & $0.29$ 			& $9.15$ & $\pm$ & $0.28$ 	& $9.99$ & $\pm$ & $0.33$\\
$\langle V_{\sun}\rangle$ & [km\,s$^{-1}$] 			& $-16.08$ & $\pm$ & $0.26$ 			& $-15.54$ & $\pm$ & $0.24$ 	& $-14.02$ & $\pm$ & $0.44$\\
$\langle W_{\sun}\rangle$ & [km\,s$^{-1}$] 			& $-5.43$ & $\pm$ & $0.26$ 			& $-5.70$ & $\pm$ & $0.15$ 	& $-5.42$ & $\pm$ & $0.27$\\
Age & [Myr] 									& $15$ & $\pm$ & $5$	 			& $15$ & $\pm$ & $5$	 	& $15$ & $\pm$ & $5$ \\
\noalign{\smallskip}
\hline
\noalign{\smallskip}
\multicolumn{2}{l}{Number of targets [sources] $^{\ddagger}$} 			& \multicolumn{3}{c}{$3$ [$7$]} 		& \multicolumn{3}{c}{$8$ [$14$]} 	& \multicolumn{3}{c}{$18$ [$29$]} \\
\multicolumn{2}{l}{Fraction of systems}			&  \multicolumn{3}{c}{$100$\,\%} 		& \multicolumn{3}{c}{$75$\,\%} 		& \multicolumn{3}{c}{$61$\,\%}\\
\multicolumn{2}{l}{Number of single stars}				& \multicolumn{3}{c}{0} 				& \multicolumn{3}{c}{2} 			& \multicolumn{3}{c}{7}\\
\multicolumn{2}{l}{Number of SB1 systems}			& \multicolumn{3}{c}{0} 				& \multicolumn{3}{c}{1} 			& \multicolumn{3}{c}{2}\\
\multicolumn{2}{l}{Number of visual binaries}			& \multicolumn{3}{c}{1} 				& \multicolumn{3}{c}{3} 			& \multicolumn{3}{c}{7}\\
\multicolumn{2}{l}{Number of multiple systems}		& \multicolumn{3}{c}{2} 				& \multicolumn{3}{c}{2} 			& \multicolumn{3}{c}{2}\\
\noalign{\smallskip}
\hline 
\end{tabular}
\label{Tab:Prop_Cepheus}
\begin{minipage}{8cm} 
\textbf{Note.} $^{\ddagger}$ We give for each group the number of targets listed in Table~\ref{Tab:List_Candidates}, while that of resolved sources in the \emph{Gaia} DR2 catalog (Tables~\ref{Tab:EWLi_Kinematics_CepSurv} and~\ref{Tab:VB_Gaia}) is bracketed. \\ 
\end{minipage}
}
\end{center}
\end{table}

\medskip
Almost all PMS stars towards the CO~Cepheus void are related to the Cepheus association (Fig.~\ref{Fig:Distribution_Cep}), except for sources \#160 and \#189 that are located at larger distances ($354\pm9$ and $320\pm97$\,pc, respectively). The former is a known member of the molecular cloud L1251 (see Sect.\,\ref{subsection:SED}), while the astrometry of the latter suffers from large uncertainties. According to the 6D phase-space data of the remaining PMS stars in this restricted area, we can divide them into three subgroups, as explained below. We summarize their properties in Table~\ref{Tab:Prop_Cepheus}.

We first identified the highest density of PMS stars in the projection of the sky positions (Fig.~\ref{Fig:Distribution_Cep}) and in the various planes in Galactic coordinates (Fig.~\ref{Fig:Kinematic_XYZUVW_cep}). Up to now, this group (henceforth Cep\,I) is composed of one visual binary (G3) and two triple hierarchical systems (\#6 and G4). It is highlighted by a hexagon in Figs.~\ref{Fig:Distribution_Cep} and~\ref{Fig:Kinematic_XYZUVW_cep}.  Moreover, source \#6 has an additional wide-separation companion (Appendix~\ref{appendix:binary-visual}).

Five additional young sources with similar properties are located in a more extended area (henceforth Cep\,II; cyan circles in Figs.\,\ref{Fig:Distribution_Cep} and \ref{Fig:Kinematic_XYZUVW_cep}). In addition to the three Cep\,I members, this group includes two single stars (\#158 and \#193), one SB1 system (\#153), and two visual binaries (\#188 and G1). 

Finally, the largest group (henceforth Cep\,III; blue circles in Figs.\,\ref{Fig:Distribution_Cep} and \ref{Fig:Kinematic_XYZUVW_cep}) is composed of five single stars (\#142, \#182, \#186, F2, F4), one SB1 system (\#156), four visual binaries (\#152, \#191, G2 and F3), and the Cep\,II members. The source~F3 has also a wide-separation companion (Table~\ref{Tab:VB_Gaia}}). 

\begin{figure*}[!t]
\centering 
\hspace{-0.5cm}
\includegraphics[width=9.2cm]{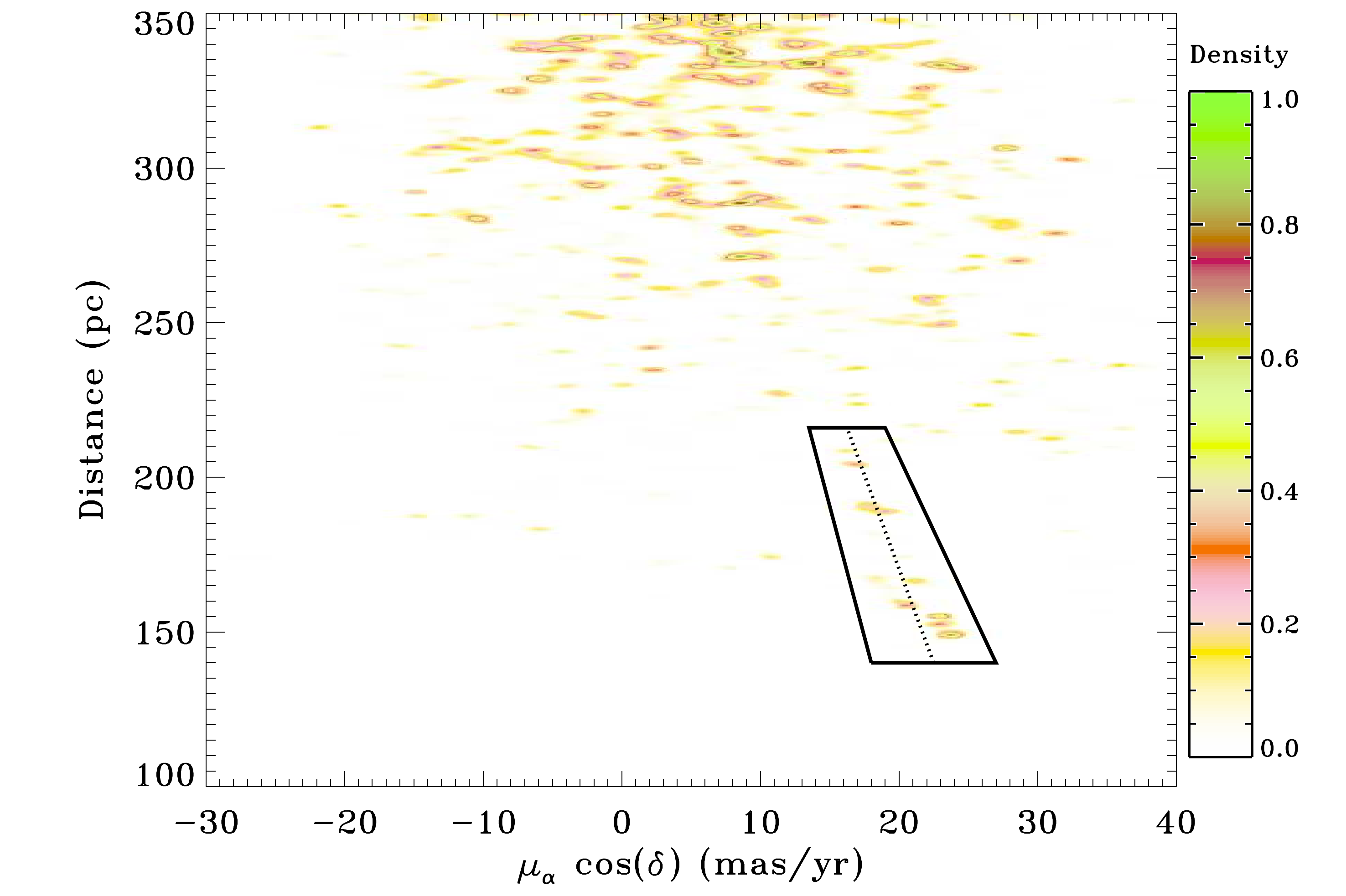}
\hspace{-0.5cm}
\includegraphics[width=9.2cm]{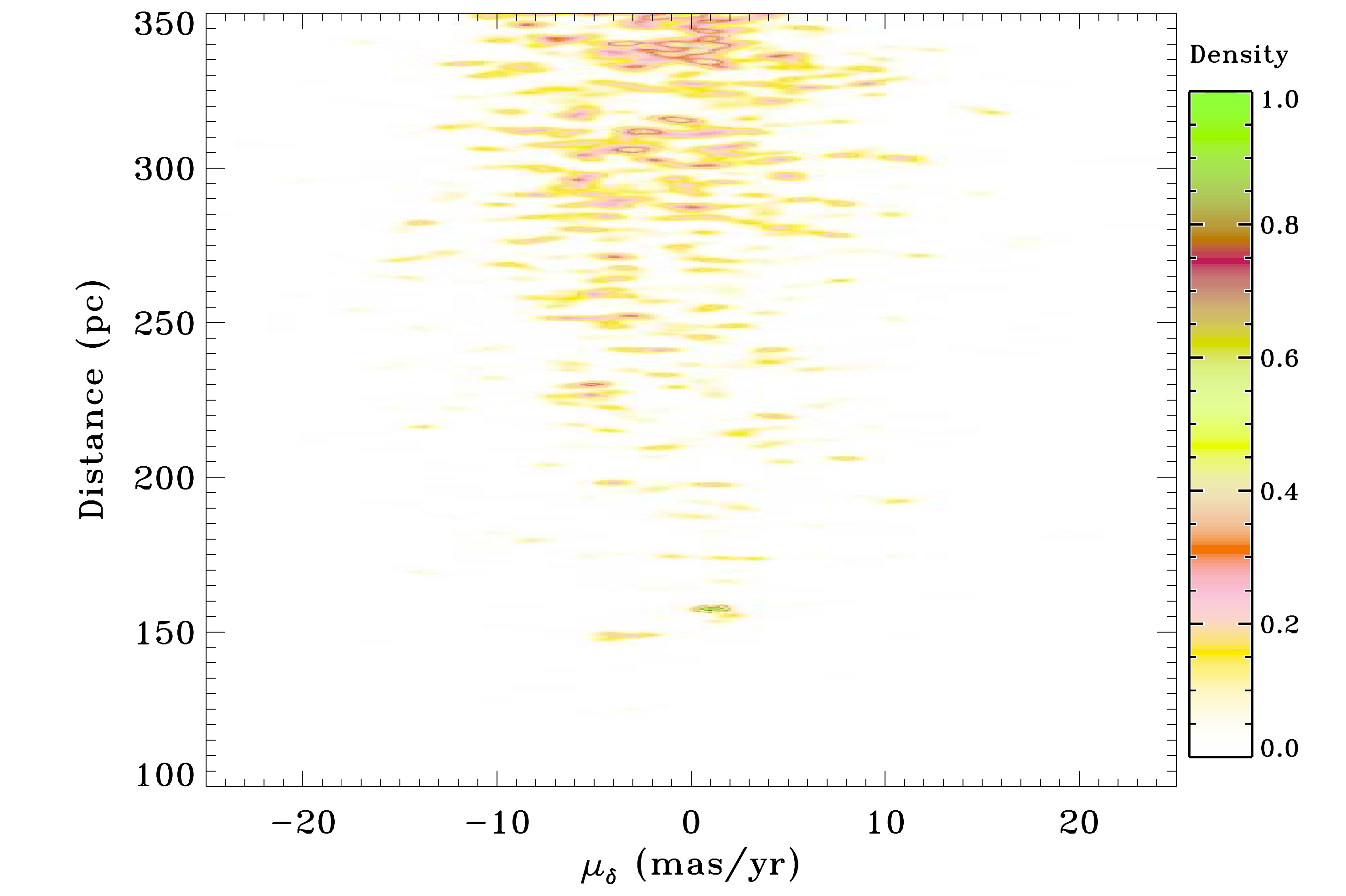}\\
\hspace{-0.5cm}
\includegraphics[width=9.2cm]{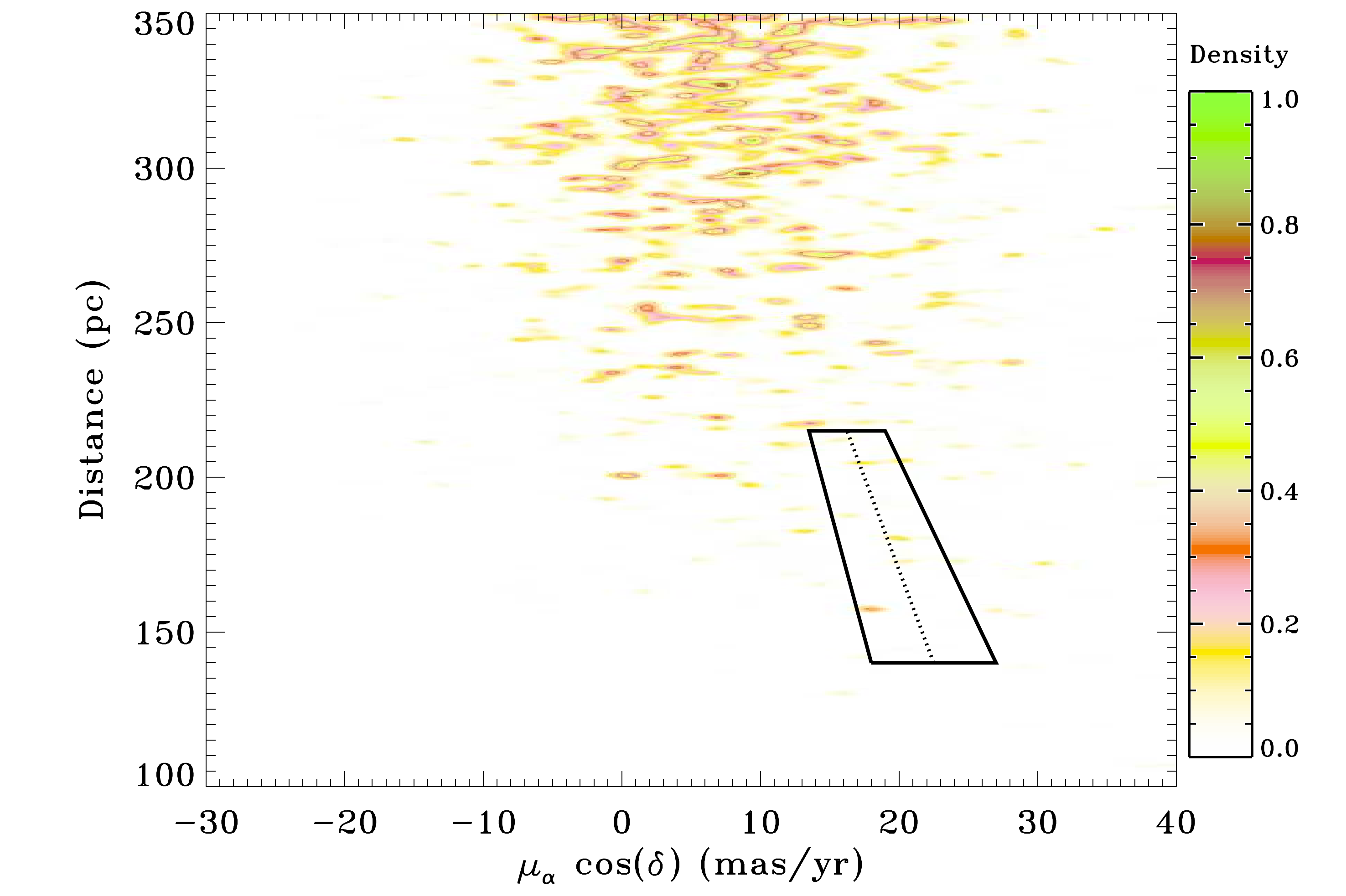}
\hspace{-0.5cm}
\includegraphics[width=9.2cm]{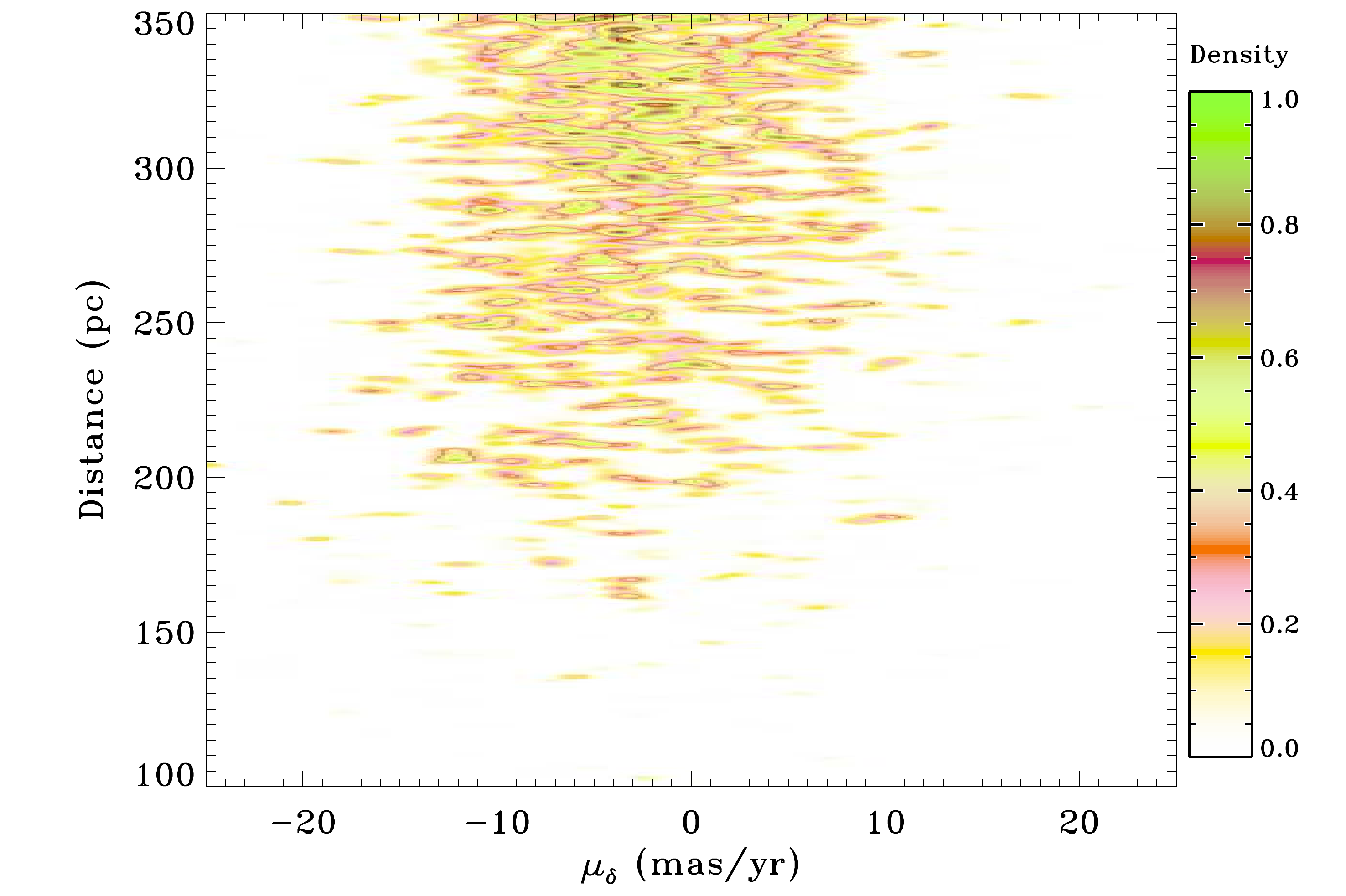}\\
\caption{Comparison of the density map obtained from the distribution of distance as a function of the proper motion in right ascension (\textit{left panels}) and in declination (\textit{right panels}), between the \emph{Gaia}-2MASS sources (\textit{upper panels}) and the simulation with the Besan\c con model (\textit{lower panels}). In the left panels, the polygonal shape denotes the region populated by the PMS-like stars belonging to Cep\,III.}
\label{Fig:Comp_dist_ppm_sources_vs_simul}
\end{figure*}

\subsection{Additional CO Cepheus void stars from \emph{Gaia}}
\label{ssec:Comp_Simulation}

To see the extent of the association in Cepheus, we built a new sample that is hereinafter referred to as \emph{Gaia}--2MASS. We selected all sources brighter than $K_{\rm s} = 12$ mag from the PPMXL catalog \citep{PPMXL2010}, using a cone search centered at (RA,~DEC) $=$ ($0$h, $+77\degr$) with a radius of $6\degr$, which covers all the sky area devoid of interstellar matter. Subsequently we matched this dataset with \emph{Gaia} DR2 (radius~$=1\arcsec$). 

\medskip
An asymmetry is clearly visible in the distribution of the proper motions in right ascension (middle panels of Fig.~\ref{Fig:Comp_dist_sources_vs_simul}), in particular in the range of proper motions of the members of the Cepheus association (delimited by vertical lines). This results in an excess of sources with positive $\mu_{\alpha} \cos\delta$ values with respect to a normal symmetric distribution. The total number of the excess \emph{Gaia}--2MASS sources is significantly larger than the number of young stars identified up to now in the CO Cepheus~void. To search for a possible link between this excess of \emph{Gaia}--2MASS stars and the Cepheus association, we performed a comparison with a simulation based on the Besan\c con model \citep{2012A&A...543A.100R,2014A&A...564A.102C,2014A&A...566A.119L}. 
 
Due to the right ascension of the members of the Cepheus association ranging from $22$h to $1$h, we have to use a rectangular area covering Galactic longitude $l$ ranging from $114\degr$ to $126\degr$ and Galactic latitude $b$ from $8\fdg5$ to $20\fdg5$, together with the same cut in K$_{s}$ magnitude that we had applied to build the \emph{Gaia}--2MASS sample. This region slightly differs from the area covered by the selected \emph{Gaia}--2MASS sources but both are consistent in terms of number of sources. The distributions of distance and proper motions are well reproduced by the simulation for both cases of the maximum distance threshold considered here ($400$ and $1000$~pc; Fig.~\ref{Fig:Comp_dist_sources_vs_simul}). Therefore, the asymmetry seen in the distribution of the $\mu_{\alpha} \cos\delta$ values is likely due to large-scale dynamical processes in the Galaxy rather than the result of the unknown supernova shock expected to be at the origin of the CO Cepheus void \citep{Grenier1989}. Thus, this excess alone cannot be used as a criterion to identify possible new members in this young association and a deeper analysis is required. 

To this end, we compared the density map obtained from the \emph{Gaia}-2MASS sources and the Besan\c con model (Fig.~\ref{Fig:Comp_dist_ppm_sources_vs_simul}). While the ($\mu_{\delta}$, distance) diagram looks the same for the \emph{Gaia}--2MASS distribution and the Besan\c con model, the ($\mu_{\alpha} \cos\delta$, distance) diagram reveals a clear over-density in the \emph{Gaia}--2MASS sample (delimited by the polygonal shape). Coupled to the location of members of the Cepheus association on this density map  (Fig.~\ref{Fig:Comp_dist_ppm_sources_zoom}), it is indisputable that the excess \emph{Gaia}--2MASS sources are attributable to the  young stars in this region. These are the best member candidates to the Cepheus association to investigate with future observations.

\subsection{Cepheus association and runaway stars}
\label{ssec:Origin_Cepheus}

\citet{Klutsch08} discovered an excess of stellar X-ray sources in Cepheus where \citet{Tachihara05}, and Papers~II and III had identified together $15$ young stars belonging to the association. The current study has added $14$ new members (i.e., six PMS stars and eight comoving companions of these $21$ young stars).  

\medskip
All but two PMS stars towards the CO Cepheus void have the same kinematic properties (Table~\ref{Tab:Prop_Cepheus}) and therefore a common origin. We also found that eight of the young stars in \citet{Tachihara05} belong to this association, confirming their assumption. The discovery of such a large number of young stars in this region is more easily explained by the in-situ model than by the runaway hypothesis. As mentioned by \citet{Tachihara05} and Paper~II, these sources are close to a faint, low density cloud that could be the remnant of the parent cloud already dissipated. 

\begin{figure}[!t]
\centering 
\includegraphics[width=8.5cm]{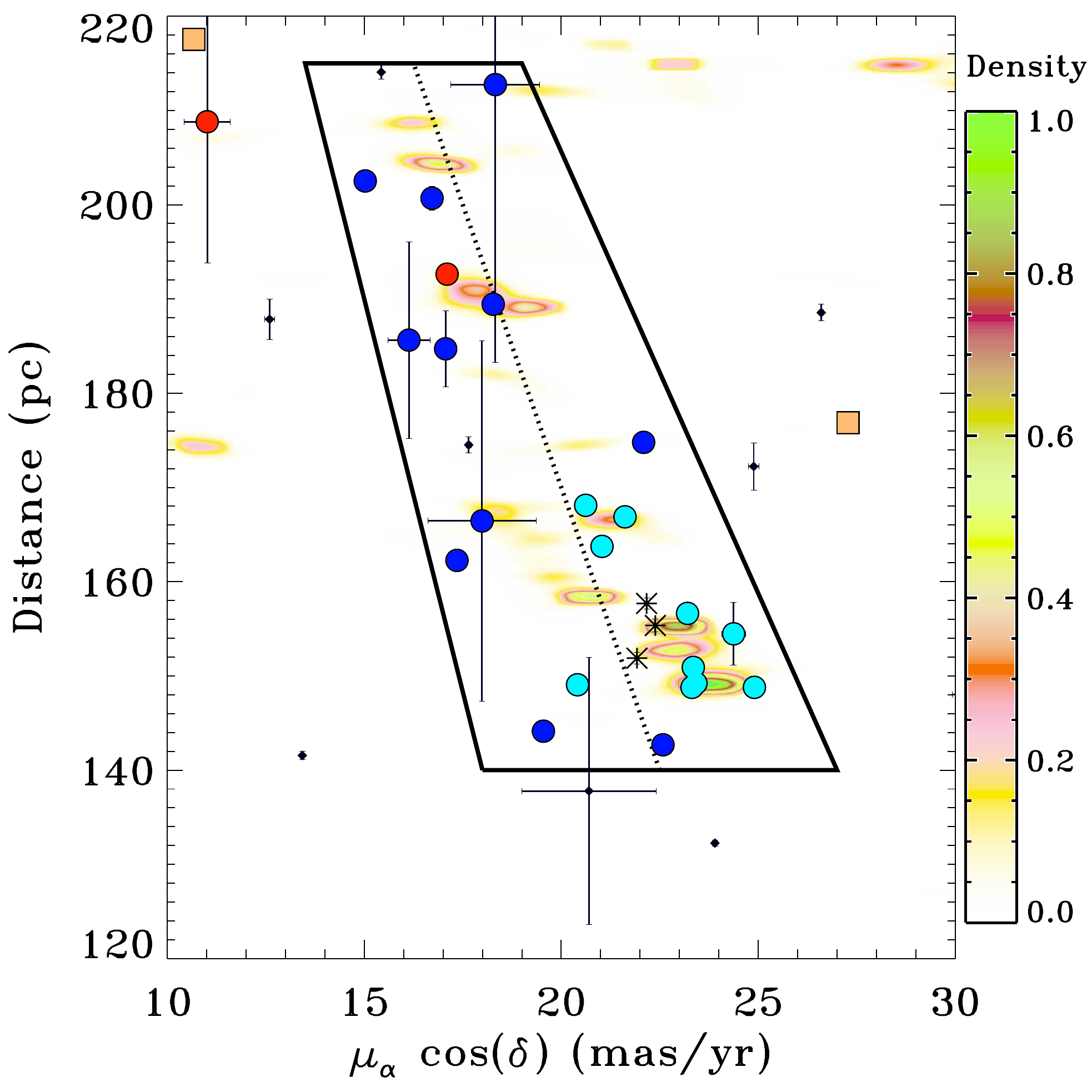}\\
\caption{($\mu_{\alpha} \cos\delta$, distance) diagram of our young star candidates located in the region populated by the members of the Cepheus association, overplotted on the \emph{Gaia}-2MASS density map (Fig.~\ref{Fig:Comp_dist_ppm_sources_vs_simul}). We use the same symbols as in Fig.~\ref{Fig:Kinematic_XYZUVW_cep}. The asterisks mark the locus of additional members reported by \citet{2017AJ....153..257O} and \citet{2018ApJ...863...91F}: \object{HIP~115764}, \object{HIP~117376}, and \object{TYC~4500-616-1} (from bottom to top).}
\label{Fig:Comp_dist_ppm_sources_zoom}
\end{figure}

\medskip
We see a decline in multiple systems and a shift from the mean Galactic position $Y$ to positive values in the more scattered group (Cep\,III) with respect to the densest group (Cep\,I). This agrees with the dispersion of the cluster members driven by the rotation of the Galaxy. This also explains the gradient found in the ($\mu_{\alpha} \cos\delta$, distance) diagram (Fig.~\ref{Fig:Comp_dist_ppm_sources_zoom}) because the members of the most extended stellar group tend to move further away from the others. Based on Table~\ref{Tab:Prop_Cepheus}, the properties of the three stellar groups could be interpreted as follows: \textit{i)} the strongest concentration of young stars (Cep\,I) could be the remnant of the past Cepheus cluster core, \textit{ii)} the sources belonging to the group Cep\,II could correspond to the members that are still gravitationally bound, and \textit{iii)} the most extended group (Cep\,III) denotes the original cluster. The latter corresponds to the Cepheus association reported in Paper~II. 

\medskip
The group~38 of \citet{2017AJ....153..257O} and \citet{2018ApJ...863...91F} actually coincides with the Cepheus association analyzed here and lists three additional members (\object{HIP~115764}, \object{HIP~117376}, and \object{TYC~4500-616-1}). Their astrometries and the $RV$ value of \object{TYC~4500-616-1} in \emph{Gaia} DR2 perfectly agree with those of the Cep\,II members (Fig.~\ref{Fig:Comp_dist_ppm_sources_zoom}), while the $RV$ value of \object{HIP~117376}  suggests a possible variation in radial velocity. The source \object{HIP~115764} would be the warmest member of the association with an effective temperature of $8020_{-140}^{+420}$\,K in \emph{Gaia} DR2. We also note that the angular separation between \object{HIP~117376} and our target \#158 is about $8\farcm30\pm0.044$~mas. This brings the number of current bona fide members to~$32$.

\medskip
We discovered six PMS stars outside the CO~Cepheus void (\#93, \#106\,c1, \#115, \#124, \#150, and F1) with space-velocity components similar to the members of the Cepheus association. These correspond to the red circles in the ($U$,~$W$) plane of Fig.~\ref{Fig:Kinematic_XYZUVW_cep}. Their distances range from $160$ to $240$~pc, except for \#150 (distance~$= 375\pm3$~pc). The five remaining sources (and their two comoving companions; Table~\ref{Tab:VB_Gaia}) could be runaway stars. This list of sources includes one of the PMS stars from Paper~III and one WTTS from \citet{Tachihara05}. 

\section{Conclusions}
\label{Sec:Conc}

We present the results of a spectroscopic survey of optical counterparts of X-ray sources in the Cepheus region (near the North Celestial pole) aimed at discovering further young objects around the four comoving stars reported by us in Paper~II. Based on multivariate analysis methods, we selected optical and infrared counterparts of ROSAT All-Sky Survey and XMM-Newton X-ray sources, which are young star candidates. The analysis of the spectra of these candidates allowed us to determine their atmospheric parameters, radial and rotational velocities, and atmospheric lithium content. These data, along with the parallaxes and proper motions from the \emph{Gaia} DR2 catalog, were used to characterize our sample and identify new young stars in this region. The main stellar populations composing this sample are young or active stars and multiple systems. We identified two distinct populations of young stars that are spatially and kinematically separated. The $18$ Pleiades-like objects with an age between $100$ and $300$~Myr are mostly projected towards the Galactic plane, while $23$ of the $37$ sources younger than $30$~Myr are located in the sky area of $8$ degree-diameter filling the CO~Cepheus void. Among these, $21$ PMS stars (including five spectroscopic binaries) and their eight comoving companions belong to the Cepheus association, which is the first nearby (distance $=157\pm 10$~pc) young (age $= 10$--$20$~Myr) stellar association found northward of $\delta = +30\degr$. 

\medskip
We provide the first comprehensive view of the Cepheus association. All studies carried out so far in this sky region have found a total of $32$~bona fide members and nine member candidates of this young association, to which $14$ ($44$\,\%) members and six ($67$\,\%) candidates are new discoveries. The kinematics of its members reveals a substantial mixture of the original cluster within the local population of the Galactic plane. The runaway hypothesis is highly improbable for explaining the formation of this homogeneous stellar group because of their kinematic properties and the identification of several new members. This raises the question of the in-situ star-formation scenario in low-mass cloud environments (as in many other SFRs). 

\medskip
Coupled to the Cepheus association properties derived from the current analysis, our \emph{Gaia}-2MASS sample should contribute to the identification of new candidates to be followed up in the future.

\begin{acknowledgements}
We thank the anonymous referee for useful suggestions. 
This work is supported by the Universidad Complutense de Madrid and Spanish \textit{Ministerio de Ciencia e Innovaci\'on y Universidades} (MICINN) under grant AYA2016-79425-C3-1-P. 
A.K. and D.M. were also supported by AstroMadrid (CAM S2009/ESP-1496), and MICINN under grants AYA2008-00695 and AYA2008-06423-C03-03. 
Part of this study is also supported by the Italian \textit{Ministero dell'Istruzione, Universit\`a e Ricerca} (MIUR), the French \textit{Centre National d'\'Etudes spatiales} (CNES), and the \textit{R\'egion Alsace}. 
This research made use of the SIMBAD database, the VIZIER catalog access and the X-Match service, which are operated at the \textit{Centre de Donn\'ees astronomiques de Strasbourg} (CDS). 
This publication makes use of the data products from ROSAT and XMM-Newton X-ray observatories, the Two Micron All Sky Survey, and the European Space Agency (ESA) missions \emph{Gaia} and \emph{Planck}. 
The ROSAT All-Sky Survey catalogs were produced by the ROSAT Scientific Data Center at the Max-Planck-Institut f\"ur Extraterrestrische Physik (MPE), Garching (Germany).
The 2XMMi-DR3 catalog is the fifth publicly released XMM-Newton X-ray source catalog produced by the XMM-Newton Survey Science Centre (SSC) consortium on behalf of ESA. The 2XMMi-DR3 is one of two incremental versions of the 2XMM catalog.
The Two Micron All Sky Survey, which is a joint project of the University of Massachusetts and the Infrared Processing and Analysis Center/California Institute of Technology, was funded by the National Aeronautics and Space Administration and the National Science Foundation. 
The data from the ESA mission \emph{Gaia} (\url{https://www.cosmos.esa.int/gaia}) was processed by the \emph{Gaia} Data Processing and Analysis Consortium (DPAC, \url{https://www.cosmos.esa.int/web/gaia/dpac/consortium}). Funding for the DPAC has been provided by national institutions, in particular the institutions participating in the \emph{Gaia} Multilateral Agreement.
\emph{Planck} (http://www.esa.int/Planck) is an ESA science mission with instruments and contributions directly funded by ESA Member States, NASA, and Canada. 
This publication used the POLLUX database (http://pollux.graal.univ-montp2.fr) operated at LUPM (\textit{Laboratoire Univers et Particules de Montpellier}, Universit\'e Montpellier~II -- CNRS, France) with the support of the French \textit{Programme National de Physique Stellaire} and \textit{Institut national des sciences de l'Univers}.\\ 
\end{acknowledgements}

\bibliographystyle{aa}
\bibliography{aa37216-19}

\Online

\begin{appendix}

\section{Extraction of spectra of binary-pair candidates with a double Gaussian profile}
\label{appendix:Extraction_2_blended_spectra}

\begin{figure}[!b]
\centering 
\hspace{-1.1cm}
\includegraphics[width=8.8cm]{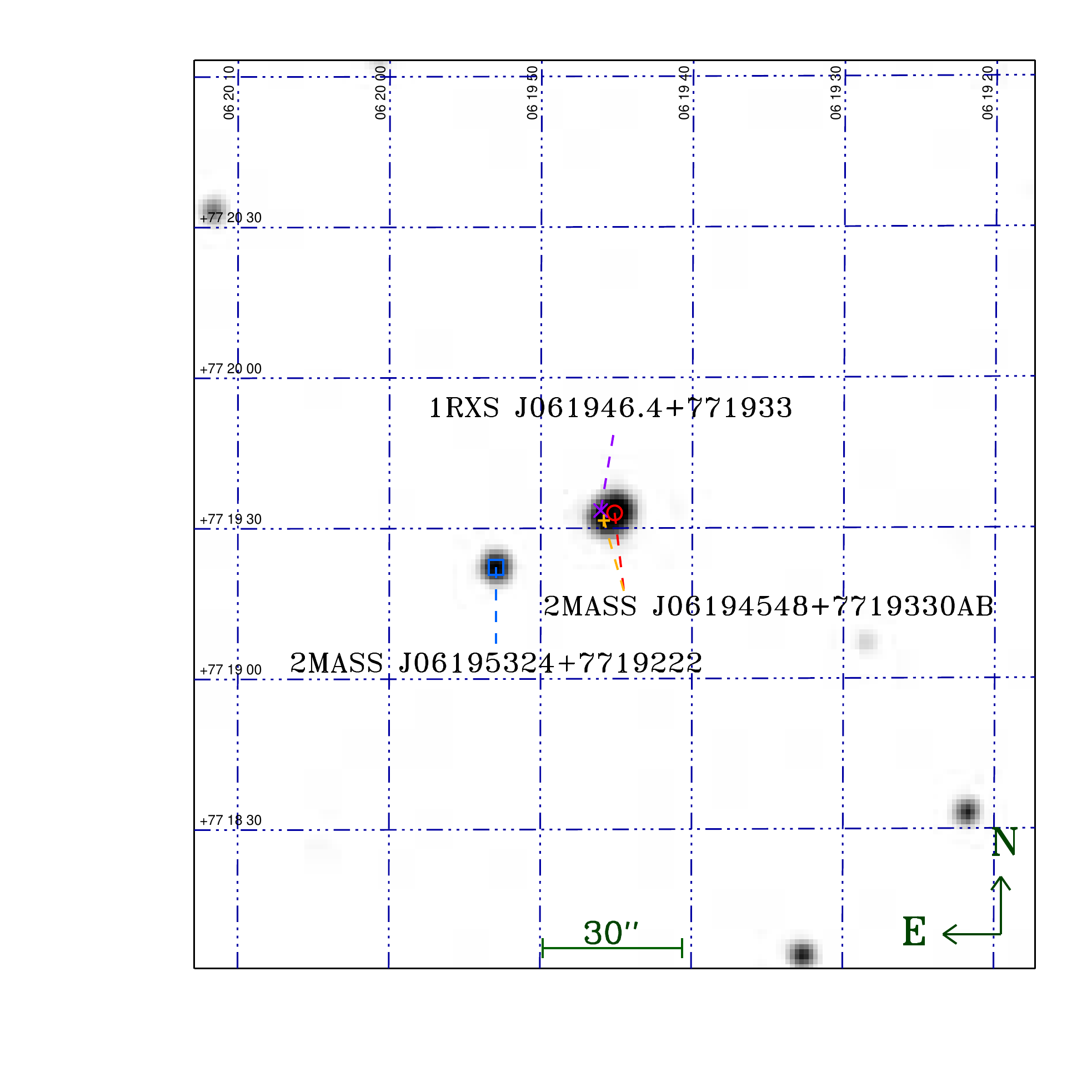} 
\vspace{-0.8cm}
\caption{2MASS K$_{s}$-band image (Epoch J1999.10), centered on the  source 2MASS~J06194548+7719330 (circle) and its nearest companion \emph{Gaia}~DR2~1116789735748819968 (``plus'' symbol). We also show the positions of the X-ray source 1RXS~J061946.4+771933 (cross) and the source 2MASS~J06195324+7719222 (square).}
\label{Fig:2MASS_0619_7719_field}
\end{figure}

To benefit from the IDS instrumental setup, we sought to identify possible binary-pair candidates within our target list before starting our observing runs. However, during the observations themselves, we found additional likely visual binaries having a lower angular separation. We did not identify them previously because the resolution of public images is not enough to resolve such binaries with sufficient confidence, before the launch of \emph{Gaia} observatory. In such cases, we oriented the slit with the position angle of the two components to simultaneously image them. Hereafter, we illustrate this work on the source \#40 (= \object{1RXS~J061946.4+771933}) because the alignment of three sources allows us to validate our work process. 

\begin{figure}[!b]
\centering 
\includegraphics[width=9.cm]{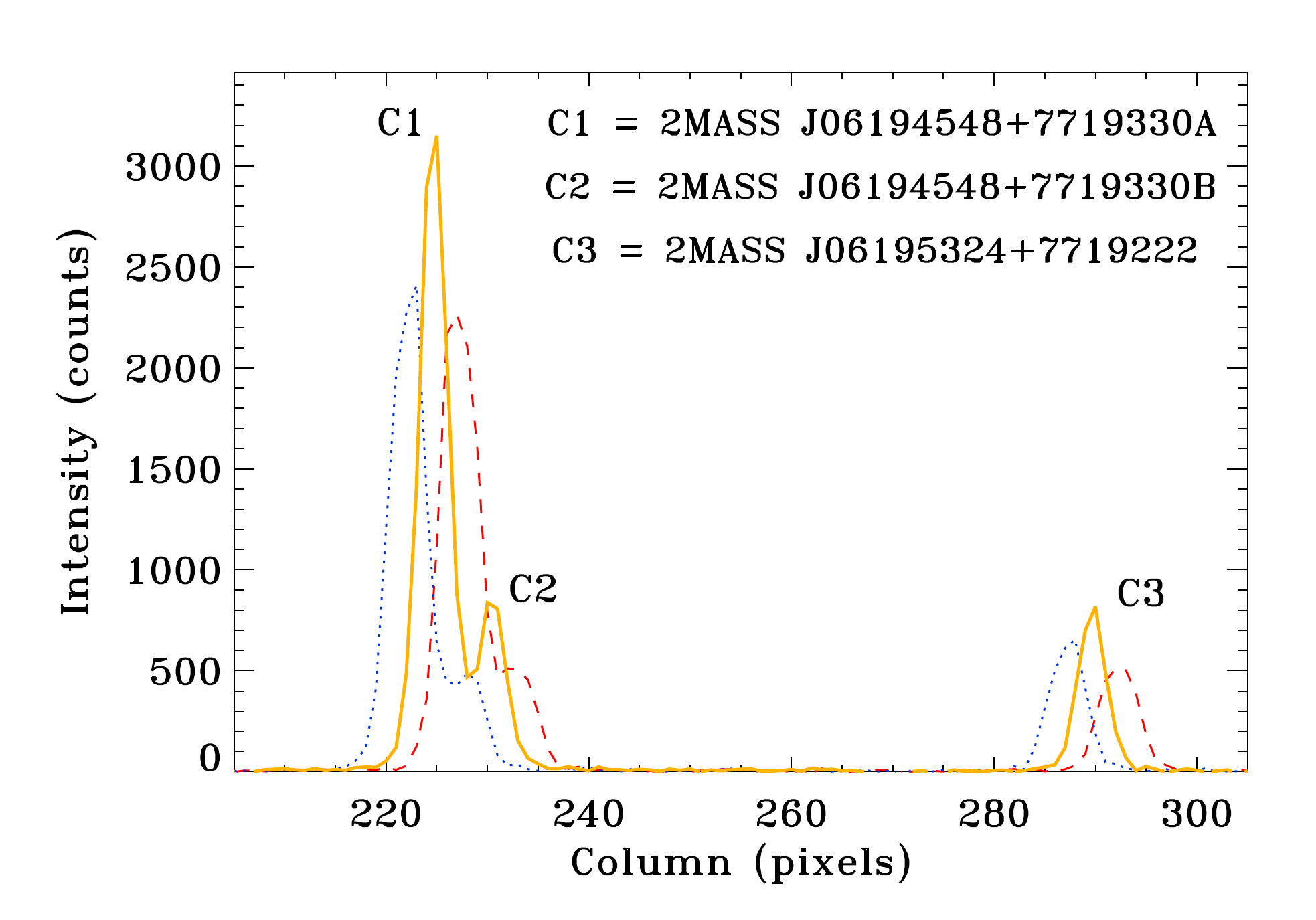} 
\caption{Plot of the line numbers $1300$ (red dashed line), $2000$ (orange solid line), and $2700$ (blue dotted line), perpendicular to the spectral direction. The three apertures C1--C3 are clearly visible.}
\label{Fig:2MASS_0619_7719_aperture}
\end{figure}

\medskip
The source 2MASS~J06194548+7719330 seems to have a weakly-elongated shape with respect to that of nearby stars (Fig.~\ref{Fig:2MASS_0619_7719_field}), which is typical of unresolved binaries and sources with a low angular separation. During our observing runs because we actually saw two sources close to the coordinates of our candidate. These turn out to be comoving stars (Table~\ref{Tab:VB_Gaia}). To position the slit on these two objects, we then set the slit position angle to about~$114\degr$. Based on the \emph{Gaia} DR2 coordinates, this angle is rather similar to the position angle $\theta$ of $115\fdg13\pm24\farcs35$ between these two sources, and of $114\fdg44\pm2\farcs95$ between the source \#40 and 2MASS J06195324+7719222. On the cuts perpendicular to the spectral direction, the aperture of the three components are clearly visible, regardless of the line used, even if the apertures C1 and C2 are partially blended (Fig.~\ref{Fig:2MASS_0619_7719_aperture}). 

\begin{figure}[!b]
\centering 
\includegraphics[width=9.cm]{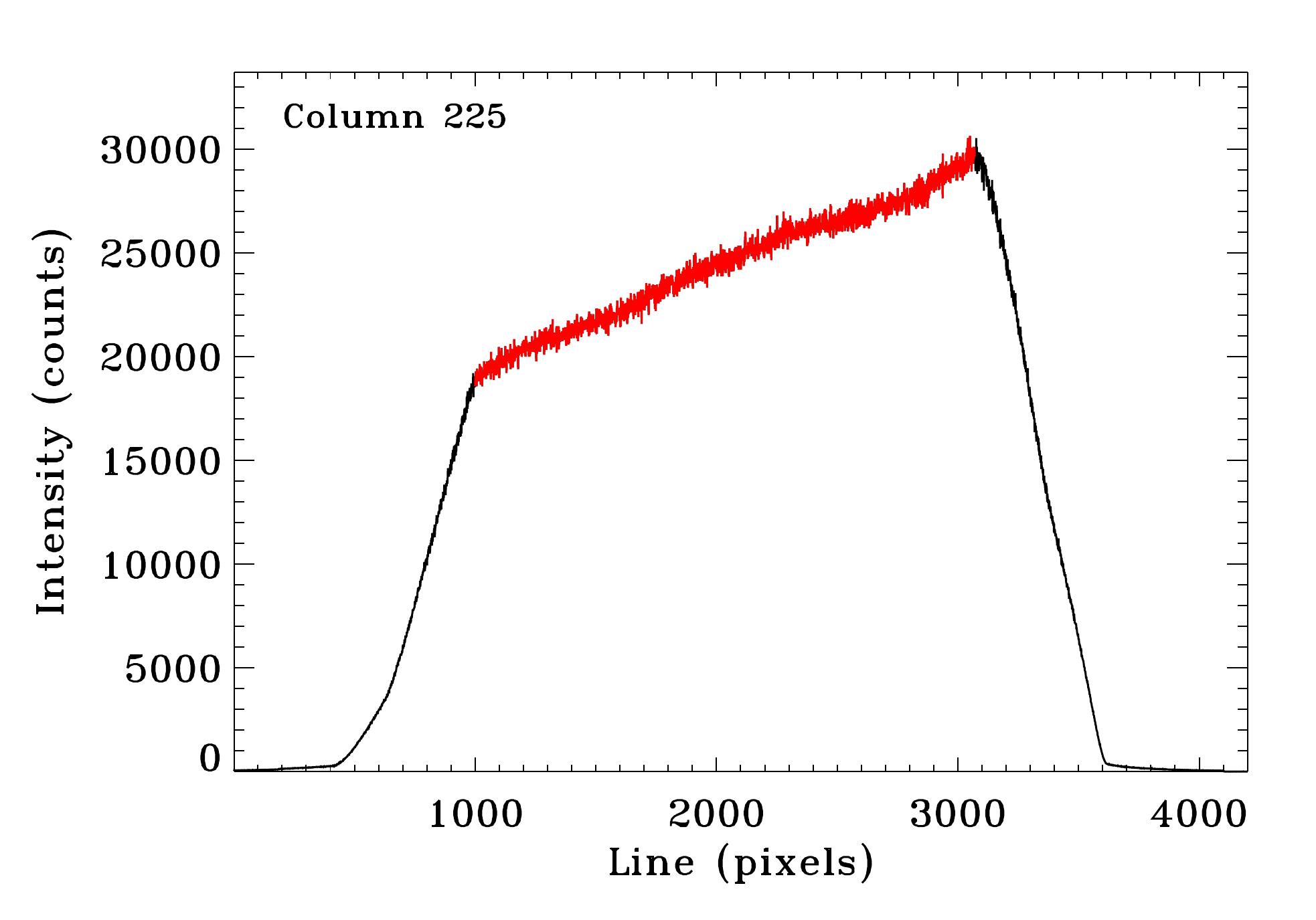} 
\caption{Plot of a lamp flat on the $235$~mm camera with the $2k$$\times$$4k$ RED+2 CCD detector mounted on IDS. The portion in red corresponds to  the approximately $2070$ CCD pixels that are clear and unvignetted.}
\label{Fig:Flat_Field}
\end{figure}

\begin{figure}[!t]
\centering 
\includegraphics[width=9cm]{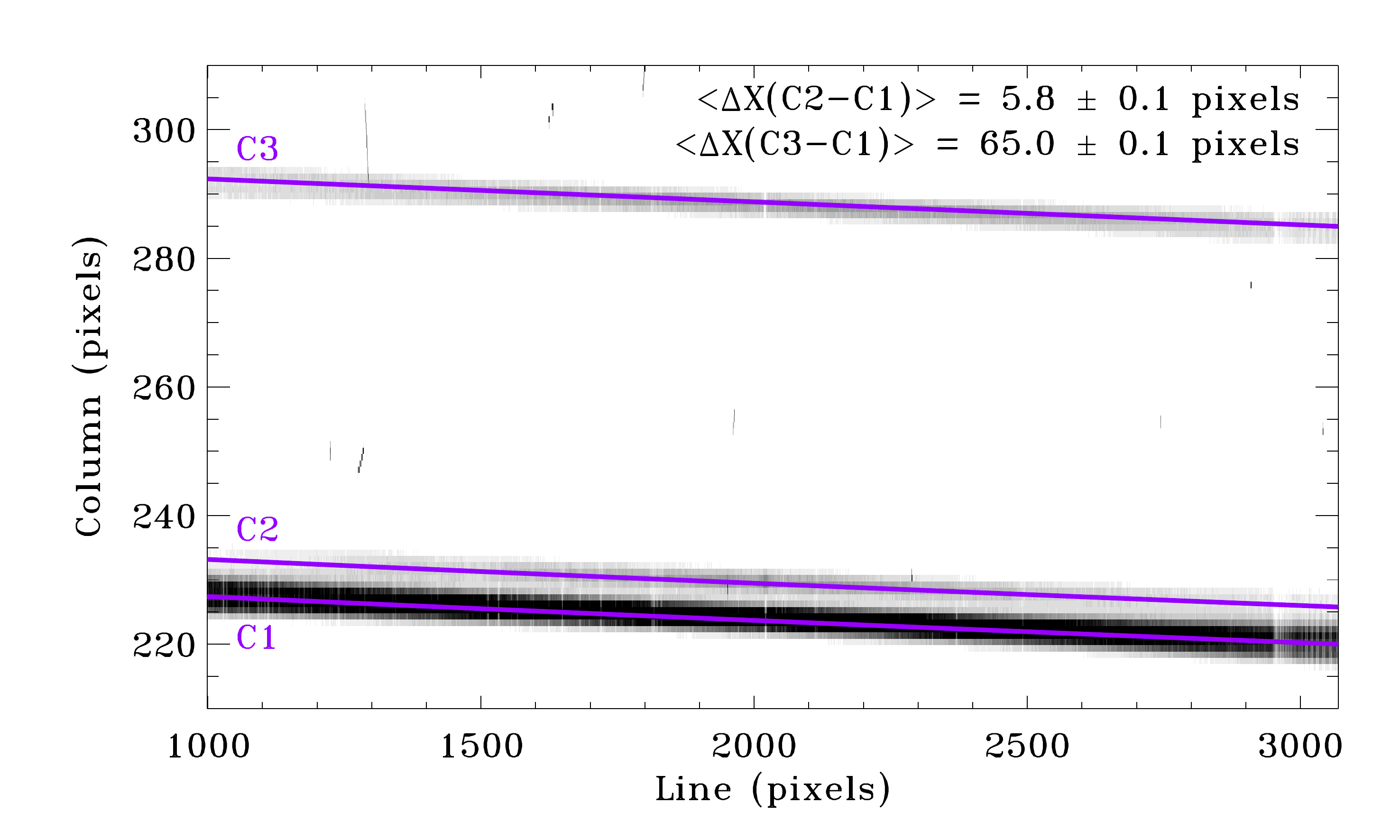} 
\caption{Zoom on the section of the image where the traces of the three sources can be used for a spectral analysis. This corresponds to all the clear and unvignetted lines (Fig.~\ref{Fig:Flat_Field}) and columns from $205$ to $305$. We also overplotted the trace (solid lines) obtained for each of the three sources. On the right upper corner, we indicate the mean separation between the main aperture (C1) and the two others (C2 and C3).}

\label{Fig:2MASS_0619_7719_trace}
\end{figure}

\medskip
We firstly used the procedure of extraction within the {\tt IRAF} environment, {\tt APALL}. As the spectrum obtained from the second aperture (C2) was contaminated by the light of the bright source (C1), we needed to perform an extraction by means of a double Gaussian profile in order to accurately fit each trace and thus substantially reduce such a contamination. In this context, we adapted the code of \citet{Frasca1997} to the IDS data. As the $235$~mm camera optics severely vignettes the outer regions of the dispersed light beam, only $2070$ of the CCD pixels are useful for the H1800V dispersion grating (Fig.~\ref{Fig:Flat_Field}). We determined the shape of each aperture by fitting all the perpendicular cuts along the spectral direction of clear and unvignetted lines. We extracted the two spectra from the partially blended apertures C1 and C2 using a double Gaussian profile, while we fitted the third aperture (C3) with an additional simple Gaussian profile. The traces for the three sources are displayed on Fig.~\ref{Fig:2MASS_0619_7719_trace}. 

\medskip
Using the 2MASS, NOMAD, and WISE coordinates, we found a spatial scale of $0.42$--$0.43$ arcsec/pixel. This agrees with the value of $0.44$ arcsec/pixel tabulated for this detector. From the clear and unvignetted lines, we found an average gap of $5.8\pm0.1$ and $65.0\pm0.1$~pixels between the main aperture (C1) and the two others (C2 and C3, respectively). These correspond to an angular separation $\rho$ of $2\farcs49 \pm 0\farcs07$ and $28\farcs0 \pm 0\farcs4$. These are consistent with the angular separation derived from the 2MASS coordinates ($\rho_{C_{1}-C_{3}} = 27\farcs707\pm0\farcs079$) and the \emph{Gaia} DR2 ones ($\rho_{C_{1}-C_{2}} = 2\farcs52\pm0.032$~mas and $\rho_{C_{1}-C_{3}} = 28\farcs39\pm0.040$~mas). 

\section{Multiple systems}
\label{appendix:multiple_targets}

The main goal of this analysis is to identify all the members of the Cepheus association among our targets and to derive the maximum information about their physical and kinematic properties. This includes their multiplicity on both small (i.e., spectroscopic systems or visual binaries with small angular separation) and large (i.e., common proper-motion binaries) scales. 

\subsection{Binary-pair candidates}
\label{appendix:binary-pair-candidates}

During our observing runs with the IDS spectrograph, it has happened that two sources are located at the coordinates of our targets. To be sure to observe the optical counterpart(s) of the unresolved X-ray source, we oriented the spectrograph slit with the position angle of such a binary-pair candidate to image them simultaneously. For the $18$ targets listed in Table~\ref{Tab:IDS_companion_slit} we acquired their long-slit spectra to characterize both of them. We subsequently determined if they are physically associated or if that is just a spurious association. We note that the sources \#6 and \#40 have two entries in this Table. Since the separation between the two apertures was large enough to avoid any contamination of the spectrum extracted from the second aperture by the light of the bright source (and for the sake of homogenization of data reduction), we used the procedure of extraction within the {\tt IRAF} environment, {\tt APALL}, with the exception of the source \#40. Nevertheless, we applied the procedure detailed in Appendix~\ref{appendix:Extraction_2_blended_spectra} to estimate the average separation in pixels between the two apertures. We then converted it into an angular separation $\rho$ according to the spatial scale indicated above.

\medskip
Based on our estimate of the angular separation and the position angle value recorded in the header of the raw data, we were able to identify each of sources located near one of our targets by relying mainly on the \emph{Gaia} DR2 catalog (Table~\ref{Tab:IDS_companion_slit}). 

\subsection{Comoving stars in \emph{Gaia} DR2}
\label{appendix:binary-visual}

To search for any common proper-motion companions to our targets, we made use of a search cone of a radius of $2\arcmin$ around each of them in \emph{Gaia} DR2. Taking advantage of the very accurate astrometry in this catalog, we found comoving stars for $46$ of our targets (Table~\ref{Tab:VB_Gaia}). In $90$\,\% of cases, we identified one companion only. Nevertheless, the sources \#123, \#125, \#192, and F3 have two comoving companions, while we are able to list three for source \#149. These five targets have one close companion ($\rho < 10\arcsec$) and at least another more distant one ($\rho > 40\arcsec$). 

For the closest systems ($\pi\geqslant4$\,mas), we also extended the radius of search cone up to $10\arcmin$. We then found that \textit{i)} the source~\#6 has another common proper-motion companion with a large angular separation ($\rho=3\farcm44 \pm 0.185$~mas) and \textit{ii)} the sources G3 and G4 are separated by $4\farcm96 \pm 0.033$~mas. 

\begin{table}
\caption{Identification of the second source that we have imaged during the acquisition of the IDS spectrum of our target.}
\smallskip
\begin{center}
\vspace{-0.4cm}
{\small
\begin{tabular}{r@{\hspace{0.1cm}}ll}
\hline
\hline
\noalign{\smallskip}
\multicolumn{2}{c}{\#} & Source name \\ 
\noalign{\smallskip}
\hline
\noalign{\smallskip}
6	& c1	& TYC 4500-1549-1	\\
6	& c3	& \emph{Gaia} DR2 564707973733134080	\\
31	& c1	& \emph{Gaia} DR2 475251432820143872	\\
37	& c2	& TYC 4354-793-1	\\
40	& c2	& \emph{Gaia} DR2 1116789735748819968	\\
40	& c3	& \emph{Gaia} DR2 1116789667029342080	\\
42	& c2	& \emph{Gaia} DR2 1116535920362524928	\\
44	& c2	& TYC 4618-329-1	\\
81	& c2	& \emph{Gaia} DR2 1723454381704389248	\\
106	& c2	& \emph{Gaia} DR2 2248508292487691008	\\
138	& c2	& \emph{Gaia} DR2 2203366330791767552	\\
140	& c2	& \emph{Gaia} DR2 2005807146670408704	\\
163	& c2	& \emph{Gaia} DR2 537608345003955840	\\
168	& c2	& \emph{Gaia} DR2 534835788997551488	\\
169	& c2	& \emph{Gaia} DR2 533784174844087168	\\
174	& c2	& \emph{Gaia} DR2 2291000598671699200	\\
184	& c2	& \emph{Gaia} DR2 2206637339215351936	\\
185	& c2	& \emph{Gaia} DR2 2226453257466755328	\\
191	& c2	& \emph{Gaia} DR2 2231866565525667712	\\
G3	& c2	& \emph{Gaia} DR2 564698451789359104	\\
\noalign{\smallskip}
\hline
\end{tabular}
\label{Tab:IDS_companion_slit}
}
\end{center}
\end{table}

\begin{table}[!t]
\caption{Spectroscopic systems among our targets. Any source whose lithium line is visible in absorption is named in bold. With respect to the profile of the \ion{Ca}{i}\,$\lambda$6717.7 line, we also highlight those for which the \ion{Li}{i}\,$\lambda$6707.8 line is \textbf{\textcolor{orange}{strong}} and \textbf{\textcolor{red}{very strong}} (see text for details).} 
\smallskip
\begin{center}
\vspace{-0.4cm}
{\footnotesize
\begin{tabular}{ll}
\hline
\hline
\noalign{\smallskip}
SB1 & 
\begin{minipage}{7.3cm}
\object{\textbf{\textcolor{orange}{1RXS~J043208.0+811627}}}, \object{1RXS~J062558.2+822124}, \object{\textbf{1RXS~J083407.4+791449}}, \object{\textbf{1RXS~J093852.0+852625}}, \object{1RXS~J193958.2+851032}, \object{\textbf{\textcolor{orange}{1RXS~J213749.2+803228}}},  \object{\textbf{\textcolor{red}{1RXS~J231616.5+784156}}}, \object{1RXS~J232209.7+575626}, \object{\textbf{\textcolor{red}{1RXS~J232647.5+770304}}}.
\end{minipage} \\
\noalign{\smallskip}
\hline 
\noalign{\smallskip}
SB1? & 
\begin{minipage}{7.3cm}
\object{1RXS~J155547.5+684014}, \object{\textbf{\textcolor{orange}{1RXS~J181048.9+701601}}}, \object{\textbf{\textcolor{red}{1RXS~J222706.6+652127}}}.
\end{minipage} \\
\noalign{\smallskip}
\hline 
\noalign{\smallskip}
SB2 & 
\begin{minipage}{7.3cm}
\object{\textbf{\textcolor{orange}{1RXS~J000142.0+773057}}}, \object{1RXS~J000806.3+475659}, \object{\textbf{\textcolor{red}{1RXS~J003904.2+791912c1}}}, \object{\textbf{\textcolor{red}{1RXS~J003941.9+790526}}}, \object{1RXS~J010929.0+683916}, \object{\textbf{1RXS~J011415.5+715933}}, \object{\textbf{1RXS~J011523.1+882923}}, \object{1RXS~J012927.4+744448}, \object{\textbf{\textcolor{orange}{1RXS~J013925.5+701853c1}}}, \object{1RXS~J023919.7+872828}, \object{1RXS~J024324.6+695320}, \object{1RXS~J025538.5+544706}, \object{1RXS~J030926.6+673238}, \object{\textbf{1RXS~J040745.1+875030}}, \object{1RXS~J061946.4+771933c2}, \object{1RXS~J085353.7+870708}, \object{1RXS~J163747.2+723937}, \object{\textbf{\textcolor{orange}{1RXS~J170526.8+743600}}}, \object{\textbf{\textcolor{orange}{1RXS~J171928.8+652227}}}, \object{1RXS~J175910.1+584300}, \object{1RXS~J183627.4+715311}, \object{1RXS~J193141.7+641951}, \object{\textbf{\textcolor{orange}{1RXS~J195758.2+664253}}}, \object{1RXS~J203857.5+580452}, \object{\textbf{\textcolor{orange}{1RXS~J212929.0+621859}}}, \object{1RXS~J214719.8+611618}, \object{1RXS~J224917.6+522634}, \object{1RXS~189583}.
\end{minipage} \\
\noalign{\smallskip}
\hline 
\noalign{\smallskip}
\begin{minipage}{0.7cm}
Likely SB2 
\end{minipage}&
\begin{minipage}{7.3cm}
\object{1RXS~J005300.8+682125}, \object{\textbf{\textcolor{red}{1RXS~J044912.7+773719}}},  \object{\textbf{\textcolor{orange}{1RXS~J071743.1+764416}}}, \object{\textbf{1RXS~J181610.9+585539}}. 
\end{minipage} \\
\noalign{\smallskip}
\hline 
\noalign{\smallskip}
SB2? &
\begin{minipage}{7.3cm}
\object{\textbf{1RXS~J000002.5+733942c1}}, \object{\textbf{1RXS~J010117.1+713114}}, \object{\textbf{\textcolor{orange}{1RXS~J045808.2+790813}}}, \object{1RXS~J050642.4+745604c2}, \object{1RXS~J064241.4+880442}, \object{1RXS~J075427.0+780633}, \object{1RXS~J161939.9+765515}, \object{1RXS~J165315.4+701554}, \object{\textbf{\textcolor{orange}{1RXS~J192127.4+611208}}}, \object{\textbf{\textcolor{orange}{1RXS~J195542.3+663207}}}, \object{\textbf{\textcolor{orange}{1RXS~J203549.9+594930}}}, \object{\textbf{\textcolor{red}{1RXS~J230822.7+790829}}}, \object{\textbf{\textcolor{orange}{1RXS~J235502.1+541516}}}. 
\end{minipage} \\
\noalign{\smallskip}
\hline 
\noalign{\smallskip}
SB3 & 
\begin{minipage}{7.3cm}
\object{1RXS~J050642.4+745604c1}, \object{\textbf{\textcolor{red}{1RXS~J185131.1+584258}}}, \object{\textbf{1RXS~J232346.4+620620}}.
\end{minipage} \\
\noalign{\smallskip}
\hline 
\noalign{\smallskip}
SB3? &
\begin{minipage}{7.3cm}
\object{\textbf{1RXS~J211232.5+741227}}.
\end{minipage} \\
\noalign{\smallskip}
\hline 
\end{tabular}
\label{tab:MultipleSyst}
}
\end{center}
\end{table}

\subsection{Spectroscopic systems}
Table~\ref{tab:MultipleSyst} lists $44$ spectroscopic systems found from our analysis of the CCF profiles (Sect.~\ref{Sec:Analysis_RV}). We also include $17$ possible spectroscopic systems for which the CCF profile is slightly asymmetric or the profile of the H$\alpha$ line is double-peaked. In the latter case, the absence of a second peak in the CCF profile can be explained by a mass ratio of the two sources much different from 1, but that the activity of the fainter star is sufficiently strong so that its H$\alpha$ line is visible in the spectrum obtained during the combined observation of the two sources. 

 \medskip
Table~\ref{tab:MultipleSyst} also highlights the lithium-rich systems. Except for the SB1 systems, the determination of $W_{\rm Li}$ value for each component of the SB2 and SB3 systems goes beyond the scope of the work presented here (see \citealt{Frasca2006} and \citealt{2008A&A...490..737K} for a detailed analysis of SB2 and SB3 systems, respectively). We therefore decided to characterize the profile of the multiple \ion{Li}{i}\,$\lambda$6707.8 lines qualitatively. To this end, we compared their intensity with respect to that of the nearby \ion{Ca}{i}\,$\lambda$6717.7 line. We selected any system with \ion{Li}{i} lines  deeper than about the half of the \ion{Ca}{i} ones. We divided them into two groups by assuming that the \ion{Li}{i} lines deeper than about the \ion{Ca}{i}~ones are very strong. The remaining sources are therefore listed as displaying a strong lithium line. Among the $44$~spectroscopic systems, the lithium line is visible in the spectrum of $22$~($50$\,\%) sources. This includes nine ($20$\,\%) and six sources ($14$\,\%) with a strong and very strong lithium line, respectively.

\begin{figure}
\centering 
\vspace{-.4cm}
\hspace{-.4cm}
\includegraphics[width=4.6cm]{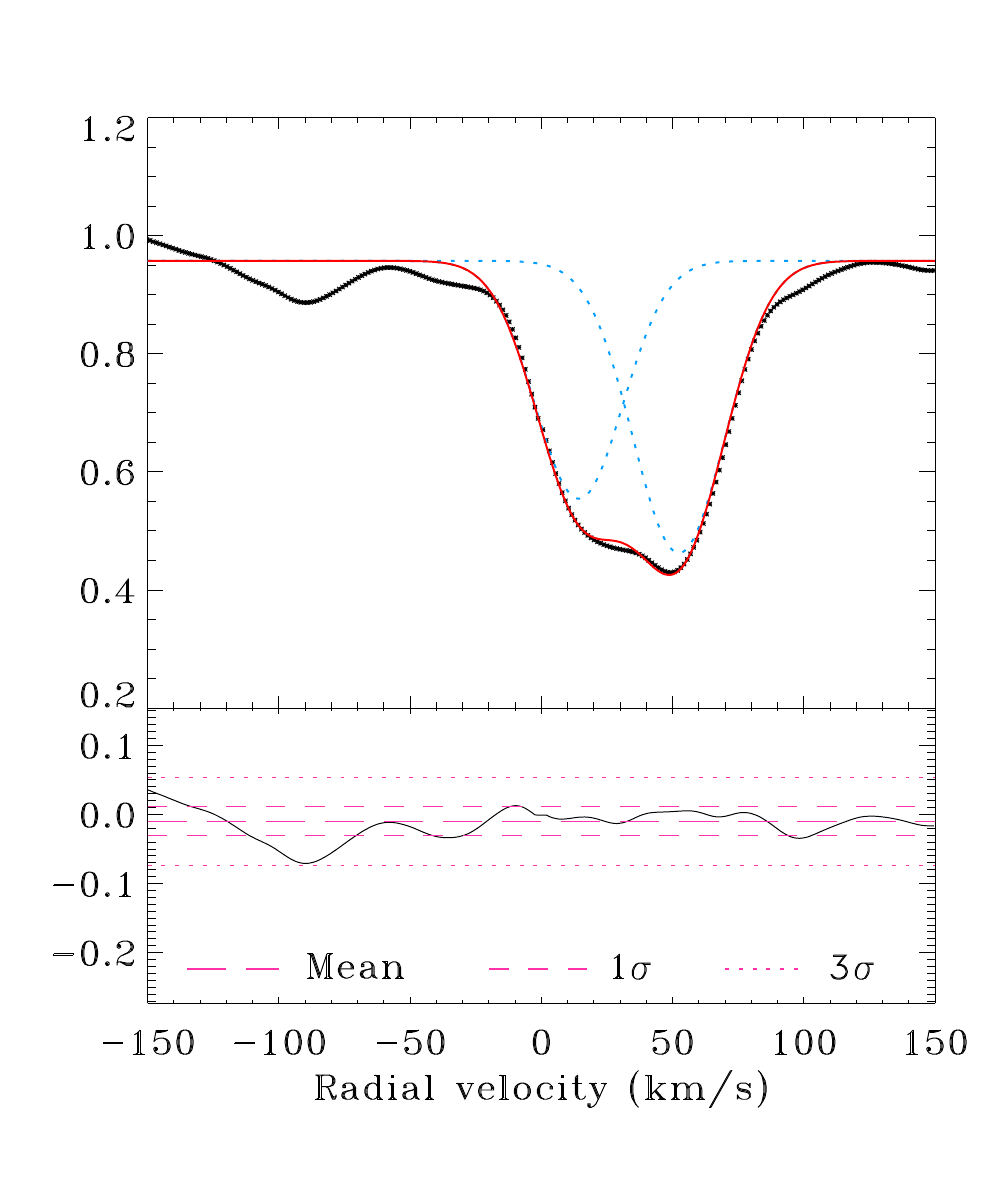} 
\hspace{-.4cm}
\includegraphics[width=4.6cm]{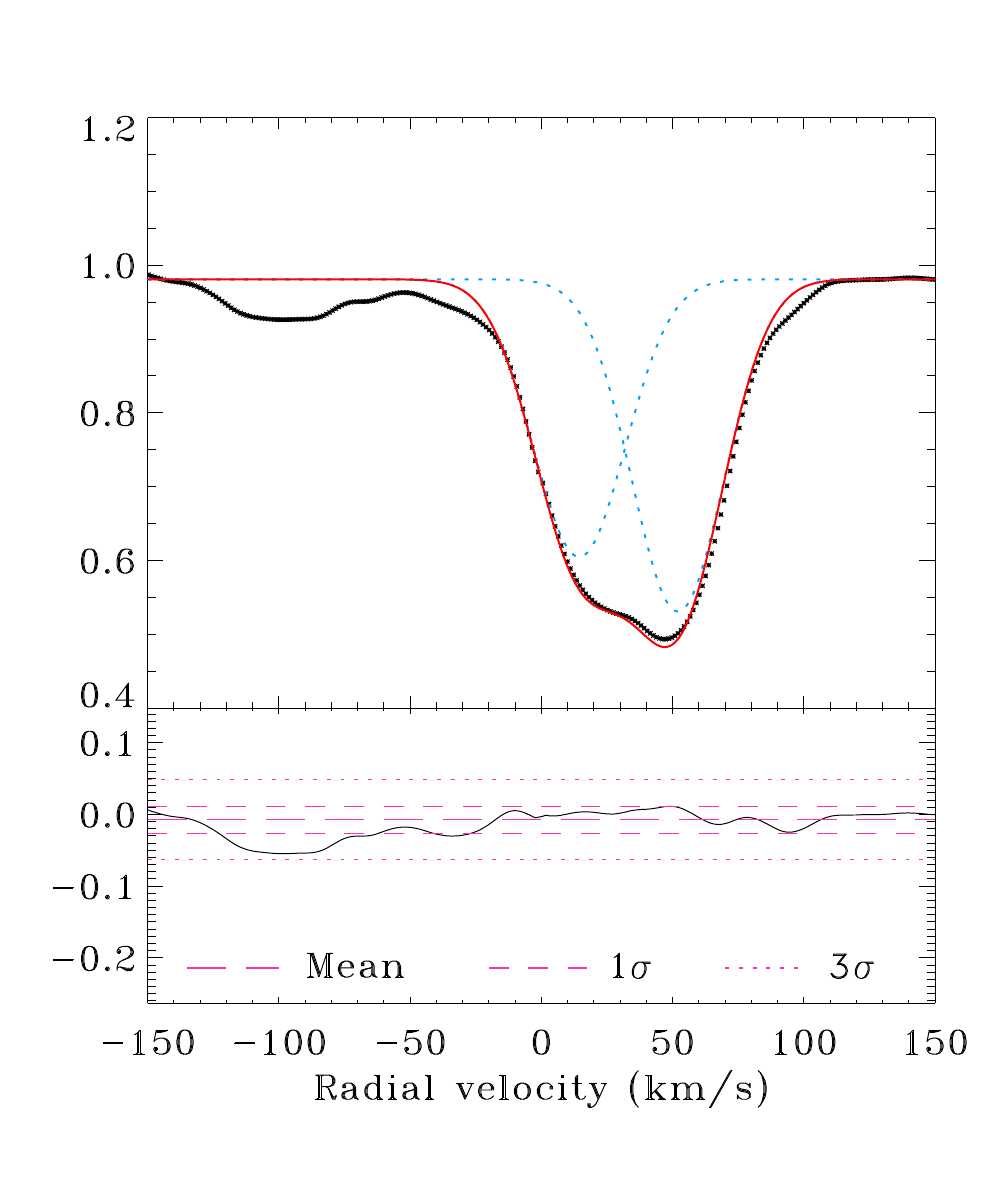} \\
\vspace{-.7cm}
\hspace{-.4cm}
\includegraphics[width=4.6cm]{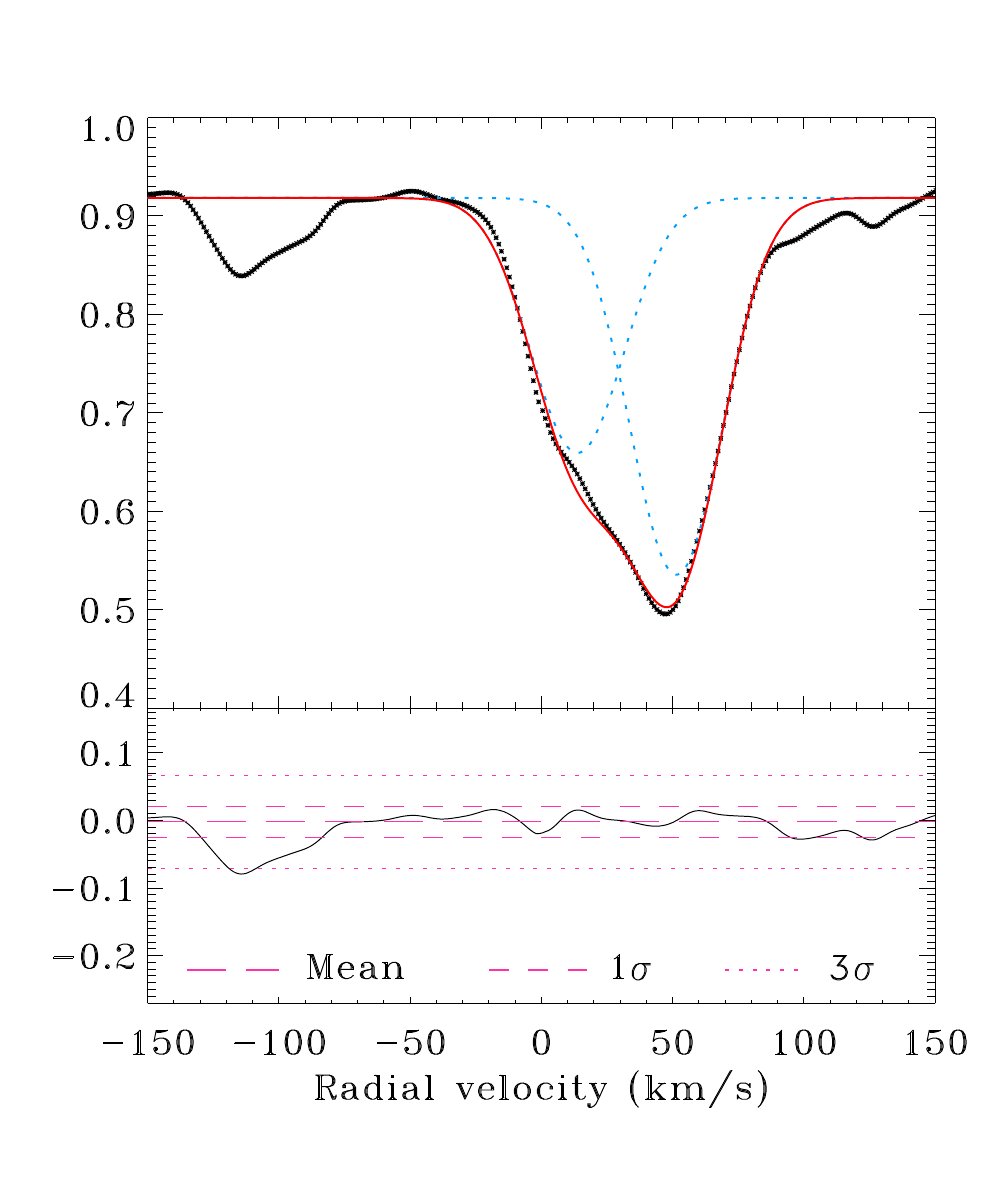} 
\hspace{-.4cm}
\includegraphics[width=4.6cm]{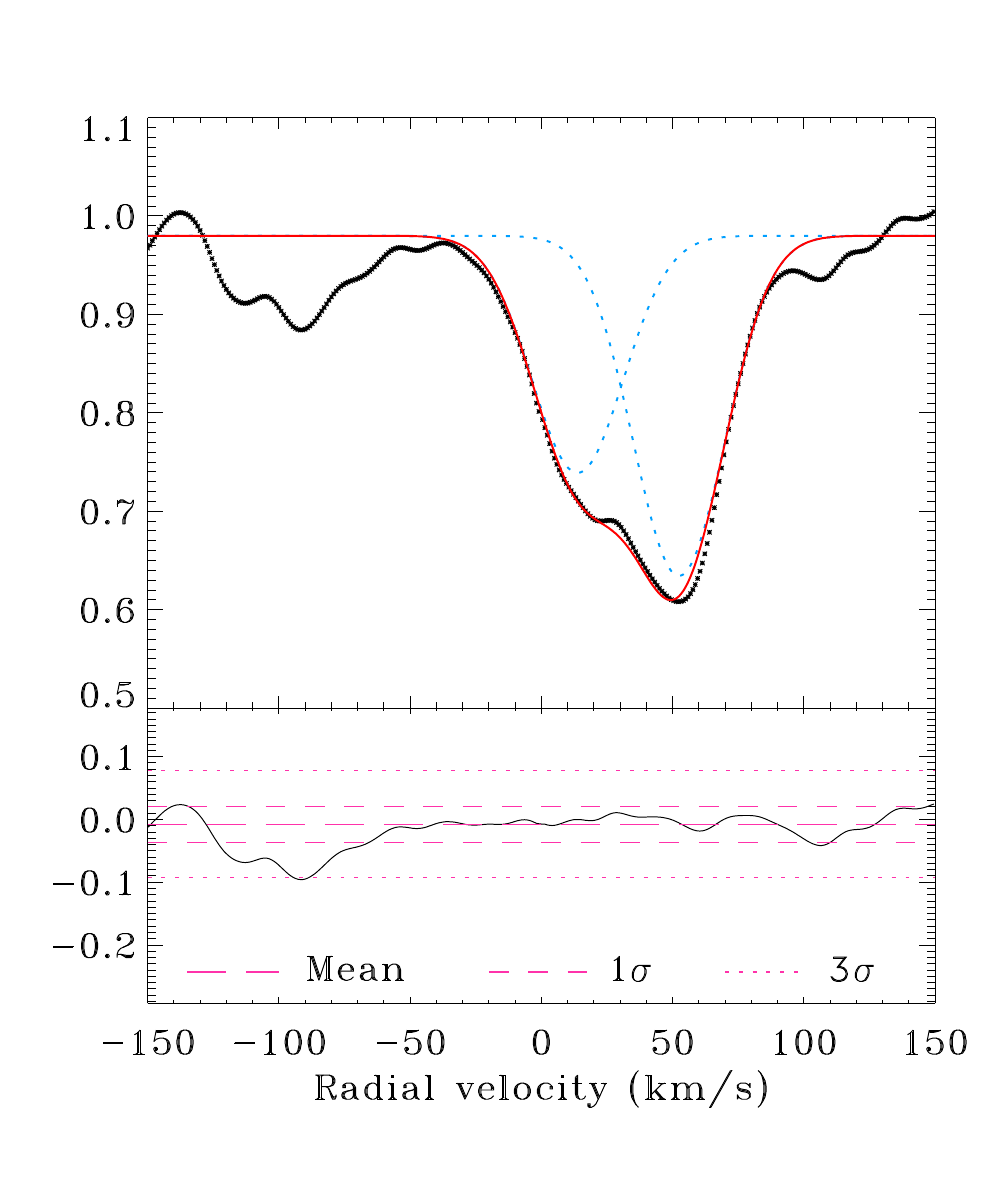} 
\caption{Cross-correlation function (asterisks) and Gaussian fitting (red solid line) for the source \#10 obtained from different orders of the SOPHIE spectrum (\textit{upper panel of each box}). We denote the Gaussian fitting of each individual peak with the blue dotted lines. We also display the CCF residuals (\textit{lower panel of each box}).}
\label{Fig:CCF_0114_7159}
\end{figure}

\medskip
Figure~\ref{Fig:CCF_0114_7159} displays the complexity of some systems and the need to carefully analyze the CCF residues after subtracting the fitting, especially when the two peaks are severely blended (i.e., one peak with a clearly asymmetric shape) on several orders. We consider the source \#10 as an SB2 system observed near the conjunction but we have also to point out the possible detection of an additional smaller peak in its CCF profile. However its amplitude remains below our threshold. 

\section{Low-mass stars}
\label{appendix:Mtype_stars}

\begin{table*}[!t]
\caption{Comparison of the spectral type derived for low-mass stars with those listed in the literature.}
\smallskip
\begin{center}
{\scriptsize
\begin{tabular}{rcccclc}
\hline
\hline
\noalign{\smallskip}
$~~~~\#$ & \multicolumn{6}{c}{Spectral type from:} \\ 
\noalign{\smallskip}
  & This work & \citet{Lepine2013} & CARMENCITA $^{a}$ & SpT vs. $V$--$J$ & Other(s) & Refs. \\ 
  &  &  &  &  relation $^{b}$ &  & \\ 
\noalign{\smallskip}
\hline
\noalign{\smallskip}
~~~~2 & M3.5V & M4V & \dots & M3.4 & M3.5+M3.6, M4V & 9, 11 \\
~~~~3 & M2.5V & \dots & M2.5V & M2.4 & M2.4V & 8 \\
~~~~5 & M4V & M4V & M4V+M & M3.6 & M4.0+M4.5, M3.7V & 7, 8 \\
~~~~6c2 & M1V & \dots & \dots & M0.4 & M2V & 6 \\
~~~~9 & M1V & M1.5V & \dots & M1.2 & \dots & \dots \\
~~12 & M2.5V & \dots & \dots & M0.9 & \dots & \dots \\
~~14 & M3.5V & \dots & \dots & M0.8 & \dots & \dots \\
~~15 & M0.5V & \dots & \dots & M3.1 & \dots & \dots \\
~~20 & \dots & \dots & \dots & M2.4 & \dots & \dots \\
~~21 & M3V & \dots & \dots & M3.6 & M2e & 5 \\
~~23 & M3.5V & \dots & \dots & M1.9 & \dots & \dots \\
~~24 & M1.5V & M2V & M1V & M2.7 & M3 & 2 \\
~~26 & M0.5V & \dots & \dots & M2.5 & \dots & \dots \\
~~29 & M0V & \dots & \dots & M1.0 & \dots & \dots \\
~~30 & M4V & \dots & \dots & M3.7 & \dots & \dots \\
~~34 & M1V & \dots & \dots & M1.8  & \dots & \dots \\
~~35 & M0.5V & \dots & \dots & M1.3 & \dots & \dots \\
~~38 & M0V & \dots & \dots & K7.85 & K5/M0 & 3 \\
~~39 & M3.5V & \dots & \dots & M3.8 & \dots & \dots \\
~~48 & \dots & M4V & M3.5V & M4.1 & M4/5 & 10 \\
~~54 & \dots & \dots & \dots & M2.8 & M3/4 & 10 \\
~~61 & \dots & \dots & \dots & M1.0& \dots & \dots \\
~~62 & \dots & M4V & \dots & M4.1& \dots & \dots \\
~~69 & \dots & \dots & \dots & M2.4& \dots & \dots \\
~~70 & M3V & M2.5V & \dots & M1.7& \dots & \dots \\
~~73 & M1V & \dots & \dots & M1.4& \dots & \dots \\
~~80 & M1.5V & \dots & \dots & M1.9& \dots & \dots \\
~~81c1 & M1V & \dots & \dots & \dots& \dots & \dots \\
~~81c2 & M2.5V & \dots & \dots & \dots& \dots & \dots \\
~~85 & M2V & \dots & \dots & M1.2 & M2V & 11 \\
~~88 & M4V & M4V & M3.5V & M3.6 & M4Ve & 4 \\
107 & M0V & \dots & \dots & M0.5& \dots & \dots \\
111 & M1V & \dots & \dots & M1.4& \dots & \dots \\
113 & M0V & M0V & \dots & M0.6& \dots & \dots \\
121 & M3.5V & \dots & \dots & M4.6& \dots & \dots \\
124 & K7V & \dots & \dots & M0.2& \dots & \dots \\
136 & K6V & \dots & \dots & M0.2& \dots & \dots \\
143 & M0V & \dots & \dots & M2.0& \dots & \dots \\
146 & M0.5V & \dots & \dots & M2.6& \dots & \dots \\
151 & M0V & M0.5V & M0V & M1.0 & M0V, M0.3V & 1, 8 \\
154 & M0.5V & \dots & \dots & M1.9& \dots & \dots \\
161 & M1.5V & \dots & \dots & M2.6& \dots & \dots \\
189 & K7V & \dots & \dots & M1.1& \dots & \dots \\
191c1 & M0V & \dots & \dots & M1.2 & K7V & 6 \\
191c2 & M2V & \dots & \dots & \dots & M3V & 6 \\
\hline 
\noalign{\smallskip}
\multicolumn{7}{l}{
\begin{minipage}{13.3cm}
\scriptsize \textbf{Notes.} \\
$^{a}$ CARMENCITA stands for ``CARMENes Cool star Information and daTa Archive'' and is a private database of the CARMENES consortium (http://carmenes.caha.es/), 
$^{b}$ We converted the $V$--$J$ color index into spectral type by using Eq.(12) of \citet{Lepine2013}. \\
\scriptsize \textbf{References.} \\
$1$ = \citet{Vyssotsky1956}; 
$2$ = \citet{Lee1984}; 
$3$ = \citet{Stephenson1986}; 
$4$ = \citet{Fleming1988}; 
$5$ = \citet{Motch1998}; 
$6$ = \citet{Tachihara05};
$7$ = \citet{Daemgen2007}; 
$8$ = \citet{Shkolnik2009};  
$9$ = \citet{Shkolnik2010};  
$10$ = \citet{Gigoyan2012}; and \\
$11$ = \citet{Janson2012}.
\end{minipage}}
\end{tabular}
\label{Tab:M_SpT}
}
\end{center}
\end{table*}

\begin{figure*}[!t]
\centering 
\includegraphics[width=8.1cm]{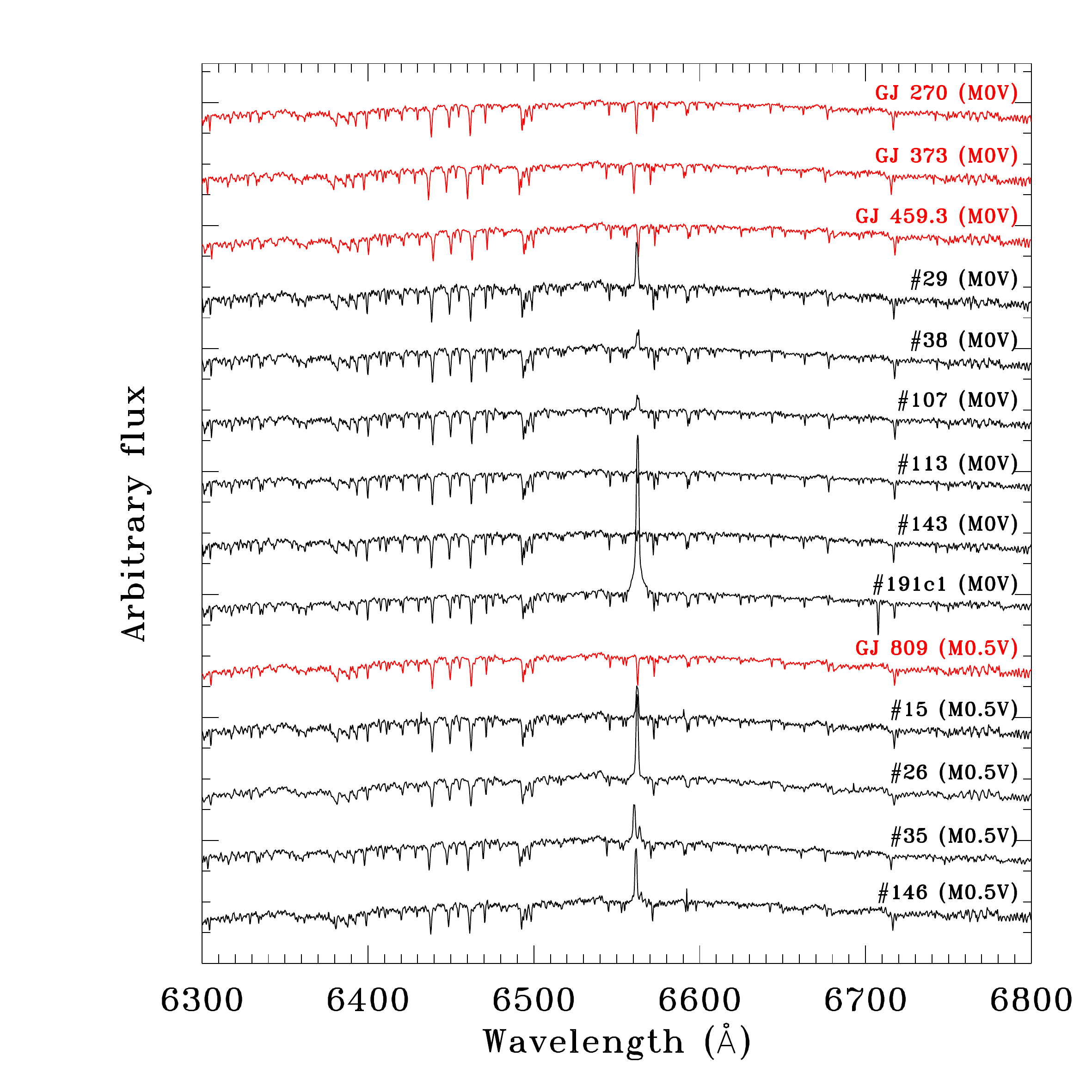}
\vspace{-0.4cm}
\hspace{-0.6cm}
\includegraphics[width=8.1cm]{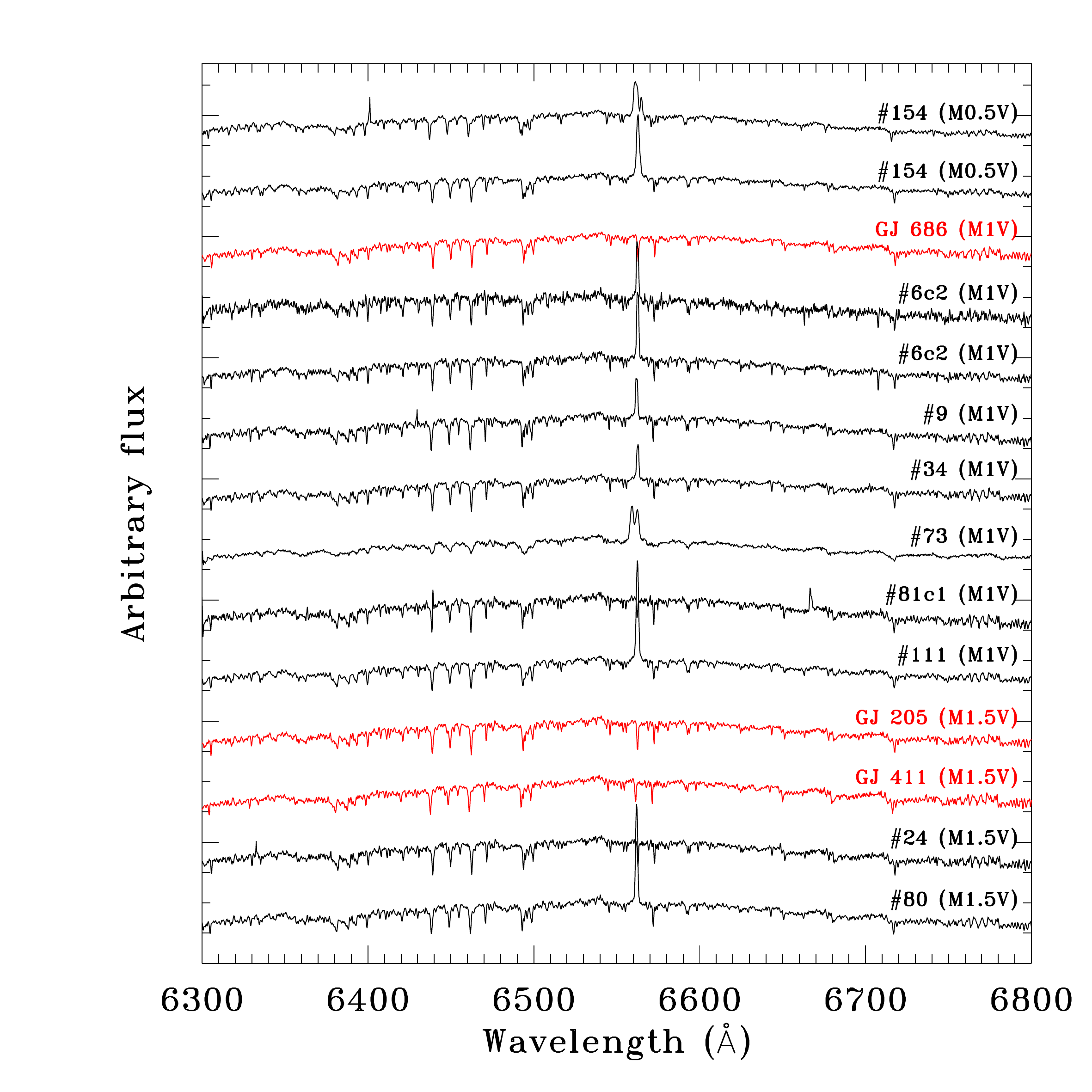} \\
\includegraphics[width=8.1cm]{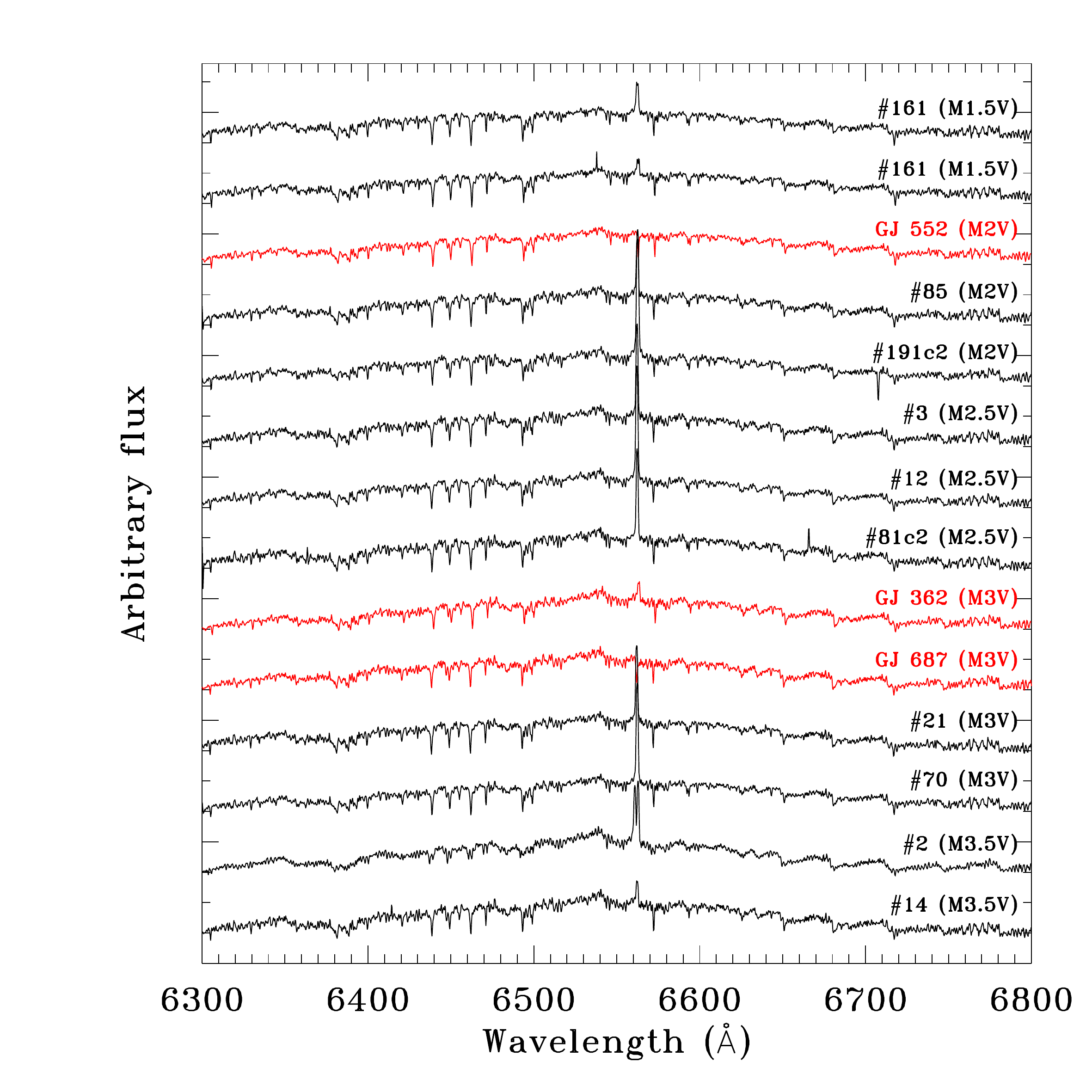}
\hspace{-0.6cm}
\includegraphics[width=8.1cm]{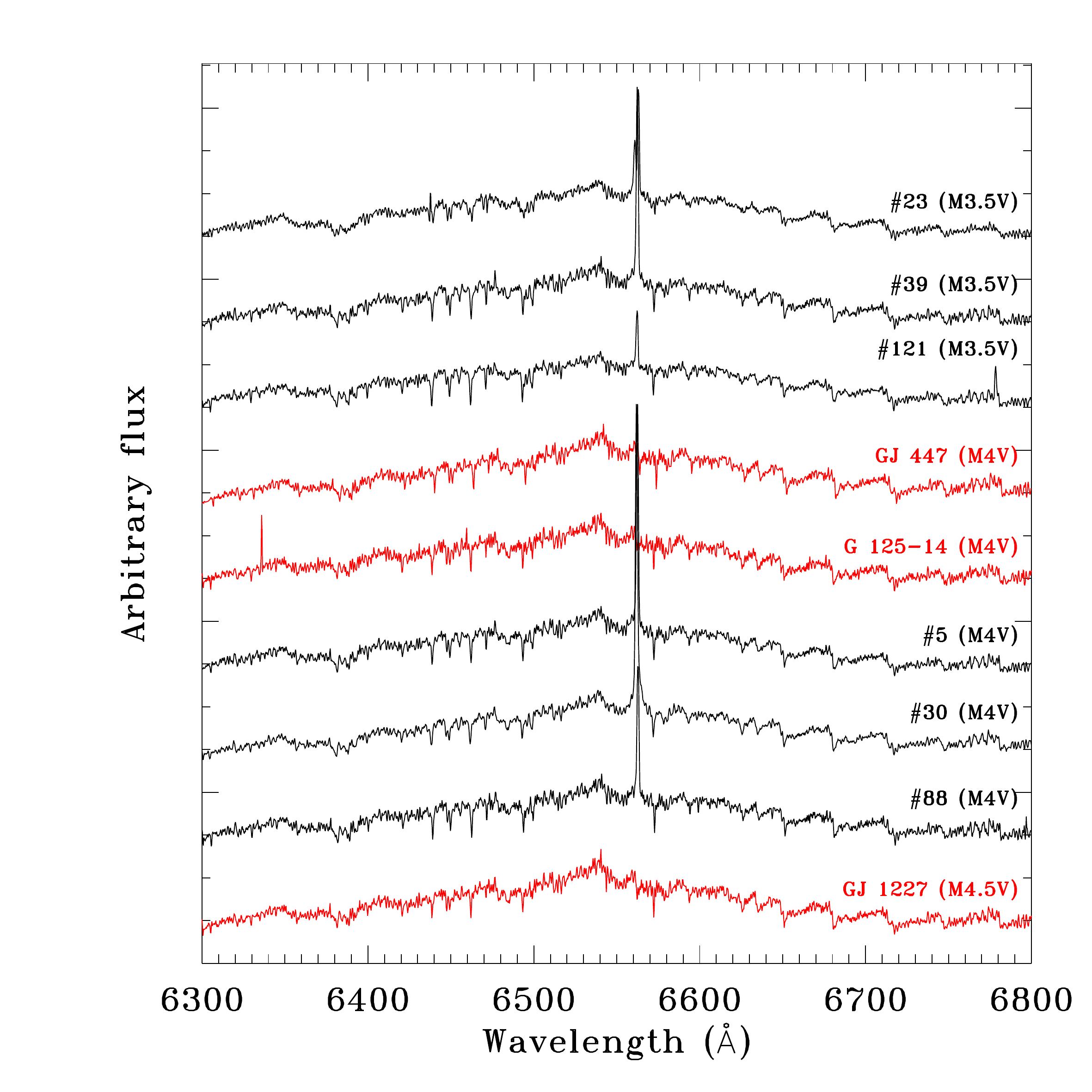}
\caption{Sequential distribution of spectra for the M-type stars. Their coloring clearly distinguishes targets (black) from standards (red).}
\label{Fig:Spectral_Typing_MStars}
\end{figure*}

We used Eq.~(12) from \citet{Lepine2013} to convert the $V$--$J$ color index listed in the Table~\ref{Tab:List_Candidates} into spectral type. We found that $45$ of our targets are likely M-type stars (Table~\ref{Tab:M_SpT}). For six of them (\#20, \#48, \#54, \#61, \#62, \#69), we were not able to acquire a spectrum to confirm this classification. 

\medskip
Due to their low luminosity and our observational strategy, the survey of low-mass stars could only be conducted with the IDS spectrograph. For these sources, we normalized the spectra in two different ways to perform their spectral typing. 

First, we defined the pseudo-continuum as a smooth line passing through the heads of molecular bands \citep[see, e.g.,][]{ZapateroOsorio2002}. In this way, all the molecular bands are fully in absorption and only the band-heads reach the 1.0 level. Such a task is impossible to be performed automatically. Normalized spectra are obtained by dividing each target spectrum by its continuum. This optimizes the comparison of our spectra with the \'Elodie library of standard stars, within the code {\tt ROTFIT}. We refer the reader to \citet{Frasca2015} for an application of this code to M-type stars. The results are listed in Table~\ref{Tab:AP_CepSurv}. 

Secondly, we divided each target spectrum by the continuum level in the spectral region near the H$\alpha$ line. This procedure allows us to perform the spectral typing of the M-type stars through a comparative analysis of the spectral shape of our targets with those of standard stars observed with the same instrumental setup (Table~\ref{tab:StandardStars}). To this end, we discarded the wavelength ranges contaminated by strong telluric lines or by chromospheric activity (e.g., the Balmer series). We looked for the best matches by means of a least-square minimization technique similar to that used by \citet{Klutsch2012} for the M-type stars from the CARMENES Input Catalog \citep[CARMENCITA,][]{Caballero2013}. Based on an internal comparison of the CARMENES results as well as with three independent works, \citet{2015A&A...577A.128A} highlighted the reliability of this approach. The spectral typing resulting from both our procedures are fully compatible with each other. Table~\ref{Tab:M_SpT} compares the spectral-type classifications resulting from our analysis with those available in the literature, and shows that they agree with each other with an accuracy of about one subclass. We finally found that three of these low-mass stars are late-K stars rather than early-M ones. Figure~\ref{Fig:Spectral_Typing_MStars} shows the spectra of the $36$ surveyed M-type stars, sorted by spectral type (from M0 to M4.5).

\section{Information about some stars}
\label{appendix:Info_targets}

\begin{figure}
\centering 
\hspace{-.5cm}
\includegraphics[width=5.4cm]{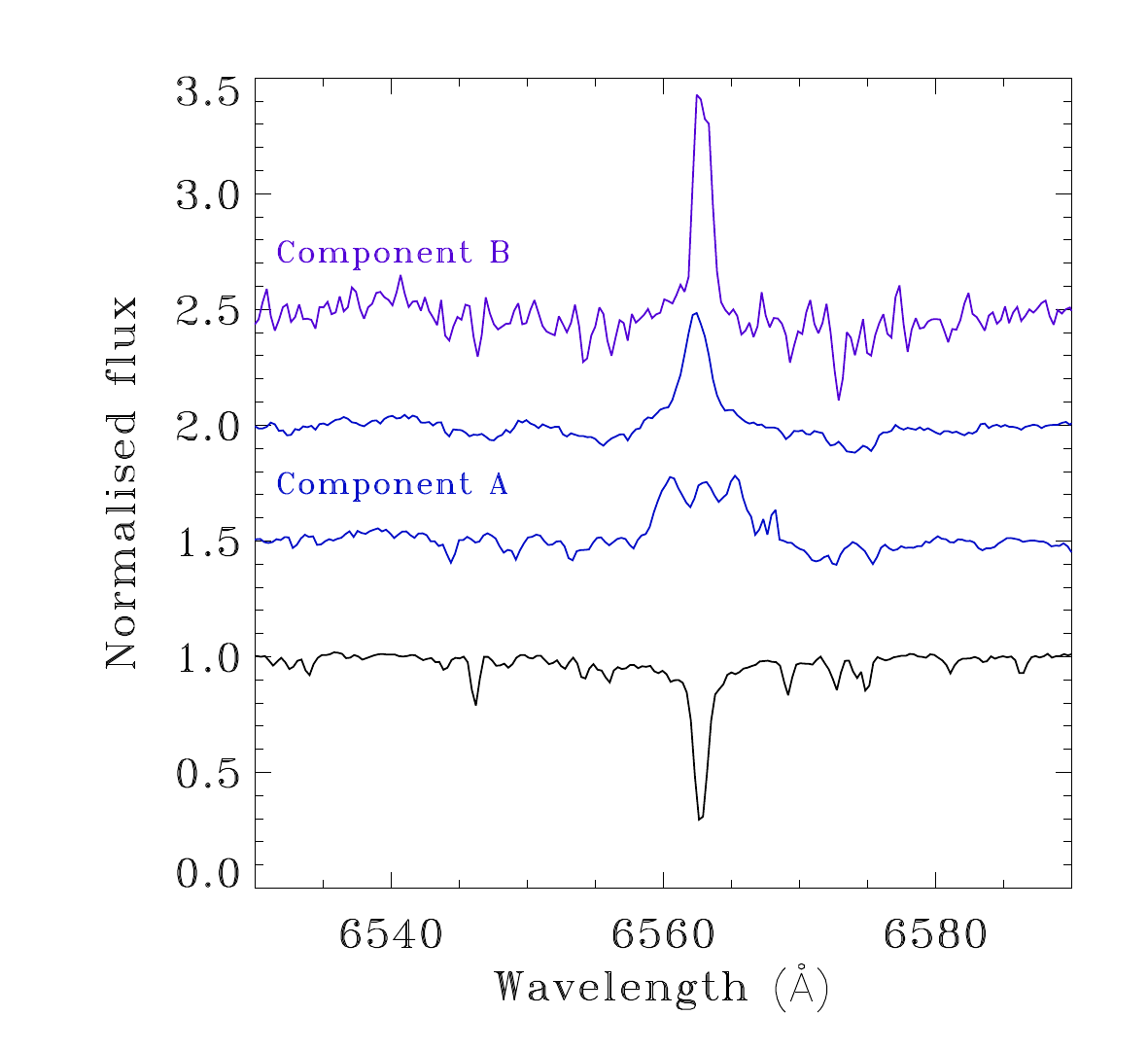}
\hspace{-1.65cm}
\includegraphics[width=5.4cm]{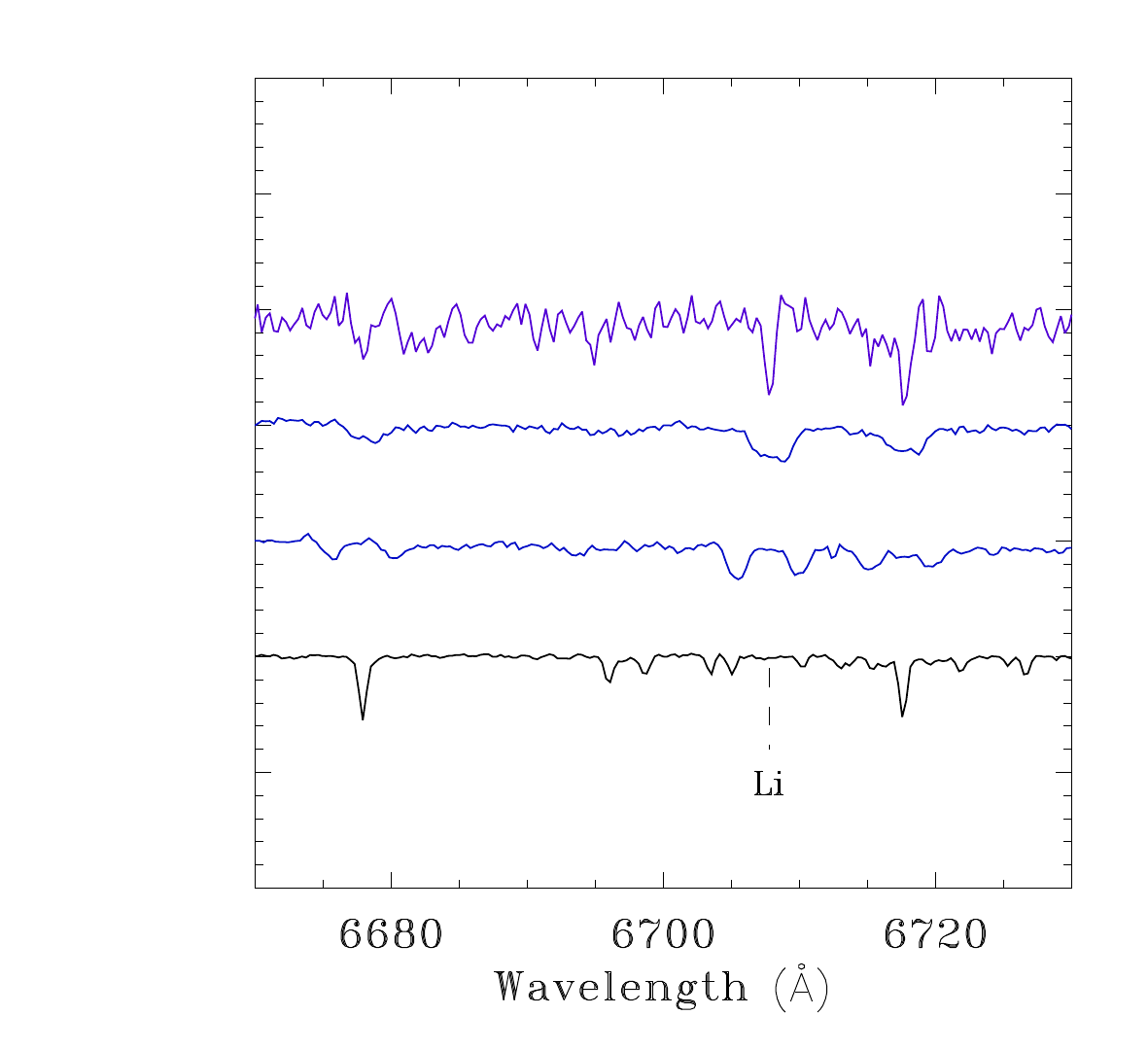}
\caption{H$\alpha$ and lithium spectra of TYC~4500-1549-1 ($\#6$) found as a visual binary. Two of spectra for the bright component are also shown.}
\label{Fig:triple_System}
\end{figure}

\begin{figure*}
\centering 
\hspace{-0.3cm}
\vspace{-0.4cm}
\includegraphics[width=7.1cm]{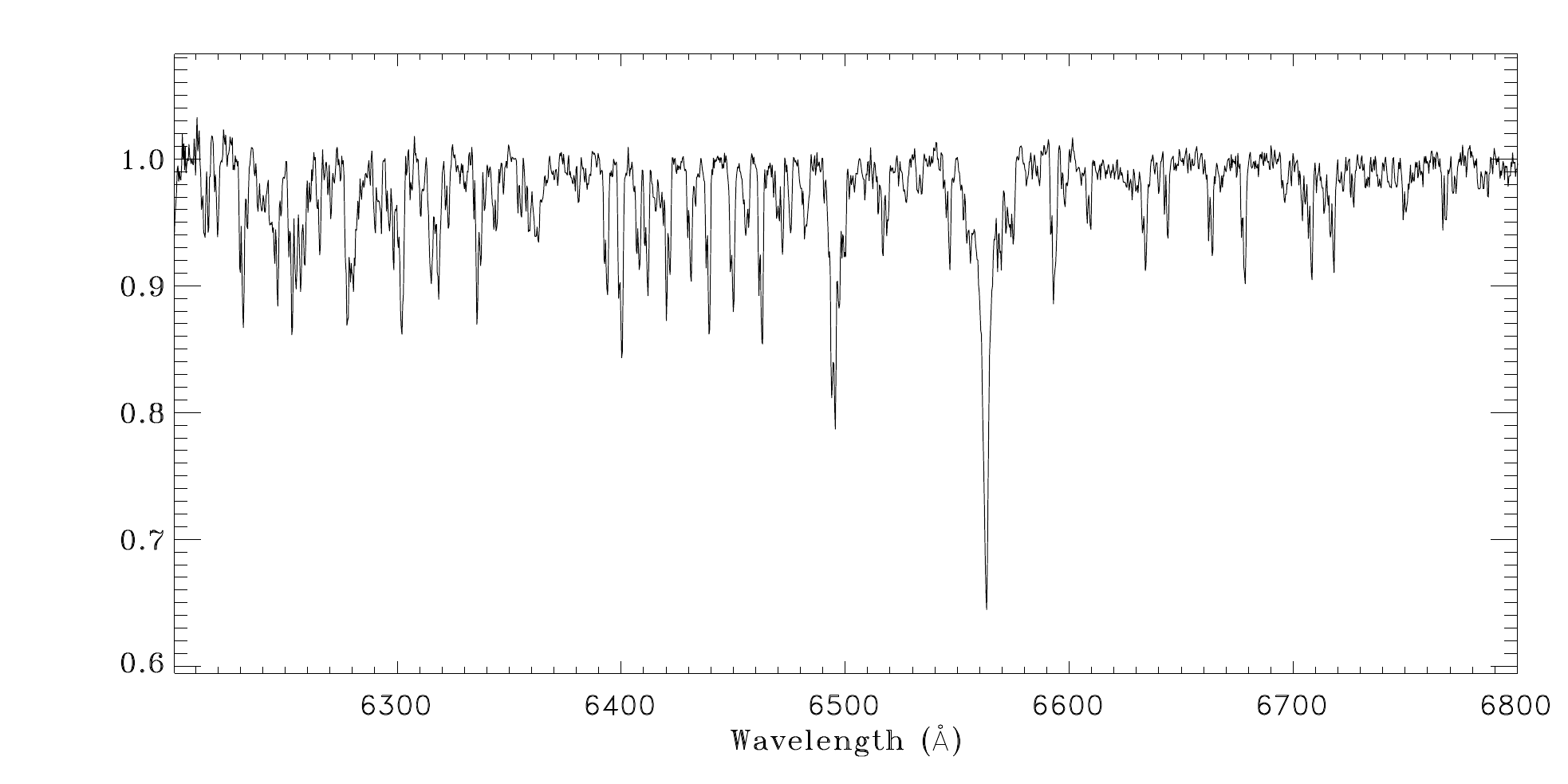}
\includegraphics[width=7.1cm]{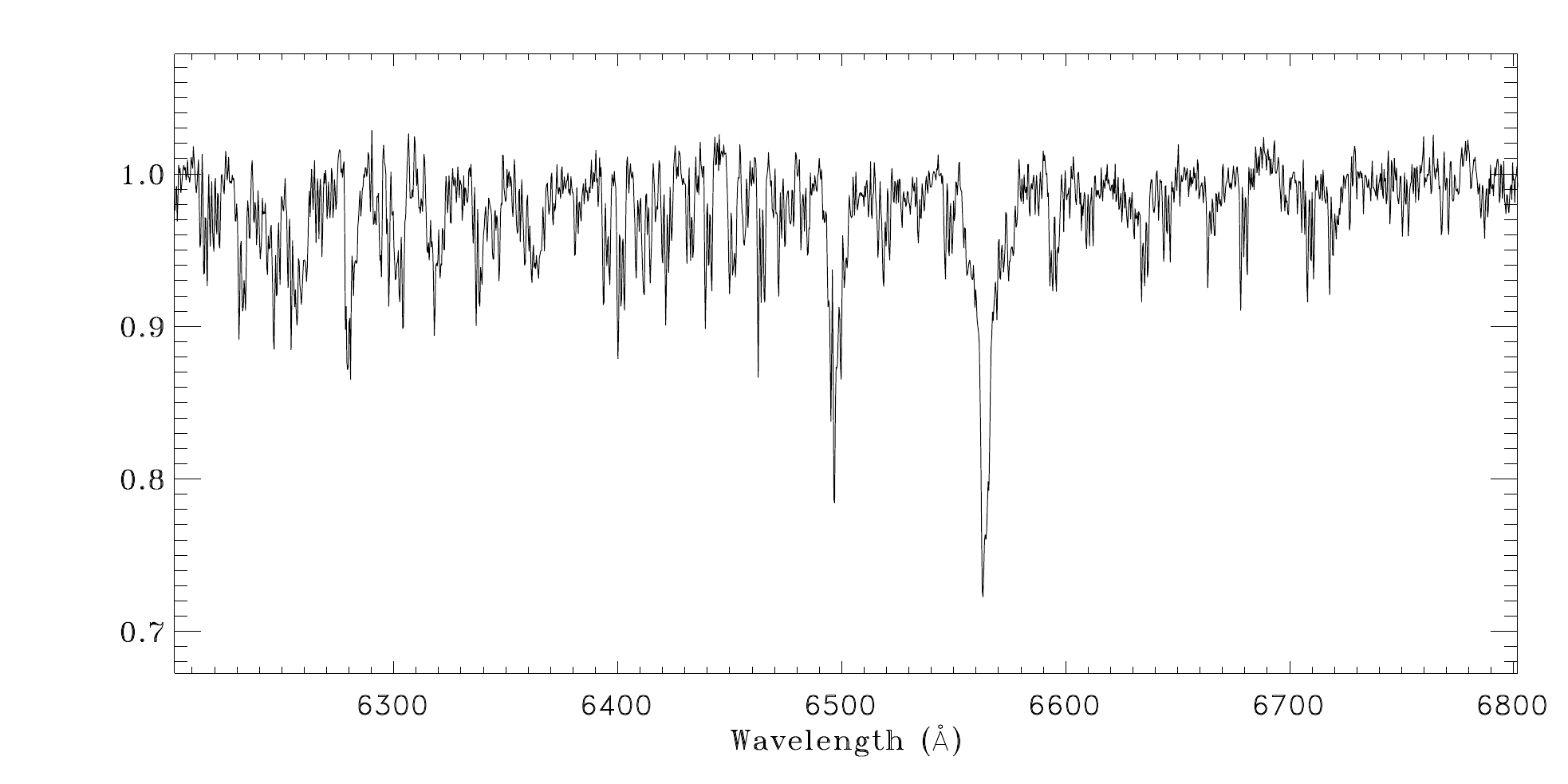} \\
\includegraphics[width=7.1cm]{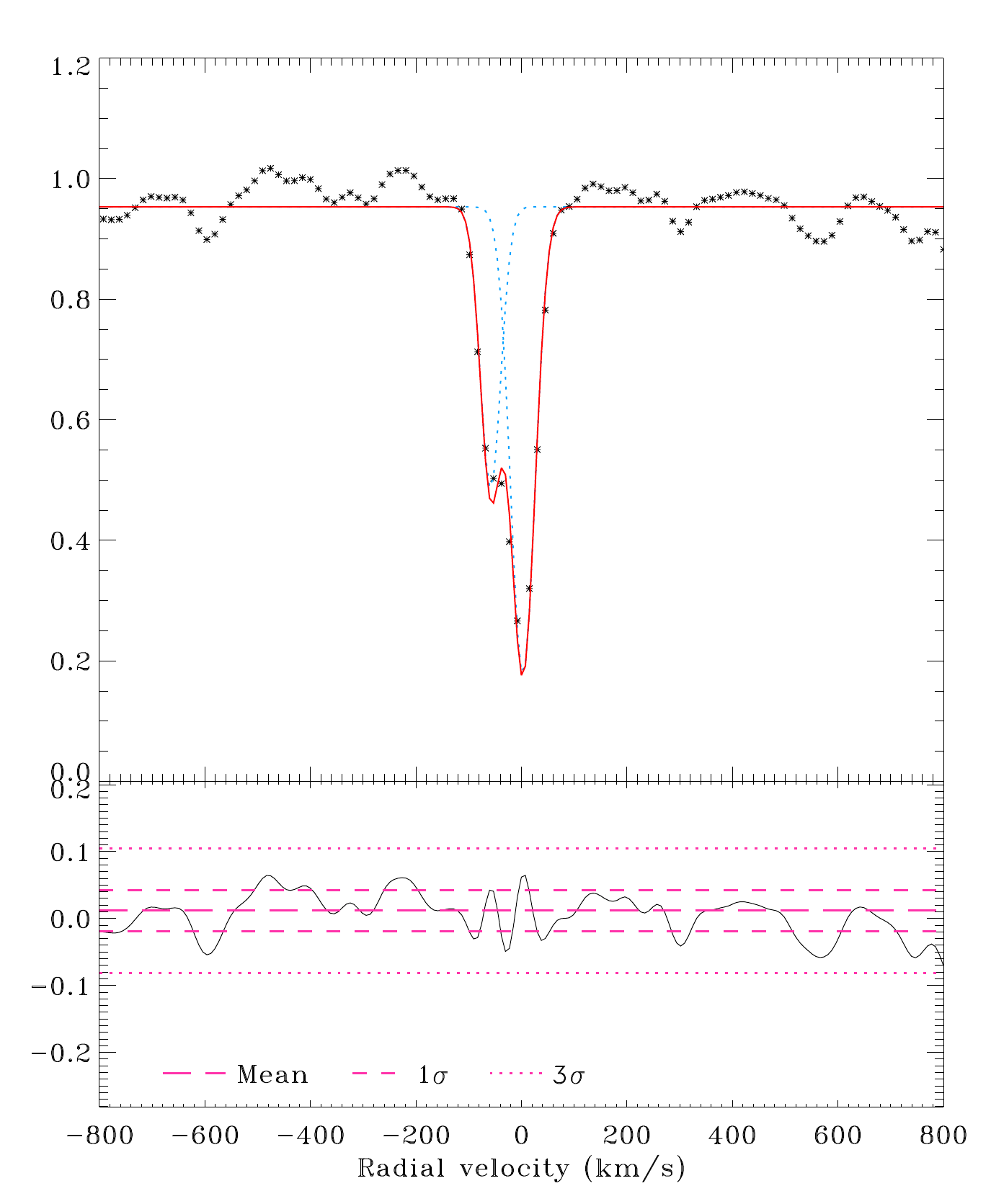}
\includegraphics[width=7.1cm]{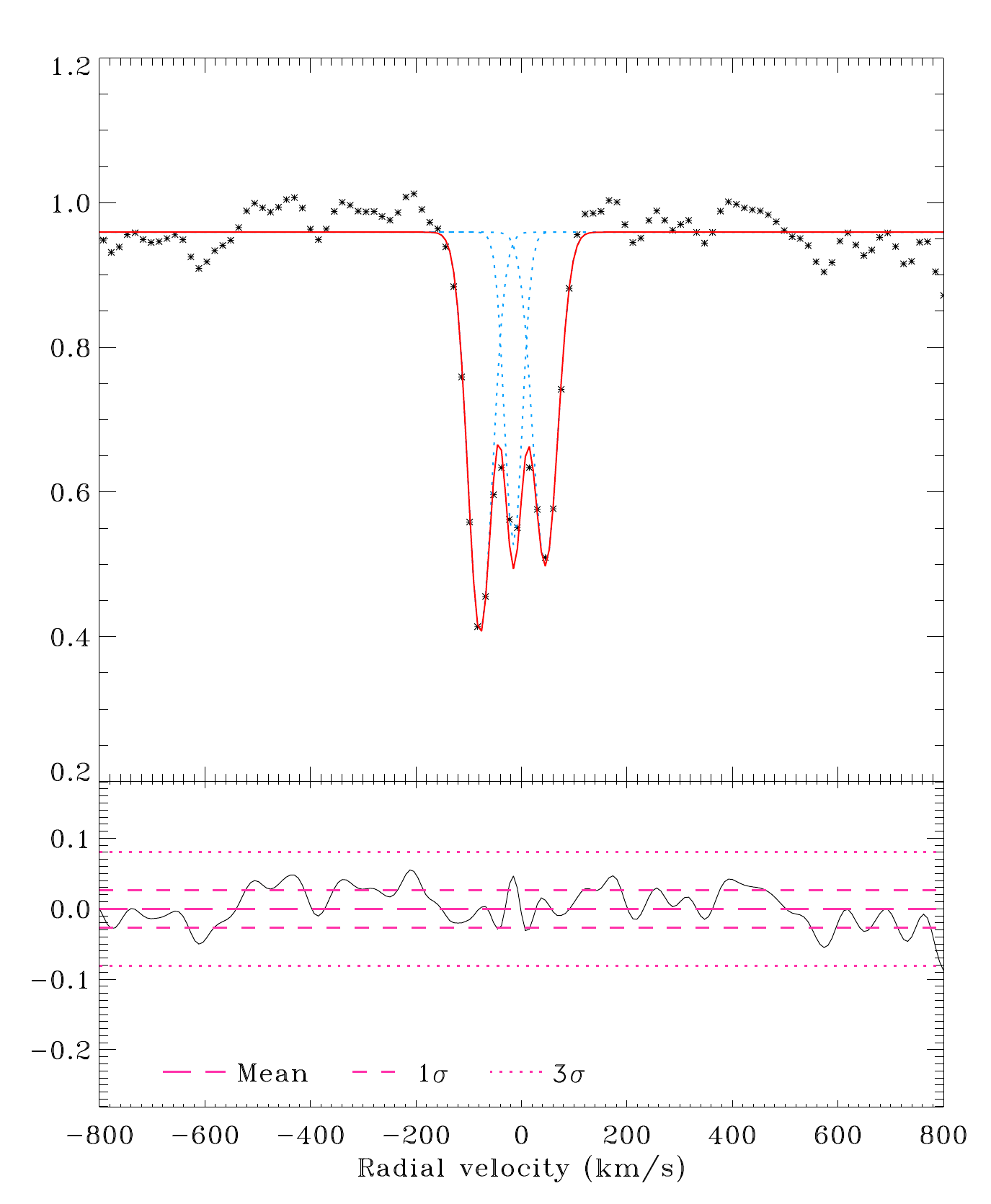}
\caption{IDS spectra of the source \#92 obtained in a time span of about 24 hours (\textit{top panels}) and their associated CCF profiles (\textit{bottom panels}).}
\label{Fig:Index92_SB2_SB3}
\end{figure*}

\begin{figure}
\centering 
\hspace{-0.5cm}
\includegraphics[width=8.5cm]{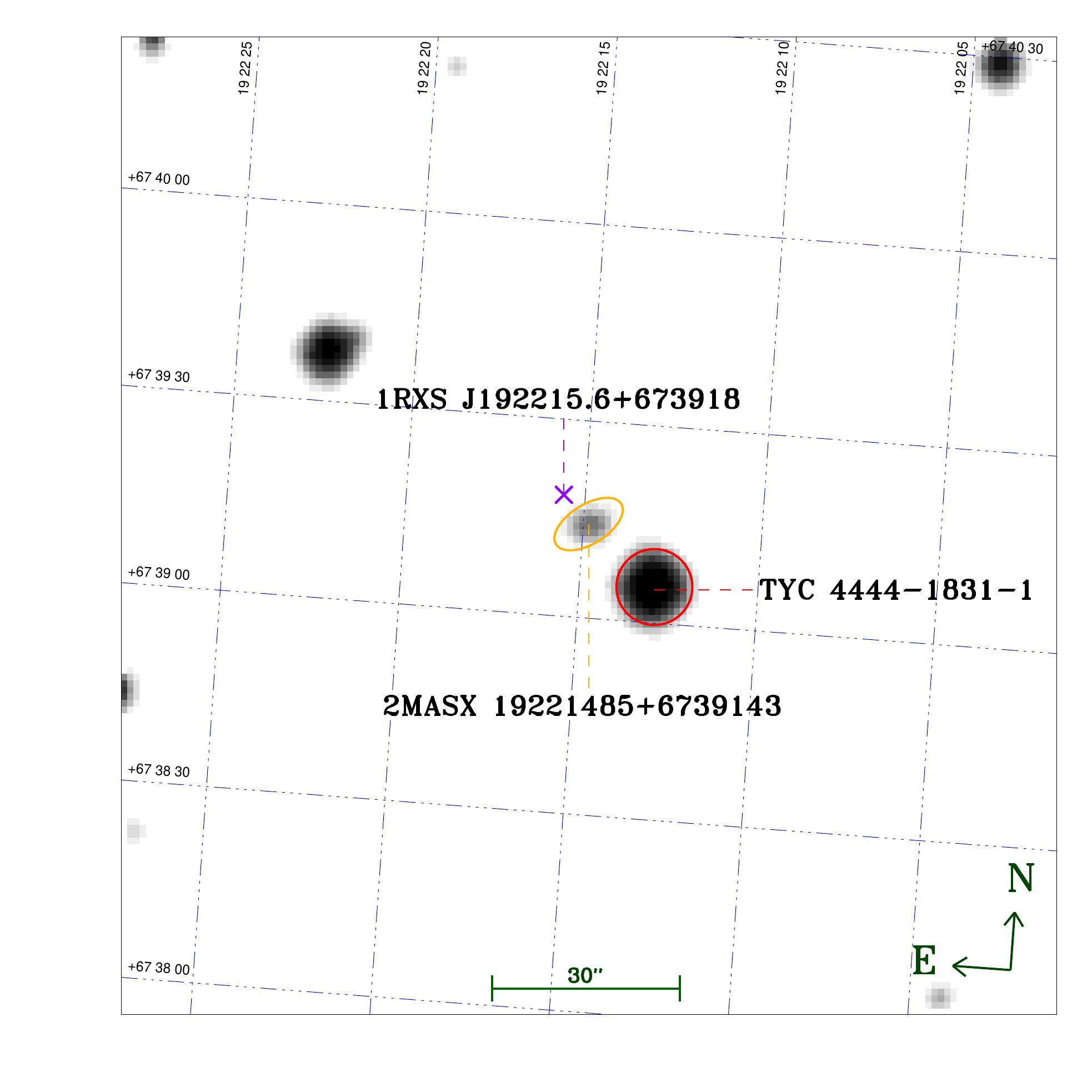} 
\caption{Photographic $B_{J}$-band image of the second Palomar Observatory Sky Survey (POSS-II, epoch J1992.57), centered on the 2MASS extended source \object{2MASX~19221485+6739143}. We also show the positions of the X-ray source \object{1RXS~J192215.6+673918} (\#$99$) and the optical source \object{TYC~4444-1831-1}.}
\label{Fig:2MASX_1922_6739_field}
\end{figure}

\begin{itemize}
\item{\object{1RXS J003904.2+791912} (Source \#6)}: 
The infrared counterpart identified for this source corresponds to the faint companion of the source \object{[TNK2005] 5} that is known as a WTTS in Cepheus \citep{Tachihara05}. Our target was only detected in photometry and classified as an M2-type companion candidate by the aforementioned authors. During one of our observing nights, both components were positioned simultaneously on the slit. These display a deep lithium absorption and the brightest component is an SB2 system (Fig.~\ref{Fig:triple_System}). Based on our analysis, this source therefore is a hierarchical triple system composed of a close inner binary plus a tertiary component in a long-period orbit. Moreover, we found an additional wide-separation companion ($\rho=3\farcm44 \pm 0.185$~mas; see Table~\ref{Tab:VB_Gaia}).  All these stars are associated to Cep\,I.

\medskip
\item{\object{1RXS J010112.8+570839} (Source \#8)}: 
\label{appendix:Giant_Dwarf}
While its intrinsic color agrees with a late-M giant (see Fig.~\ref{Fig:2mass_photometry_color}, along with Fig.~2 of \citealt{JimenezEsteban2012}), we classified it as a K5 giant with a large uncertainty on its surface gravity ($\log g = 4.20 \pm 0.95$ dex). This is consistent with the extinction of $A_V\sim0.94$~mag derived from our SED analysis (Sect.~\ref{subsection:SED}). Moreover, this source is more than $2$~mag fainter compared to giants included in the LSPM-North catalog \citep{LS2005}, with a similar $J$--$K_{\rm s}$ color index. 

\medskip
\item{\object{1RXS J174104.6+842458} (Source \#81)}: 
This visual binary is composed of an M1 and an M2.5-type star, with a small angular separation (Table~\ref{Tab:VB_Gaia}). These are the only M-type stars displaying the \ion{He}{i} $\lambda$6678\,\AA~line in emission (Fig.~\ref{Fig:Spectral_Typing_MStars}).

\medskip
\item{\object{1RXS J185131.1+584258} (Source \#92)}: 
\label{appendix:SB2_SB3}
We observed this source thrice. At the first epoch, it appears as an SB3 system. We then re-observed this star during our IDS runs in order to learn more about the temporal variation of this system. Since the source turned out to be an SB2 system, we repeated the observation. Within this time span of about one day its spectrum has drastically changed and a third peaks is clearly visible in the CCF profile (Fig.~\ref{Fig:Index92_SB2_SB3}). At this stage, it is complicated to put forward any hypothesis about the dynamics of this system, but the radial velocity of the three components seem to change quickly. This is in contrast with the three SB3 systems analyzed by \citet{2008A&A...490..737K}.

\medskip
\item{\object{1RXS J192215.6+673918} (Source \#99)}: 
\label{appendix:galaxy}
We selected the source 2MASS J19221478+6739142 as its infrared counterpart. With an angular separation of $\sim$$6\farcm42$, this source fulfilled all our criteria. Its infrared colors suggest a late M-type star according to \citet{West08}. At this position, however, we observed a giant star ($T_{\rm eff}~=~4835\pm102$\,K, $\log g~=~2.80\pm0.23$\,dex, and spectral type = G9\,III) with a small lithium absorption line: $EW$(Li)  $=16 \pm 11$\,m$\AA$, and $\log$~N(Li)~$\sim 0.9$\,dex. Seeking to understand the inconsistency between the photometric and spectroscopic data, we found that  \object{2MASX~19221485+6739143} from the catalog of the 2MASS extended sources \citep{Cat2MASX03} has similar coordinates. Therefore the infrared counterpart of this target is indeed the galaxy \object{GALEX~J192214.8+673914}. 

\smallskip
It appears that we observed the source \object{TYC~4444-1831-1} (or~\object{2MASS~J19221289+6739053}) that is somewhat more distant with a separation of  $\sim$$20\farcm42$ to the X-ray source. We illustrated the space configuration of these three sources on Fig.~\ref{Fig:2MASX_1922_6739_field}. Moreover, its photometry is consistent with a late-G or early-K star and its small proper motions are in agreement with the expectation for a giant star. The probabilities of \citet{Flesch10a} are rather close that the radio/X-ray association is erroneous ($51$\,\%) and this X-ray source is a star ($42$\,\%), while the probability to be a galaxy is only of $7$\,\%. 

\medskip
\item{\object{1RXS J221055.6+632339} (Source \#139)}: 
\label{appendix:Id139}
As in \citet{Motch1997c}, we consider the early-type star \object{STF~2879\,C} (Fig.~\ref{Fig:s139_spec}) as being its optical counterpart due to its $B$--$V$ color index of about $0.75\pm0.44$~mag from the GSC\,II photometric data. In contrast, the APASS catalog reported a more reliable color index ($B$--$V=0.320\pm0.031$~mag). This source has probably passed through the cracks because of its proximity to the B5 binary \object{V447~Cep} whose extended halo of light could affect the entire photometry of our target (Fig.~\ref{Fig:s139_field}). Nevertheless, the particularity of this source is not limited to this aspect. Its magnitudes quoted in the final release of the WISE all-sky survey catalog are also atypical. With a $W1$ magnitude of $10.819\pm0.023$~mag and a $W1$--$W4$ of $4.691\pm0.050$~mag, this source falls in the area mainly populated by T\,Tauri stars displaying an infrared excess (Guillout, priv. com.). 

\begin{figure}
\centering 
\hspace{-0.6cm}
\includegraphics[width=9.5cm]{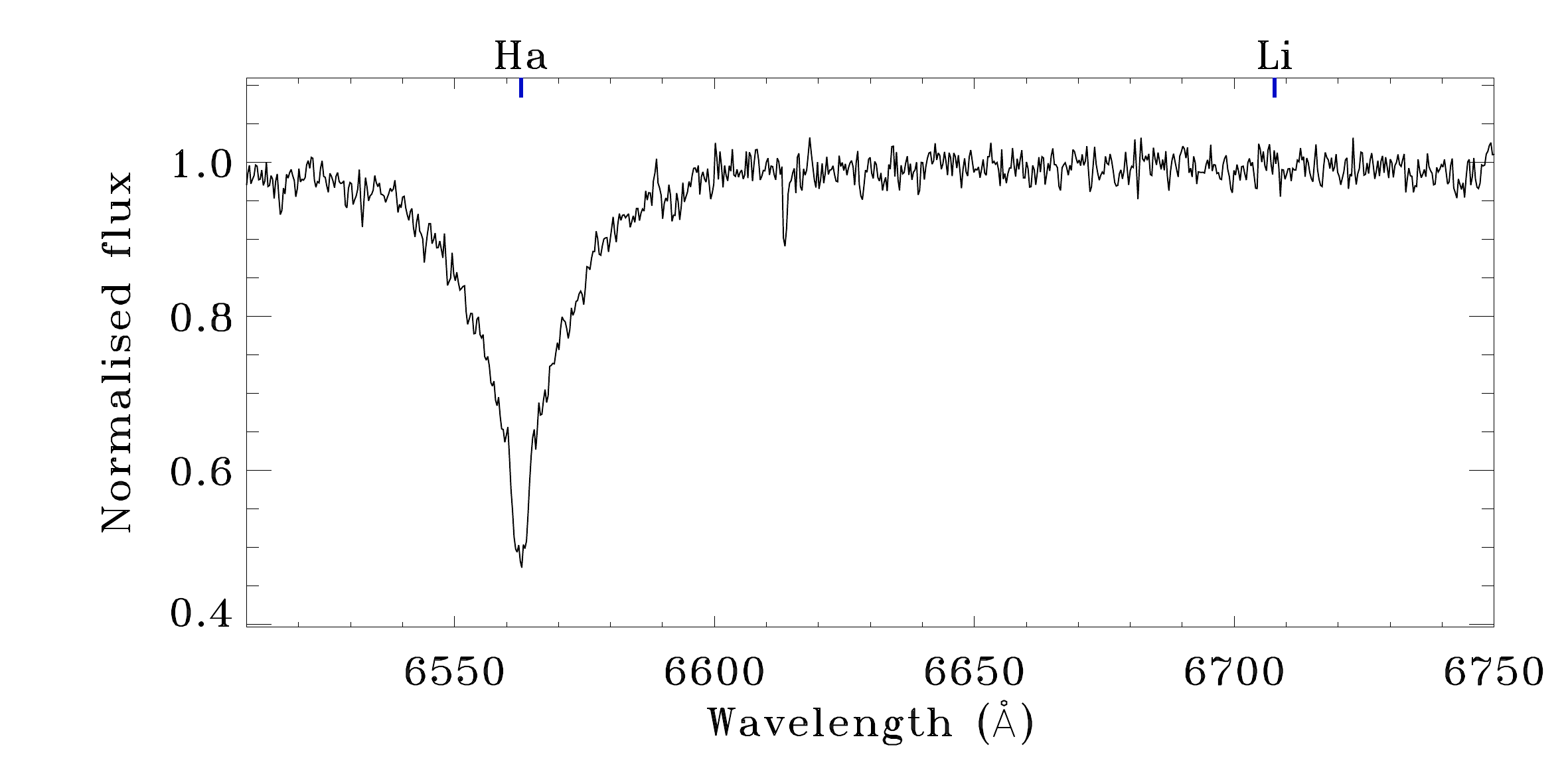}
\caption{IDS spectrum of source \#$139$.}
\label{Fig:s139_spec}
\end{figure}

\begin{figure}
\centering 
\hspace{-0.5cm}
\includegraphics[width=8.5cm]{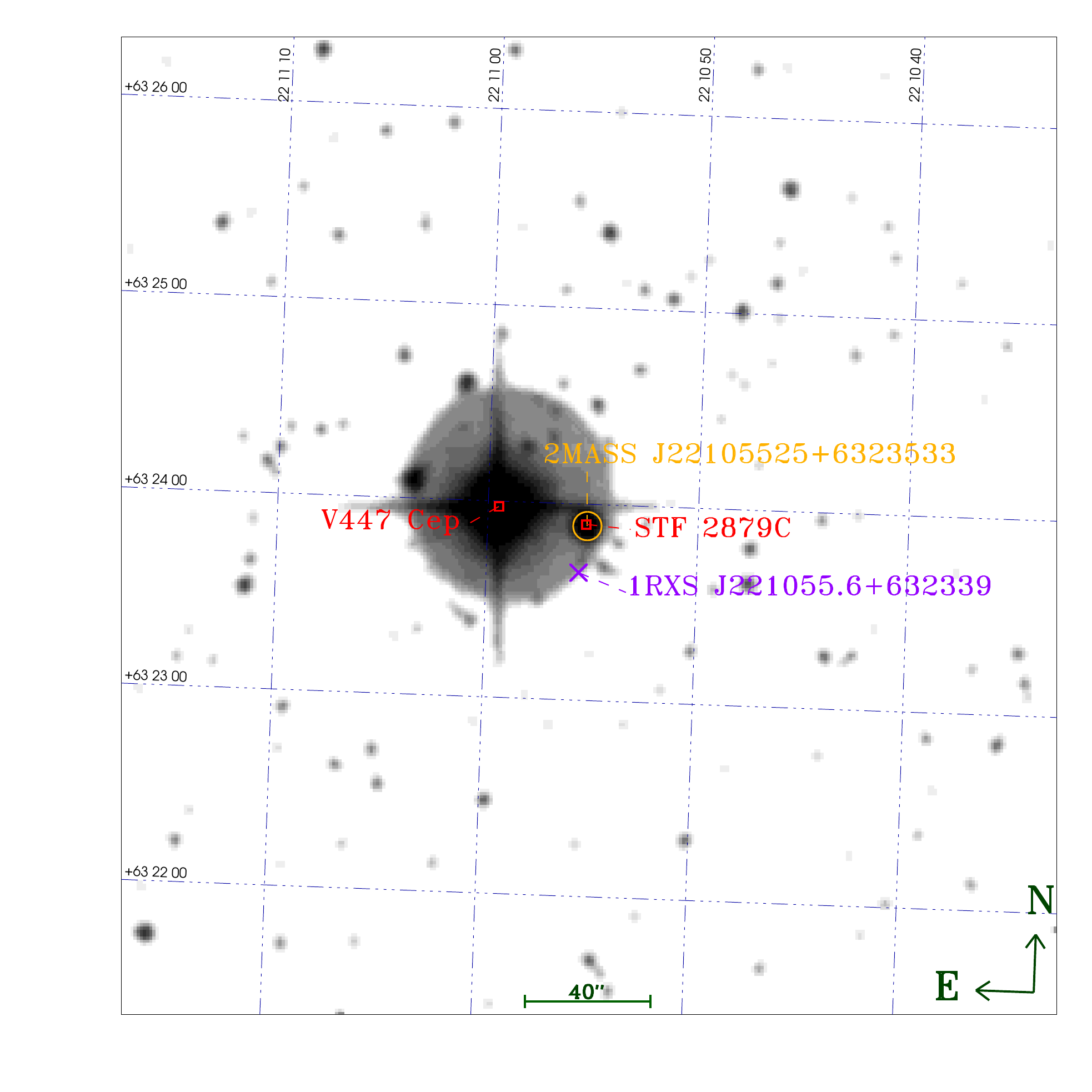} 
\caption{Photographic $B_{J}$-band image of the second Palomar Observatory Sky Survey (POSS-II), centered on the infrared source 2MASS~J22105525+6323533. We mark the locus of the X-ray source 1RXS~J221055.6+632339 (\#$139$) and its possible optical counterpart STF~2879~C, which is the companion of the bright B5 binary V447~Cep.}
\label{Fig:s139_field}
\end{figure}

\medskip
\item{\object{1RXS\,184257} (Source \#160)}:
This source \citep[=~\object{[KP93] 2$-$43} in][]{Kun1993} is too faint in optical to acquire a spectrum with the instruments at our disposal in a reasonable exposure time. It is known as a visual binary whose primary component is a young star ($1$-Myr-old WTTS in \citealt{Kun2009} or a CTTS in \citealt{Simon2009}) and one of the strongest X-ray emitters during the XMM-Newton and Chandra observations of the L1251 cloud \citep{Simon2006, Simon2009}. 

\medskip
\item{\object{1RXS J003941.9+790526} (Source G4)}:
We followed a procedure similar to \citet{Frasca2006} to derive its orbital parameters. For this purpose we combined the radial velocities that we previously reported in Paper~III with those obtained during the observing runs analyzed in the current paper. However it turns out that most measurements (HJD~$=2455077$--$2455111$) only cover half of the orbit of this system. From our preliminary analysis the period would be estimated at $\sim$$70$ days.

\end{itemize}

\section{Additional materials online}
\label{appendix:Spec_regions_online}

\begin{table*}[!t]
\caption{Basic parameters of reference stars observed during our runs. We adopted the spectral type, atmospheric parameters, and projected rotational velocities found in the literature (see references below). 
We also determined the mean $RV$ values from those listed in the online \'Elodie archive, along with the number of observations used o$_{RV}$.} 
\smallskip
\begin{center}
{\scriptsize
\begin{tabular}{llccccccccc}
\hline
\hline
\noalign{\smallskip}
 Star name & SpT $^{a}$ & Ref. & $T_{\rm eff}$ $ ^{b}$  & $\log g$ $ ^{b}$  & [Fe/H] $ ^{b}$  & Ref. & $v \sin i $ $ ^{c}$ & Ref. & $\langle RV \rangle$ $^{d}$& o$_{RV}$ \\
\noalign{\smallskip}
 &  &  & (K) & (cm$^{2}$ s$^{-1}$) & (dex) &  & (km s$^{-1}$) &  & (km s$^{-1}$) \\
\noalign{\smallskip}
\hline
\noalign{\smallskip}
\object{HD 187691} 	& F8\,V 		& A3 & 	$6173$	&	$4.25$	&	~~$0.04$	&	B15		&	$3.6$		&	C3~~	&	 ~~~~~~$0.03$				&	$6$	\\
\object{HD 19373} 	& F9.5\,V 		& A4 & 	$6008$	&	$4.33$	&	~~$0.15$	&	B12		&	$4.5$		&	 B11		&	~~~~~$49.4$				&	$1$	\\
\object{HD 22879} 	& G0\,V 		& A3 & 	$5759$ 	&	$4.25$	&	$-0.85$ 	&	B14		&	$2.3$		&	C3~~	&	~~~$120.31$				&	$4$	\\
\object{HD 10307} 	& G1\,V 		& A4 & 	$5859$	&	$4.27$	&	~~$0.04$	&	B15		&	$2.3$		&	C3~~	&	~~~~~~$4.97$				&	$10$	\\
\object{HD 159222} 	& G1\,V 		& A4 & 	$5851$	&	$4.41$	&	~~$0.16$	&	B12		&	$3.3$		&	B11		&	~~$-51.62$				&	$15$	\\
\object{HD 196850} 	& G1\,V 		& A4 & 	$5838$	&	$4.37$	&	$-0.09$ 	&	B16		&	$2.0$		&	B02		&	~~$-21.07$				&	$9$	\\
\object{HD 197076} 	& G1\,V 		& A4 & 	$5828$	&	$4.45$	&	$-0.10$	&	B15		&	$2.7$		&	C7~~	&	~~$-35.44$				&	$13$	\\
\object{HD 19445} 	& G2\,V Fe-3 	& A4 & 	$5918$	&	$4.41$	&	$-1.89$	&	B08		&	$4.1$		&	C2~~	&	$-139.84$					&	$2$	\\
\object{HD 193664} 	& G3\,V 		& A5 & 	$5886$	&	$4.48$	&	$-0.09$	&	B02		&	$1.3$		&	B02		&	~~~~$-4.51$	 			&	$28$	\\
\object{HD 65583} 	& G8\,V 		& A1& 	$5279$	&	$4.76$	&	$-0.69$	&	B02		&	$3.3$		&	C8~~	&	~~~~~~~~$14.89$ $ ^{3}$		&	$20$	\\
\object{HD 10700} 	& G8.5\,V 		& A6 & 	$5290$	&	$4.46$	&	$-0.48$	&	B11		&	$0.9$		&	C3~~	&	~~$-16.65$				&	$1$	\\
\object{HD 10780} 	& G9\,V 		& A4 & 	$5309$	&	$4.56$	&	~~$0.00$	&	B04		&	$1.3$		&	B01		&	~~~~~~$2.72$				&	$33$	\\
\object{HD 12051} 	& G9\,V 		& A4 & 	$5458$	&	$4.55$	&	~~$0.24$	&	B07		&	$0.5$		&	B06		&	~~~-$35.22$				&	$4$	\\
\object{HD 185144} 	& G9\,V 		& A4 & 	$5204$	&	$4.37$	&	$-0.26$	&	B15		&	$0.8$		&	C3~~	&	~~~~~$26.61$				&	$1$	\\
\object{HD 182488} 	& G9+\,V 		& A4 & 	$5393$	&	$4.55$	&	~~$0.22$	&	B05		&	$1.2$		&	C3~~	&	~~$-21.57$				&	$12$	\\
\object{HD 3651} 	& K0\,V 		& A4 & 	$5218$	&	$4.52$	&	~~$0.20$	&	B05		&	$1.2$		&	C2~~	&	~~$-33.08$				&	$2$	\\
\object{HD 38230} 	& K0\,V 		& A4 & 	$5174$	&	$4.53$	&	$-0.08$	&	B02		&	$0.9$		&	C3~~	&	~~$-29.22$				&	$33$	\\
\object{HD 10476} 	& K1\,V 		& A6 & 	$5173$	&	$4.59$	&	$-0.08$	&	B04		&	$1.7$		&	B01		&	~~$-33.76$				&	$3$	\\
\object{HD 4628} 	& K2\,V 		& A4 & 	$4905$	&	$4.60$	&	$-0.36$	&	B07		&	$1.5$		&	B06		&	~~$-10.35$				&	$1$	\\
\object{HD 73667} 	& K2\,V 		& A4 & 	$4884$	&	$4.40$	&	$-0.58$	&	B07		&	$1.2$		&	B01		&	~~$-12.12$				&	$21$	\\
\object{HD 166620} 	& K2\,V		& A4 & 	$5007$	&	$4.62$	&	$-0.24$	&	B04		&	$0.6$		&	C3~~	&	~~$-19.59$				&	$3$	\\
\object{HD 16160} 	& K3\,V		& A6 & 	$4829$	&	$4.60$	&	$-0.16$	&	B07		&	$0.9$		&	B06		&	~~~~~$25.73$				&	$1$	\\
\object{HD 219134} 	& K3\,V 		& A4 & 	$4835$	&	$4.56$	&	~~$0.12$	&	B02		&	$1.8$		&	B01		&	~~$-18.69$				&	$4$	\\
\object{HD 190007} 	& K4\,V 		& A4 & 	$4786$	&	$4.31$	&	$-0.02$	&	B10		&	$0.9$		&	B10		&	~~$-30.43$				&	$10$	\\
\object{HD 201091} 	& K5\,V 		& A4 & 	$4236$	&	$4.50$	&	$-0.03$	&	B07		&	$3.8$		&	B06		&	~~$-65.51$				&	$5$	\\
\object{HD 221503} 	& K6\,V 		& A6 & 	$4270$	&	$4.99$	&	~~$0.02$	&	B13		&	$2.5$		&	C4~~	&	~~~~~~$-0.96$ $ ^{5}$		&	$6$	\\
\object{HD 201092} 	& K7\,V 		& A4 & 	$4200$	&	$4.60$	&	$-0.63$	&	B09		&	$1.6$		&	C3~~	&	~~$-64.15$				&	$5$	\\
\object{GJ 270} 	& M0\,V 		& A9 & 	$3856$	&	$4.69$	& 	\dots		& 	B17		& 	$\leqslant 3.0$	&	C9~~	&	~~~~~$-69.78$ $^{4}$		&	$1$	\\
\object{GJ 373} 	& M0\,V 		& A9 & 	$3820$	&	$4.68$	& 	\dots		& 	B17		& 	$\leqslant 2.0$	&	C9~~	&	~~~~~~~$15.13$ $^{4}$		&	$4$	\\
\object{GJ 459.3} 	& M0\,V 		& A9 & 	$3852$	&	$4.66$	& 	\dots		& 	B17		& 	$\leqslant 3.0$	&	C9~~	&	~~~~$-0.64$				&	$1$	\\
\object{GJ 809} 	& M0.5\,V 		& A9 & 	$3720$	&	$4.67$	&	$-0.13$	&	B01		&	$\leqslant 2.0$	&	C9~~	&	~~$-17.55$				&	$10$	\\
\object{GJ 686} 	& M1\,V 		& A9 & 	$3611$	&	$4.84$	&	$-0.44$	&	A8~~	&	$\leqslant 2.0$	&	C9~~	&	~~~~$-9.87$				&	$11$	\\
\object{GJ 205} 	& M1.5\,V 		& A9 & 	$3626$	&	$4.80$	&	~$0.60$	&	B09		&	$\leqslant 2.0$	&	C9~~ 	&	~~~~~~~$8.25$			&	$15$	\\
\object{GJ 411} 	& M1.5\,V 		& A9 & 	$3671$	&	$4.89$	& 	\dots		& 	B17 		& 	$\leqslant 2.0$	&	C9~~	&	~~~~~~$-84.71$ $^{4}$		&	$7$	\\
\object{GJ 552} 	& M2\,V 		& A9 & 	$3574$	&	$4.79$	& 	\dots		& 	B17		& 	$\leqslant 2.0$	&	C9~~	&	~~~~~~~~~$7.36$ $^{4}$		&	$2$	\\
\object{GJ 362} 	& M3\,V 		& A9 & 	$3430$	&	$4.85$	& 	\dots		& 	B17		& 	$\leqslant 2.0$	&	C9~~ 	&	~~~~~~$6.21$				&	$2$	\\
\object{GJ 687} 	& M3\,V 		& A9 & 	$3340$	&	$4.82$	&	~$0.15$	&	B01		&	$\leqslant 2.5$	&	C9~~	&	~~$-29.16$				&	$9$	\\
\object{GJ 447} 	& M4\,V 		& A9 & 	$3192$	&	$5.07$	& 	\dots		& 	B17		& 	$\leqslant 2.0$	&	C9~~	&	~~~~~$-30.86$ $^{4}$		&	$1$	\\
\object{GJ 1227} 	& M4.5\,V 		& A8 & 	$3072$	&	$5.01$	&	\dots		& 	B17		&	$\leqslant 2.0$	&	C9~~	&	~~$-14.04$				&	$6$	\\
\object{HD 84737} 	& G0\,IV-V 	& A4 & 	$5934$	&	$4.16$	&	~$0.16$	&	B12		&	$3.0$		&	B11		&	~~~~~~~~~$4.79$ $ ^{1}$		&	$25$	\\
\object{HD 51000} 	& G5\,III 		& A2 & 	$5180$	&	$3.05$	&	$-0.04$	&	B03		&	$2.3$		&	B03		&	~~~~$-9.36$				&	$1$	\\
\object{HD 62509} 	& G9\,III 		& A7 & 	$4955$	&	$3.07$	&	~$0.16$	&	B15		&	$2.4$		&	C3~~	&	~~~~~~$3.24$				&	$2$	\\
\object{HD 124897} 	& K0\,III 		& A4 & 	$4280$	&	$1.69$	&	$-0.52$	&	B06		&	$2.5$		&	C3~~	&	~~~~$-5.30$				&	$8$	\\
\object{HD 12929} 	& K1\,III 		& A4 & 	$4546$	&	$2.40$	&	$-0.24$	&	B04		&	$1.6$		&	C3~~	&	~~~~~$-14.64$ $ ^{2}$		&	$170$	\\
\object{HD 26162} 	& K2\,III 		& A2 & 	$4800$	&	$2.90$	&	$0.06$	&	B03		&	$4.0$		&	B03		&	~~~~$24.74$				&	$7$	\\
\object{HD 29139} 	& K5\,III 		& A6 & 	$3891$	&	$1.20$	&	$-0.15$	&	B06		&	$2.6$		&	C3~~	&	~~~~$54.02$				&	$8$	\\
\noalign{\smallskip}
\hline 
\noalign{\smallskip}
\multicolumn{11}{l}{
\begin{minipage}{12.5cm}
\scriptsize \textbf{References.} \\
$^{a}$~For the G- and K-type sources, we list the spectral type most often cited in the last release of the catalog of stellar spectral classification \citep{Skiff2010}: \\
A1 = \citet{Harlan1969}, 
A2 = \citet{Cowley1979}, 
A3 = \citet{Gray2001}, 
A4 = \citet{Gray2003}, 
A5 = \citet{Abt2006}, 
A6 = \citet{Gray2006},
A7 = \citet{Abt2008}, and
A8 = \citet{Jenkins2009}. 

For the M-type stars, we usually adopt the spectral type from \citet{2015A&A...577A.128A}, labeled A9. \\
$^{b}$~The atmospheric parameters come from: 
B01 = \citet{Woolf2005}, 
B02 = \citet{Valenti2005}, 
B03= \citet{Hekker2007}, 
B04 = \citet{Ramirez2007}, 
B05 = \citet{Fuhrmann2008}, 
B06 = \citet{Melendez2008}, 
B07 = \citet{Mishenina2008}, 
B08 = \citet{Reddy2008}, 
B09 = \citet{Frasca2009}, 
B10 = \citet{Guillout09}, 
B11 = \citet{Bruntt2010}, 
B12 = \citet{Gonzalez2010}, 
B13 = \citet{Casagrande2010},
B14 = \citet{Nissen2010},  
B15 = \citet{daSilva2011}, 
B16 = \citet{Casagrande2011}, and
B17 = \citet{2018AJ....156..102S}.
Most of them are available in the PASTEL catalog \citep{Pastel_Cat2010}. \\
 $^{c}$~$v \sin i$ values from:
C1 = \citet{Delfosse1998}, 
C2 = \citet{Fischer2005},
C3 = the catalog of stellar rotational velocities of \citet{Glebocki2005}, 
C4 = \citet{Torres06}, 
C5 = \citet{Houdebine2010}, 
C6 = \citet{Houdebine2011},
C7 = \citet{Herrero2012}, 
C8 = \citet{2014AJ....148...70M} , and 
C9 = \citet{2018A&A...614A..76J}.
Some of these publications contain also an AP determination. \\
$^{d}$~We also made use of the $RV$ values from $^{1}$~\citet{Soubiran1998}, $^{2}$~\citet{Famaey2005}, $^{3}$~\citet{2013AA...552A..64S}, $^{4}$~\citet{2018A&A...614A..76J}, and $^{5}$~\citet{2018A&A...616A...7S}. \\
\end{minipage}}
\end{tabular}
\label{tab:StandardStars}
}
\end{center}
\end{table*}

All long online tables are available at the CDS. Table~\ref{Tab:List_Candidates} summarizes the optical and infrared names of each X-ray source, along with those appearing in Simbad and some main parameters coming from the literature. The three additional online tables present our results. Table~\ref{Tab:AP_CepSurv} reports the radial velocity of all the stars, including the measurements for each component of a spectroscopic system, along with the rotational velocity, atmospheric parameters, and lithium equivalent widths derived for the targets identified as single stars or SB1 systems. Table~\ref{Tab:EWLi_Kinematics_CepSurv} provides the astrometry in \emph{Gaia} DR2, the Galactic positions, and the extinction estimate of our targets, as well as the space velocities of those identified as single stars. Table~\ref{Tab:VB_Gaia}  lists all the sources that are comoving with one of our targets. 

\medskip
To derive the radial and rotational velocities of our targets and to perform their spectral typing, we made use of a smaller library of template spectra taken with FOCES and IDS during our observing runs. Their spectral type, atmospheric parameters, and radial and rotational velocities are given in Table~\ref{tab:StandardStars}. 

\begin{figure}[t]
\centering 
\hspace{-0.5cm}
\vspace{-.4cm}
\includegraphics[width=9cm]{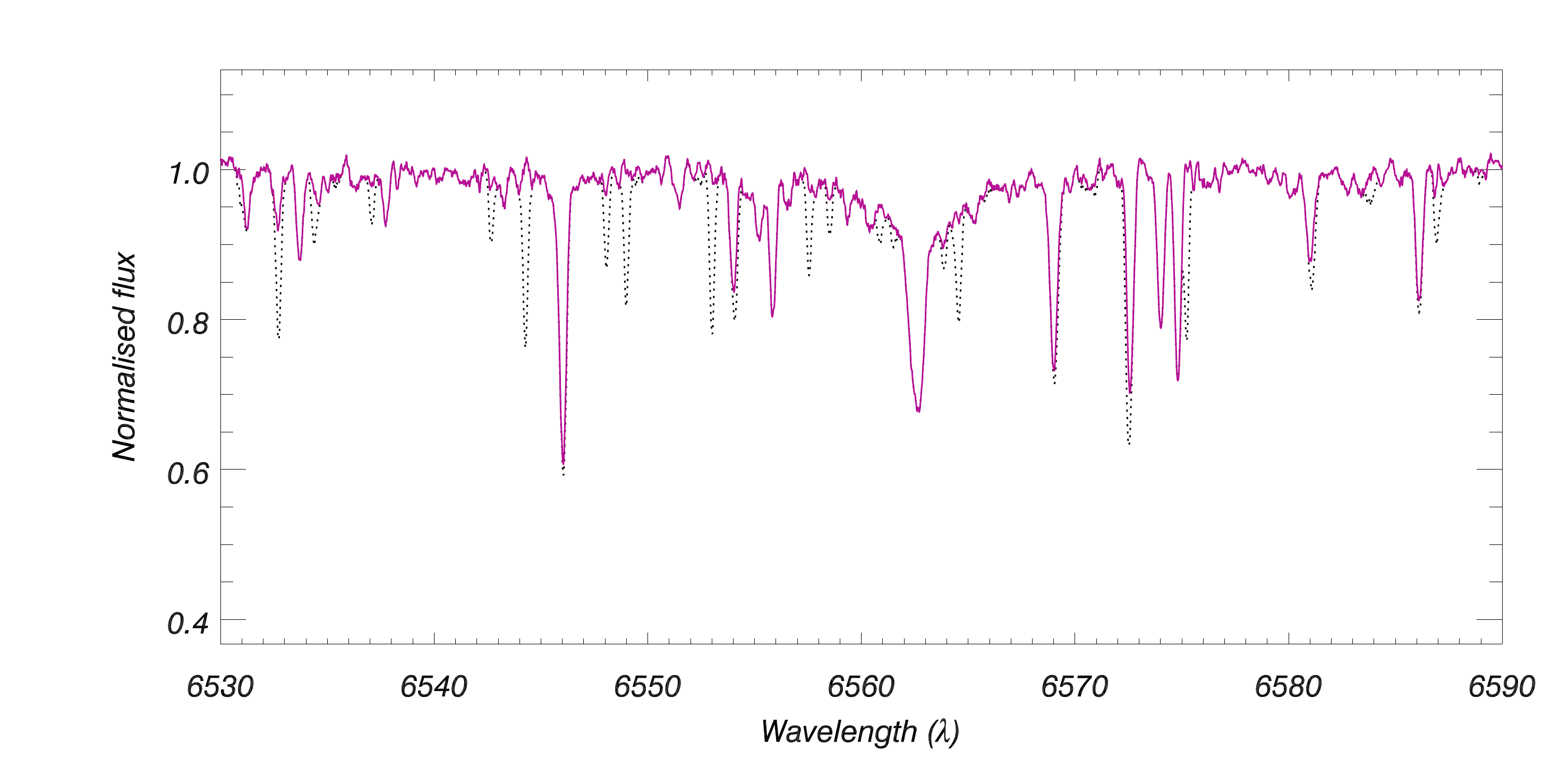} \\
\hspace{-0.5cm}
\includegraphics[width=9cm]{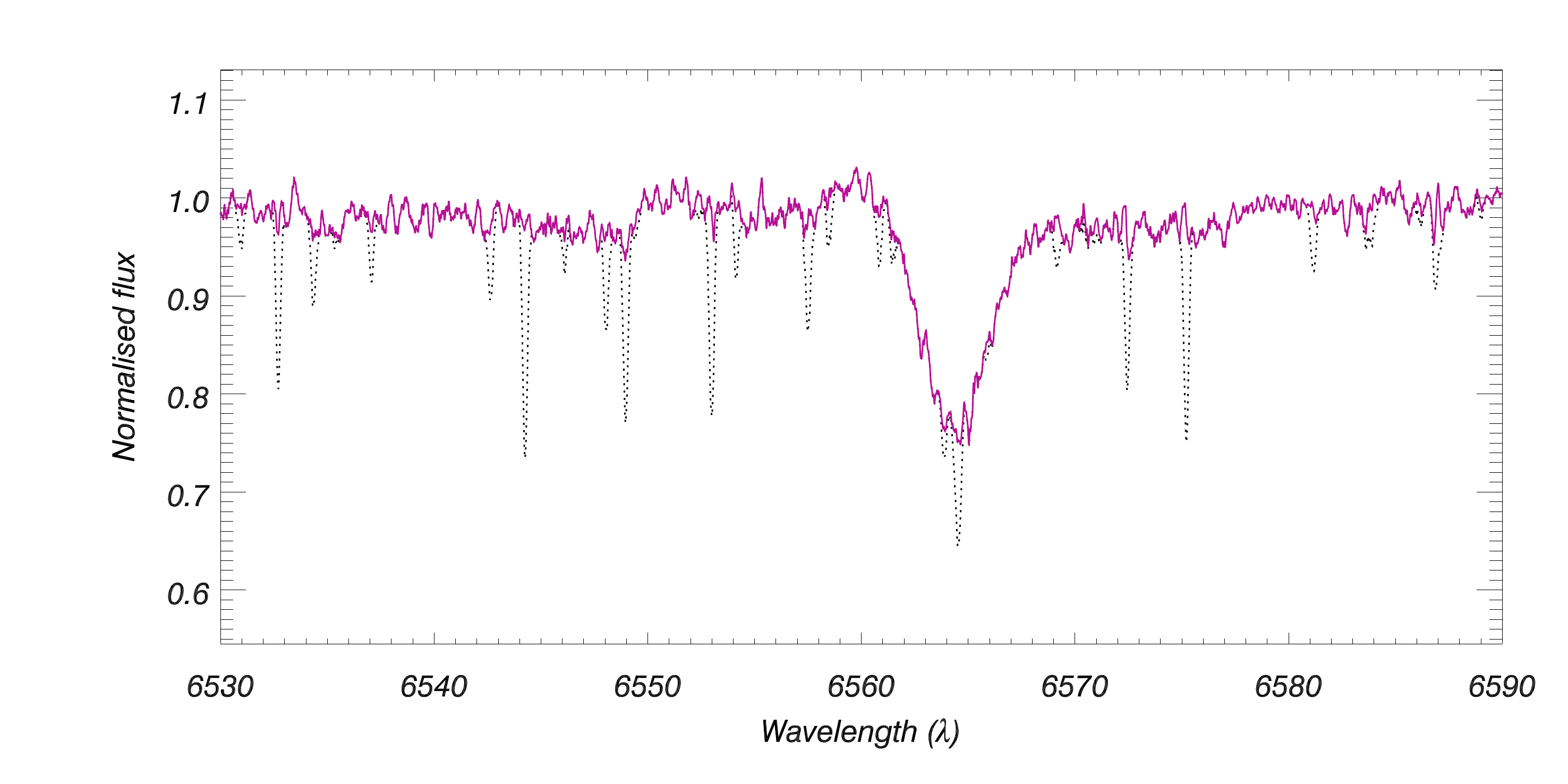} 
\caption{Results of our telluric line removal on SOPHIE spectra for a slow-rotating source (\textit{top panel}) and a fast rotator (\textit{bottom panel}). On each panel, the black dotted lines and purple solid lines show the spectra before and after the application of our procedure, respectively.}
\label{Fig:Telluric_Removal_Sophie}
\end{figure}

\medskip
We show the efficiency of our procedure for removing telluric lines in the case of a few SOPHIE spectra (Fig.~\ref{Fig:Telluric_Removal_Sophie}). For each target observed during our survey, we display only one spectrum in the region around the \ion{Li}{i} $\lambda$6707.8 line (Fig.~\ref{Fig:HaLiSpec_CepSurv}). 

\begin{figure*}[t]
\centering 
\includegraphics[angle=90,width=17cm]{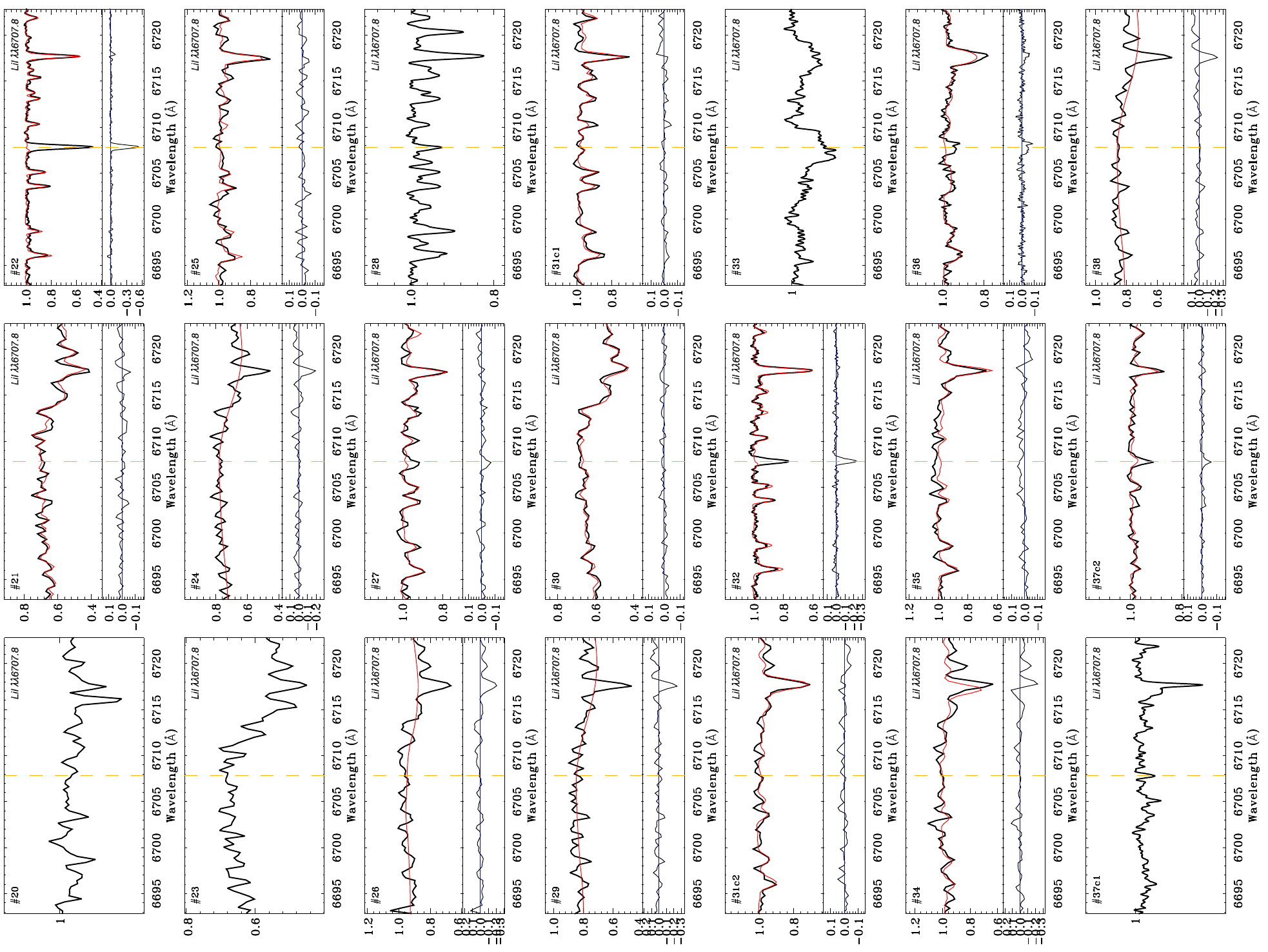}
\includegraphics[angle=90,width=17cm]{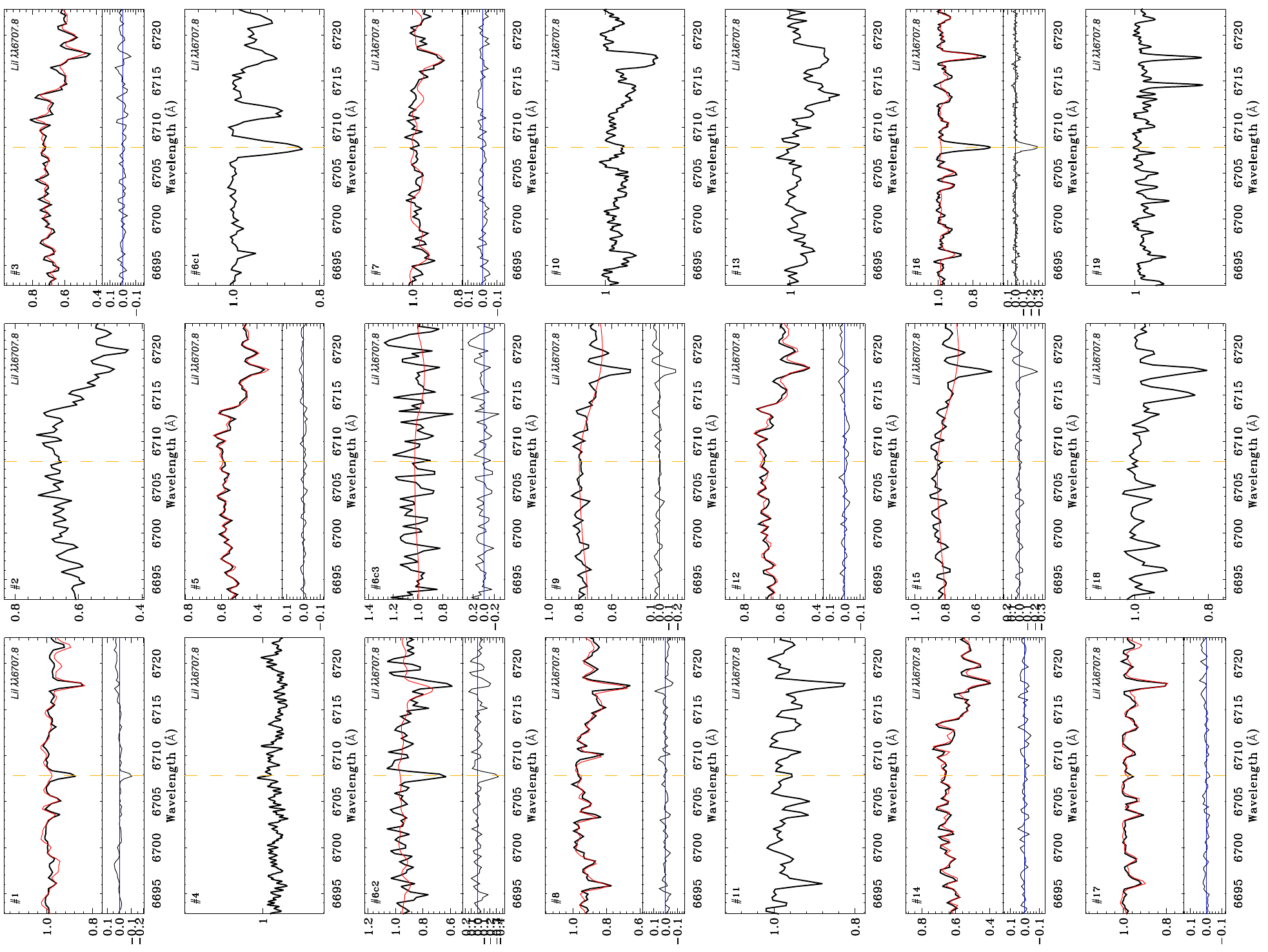}
\caption{Target spectra (black line) around the lithium line (vertical dashed line). We display the best match rotationally broadened to the $v \sin i$ obtained with {\tt ROTFIT} with the red line only for single stars and SB1 systems (\textit{upper panel}), while the difference (target--template) spectrum shows their lithium content (\textit{lower panel}). The number of the source appears in the upper left corner.}
\label{Fig:HaLiSpec_CepSurv}
\end{figure*}

\setcounter{figure}{1}
\begin{figure*}[t]
\centering 
\includegraphics[angle=90,width=17cm]{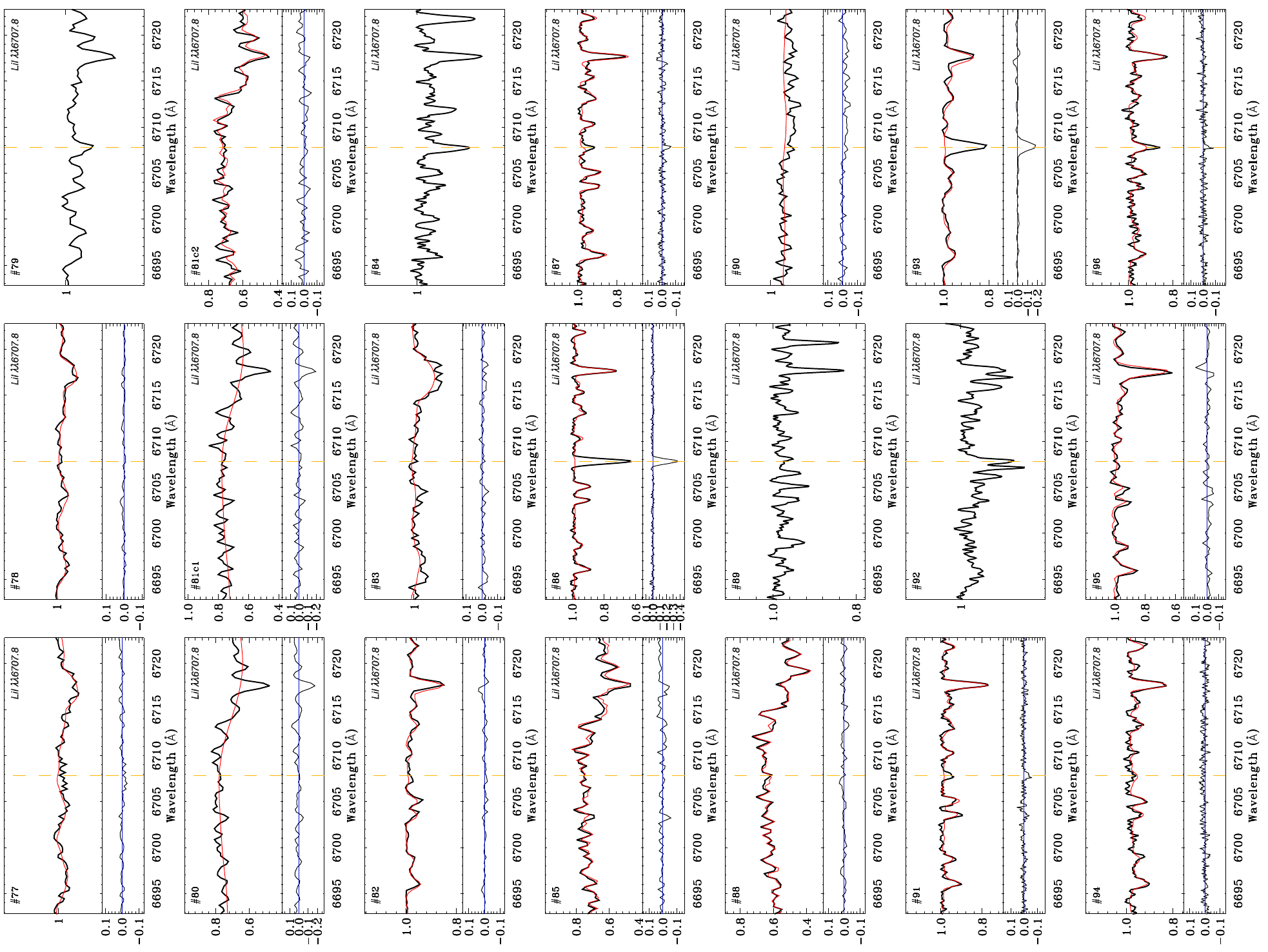}
\includegraphics[angle=90,width=17cm]{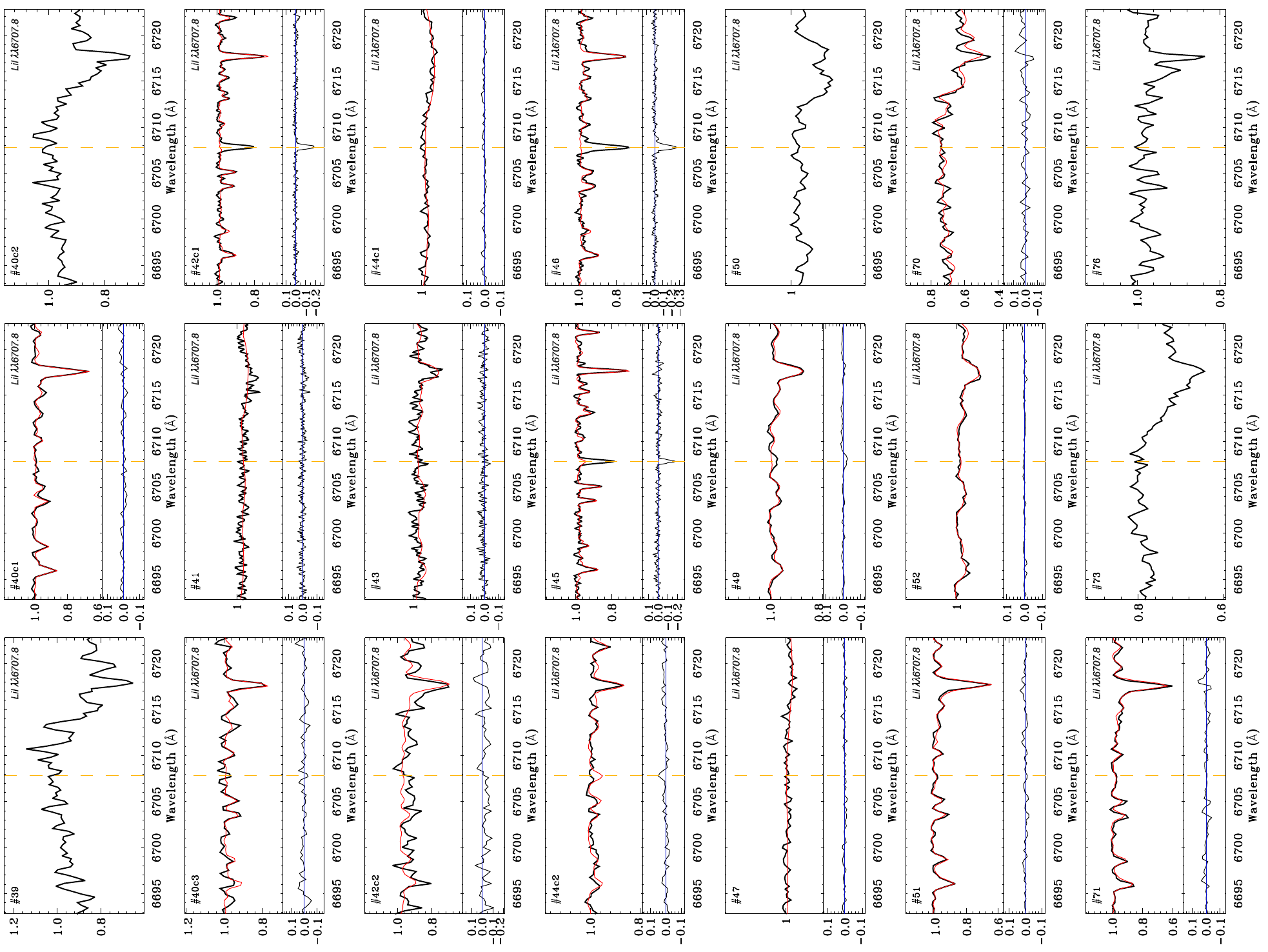}
\caption{Continued}
\end{figure*}

\setcounter{figure}{1}
\begin{figure*}[t]
\centering 
\includegraphics[angle=90,width=17cm]{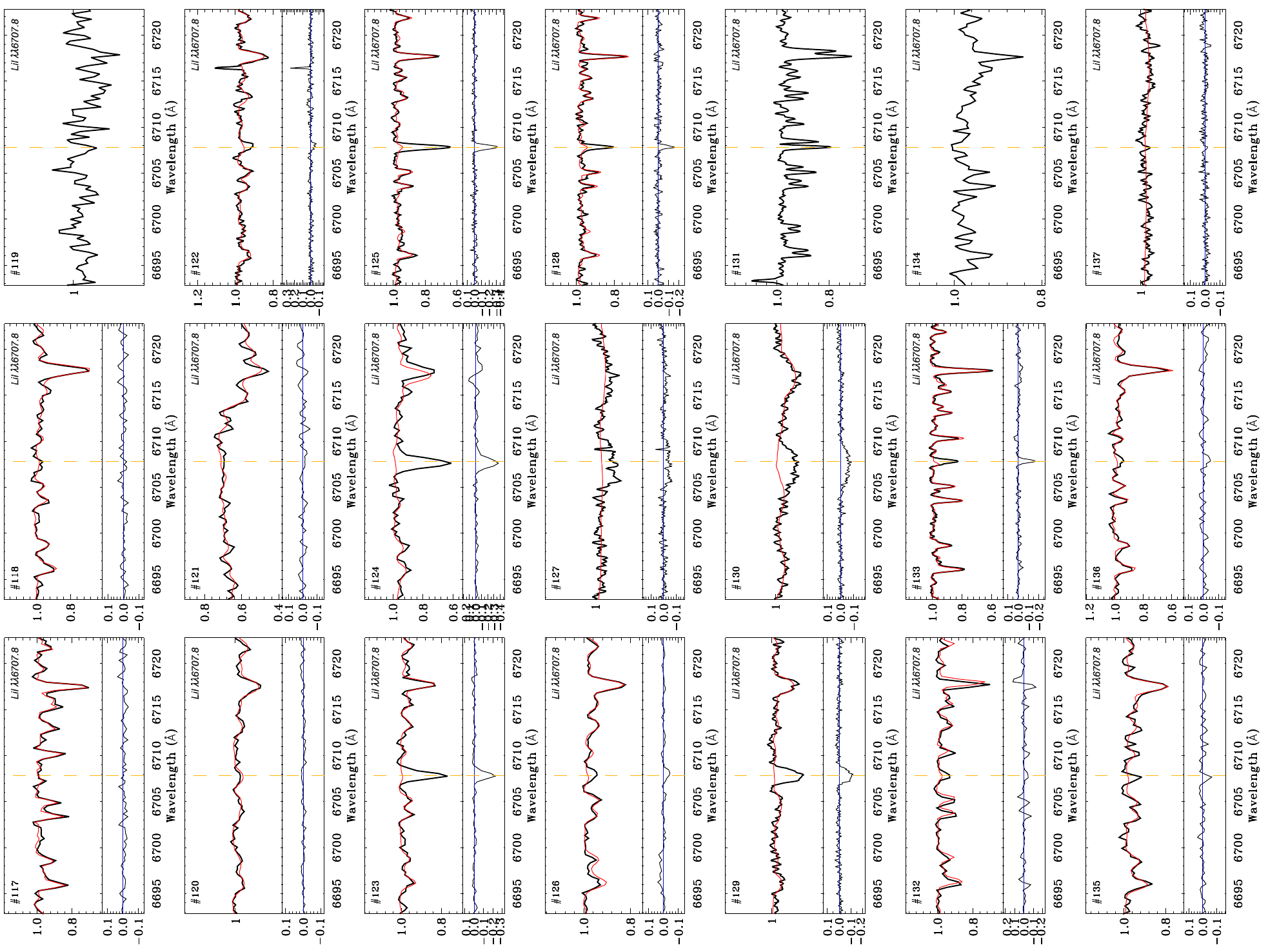}
\includegraphics[angle=90,width=17cm]{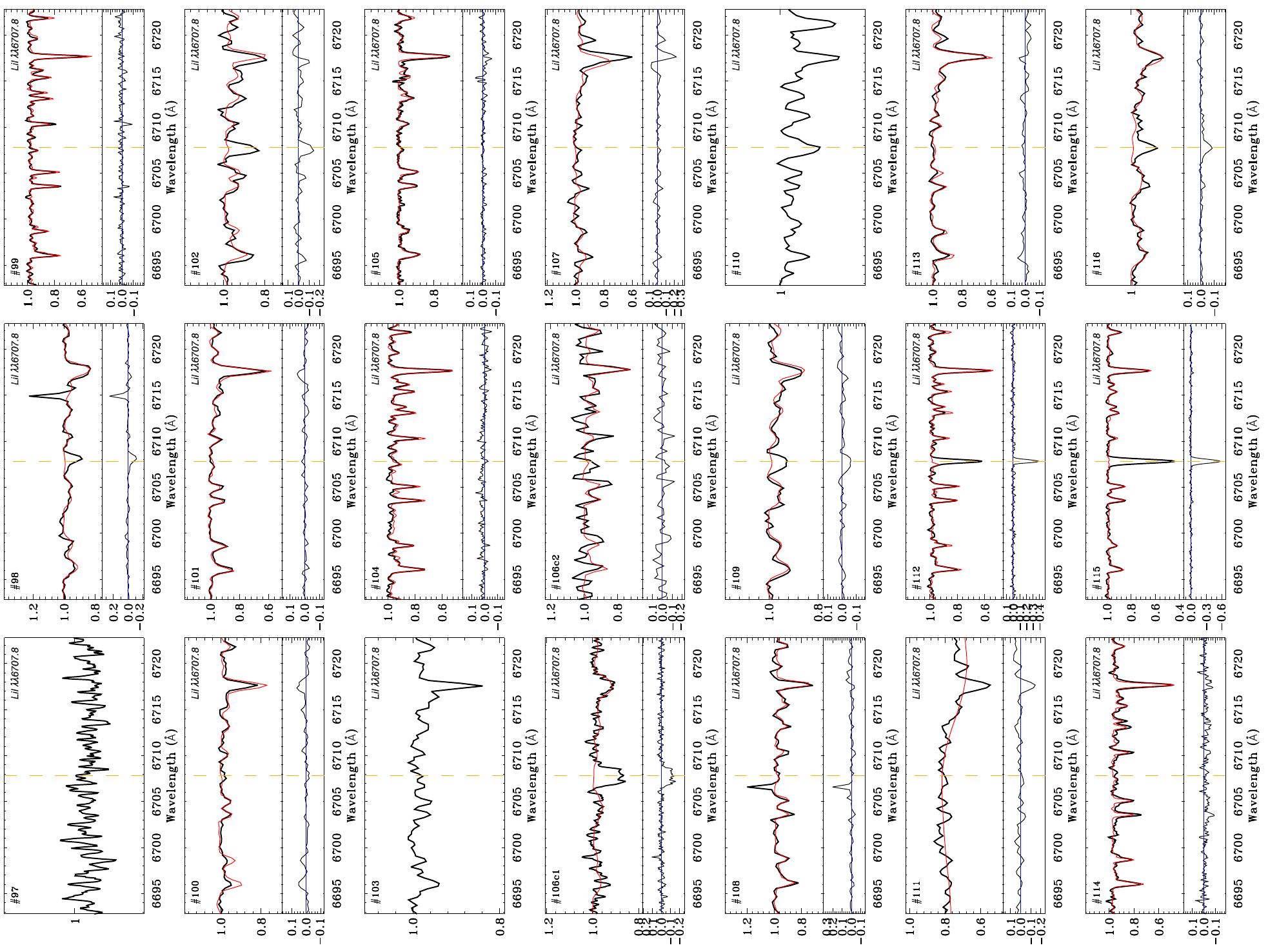}
\caption{Continued}
\end{figure*}

\setcounter{figure}{1}
\begin{figure*}[t]
\centering 
\includegraphics[angle=90,width=17cm]{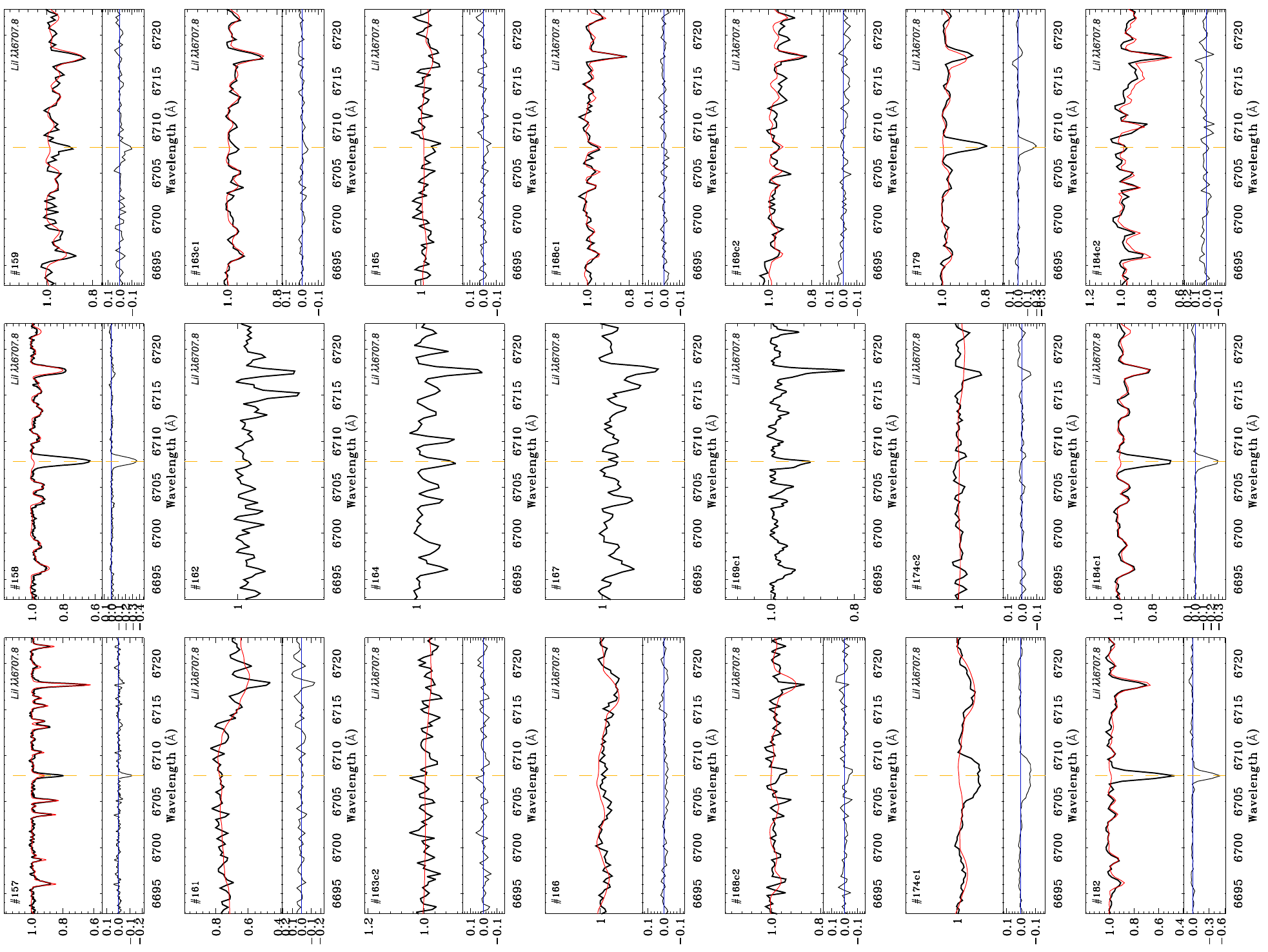}
\includegraphics[angle=90,width=17cm]{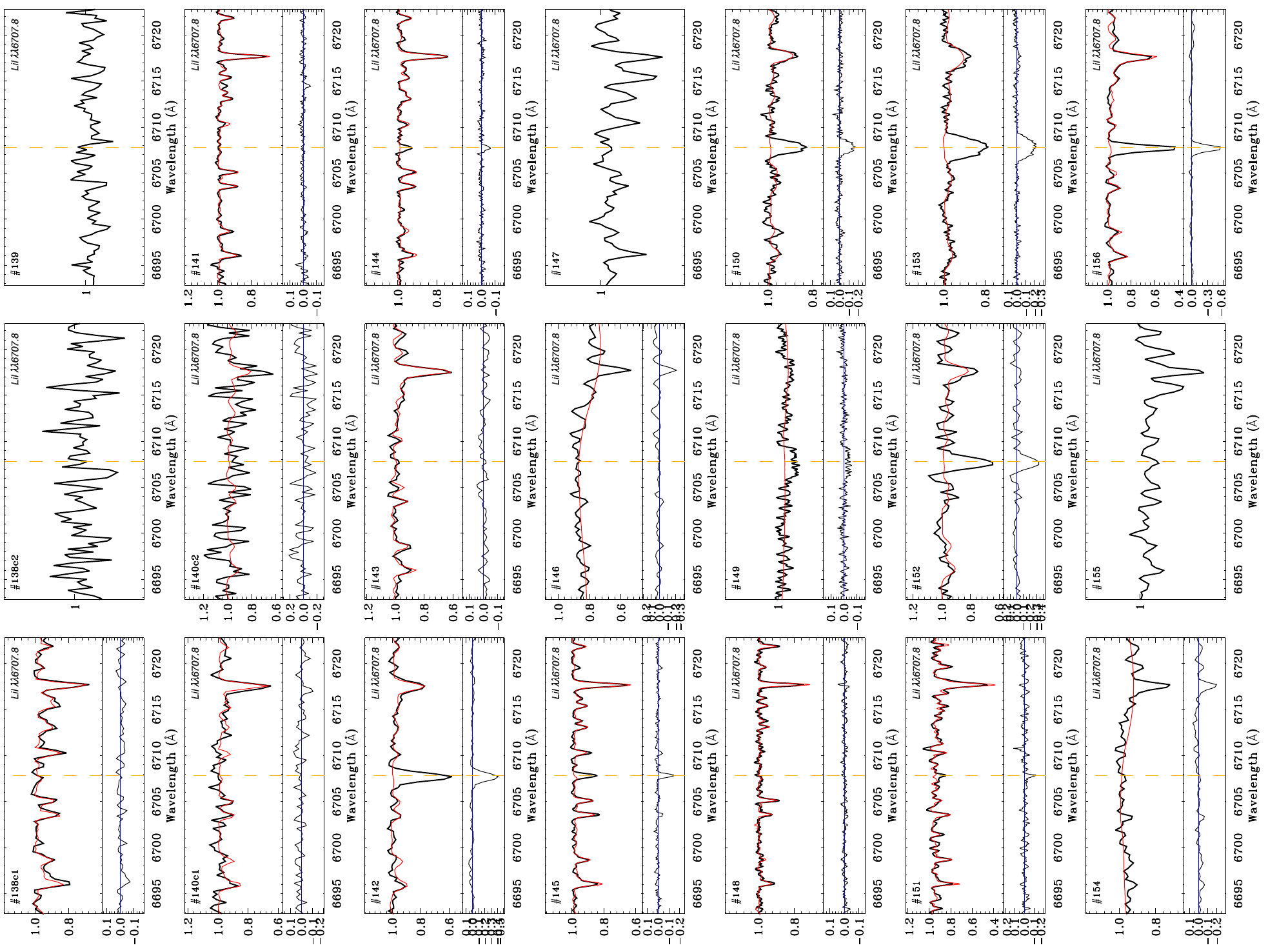}
\caption{Continued}
\end{figure*}

\setcounter{figure}{1}
\begin{figure*}[t]
\centering 
\includegraphics[angle=90,width=17cm]{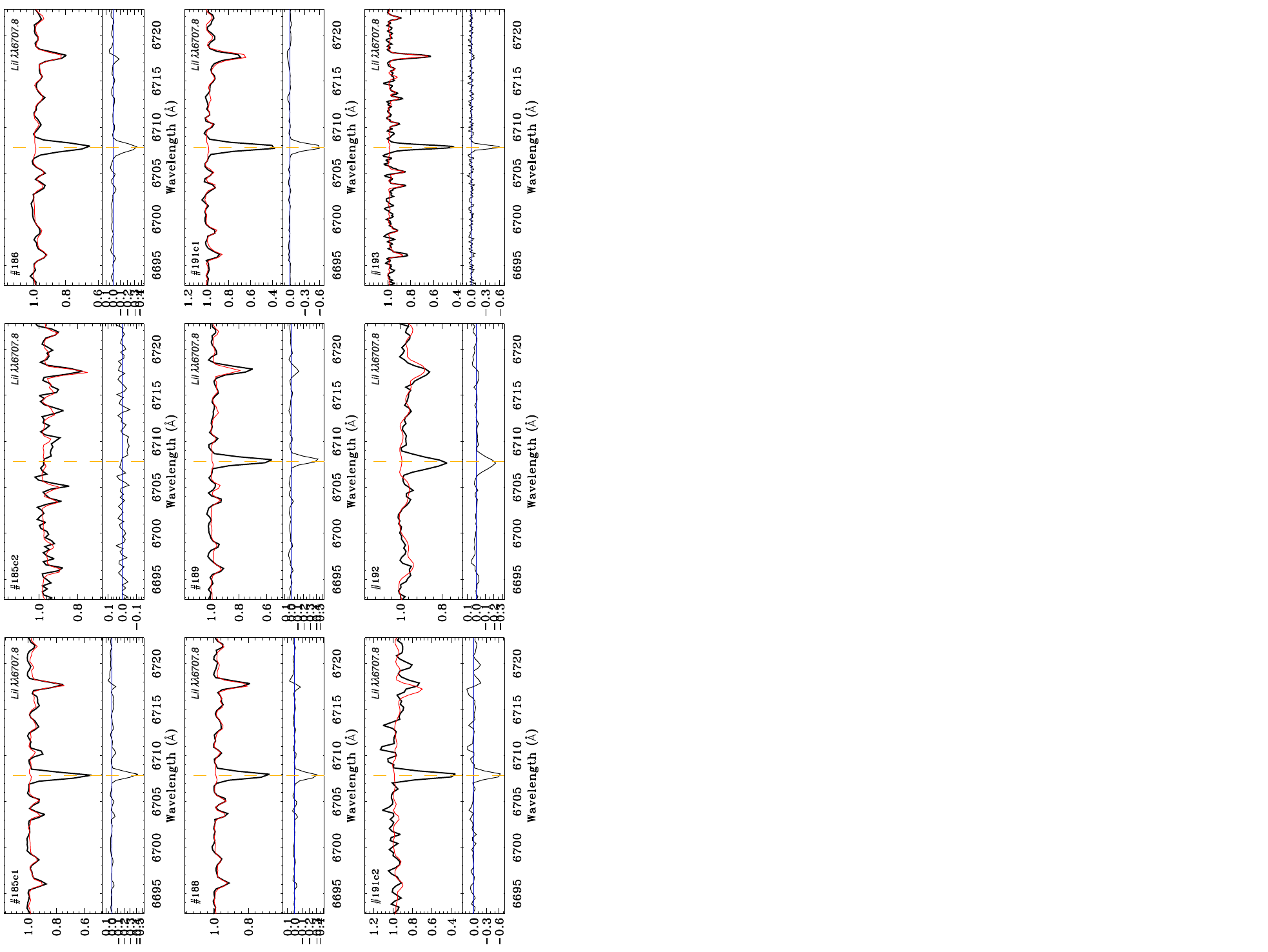}
\caption{Continued}
\end{figure*}

\end{appendix}

\end{document}